\documentclass[a4paper,12pt]{book}
\usepackage[ margin=1in]{geometry}
\usepackage[nottoc,notlot,notlof]{tocbibind} 
\usepackage[sort&compress,numbers,super]{natbib}
\usepackage{amsmath, amssymb, bm}
\usepackage{amsthm}
\usepackage{setspace} 
\usepackage{physics}
\usepackage{textcomp, gensymb}
\usepackage{graphicx}
\usepackage[bf,small]{caption}
\usepackage[english]{babel}
\usepackage[autostyle]{csquotes}
\usepackage{listings}

\usepackage{multirow}
\usepackage{tabularray}

\usepackage{array}
\usepackage{fontawesome}
\usepackage{enumitem}
\SetLabelAlign{center}{\strut\smash{\parbox[t]\labelwidth{\centering#1}}}
\usepackage{pdfpages}
\usepackage{bookmark} 

\usepackage{xcolor}

\definecolor{darkblue}{rgb}{0,0,0.55}
\definecolor{newmidnightblue}{rgb}{0,0,0.61}
\definecolor{darkslategray}{rgb}{0.18,0.31,0.31}

\hypersetup{colorlinks,linkcolor=blue,citecolor=newmidnightblue,urlcolor=magenta}

\usepackage[page,toc,titletoc,title]{appendix}

\usepackage[Sonny]{fncychap}
\usepackage{fancyhdr}

\ChNumVar{\fontsize{100}{42}\usefont{OT1}{ptm}{m}{n}\selectfont}
\usepackage{easyReview} 



\newtheorem*{conjecture*}{Conjecture}

\makeatletter
\renewcommand*{\cleardoublepage}{\clearpage\if@twoside \ifodd\c@page\else
\hbox{}%
\thispagestyle{empty}%
\newpage%
\if@twocolumn\hbox{}\newpage\fi\fi\fi}
\makeatother

\begin{document}

\frontmatter
	\pagestyle{headings}

\begin{titlepage}
	\centering
	{\vskip 1.0cm
\huge \textbf{Superconductivity in \textit{a}-MoGe thin films: effect of phase fluctuations with decreasing thickness and study of vortex dynamics in presence of low-frequency ac excitation}
\vskip 1.0cm
}
\vskip 0.5cm 
{\large A Thesis \\
\vskip 1.01cm 
Submitted to the\\
\vskip 0.01cm
Tata Institute of Fundamental Research, Mumbai\\
\vskip 0.01cm
for the Degree of Doctor of Philosophy  \\
in Physics\\
\vskip 1cm
by\\
\vskip 0.5cm 
{\Large \textbf{Soumyajit Mandal}}\\

\vskip 1cm 
Department of Condensed Matter Physics and Materials Science\\
Tata Institute of Fundamental Research \\
\vskip 0.02 cm
 Mumbai, India\\
\vskip 0.2 cm

\vspace{0.02\textheight}
\includegraphics[scale=0.35]{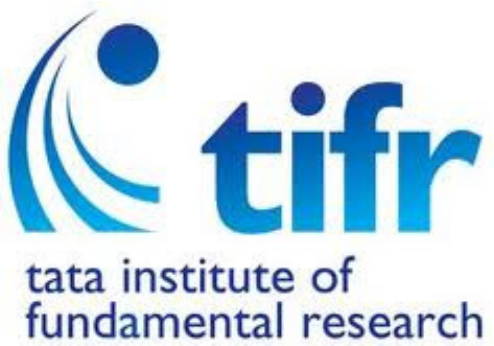}
\vskip 0.5 cm
July, 2023
\vskip 0.1 cm

{Final Version Submitted in January, 2024}

}
\end{titlepage}



        \cleardoublepage
        \phantomsection
        \includepdf[pages=1]{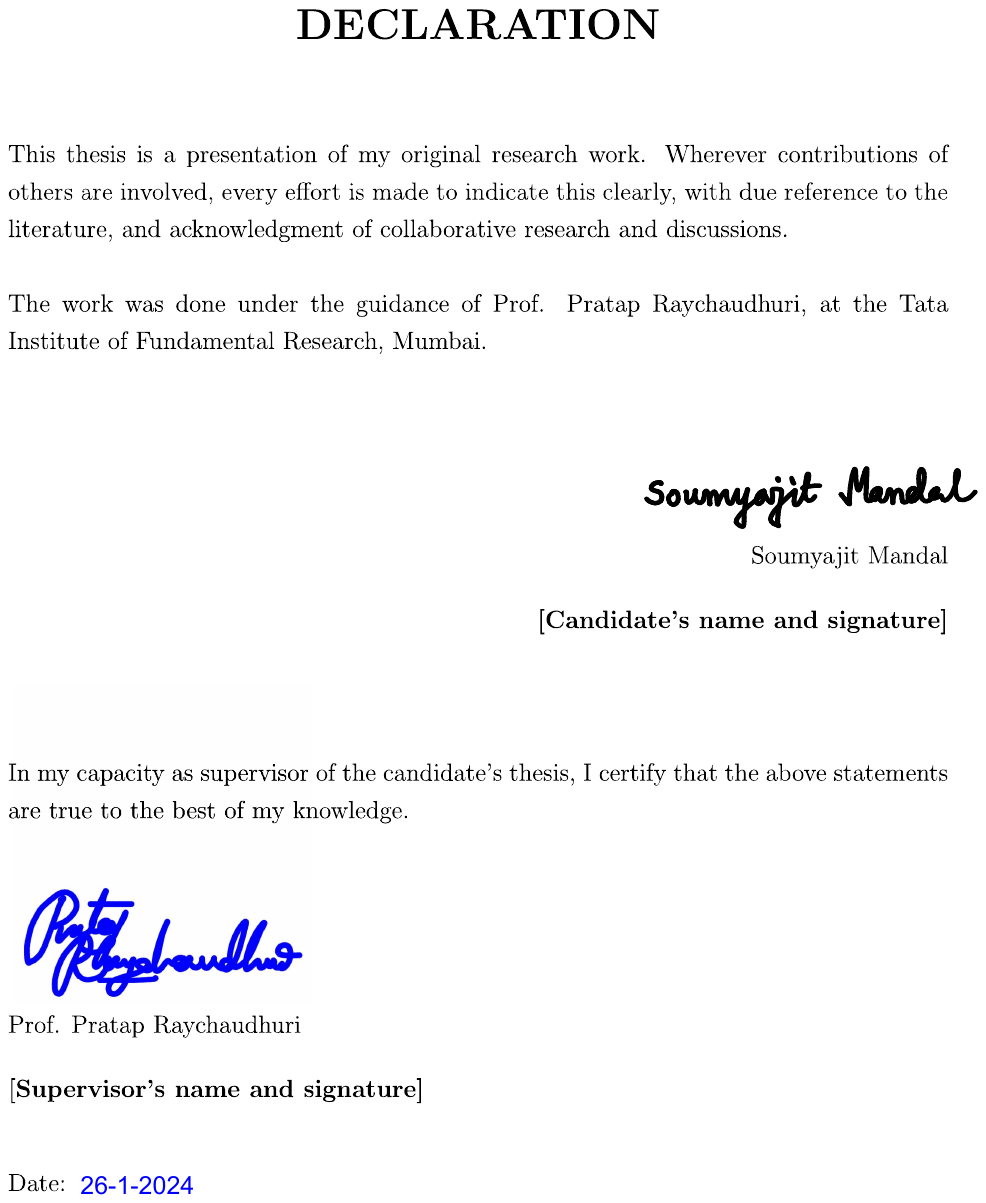}
        
        \cleardoublepage
        \phantomsection

\begin{center}

	\vspace*{6cm}
	
	{\LARGE \bf \it {To Maa, Baba and Bhai}}

	\vspace{-0.6cm}

\end{center}

\thispagestyle{empty}

        \clearpage
        
        \pagestyle{fancy}
        
\chapter*{Acknowledgement}

First and foremost, I express my deepest gratitude to my supervisor, Prof. Pratap Raychaudhuri, for his unwavering guidance and immense support throughout my Ph.D. journey. Without his invaluable mentorship, this work would not have been possible. His perseverance and enthusiasm for exploring new ideas have been a constant source of inspiration for me. I will forever cherish the numerous insightful discussions we had over the years.

I extend heartfelt appreciation to Prof. Lara Benfatto for her invaluable theoretical collaborations and valuable input, which enriched the quality of this research.

I am grateful to Prof. Vikram Tripathi and Prof. Arnab Bhattacharya for dedicating their time and expertise as members of the thesis monitoring committee, offering valuable insights that greatly contributed to the refinement of this work.

My sincere thanks go to my esteemed colleagues, both past and present. Dr. Somesh Chandra Ganguli, Dr. Rini Ganguly, Dr. Harkirat Singh, Dr. Aditya Narayan Roy Choudhury, Dr. Pradnya Parab, Dr. Ruchi Tomar, Dr. Indranil Roy, Dr. Surajit Dutta, John Jesudasan, Vivas Bagwe, Somak Basistha, Rishabh Duhan, Srijita Das, Pritam Das, Arghya Dutta, Sulagna Dutta, Sitakanta Behera and Dr. Subhamita Sengupta, your collaboration, unwavering support, and camaraderie have been instrumental in the success of this research.

I am grateful to Subash Pai, Supriya, and the team at Excel Instruments for their prompt and indispensable technical support. My heartfelt thanks also go to Atul Raut and the TIFR Central Workshop for their invaluable technical assistance. I am deeply appreciative of Ganesh Jangam and Prof. P. L. Paulose for their assistance with SQUID magnetization measurements. I extend my thanks to Bhagyashree Chalke, and Rudhir Bapat for characterizing the samples using EDX and SEM, and Jayesh B. Parmar for their invaluable help in TEM sample-surface preparations. Moreover, I extend my deepest appreciation to the Low Temperature Facility team of TIFR: Srinivasan, Vijay Arolkar, Bosco, Arvind, Jaison, Purao, and Nikhil for their unfaltering support in providing liquid helium and nitrogen, without which these experiments would not have been possible.

I am grateful for the invaluable friendship and camaraderie extended to me by my fellow researchers at TIFR. The constant support of Triparno, Nairit, Aritra, Manibrata, Chandradoy, Sabir, Saikat, Amrita and other seniors, as well as Anupam, Rabisankar, Anirban, Soumen, Sawani, Sanat, Anindya, Dibysankar, Souvik, Deepjyoti, Sunil, Arnab, and other juniors, Sudip, Indranil, Pratap, Sandip, Soham, Saikat, Soumen, Anurag, Haritha, Arusha, Kishor, Vishal, Dibyendu, Sumeru, Arkadipta, Sazedur, Suman, and other wonderful friends and batch-mates have been a delightful part of this voyage.
During challenging times towards the conclusion of my tenure, I would like to acknowledge the unwavering support from Mintu, Sanat, Aman, Anirban Bala, Anindya, Dibyasankar, and many others. Beyond the TIFR community, I have been fortunate to have friends like Madhumita, Semanti, Srijoyee, Shreya, and Debamitra during various phases of my Ph.D. journey. My sincere appreciation goes to Aditi for being a companion in sharing our daily life experiences — some enthralling, others mere moments of shared frustration.

I consider myself fortunate to have forged meaningful friendships throughout my academic journey, spanning from school to bachelors and masters. While it is impossible to name everyone, I want to express my gratitude to friends like Arunava, Subir, Dipanjan, Aritra, and others from school. In my college days, I cherish the camaraderie of Poulab, Bodhisattwa, Snehashish, Rahul, Jui, Amartya, Rajesh, Supriyo, Durgesh, Santu, Shirsendu, and many others. Each of these friendships has enriched my life in unique ways.

It would be remiss of me not to acknowledge the impact of my fellow Duffianz from Scottish Church College. Mustak, Subhaskar, Ankur, Mithun, Soumyadeep, Subhranil, Suman, Sumit – you all are truly incredible. Our circle of friends was a rare haven where I could freely share both the frustrations and joys of my life without any hesitation.

I am deeply grateful for the continuous encouragement and unwavering support I have received from my family members, including dida, dadu, masi, mama, mami, and my younger cousins, throughout my academic journey.

Lastly, I would like to acknowledge my beloved family - baba, maa, and bhai for their unconditional love, unwavering encouragement, and unyielding support throughout this academic journey. Their belief in me has been my greatest motivation, and I am forever indebted to them for shaping the person I have become today.


	\clearpage


        
 
 
	\setcounter{tocdepth}{2}
        \setcounter{secnumdepth}{3}
	\tableofcontents
	\clearpage

        \cleardoublepage
        \phantomsection         
        
\chapter*{Abstract}
\addcontentsline{toc}{chapter}{Abstract}

We have studied the evolution of superconductivity in amorphous Molybdenum Germanium (\textit{a}-MoGe) thin films. The work can be broken down into two parts. In the first part, we investigate the effect of decreasing thickness on the suppression of superconductivity in \textit{a}-MoGe thin films. Thick \textit{a}-MoGe thin film is a typical type-II superconductor and follows the conventional Bardeen-Cooper-Schrieffer (BCS) equation. Conventionally, it is believed that decreasing thickness will decrease the effective attractive pairing interaction because of the gradual loss of screening which holds true for large thicknesses in \textit{a}-MoGe. But for lower thicknesses, a new mechanism comes into the picture where superfluid density is suppressed making the superconductor vulnerable to phase fluctuations. This is known as the Bosonic mechanism, where superconductivity is destroyed due to the loss of phase coherence of the superconducting state even though the attractive pairing amplitude remains finite above the transition temperature. \textcolor{black}{The study unveils that the impact of phase fluctuations is significantly more potent than anticipated. Specifically, these fluctuations suppress the critical temperature ($T_c$) of the thinnest films, where both superfluid density and resistivity vanish, well below the predicted conventional Kosterlitz-Thouless-Berezinskii transition}. In the second part, we explore the electromagnetic response of vortices in \textit{a}-MoGe thin film by low-frequency two-coil mutual inductance technique. Penetration depth measured from the two-coil technique was earlier used to determine superfluid density. However in the present work, by analyzing the in-field penetration depth data with the help of a mean-field model proposed by Coffey and Clem, we have demonstrated a procedure of extraction of vortex parameters such as pinning restoring force constant or Labusch parameter, vortex lattice drag coefficient and pinning potential barrier for the thermally activated motion of vortices. The temperature variation of vortex parameters suggests the dominant effect of thermal fluctuations.

        \pagestyle{fancy}

	\clearpage
        \markboth{}{}
        \cleardoublepage
        \phantomsection
        \pagestyle{fancy}
	\chapter*{List of Publications}
\phantomsection
\addcontentsline{toc}{chapter}{List of Publications}

\section*{Publications related to this thesis}
\begin{itemize}
	\item \textbf{Soumyajit Mandal}, Surajit Dutta, Somak Basistha, Indranil Roy, John Jesudasan, Vivas Bagwe, Lara Benfatto, Arumugam Thamizhavel, and Pratap Raychaudhuri; \textit{Destruction of superconductivity through phase fluctuations in ultrathin a-MoGe films}; \href{https://doi.org/10.1103/PhysRevB.102.060501}{\textit{Physical Review B}, \textbf{102}, 060501(R) (2020).}
 
	\item \textbf{Soumyajit Mandal}, Somak Basistha, John Jesudasan, Vivas Bagwe, and Pratap Raychaudhuri; \textit{Study of vortex dynamics in an a-MoGe thin film using low-frequency two-coil mutual inductance measurements};  \href{https://doi.org/10.1088/1361-6668/aca62c} {\textit{Superconductor Science and Technology}, \textbf{36}, 014004 (2023).}
\end{itemize}
\vspace{0.5cm}
\section*{Other publications}
\begin{itemize}
	\item Indranil Roy, Surajit Dutta, Aditya N. Roy Choudhury, Somak Basistha, Ilaria Maccari, \textbf{Soumyajit Mandal}, John Jesudasan, Vivas Bagwe, Claudio Castellani, Lara Benfatto, and Pratap Raychaudhuri; \textit{Melting of the vortex lattice through intermediate hexatic fluid in an a-MoGe thin film}; \href{https://doi.org/10.1103/PhysRevLett.122.047001}{\textit{Physical Review Letters}, \textbf{122}, 047001 (2019).}
 
		\item Surajit Dutta, Indranil Roy, \textbf{Soumyajit Mandal}, John Jesudasan, Vivas Bagwe, and Pratap Raychaudhuri; \textit{Extreme sensitivity of the vortex state in a-MoGe films to radio-frequency electromagnetic perturbation}; \href{https://doi.org/10.1103/PhysRevB.100.214518}{\textit{Physical Review B}, \textbf{100}, 214518 (2019).}
  
  	\item Surajit Dutta, Indranil Roy, Somak Basistha, \textbf{Soumyajit Mandal}, John Jesudasan, Vivas Bagwe and Pratap Raychaudhuri; \textit{Collective flux pinning in hexatic vortex fluid in a-MoGe thin filmn}; \href{https://doi.org/10.1088/1361-648X/ab52b8}{\textit{Journal of Physics: Condensed Matter}, \textbf{32}, 075601 (2020).}
   
    	\item Somak Basistha, Vivas Bagwe, John Jesudasan, Gorakhnath Chaurasiya, \textbf{Soumyajit Mandal}, Surajit Dutta, and Pratap Raychaudhuri; \textit{Growth and Characterization of amorphous Molybdenum Germanium (a-MoGe) and amorphous Rhenium Zirconium (a-Re$_6$Zr) superconducting thin films using Pulsed Laser Deposition (PLD) technique}; \href{https://www.smcindia.org/smc-bulletin.php}{\textit{Society for Materials Chemistry Bulletin}, \textbf{11}(3), 150 - 159 (Dec 2020).}

     	\item Somak Basistha,\hspace{0.1cm} \textbf{Soumyajit Mandal},\hspace{0.1cm} John Jesudasan,\hspace{0.1cm} Vivas Bagwe, and Pratap Raychaudhuri;\hspace{0.1cm} \textit{Low\hspace{0.1cm} frequency\hspace{0.1cm} electrodynamics\hspace{0.1cm} in\hspace{0.1cm} the\hspace{0.1cm} mixed\hspace{0.1cm} state\hspace{0.1cm} of superconducting\hspace{0.1cm} NbN\hspace{0.1cm} and \textit{a}-MoGe\hspace{0.1cm} films\hspace{0.1cm} using\hspace{0.1cm} two-coil\hspace{0.1cm} mutual inductance technique}; \href{https://doi.org/10.48550/arXiv.2311.03752}{arXiv:2311.03752}. (submitted) 
\end{itemize}

	\cleardoublepage
 
	\pagestyle{fancy}
	\phantomsection
	\listoffigures
	\addcontentsline{toc}{chapter}{\listfigurename}
        
        \cleardoublepage
        \phantomsection
        \listoftables
        \addcontentsline{toc}{chapter}{\listtablename}

        \cleardoublepage
        \phantomsection
        
\chapter*{Preamble}
\addcontentsline{toc}{chapter}{Preamble}
This thesis explores the evolution of superconductivity in amorphous Molybdenum Germanium (\textit{a}-MoGe) thin films, focusing on the effect of phase fluctuations on the suppression of superconductivity with decreasing thickness. Additionally, it investigates the electromagnetic response of vortices in these films, extracting vortex parameters and examining the impact of thermal fluctuations.
\par
The layout of the thesis is as follows:
\begin{itemize}
    \item Chapter~\ref{ch:intro} discusses the basic concepts of superconductivity, including BCS and Ginzburg-Landau theory. Additionally, necessary theoretical concepts and literature surveys required for the ensuing work are also explored and presented.
    \item Chapter~\ref{ch:exp-methods} focuses on the discussion of sample preparation and measurement techniques.
    \item Chapter~\ref{ch:MoGe-sample} describes the sample details and characterization of the sample. Additionally, it includes a discussion of some recent results that have not been covered in the later chapters.
    \item Chapter~\ref{ch:phase_fluctuations} delves into the effect of decreasing thickness in \textit{a}-MoGe thin films and how phase fluctuations affect the superconductivity at lower thicknesses.
    \item Chapter~\ref{ch:vortex-par-chap} examines the vortex dynamics in \textit{a}-MoGe in the presence of low-frequency small ac excitation and demonstrates the extraction of vortex parameters.
    \item Chapter~\ref{ch:summary} provides a summary of our work and presents an outlook for future studies.
\end{itemize}

        \cleardoublepage
        \phantomsection
	
\chapter*{Statement of Joint Work}
\addcontentsline{toc}{chapter}{Statement of Joint Work}
\vspace{-1cm}
The experiments reported in this thesis were conducted in the Department of Condensed Matter Physics and Materials Science at the Tata Institute of Fundamental Research under the guidance of Prof. Pratap Raychaudhuri. A significant portion of the results presented here has already been published in refereed journals.

While I have conducted the majority of the experiments and analyses presented in this thesis, I have also included some collaborative work carried out by others. The details of these collaborations are as follows:
\begin{itemize}
    \item The \textit{a}-MoGe thin films for two-coil mutual inductance measurements were grown in collaboration with Somak Basistha and John Jesudasan. The MoGe target for sample deposition in Pulsed Laser Deposition (PLD) was grown in collaboration with Prof. Arumugam Thamizhavel.
    \item In the investigation of the effect of phase fluctuations on ultrathin $a$-MoGe films, I conducted the penetration depth measurements and analyses. Scanning Tunneling Spectroscopy (STS) measurements and analyses were performed by Surajit Dutta and Indranil Roy. Transport measurements were carried out by Surajit Dutta and Somak Basistha. Theoretical inputs were provided by Prof. Lara Benfatto.
    \item For the study of vortex dynamics in $a$-MoGe, I collaborated with Somak Basistha for the penetration depth measurements and analyses.
    \item SQUID-VSM measurements on $a$-MoGe were performed with assistance from Ganesh Jangam.
    \item Broadband microwave measurements on \textit{a}-MoGe films were conducted in collaboration with Somak Basistha.
    \item The study of the movement of vortices in \textit{a}-MoGe using STS and subsequent analyses were done in collaboration with Indranil Roy. 
\end{itemize}

	\mainmatter
		
	\chapter{Introduction}\label{ch:intro}
Superconductivity, discovered by Heike Kamerlingh Onnes~\cite{onnes_notitle_1911,onnes_through_1990} in 1911, has been a subject of intense research for over a century. While the practical applications of superconductivity have been a driving force behind extensive research, understanding its fundamental aspects has also been crucial. \;In the 1950s, Bardeen, Cooper, and Schrieffer~\cite{bardeen_theory_1957} ( BCS ) revolutionized our understanding of superconductivity, providing a microscopic explanation for the phenomenon and paving the way for practical applications. In another major breakthrough, Vitaly Ginzburg and Lev Landau introduced a macroscopic framework within Landau's general theory~\cite{landau_theory_phase_1965} of second-order phase transitions, known as Ginzburg Landau theory~\cite{Ginzburg:1950sr,landau_theory_1965} ( GL ), which further deepened our comprehension of superconducting materials and their behavior near the critical temperature. The discovery of high-temperature superconductors by Bednorz and Müller~\cite{bednorz_possible_1986} further revitalized the field, although the basic microscopic mechanism in these new superconductors remains an open question to date.  However, their overall phenomenology follows the patterns observed in conventional superconductors. Hence, a common strategy is often adopted to explore several high-temperature phenomena, including the presence of a pseudogap phase~\cite{alloul_89mathrmy_1989,loram_electronic_1993}, quantum fluctuations of vortices~\cite{blatter_vortices_1994}, and phase fluctuations~\cite{emery_importance_1995} in conventional superconductors with low superfluid density, in order to uncover similarities and draw connections between the behavior of high-temperature superconductors and that of traditional superconductors.
\par
The superconducting state is primarily dependent on two phenomena: (i) weak attractive interaction between a pair of electrons with opposite momenta and opposite spins forming a spin-zero Boson-like object called Cooper pairs and (ii) condensation of all such Cooper pairs into a phase-coherent macroscopic quantum state. The extent of phase coherence depends on the phase rigidity of the superconducting state and it gives rise to two of the most defining properties of the superconductor: zero resistance and perfect diamagnetism. The response to the external magnetic field further classifies the superconductors- type-I, and type-II. Type-I superconductors behave like a perfect diamagnet in the presence of an external field until it reaches a critical field ( $H_c$ ) characteristic of each material. On the other hand, at low fields type-II superconductors are in perfect diamagnetic or Meissner state up to the lower critical field $H_{c1}$, above which quantized flux lines ( vortices ) penetrate the sample before going to the normal state above the upper critical field $H_{c2}$.
\par
We have studied the evolution of superconductivity in amorphous Molybdenum Germanium ( \textit{a}-MoGe ) thin films. The work can be broken down into two parts. In the first work, we have studied the effect of disorder on the superconductivity in the absence of a magnetic field in \textit{a}-MoGe thin films and the interplay of disorder and phase fluctuations on the superconducting state. In the second work, we have explored the response of vortices to small ac excitation ( $kHz$ frequency range ) in the weakly-pinned \textit{a}-MoGe film.
\par
In this chapter, I shall first introduce the basic theoretical understanding and the phenomenology of superconductivity including BCS and GL theory. Then I shall provide a brief literature review of the effect of disorder on superconductivity as an introduction for the first work ( Chapter~\ref{ch:phase_fluctuations} ). Also, I shall briefly discuss the current understanding of the effect of\hspace{0.1cm} ac excitation\hspace{0.1cm} on the vortex lattice as an\hspace{0.1cm} introduction to the\hspace{0.1cm} second work ( Chapter~\ref{ch:vortex-par-chap} ).
\section{Basics of Superconductivity}
The two fundamental properties distinguishing a superconductor from a normal metal are its zero resistance and perfect diamagnetism.
\subsection{Zero resistance}
It was first observed by Kamerlingh Onnes~\cite{onnes_notitle_1911,onnes_through_1990} in 1911 that the electrical resistance of various metals such as mercury, tin, and lead drops below measurable limits below a certain critical temperature, $T_c$, characteristic of the corresponding material. Later this phenomenon was observed in a lot of other materials. This property is most successfully demonstrated by the persistent current in superconducting loops which has the characteristic decay time of more than $10^5$ years~\cite{tinkham_introduction_2004}. This perfect conductivity has made the superconductors the potential candidates for applications in the transmission of high current and high field magnets.
\subsection{Perfect diamagnetism}
The other defining property of a superconductor, discovered by Meissner and Oschsenfeld~\cite{meissner_neuer_1933} in 1933, is perfect diamagnetism. When a superconductor is cooled below its characteristic transition temperature $T_c$, it repels the external magnetic field irrespective of its magnetic history, whether the field was applied before or after cooling. In contrast, a perfect conductor would trap the flux inside if they are cooled in the field. This perfect diamagnetism in superconductors is known as the Meissner effect.
\subsection{Critical field}
While superconductors exhibit perfect diamagnetism in weak magnetic fields, it is important to note that their superconductivity can be disrupted by strong magnetic fields. When the magnetic field surpasses a certain threshold known as the critical field, $H_c$, most elemental superconductors lose their diamagnetic properties and start behaving like normal metals. It reflects in the increase in the free energy of the superconducting state in the presence of an external magnetic field, also known as condensation energy, $\epsilon_{cond}$:
\begin{equation}\label{eq:condens_Hc}
    \epsilon_{cond}=f_s(H_c)-f_s(0)=\frac{1}{2} \mu_0 H_c^2.
\end{equation}
$H_c$ is a parameter dependent on the superconducting material. Experimental findings have revealed that the temperature dependence of the critical fields in different superconductors can be approximated by the following empirical relation~\cite{rose-innes_introduction_1994},
\begin{equation}
	H_c(T) = H_c(0)\left[1-(T/T_c)^2\right].
\end{equation}
\subsection{London equation}
Strictly speaking, the perfect diamagnetism mentioned above is only valid for bulk samples. For finite-size superconductors, the magnetic field does penetrate the superconductor to a finite distance from the surface, known as London penetration depth ( $\lambda_L$ ). Fritz and Heinz London \cite{london_electromagnetic_1935} in 1935 came up with a phenomenological model to explain the two basic electrodynamic properties of superconductors mentioned above. They proposed that the microscopic electric and magnetic fields of the superconducting state follow the following two equations:
\begin{equation}\label{eq:london_eqn_E}
	\bm{E}=\frac{\partial (\Lambda \bm{J_s})}{\partial t},
\end{equation}
\begin{equation}\label{eq:london_eqn_B}
	\bm{B}=\bm{\nabla}\times\bm{A}=-\bm{\nabla}\times(\Lambda \bm{J_s}),
\end{equation}
Here, $\Lambda=m_e/n_se^2$ is a phenomenological parameter ( $m_e$ is the effective electron mass, $\mu_0$ is vacuum permeability, $n_s$ is superfluid density and $e$ is electron charge ), which is related to the London penetration depth by the following relation:
\begin{equation}\label{eq:london-pen-depth-ns}
	\lambda_L=\sqrt{\frac{m_e}{\mu_0 n_se^2}}=\sqrt{\frac{\Lambda}{\mu_0}}
\end{equation}
Eq.~\ref{eq:london_eqn_E} pertains to the acceleration of electrons, rather than their ability to sustain velocity in the presence of resistance caused by electric fields. As such, it signifies the perfect conductivity exhibited by superconductors, where electrons can move without hindrance.
Eq.~\ref{eq:london_eqn_B} combined with the Maxwell’s equation, $\bm{\nabla}\times \bm{B}=\mu_0\bm{J_s}$, gives rise to the following equation:
\begin{equation}\label{eq:london_maxwell_eqn_combined}
    \nabla^2\bm{B}=\frac{\bm{B}}{\lambda_L^2}
\end{equation}
Let us write Eq.~\ref{eq:london_maxwell_eqn_combined} in a simple case where a superconductor is placed in a uniform applied magnetic field $\bm{B_a}=B_0 \bm{\hat{z}}$. Hence, the magnetic field outside the surface of the superconductor is, $\bm{B}(x=0)=B_0\bm{\hat{z}}$ where $\bm{B}(x)$ is the magnetic induction inside the superconductor and x is the distance from the surface. Thus Eq.~\ref{eq:london_maxwell_eqn_combined} is simplified as:
\begin{equation}\label{eq:london_maxwell_combined_simple}
    \frac{\partial^2 B(x)}{\partial x^2}=\frac{B(x)}{\lambda_L^2}
\end{equation}
Solution to Eq.~\ref{eq:london_maxwell_combined_simple} is given by:
\begin{equation}
    B(x)=B_0e^{-x/\lambda_L}
\end{equation}
This means that magnetic induction $B$ decays exponentially inside the superconductor with the characteristic length $\lambda_L$. Additionally, it follows from Eq.~\ref{eq:london_eqn_B} that the current also decays exponentially inside the superconductor.
\subsection{Pippard's Coherence length}
In 1953, Pippard~\cite{pippard_experimental_1953} introduced a crucial concept regarding superconducting electrons. He argued that the density of superconducting electrons in space cannot abruptly change to zero. The shortest length scale, over which the superconducting electron density can be modified to transition from a superconductor to a normal metal, is known as the coherence length, denoted by $\xi$. However, the presence of impurities in the material can reduce this coherence length. For impure materials or alloys, $\xi$ is given by $(\xi_0l)^{1/2}$, where $\xi_0$ is the coherence length of the pure material and $l$ is the electronic mean free path of the material. Pippard employed the position-momentum uncertainty argument to demonstrate that $\xi_0\sim a\frac{\hbar v_F}{k_B T_c}$, where $a$ is a constant of order 1, $v_F$ is the Fermi velocity, $\hbar$ is the reduced Planck constant, $k_B$ is the Boltzmann constant, and $T_c$ is the transition temperature.
\subsection{Type-I and Type-II superconductors}\label{sec:typeI-II-SC}
The response to the external magnetic field classifies the superconductors into two categories - type-I and type-II. Type-I superconductors exhibit perfect diamagnetism (Meissner effect) up to a critical magnetic field, $H_c$, beyond which they transition to a normal state. Except for Nb, elemental superconductors fall in the type-I class of superconductors.
\par
On the other hand, in type-II superconductors, which include nearly all superconducting alloys and compounds, the magnetic response to the applied field is relatively complex. It shows perfect diamagnetism up to a lower characteristic field, $H_{c1}$, where the magnetic flux lines do not penetrate the superconductor. However, when the applied magnetic field exceeds $H_{c1}$, quantized magnetic flux lines start to penetrate inside the superconductor, giving rise to a vortex state where superconducting regions coexist with normal regions, and the flux lines pass through the normal regions. With further increase in the magnetic field beyond a second characteristic critical field, $H_{c2}$, the superconductivity is completely destroyed, and the material transitions to a normal state. $H_{c2}$ is known as the upper critical field.
\section{BCS theory}
The Bardeen-Cooper-Schrieffer ( BCS ) theory of superconductivity revolutionized our understanding of the behavior of electrons in materials at low temperatures. The theory, proposed in 1957 by John Bardeen, Leon Cooper, and John Robert Schrieffer~\cite{bardeen_theory_1957}, provided a comprehensive explanation for the emergence of superconductivity, the phenomenon in which certain materials can conduct electricity with zero resistance at very low temperatures. According to the theory, even a weak attractive interaction between the electrons can lead to the formation of a macroscopic quantum state, in which many electrons behave as a single entity. This state arises due to the interaction between electrons and the crystal lattice and is characterized by the presence of an energy gap in the density of states of electrons. The BCS theory has since been confirmed by numerous experiments and has led to many important technological applications, such as magnetic resonance imaging ( MRI ) and particle accelerators.
\par
In this section, I shall give a brief overview of this elegant theory and its novelty to describe different properties of superconductors.
\subsection{Cooper pairs}
In 1950, Fröhlich~\cite{frohlich_theory_1950} suggested that electrons can experience an attractive interaction mediated by phonons, contrary to the commonly held belief that electrons always repel each other. This occurs because when an electron travels through the lattice, it creates a positive charge imbalance that attracts other electrons.
\par
Six years later, Cooper~\cite{cooper_bound_1956} showed that even a weak attractive interaction can cause two electrons with opposite momenta and spin to form a bound pair known as a Cooper pair. To investigate the properties of a superconducting state, Cooper considered a two-particle wave function for the Cooper pair, assuming that the two electrons did not interact with the rest of the Fermi sea. For simplicity, he assumed a momentum-independent attractive potential ($-V$). By solving Schrodinger's equation, Cooper was able to obtain the eigenvalues of the wave function for this state as the following:
\begin{equation}\label{eq:cooper_negative_E}
    E\approx 2E_F - 2\hbar\omega_D e^{-2/N(0)V},
\end{equation}
where $N(0)$ represents the density of states ( DOS ) at the Fermi sea for electrons of one spin, while $E_F$ is the Fermi energy and $\omega_D$ is the Debye frequency of the material. Eq.~\ref{eq:cooper_negative_E} indicates that even if the net attractive potential $V$ is very small, the eigenvalues of the wave function are negative with respect to the Fermi energy, resulting in the formation of bound states of electrons with kinetic energy greater than $E_F$. This analysis assumes that the coupling between the electrons and the lattice is weak, i.e., $N(0)V \ll 1$ ( weak-coupling approximation )~\cite{tinkham_introduction_2004}.
\subsection{BCS ground state}\label{sec:BCS-gnd-state}
Cooper's~\cite{cooper_bound_1956} approach of considering a single pair of electrons was not sufficient to describe real-world systems. Dealing with multiple Cooper pairs in a state was a challenging task. To tackle this issue, Bardeen, Cooper, and Schrieffer~\cite{bardeen_theory_1957} proposed a novel expression for the ground state of a superconductor:
\begin{equation}\label{eq:BCS_wave_fn}
    \ket{\Psi_{BCS}}=\prod_{\bm{k}}(u_{\bm{k}}+v_{\bm{k}}e^{i\phi_{\bm{k}}}c^{\dagger}_{\bm{k}\uparrow}c^{\dagger}_{-\bm{k}\downarrow})\ket{0},
\end{equation}
where $\ket{0}$ is the vacuum state in the absence of any particle. $c^{\dagger}_{\bm{k}\uparrow} (c_{\bm{-k}\downarrow})$ is the creation (annihilation) operator of an electron of momentum $\bm{k} (-\bm{k})$ and spin up (down). The probability of having a bound pair of electrons ( Cooper pair ) is $v^2_{\bm{k}}$ and not having is $u^2_{\bm{k}}$, with the normalization condition: $u^2_{\bm{k}}+v^2_{\bm{k}}=1$. Here, $u_{\bm{k}}$ and $u_{\bm{k}}$ are chosen to be real while the phase part is absorbed in $\phi_{\bm{k}}$, which is the phase of each Cooper pair. The use of a wavefunction expressed as a product of individual wavefunctions shown in Eq.~\ref{eq:BCS_wave_fn}, rather than a many-body wavefunction composed of Slater determinants of momentum eigenfunctions, is only justified when dealing with a large number of particles.
\par
We first consider the ground state of the superconductor. We can determine the coefficients $u_{\bm{k}}$, $v_{\bm{k}}$, and $\phi_{\bm{k}}$ by minimizing the ground state energy $E$ by variational method,
\begin{equation}\label{eq:E-BCS-variational}
    \delta E=\delta \expval{H_{pair}}{\Psi_{BCS}}=0,
\end{equation}
where $H_{pair}$ is the BCS pairing Hamiltonian denoted by,
\begin{equation}\label{eq:BCS_hamiltonian}
\begin{split}
    H_{pair}
    &=\sum_{\bm{k},\sigma} (\epsilon_{\bm{k}}-\mu) n_{\bm{k},\sigma}+\sum_{\bm{k},\bm{l}}V_{\bm{k,l}}c_{\bm{k}\uparrow}^{\dagger}c_{-\bm{k}\downarrow}^{\dagger}c_{-\bm{l}\downarrow}c_{\bm{l}\uparrow}\\
    &=\sum_{\bm{k},\sigma} \xi_{\bm{k}} c^{\dagger}_{\bm{k},\sigma}c_{\bm{k},\sigma}+\frac{1}{2}\left(\sum_{\bm{k},\bm{l}}V_{\bm{k,l}}c_{\bm{k}\uparrow}^{\dagger}c_{-\bm{k}\downarrow}^{\dagger}c_{-\bm{l}\downarrow}c_{\bm{l}\uparrow}+\sum_{\bm{k},\bm{l}}V_{\bm{l,k}}c_{\bm{l}\uparrow}^{\dagger}c_{-\bm{l}\downarrow}^{\dagger}c_{-\bm{k}\downarrow}c_{\bm{k}\uparrow}\right).
\end{split}
\end{equation}
The first term in the right-hand side of Eq.~\ref{eq:BCS_hamiltonian} corresponds to the kinetic energy of noninteracting electrons, where $\epsilon_{\bm{k}}$ is the kinetic energy of a single electron, $\mu$ is the chemical potential and $n_{\bm{k},\sigma}$ is called particle number operator defined by $n_{\bm{k},\sigma}=c^{\dagger}_{\bm{k},\sigma}c_{\bm{k},\sigma}$ with momentum $\bm{k}$ and spin $\sigma$. The chemical potential $\mu$ ( or Fermi energy $E_F$ ) is included in Eq.~\ref{eq:BCS_hamiltonian} under the framework of grand canonical ensemble, where the mean number of particles is kept fixed instead of the total number of particles, $N$. This choice of the ensemble is justified in the large $N$ limit, where fractional uncertainty in $N$ becomes negligible. In the second equality, $\xi_{\bm{k}}$ is defined as the kinetic energy of a single electron with respect to Fermi level ( $E_F$ or $\mu$ ), $\xi_{\bm{k}}=\epsilon_{\bm{k}}-\mu$.
\par
On the other hand, the second term in the right-hand side of Eq.~\ref{eq:BCS_hamiltonian} pertains to the pairing interaction due to second-order phonon scattering. In the second equality, the scattering term has been split between both the scattering possibilities for the combination of momenta $\bm{k}$ and $\bm{l}$, taking into account the symmetry of the term. Additionally, following Cooper's approach~\cite{cooper_bound_1956}, BCS adopted a simplified form for the electron-phonon scattering matrix element $V_{\bm{k},\bm{l}}$ as follows,
\begin{equation}\label{eq:V-scattering-BCS}
    V_{\bm{k},\bm{l}}=
    \begin{cases}
    -V & \text{for $\abs{\xi_{\bm{k}}}, \abs{\xi_{\bm{l}}}\le\hbar\omega_D$}\\
    0 &\text{otherwise}.
    \end{cases}
\end{equation}
\par
By utilizing Eq.~\ref{eq:BCS_wave_fn}, \ref{eq:BCS_hamiltonian}, and \ref{eq:V-scattering-BCS}, we can calculate the expectation value of energy at the ground state as follows,
\begin{equation}\label{eq:E-BCS-gnd-simplified}
\begin{split}
    E
    &=\expval{H_{pair}}{\Psi_{BCS}}\\
    &=2 \sum_{\bm{k}} \xi_{\bm{k}}v^2_{\bm{k}}+\frac{1}{2}\sum_{\bm{k,l}}(-V) u_{\bm{k}}v_{\bm{k}}u_{\bm{l}}v_{\bm{l}}\left[e^{-i(\phi_{\bm{k}}-\phi_{\bm{l}})}+e^{i(\phi_{\bm{k}}-\phi_{\bm{l}})}\right].
\end{split}
\end{equation}
Now, to minimize $E$ in Eq.~\ref{eq:E-BCS-gnd-simplified} by varying $\phi_{\bm{k}}$, we obtain,
\begin{equation}
\begin{split}
    \frac{\partial E}{\partial \phi_{\bm{k}}}
    &=\frac{1}{2}\sum_{\bm{k,l}}(-V) u_{\bm{k}}v_{\bm{k}}u_{\bm{l}}v_{\bm{l}}\left[-ie^{-i(\phi_{\bm{k}}-\phi_{\bm{l}})}+ie^{i(\phi_{\bm{k}}-\phi_{\bm{l}})}\right]=0\\
    &\Rightarrow V \sum_{\bm{k,l}} u_{\bm{k}}v_{\bm{k}}u_{\bm{l}}v_{\bm{l}}\left[\frac{e^{i(\phi_{\bm{k}}-\phi_{\bm{l}})}-e^{i(\phi_{\bm{k}}-\phi_{\bm{l}})}}{2i}\right]=0\\
    &\Rightarrow V \sum_{\bm{k,l}} u_{\bm{k}}v_{\bm{k}}u_{\bm{l}}v_{\bm{l}}\sin{(\phi_{\bm{k}}-\phi_{\bm{l}})}=0.
\end{split}
\end{equation}
Now, excluding a trivial solution for $u_{\bm{k}}(u_{\bm{l}})$ or $v_{\bm{k}}(v_{\bm{l}})$ for the ground state energy, we get to the condition,
\begin{equation}
    \begin{split}
        &\sin{(\phi_{\bm{k}}-\phi_{\bm{l}})}=0\\
        &\Rightarrow \phi_{\bm{k}}=\phi_{\bm{l}}.
    \end{split}
\end{equation}
Thus, $\phi_{\bm{k}}$ is independent of momentum $\bm{k}$, indicating that the phase $\phi$ is the same for all the Cooper pairs, implying phase coherence in the superconducting ground state. However, this condition is only true for $T=0$. At finite temperatures, the effects of phase fluctuations come into play, although they have a negligible impact on conventional superconductors.
\par
Similar to $\phi_{\bm{k}}$, $u_{\bm{k}}$ and $v_{\bm{k}}$ can also be determined by minimizing the ground state energy as follows~\cite{tinkham_introduction_2004,ketterson_superconductivity_1999},
\begin{equation}
    v_{\bm{k}}^2=\frac{1}{2} \left(1-\frac{\xi_{\bm{k}}}{E_{\bm{k}}}\right)=\frac{1}{2}\left[1-\frac{\xi_{\bm{k}}}{\left(\Delta^2+\xi_{\bm{k}}\right)^{1/2}}\right],
\end{equation}
and
\begin{equation}
    u_{\bm{k}}^2=\frac{1}{2} \left(1+\frac{\xi_{\bm{k}}}{E_{\bm{k}}}\right)=1-v_{\bm{k}}^2.
\end{equation}
\par
Once we get the values of the coefficients $u_{\bm{k}}$ and $v_{\bm{k}}$, both $\ket{\Psi_{BCS}}$ and ground state energy can be determined. It is found that the energy difference between the ground state superconducting state and normal state is given by the following,
\begin{equation}
    \epsilon_{cond}=f_n(0)-f_s(0)=\frac{1}{2}N(0)\Delta^2(0),
\end{equation}
where $f_n$ and $f_s$ are the free energy per unit volume at normal and superconducting states respectively and $\Delta(0)$ is the superconducting gap which is the manifestation of the pairing energy in the electronic spectrum. This difference is equivalent to the condensation energy $\epsilon_{cond}$ stated in Eq.~\ref{eq:condens_Hc}.
$\Delta(0)$ at ground state ( or zero temperature ) is determined in the BCS theory by the following expression~\cite{tinkham_introduction_2004}:
\begin{equation}\label{eq:gap_BCS_eqn}
    \Delta(0)=2\hbar\omega_D e^{-1/N(0)V}
\end{equation}
\begin{figure*}[hbt]
	\centering
	\includegraphics[width=8cm]{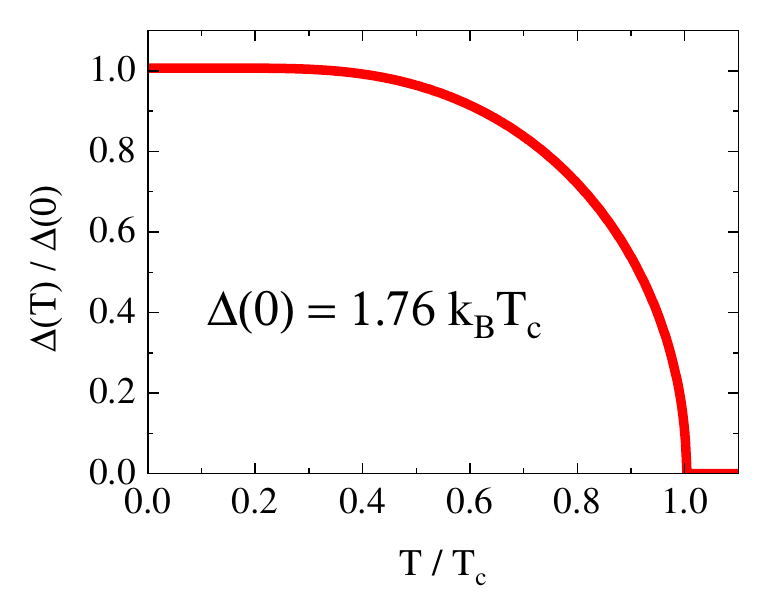}
	\caption[Universal BCS superconducting gap variation]{
    BCS variation of normalized $\Delta$ as a function of $T/T_c$ (universal BCS gap variation) obtained by numerically solving Eq.~\ref{eq:gap_var_BCS} under weak-coupling approximation~\cite{tinkham_introduction_2004} ($N(0)V\approx0.3$).}
 \label{fig:gap-variation-BCS} 
\end{figure*}
\subsection{Elementary excitations and BCS gap variation at finite T}
At finite temperatures, the breaking of Cooper pairs gives rise to electron and hole-like quasi-particles, which are the elementary excitations in superconductors. While the variational method can be used to deal with excited states, a more elegant and sophisticated approach is the Bogoliubov-Valatin~\cite{bogoliubov_new_1958,bogoljubov_new_1958,valatin_comments_1958} self-consistent method. The energy of the quasi-particles is given by~\cite{tinkham_introduction_2004,ketterson_superconductivity_1999},
\begin{equation}
    E_{\bm{k}}=(\xi^2_{\bm{k}}+\Delta^2)^{1/2}.
\end{equation}
By solving the Bogoliubov-Valatin self-consistent equation, we get the following equation:
\begin{equation}\label{eq:gap_var_BCS}
    \frac{1}{N(0)V}=\int_0^{\hbar \omega_D}\frac{\tanh{\frac{\beta}{2} (\xi^2 +\Delta^2)^{1/2}}}{(\xi^2 +\Delta^2)^{1/2}} d\xi.
\end{equation}
The temperature variation of superconducting gap $\Delta$ can be obtained by numerically solving Eq.~\ref{eq:gap_var_BCS} ( shown in Fig.~\ref{fig:gap-variation-BCS} ). According to the BCS theory, $\Delta(T)$ or the pairing energy of the Cooper pairs goes to zero at transition temperature $T_c$. Hence, at $T_c$ the Eq.~\ref{eq:gap_var_BCS} translates into the expression:
\begin{equation}\label{eq:Tc_BCS_eqn}
    k_BT_c=1.13\hbar\omega_D e^{- 1/N(0)V}.
\end{equation}
Now, Eq.~\ref{eq:Tc_BCS_eqn} combined with Eq.~\ref{eq:gap_BCS_eqn} gives rise to the BCS prediction about the relation between $\Delta$ and $T_c$:
\begin{equation}
    \frac{\Delta(0)}{T_c}=1.764.
\end{equation}
This signifies that the superconducting gap vanishes at $T_c$ according to BCS theory, which is true for conventional superconductors.
\par
However, it is worth noting that BCS did not consider the effect of phase fluctuations at finite temperatures. To date, a comprehensive microscopic theory that can effectively address the entire problem, accounting for both quasi-particle excitations and phase fluctuations on equal terms, remains elusive. Nevertheless, the effect of phase fluctuations can be formulated in the framework of a phase-only Hamiltonian, which will be discussed in Section~\ref{sec:phase-fluc-intro}.
\subsection{BCS density of states}
\begin{figure*}[hbt]
	\centering
	\includegraphics[width=16cm]{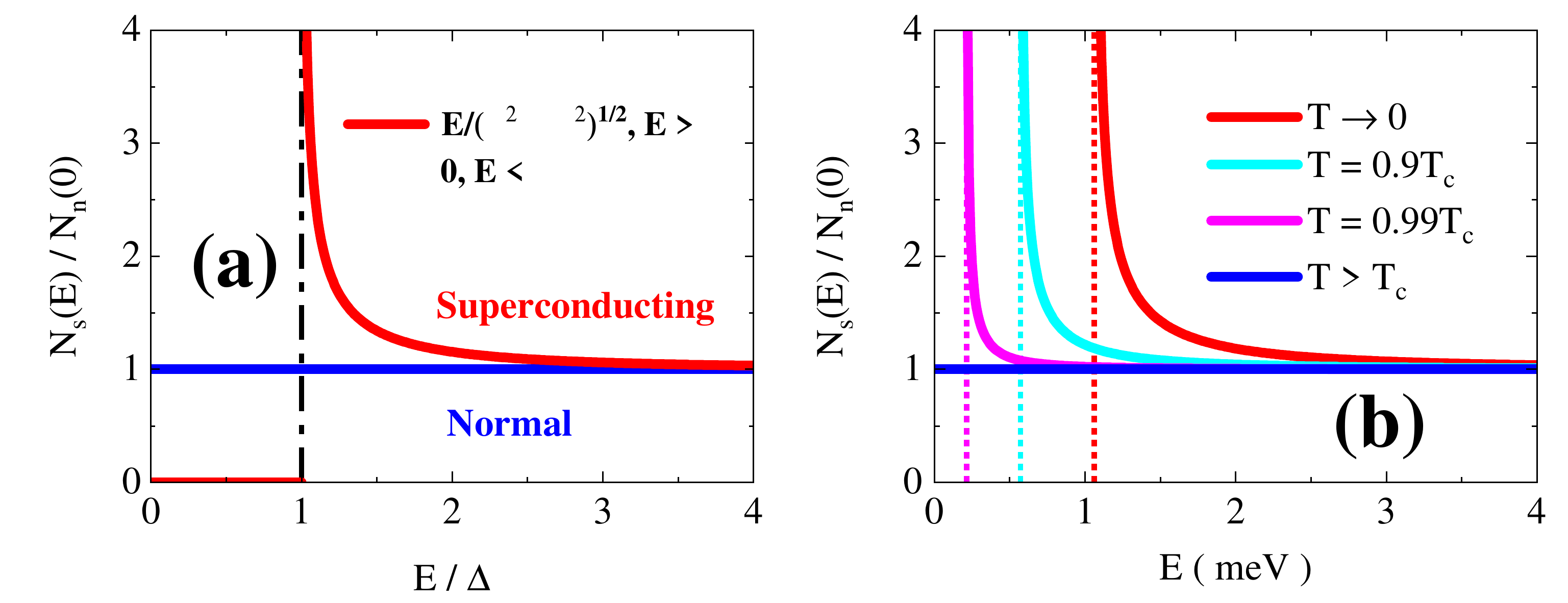}
	\caption[Density of states in superconducting and normal state]{
\textbf{(a)} The comparison of the BCS density of states (DOS) (red line) with the normal metal DOS (blue line). At $E=\Delta$ (indicated by the vertical dotted line), the superconducting DOS diverges. Both DOS are normalized with respect to the DOS of the normal metal at $E=E_F$. \textbf{(b)} The variation of the normalized superconducting DOS is shown at intermediate temperatures ($T_{int}$), where the superconducting DOS diverges at $E=\Delta(T_{int})$ - derived from the corresponding temperature profile of $\Delta$ in Fig.~\ref{fig:gap-variation-BCS}. All the plots were simulated using Eq.~\ref{eq:NsbyN0}.
}
 \label{fig:Ns-vs-E-variation-BCS} 
\end{figure*}

The quasi-particle excitation is Fermionic in nature and the quasi-particles behave like normal electrons. Also, the Bogoliubov creation operator which creates the quasi-particles is in one-to-one correspondence with the creation operator $c^*_{\bm{k}}$ for normal electrons. Hence, we can get the superconducting DOS, $N_s(E)$ by equating with the normal DOS, $N_n(\xi_n)$,
\begin{equation}
 N_s(E)dE=N_n(\xi_n)d\xi_n. 
\end{equation}
Since we are considering $\xi_n$ only few meV around the Fermi level, we can assume normal DOS to be constant in that range so that $N_n(\xi_n)\approx N_n(0)$.
Hence, we get the superconducting DOS by the following relation,
\begin{equation}\label{eq:NsbyN0}
    \frac{N_s(E)}{N_n(0)}=\frac{d\xi}{dE}=
    \begin{cases}
    \frac{E}{(E^2-\Delta^2)^{1/2}} & (E>\Delta)\\
    0 & (E<\Delta)
    \end{cases}
\end{equation}
The comparison of superconducting and normal density of states (DOS) is illustrated in Fig.~\ref{fig:Ns-vs-E-variation-BCS}(a) ( showing only the positive half of the quadrant ). At $E=\Delta$, as also evident from Eq.~\ref{eq:NsbyN0}, the superconducting DOS diverges. In Fig.~\ref{fig:Ns-vs-E-variation-BCS}(b), the superconducting DOS is plotted for different temperatures. At temperatures very close to zero, it diverges in proximity to the superconducting gap at zero temperature, $\Delta(0)$. These points of divergence at $E=\pm \Delta$ are known as coherence peaks. With an increase in temperature, the coherence peak shifts to lower values of $\Delta$, identifiable from the temperature-dependent gap values curated in Fig.~\ref{fig:gap-variation-BCS}. Beyond $T_c$, it becomes similar to the normal DOS.

\subsection{Broadening of coherence peaks: Dynes parameter}\label{sec:dynes-par}
In the case of disordered superconductors, the coherence peaks in the superconducting DOS are comparatively broader than expected from BCS theory. To address this broadening, Dynes et al.~\cite{dynes_direct_1978} proposed to incorporate a phenomenological parameter $\Gamma$ in the BCS DOS such that Eq.~\ref{eq:NsbyN0} becomes,
\begin{equation}\label{eq:NsbyN0_Dynes}
    N_s(E)=Re\left(\frac{E+i\Gamma}{\sqrt{\left(E+i\Gamma\right)^2-\Delta^2}}\right).
\end{equation}
The parameter $\Gamma$, named after Dynes, is integral to our discussion. The modified density of states ( DOS ), as defined in Eq.~\ref{eq:NsbyN0_Dynes}, has proven effective in accurately modeling tunneling data in both zero-field~\cite{white_destruction_1986,plecenik_finite-quasiparticle-lifetime_1994,raychaudhuri_evidence_2004,chockalingam_tunneling_2009,sherman_higgs_2015,szabo_fermionic_2016} environments and in the presence of a magnetic field~\cite{mukhopadhyay_magnetic-field_2005,sirohi_multiband_2019,le_single_2019}. In fact, the Dynes formula for the density of states ( DOS ) has been applied to a wide variety of superconductors, encompassing tunneling spectra in bulk~\cite{dynes_tunneling_1984} and thin film~\cite{white_destruction_1986,chockalingam_tunneling_2009,szabo_fermionic_2016} superconductors, superconducting nanoparticles~\cite{bose_observation_2010}, two-band superconductors like MgB\textsubscript{2}~\cite{iavarone_directional_2003}, and novel superconductors such as CaC\textsubscript{6}~\cite{bergeal_scanning_2006, kurter_large_2007}. Additionally, it has been utilized in photoemission studies of the superconducting h-ZrRuP~\cite{matsui_ultrahigh-resolution_2005} and the filled skutterudite superconductor LaRu\textsubscript{4}P\textsubscript{12}~\cite{tsuda_superconducting_2006}.
\par
Dynes et al.~\cite{dynes_direct_1978, dynes_tunneling_1984} initially interpreted the observed broadening as arising from the finite lifetime of quasi-particles at the superconducting gap edge. However, this interpretation has faced criticism~\cite{mitrovic_correct_2007, dentelski_tunneling_2018} on the basis that the parameter is not directly derived from a microscopic pair-breaking mechanism and alternative forms of the single-particle superconducting density of states ( DOS ) have been proposed, each with its own set of microscopic justifications. Rather than introducing the parameter $\Gamma$ in Eq.~\ref{eq:NsbyN0}, Mitrović et al.~\cite{mitrovic_correct_2007} considered the gap to be complex. On the other hand, Dentelski et al.~\cite{dentelski_tunneling_2018} employed a theory of free electrons coupled to superconducting fluctuations with finite-range correlations to fit the tunneling spectra. Nevertheless, in recent years, some researchers~\cite{herman_microscopic_2016, boschker_microscopic_2020} have proposed microscopic interpretations for the utilization of $\Gamma$. Interestingly, for addressing the broadening in the presence of magnetic scatterers, Herman et al.~\cite{herman_microscopic_2016} utilized the following elegant transformation suggested by Mikhailovsky et al.~\cite{mikhailovsky_thermal_1991},
\begin{equation}\label{eq:Delta_BCS_to_Dynes}
    \Delta'(E)=\frac{E\Delta}{E+i\Gamma},
\end{equation}
which, upon substituting $\Delta\rightarrow\Delta'(E)$, transforms Eq.~\ref{eq:NsbyN0} into Eq.~\ref{eq:NsbyN0_Dynes}.
\par
\begin{figure*}[hbt]
	\centering
	\includegraphics[width=8cm]{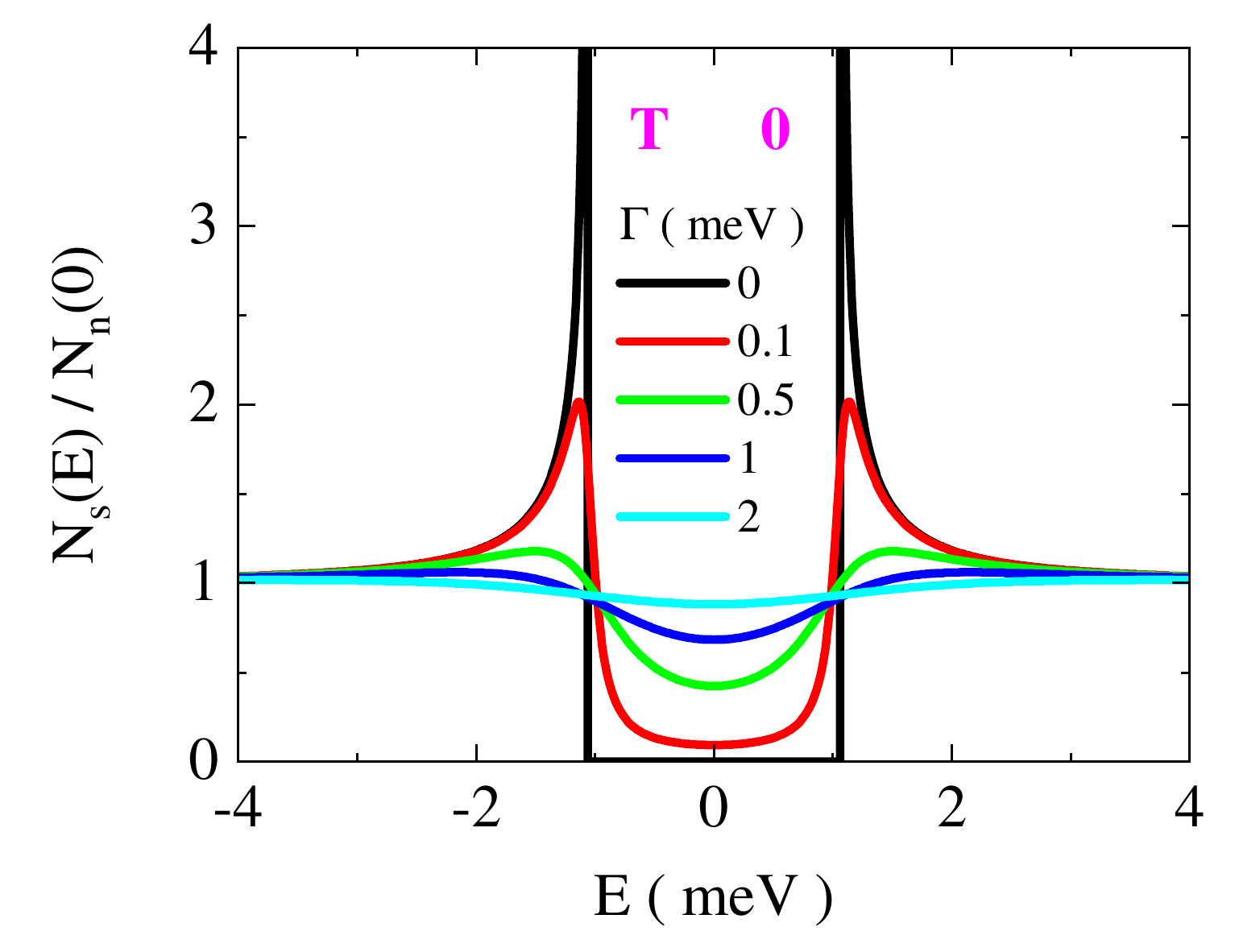}
	\caption[Variation of normalized density of states with different Dynes parameter values]{
    Normalized density of states spectra ($N_s/N_n$ vs $E$) are presented for various Dynes parameter values ($\Gamma$), simulated using Eq.~\ref{eq:NsbyN0_Dynes} as $T\rightarrow 0$. For $\Gamma=0$, a sharp divergence occurs at $E=\pm \Delta(0)$, where $\Delta(0)=1.76k_BT_c=1.0626$ meV with $T_c=7$ K. In the intermediate region, $N_s$ is zero for $\Gamma=0$, as indicated by Eq.~\ref{eq:NsbyN0}. As we increase $\Gamma$, the coherence peaks flatten, leading to a finite density of states (DOS) in the intermediate gap region.}
 \label{fig:Ns-vs-E-diff-Gamma-BCS} 
\end{figure*}
Dyne broadening can be attributed to various factors, encompassing spatial inhomogeneity within the electron system~\cite{boschker_microscopic_2020} and the averaging effect over the anisotropic gap distribution in a superconductor~\cite{raychaudhuri_evidence_2004}. Another significant contributor to the broadening of quasiparticle peaks in $N_s(E)$ stems from pair-breaking scattering, limiting the lifetime of quasiparticle states~\cite{boschker_microscopic_2020}. Cooper pair-breaking can occur due to diverse factors, such as external magnetic fields, current, rotation, spin exchange, hyperfine fields, and magnetic impurities~\cite{tinkham_introduction_2004}. Thermal phase fluctuations represent another such pair-breaking mechanism~\cite{lee_electron-electron_1989}, holding\hspace{0.2cm} particular relevance\hspace{0.2cm} for our system\hspace{0.2cm} ( refer\hspace{0.2cm} to Chapter~\ref{ch:phase_fluctuations} for further details ). Abrikosov and Gor'kov~\cite{abrikosov_problem_1961} initially demonstrated the effect of pair-breaking on the depression of $T_c$ and the modification of BCS density of states in the presence of magnetic impurities. Later, Maki~\cite{maki_persistent_1963,maki_behavior_1964,maki_pauli_1964,maki_equivalence_1965} and de Gennes et al.~\cite{de_gennes_relations_1963,de_gennes_behavior_1964,de_gennes_magnetic_1964} extended these results to describe the effects of other pair-breaking perturbations, i.e., those that destroy the time-reversal degeneracy of the paired states.
\subsection{Measurement of superconducting gap by normal metal-to-superconductor tunneling}\label{sec:tunneling}
According to the BCS theory, an energy gap, $\Delta$, which is equivalent to the pairing energy, emerges in the electronic density of states ( DOS ) during the transition from the normal to the superconducting state. In 1960, Ivar Giaever~\cite{giaever_energy_1960,giaever_electron_1960,giaever_electron_1974} conducted direct measurements of $\Delta$ for superconductors like Al, Pb, and others. He employed tunneling experiments in Superconductor-Insulator-Normal ( SIN ) tunnel junctions for this purpose. I shall briefly explain the basics of normal-to-superconductor tunneling.
\par
Tunneling is a quantum mechanical phenomenon that allows particles to pass through a potential barrier with a non-zero transmission probability, which is forbidden in classical physics. When a normal metal and a superconductor are brought close to each other while maintaining a thin insulating barrier between them, electron tunneling occurs and reveals the density of states of the superconductor.
\par
\begin{figure*}[hbt]
	\centering
	\includegraphics[width=8cm]{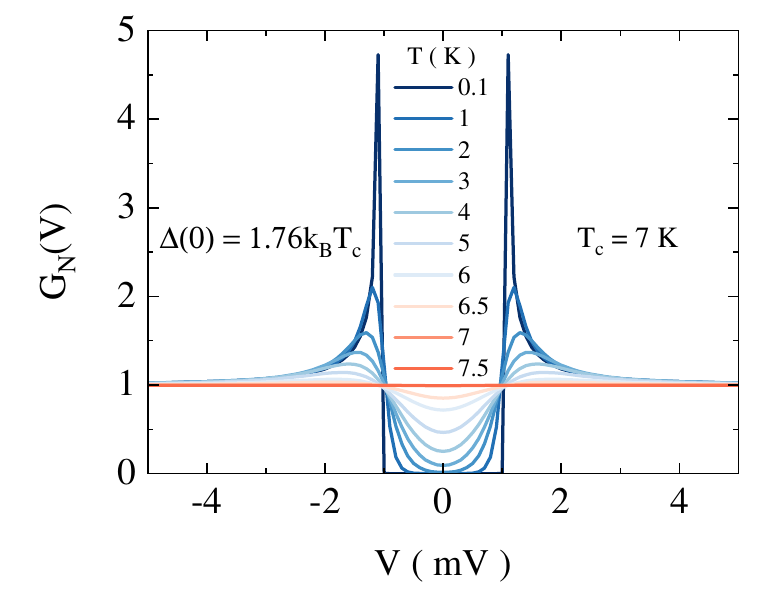}
	\caption[Temperature variation of simulated normalized differential conductance spectra for normal-to-superconductor tunneling.]{
    Normalized differential conductance ($G_N(V)$ vs $V$) spectra for normal-to-superconductor tunneling for different temperatures simulated from Eq.~\ref{eq:tunneling_conductance} assuming $\Delta(0)=1.76k_BT_c$ where $T_c=7$ $K$.}
 \label{fig:G-vs-E-diff-T-BCS} 
\end{figure*}
In the absence of an externally applied voltage, the Fermi levels of the normal metal and superconductor align themselves at the same energy in accordance with thermodynamic principles, resulting in a single chemical potential for the combined system. However, due to the presence of a single-particle energy gap ( $\Delta$ ) in the superconductor across the Fermi level, tunneling from the normal metal to the superconductor at the same energy is prohibited. Therefore, in the absence of an external bias, no steady-state tunneling current is observed.
\par
Tunneling becomes possible only when the applied bias voltage ( $V$ ) exceeds the energy gap ( $\Delta$ ), specifically when $eV>\Delta$. By observing the tunneling current under applied bias, the superconducting energy gap can be determined accurately.
\par
The tunneling current from the normal to the superconducting side takes the form:
\begin{equation}
    I_{ns}=A\abs{T}^2N_n(0)\int_{-\infty}^\infty N_s(E)[f(E)-f(E+eV)]dE,
\end{equation}
where $T$ is the tunneling matrix element, $f$ is the Fermi-Dirac distribution, $N_n$ and $N_s$ are the electronic density of states of normal and superconducting state respectively.
\par
The differential conductance of such SIN tunnel junctions, as a function of the bias voltage, $V$, can be accurately described using Fermi's golden rule:
\begin{equation}\label{eq:tunneling_conductance}
    G_{ns}=\frac{dI_{ns}}{dV} = G_{nn} \int_{-\infty}^{\infty}\frac{N_s(E)}{N_n(0)} \left[-\frac{\partial f(E+eV)}{\partial (eV)} \right]dE,
\end{equation}
where $G_{nn}$ is the differential conductance of the junction above the transition temperature when the superconductor becomes normal. Now at the limit $T=0$, $\partial f(E+eV)/\partial (eV)$ becomes a $\delta$ function at $E=eV$ and the differential conductance becomes, 
\begin{equation}
    G_{ns}|_{T=0} =G_{nn}\frac{N_s(e\abs{V})}{N_n(0)}=G_{nn}\frac{e\abs{V}}{(e\abs{V}^2-\Delta^2)^{1/2}}.
\end{equation}
We employ this normal-to-superconductor tunneling phenomenon to directly measure the superconducting DOS with the help of a scanning tunneling microscope where the \enquote{insulator} in the superconductor-insulator-normal metal junction is the vacuum between the STM tip (metal) and the sample (superconductor).
\subsection{Temperature variation of penetration depth}
In the clean limit ( $l > \xi_0$ ), the BCS variation of penetration depth ( $\lambda_L$ ) with temperature is as follows~\cite{tinkham_introduction_2004}, 
\begin{equation}\label{eq:lambda-bcs-clean}
   \left. \frac{\lambda_{L}^{-2}(T)}{\lambda_{L}^{-2}(0)} \right\vert_{Clean}=\left[1-2\int_{\Delta}^{\infty}\left(-\frac{\partial f}{\partial E}\right) \frac{E}{(E^2-\Delta^2)^{1/2}} dE\right],
\end{equation}
where $\lambda_{L}^{-2}(0)$ in the clean limit is determined from Eq.~\ref{eq:london-pen-depth-ns} where $n_s$ becomes the total carrier density $n$ in the limit $T\rightarrow 0$,
\begin{equation}
     \lambda_{L,Clean}^{-2}(0)=\frac{\mu_0 n e^2}{m_e}.  
\end{equation}
On the other hand, the variation of $\lambda_L$ In the dirty limit ( $l < \xi_0$ ) is given by~\cite{abrikosov_superconducting_1959},
\begin{equation}\label{eq:lambda-bcs-dirty}
    \left. \frac{\lambda_L^{-2}(T)}{\lambda_L^{-2}(0)}\right\vert_{Dirty}=\frac{\Delta(T)}{\Delta(0)}\tanh{\frac{\Delta(T)}{2k_BT}},
\end{equation}
where the zero-temperature dirty limit BCS estimate of $\lambda_L$ is given by the following relation~\cite{abrikosov_theory_1959},
\begin{equation}\label{eq:lambda0_dirty}
     \lambda_{L,Dirty}^{-2}(0)=\frac{\pi \mu_0 \Delta(0)\sigma_n}{\hbar},
\end{equation}
which can be estimated from the sum rule for the suppression of superfluid density in the presence of disorder~\cite{mondal_enhancement_2013,dutta_superfluid_2022} ( $\sigma_n$ is the normal state conductivity ). It should be noted that a comprehensive expression of the temperature variation of $\lambda_L$ as a function of disorder has been derived in Ref.~\citenum{dutta_superfluid_2022}.
\begin{figure*}[hbt]
	\centering
	\includegraphics[width=8cm]{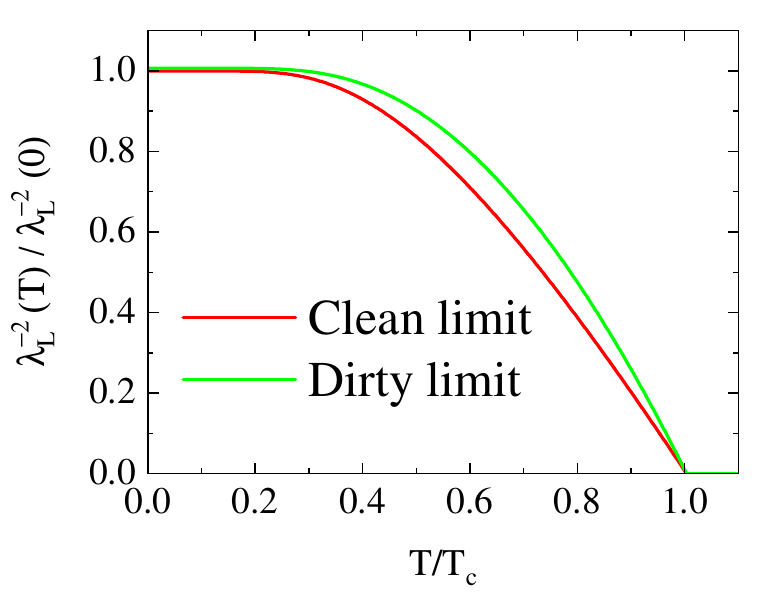}
	\caption[Comparison of the temperature variation of $\lambda_L^{-2}$ in the clean and dirty limit.]{
    Temperature variation of $\lambda_L^{-2}$ in the clean and dirty limit simulated from Eq.~\ref{eq:lambda-bcs-clean} and \ref{eq:lambda-bcs-dirty} respectively.}
 \label{fig:invlambda2-vs-T-BCS-clean-dirty} 
\end{figure*}
\section{Ginzburg-Landau theory}
The microscopic BCS theory offers excellent explanations for various characteristics of superconductors, such as nuclear relaxation, energy gap, and elementary excitations, assuming a constant energy gap throughout space. However, in the case of inhomogeneous superconductors where the gap varies spatially and involves fluctuations, the BCS theory becomes highly complex. In such scenarios, an intriguing theoretical framework proposed by Ginzburg and Landau~\cite{Ginzburg:1950sr,landau_theory_1965,ginzburg_theory_1955} ( GL ) in 1950, predating the development of the BCS theory, presents an elegant depiction of superconductivity near $T_c$. Although this theory originated as a phenomenological assumption, Gor'kov~\cite{gorkov_microscopic_1959} demonstrated that the Ginzburg-Landau theory can be deduced from the BCS theory.
\subsection{Order parameter and GL free energy}\label{sec:order-par-GL}
Taking a cue from the general theory on second-order phase transition by Landau~\cite{landau_theory_phase_1965}, Ginzburg and Landau proposed a pseudo-wavefunction in the form of a complex order parameter, $\psi=\abs{\psi}e^{i\phi}$ associated with the superconducting state. Local superconducting electron density, $n_s(\bm{r})$, is assumed to be related to the order parameter in the following way:
\begin{equation}
    n_s(\bm{r})=\abs{\psi(\bm{r})}^2
\end{equation}
Assuming small $\psi$ and slow variation in space, free energy density $f$ can be expanded in a power series of $\abs{\psi}^2$ and $\abs{\nabla\psi}^2$ as follows~\cite{tinkham_introduction_2004}:
\begin{equation}\label{eq:GL-free-energy}
    f=f_{n0}+\alpha\abs{\psi}^2+\frac{\beta}{2}\abs{\psi}^4+\frac{1}{2m^*}\abs{\left(\frac{\hbar}{i}\bm{\nabla}-e^*\bm{A}\right)\psi}^2+\frac{1}{2}\mu_0H^2,
\end{equation}
where $f_{n0}$ is the normal state free energy, $\alpha$ and $\beta$ are phenomenological coefficients, $m^*$ and $e^*$ are the effective mass and effective charge of a pair of electrons respectively, $\bm{A}$ is the vector potential. 
\par
Similar to the Landau theory of phase transitions~\cite{landau_theory_phase_1965}, temperature variation of $\alpha$ and $\beta$ across the transition is taken as:
\begin{equation}
    \alpha=\alpha_0(T/T_c-1),
\end{equation}
and,
\begin{equation}
    \beta=\beta_0.
\end{equation}
where $\alpha_0,\beta_0>0$
\subsection{Ginzburg-Landau equations}
By minimizing the free energy expression in Eq.~\ref{eq:GL-free-energy} by applying variational method, we get the celebrated Ginzburg-Landau equations: 
\begin{equation}\label{eq:GL-eqn-1}
    \alpha\psi+\beta\abs{\psi}^2\psi+\frac{1}{2m^*}\left(\frac{\hbar}{i}\bm{\nabla}-e^*\bm{A}\right)^2\psi=0,
\end{equation}
and
\begin{equation}\label{eq:GL-eqn-2}
    \bm{J}=\frac{e^*}{m^*}\abs{\psi}^2(\hbar\bm{\nabla}\phi-e^*\bm{A})=e^*\abs{\psi}^2\bm{v_s},
\end{equation}
where $\bm{J}$ is the supercurrent density and $\bm{v_s}$ is interpreted as the superelectron velocity given by:
\begin{equation}
    \bm{v_s}=\frac{1}{m^*}(\hbar\bm{\nabla}\phi-e^*\bm{A})
\end{equation}
\subsection{GL coherence length}
In zero field, the first GL equation~\ref{eq:GL-eqn-1} can be written in one dimension as follows,
\begin{equation}\label{eq:GL-eqn-1d}
    \frac{\hbar^2}{2m^*\abs{\alpha}}\frac{d^2f}{dx^2}+f-f^3=0,
\end{equation}
where $f=\psi/\psi_{\infty}$ is a normalized wave function with $\psi_{\infty}=-\alpha/\beta>0$. A new characteristic length scale $\xi_{GL}$ can be defined as evident in Eq.~\ref{eq:GL-eqn-1d}:
\begin{equation}
    \xi^2_{GL}(T)=\frac{\hbar^2}{2m^*\abs{\alpha}}=\frac{\Phi_0}{2\sqrt{2}H_c(T)\lambda_L(T)}.
\end{equation}
$\xi_{GL}$ is called Ginzburg-Landau coherence length which is the characteristic length scale over which a small disturbance of $\psi$ with respect to $\psi_{\infty}$ will decay spatially.
\subsection{GL parameter and Classification of superconductors}
Ginzburg-Landau parameter is defined as the ratio of two characteristic lengths of the system, namely, London penetration depth ( $\lambda_L$ ) and GL coherence length ( $\xi_{GL}$ ):
\begin{equation}
    \kappa=\frac{\lambda_L(T)}{\xi_{GL}(T)}.
\end{equation}
In their original paper~\cite{Ginzburg:1950sr,landau_theory_1965}, Ginzburg and Landau showed by numerical integration that there is a crossover from positive to negative surface energy at $\kappa=1/\sqrt{2}$. But, the enormity of this crossover was later realized in the work of Abrikosov's seminal work~\cite{abrikosov_vliyanie_1952,abrikosov_magnetic_1957-1,abrikosov_magnetic_1957}, where he demonstrated that the presence of negative surface energy leads to the division of flux-bearing normal regions, eventually reaching a quantum limit where each quantum of flux $\Phi_0=h/2e$ penetrates the sample as an individual flux tube. This new class of superconductors was named \enquote{superconductors of the second
group} by Abrikosov and later known as type-II superconductors. We have already introduced the type-I and type-II superconductors in section~\ref{sec:typeI-II-SC} which are classified on the basis of the following relation:
\begin{equation}
    \kappa
    \begin{cases}
        <\frac{1}{\sqrt{2}} & \text{for type-I}\\
        >\frac{1}{\sqrt{2}} & \text{for type-II}.
    \end{cases}
\end{equation}
\subsection{Fluxoid quantization}
While investigating the state of a multiply connected superconductor in the presence of a magnetic field, F. London~\cite{london_superfluids_1950} put forward the concept of the fluxoid associated with each hole ( or normal region ) passing through the superconductor. Fluxoid $\Phi'$ is defined by,
\begin{equation}\label{eq:fluxoid-def}
    \Phi'=\Phi+\mu_0\lambda^2 \oint \bm{J_s}\cdot d\bm{l}=\Phi+\frac{m^*}{e^*}\oint \bm{v_s} \cdot d\bm{l},
\end{equation}
where $m^*$ and $e^*$ are twice the mass $m$ and charge $e$ of electron respectively, and $\Phi$ is the ordinary magnetic flux passing through the integration loop expressed as,
\begin{equation}
    \Phi=\iint\displaylimits_S \bm{B} \cdot d \bm{S}=\oint \bm{A} \cdot d\bm{l}.
\end{equation}
In the framework of GL theory, it is possible to explain the quantization of fluxoids. Using the Bohr-Sommerfeld quantum condition, it can be shown that:
\begin{equation}~\label{eq:fluxoid-quantization}
    \Phi'=\frac{1}{e^*}\oint(m^*\bm{v_s}+e^*\bm{A})\cdot d\bm{s}=\frac{1}{e^*}\oint\bm{p}\cdot d\bm{s}=n\frac{h}{e^*}=n\Phi_0,
\end{equation}
where $\Phi_0=h/2e=2.07\times10^{-15}$ $Wb$ is the flux quantum ( $h$ is the Planck constant ).
\subsection{Little-Parks experiment}\label{sec:Little-Parks-intro}
Little and Parks~\cite{little_observation_1962,parks_fluxoid_1964} conducted a remarkable experiment to demonstrate the quantization of the fluxoid. They utilized a thin-walled superconducting cylinder placed in an axial magnetic field and observed shifts $\Delta T_c(H)$ in the critical temperature corresponding to the magnetic flux enclosed by the cylinder. 
\par
For a thin-walled cylinder of radius $R$ in the presence of applied field $H$, flux $\Phi$ is given by,
\begin{equation}\label{eq:flux-cylinder}
    \Phi=\pi R^2 H.
\end{equation}
Now, Putting Eq.~\ref{eq:flux-cylinder} and the fluxoid quantization relation $\Phi'=n\Phi_0$ into Eq.~\ref{eq:fluxoid-def}, we get the expression of supercurrent density,
\begin{equation}
    v_s=\frac{\hbar}{m^*R}\left(n-\frac{\Phi}{\Phi_0}\right).
\end{equation}
Also, by minimising GL equation (Eq.~\ref{eq:GL-free-energy}) for a uniform current density in a thin cylinder, we get,
\begin{equation}
    \abs{\psi}^2=\psi^2_{\infty}\left[1-\left(\frac{\xi_{GL} m^* v_s}{\hbar}\right)^2\right],
\end{equation}
where $\psi_{\infty}=-\alpha/\beta>0$. Now, at the transition $\abs{\psi}^2=0$, which leads us to the condition,
\begin{equation}\label{eq:little-parks-xi-var}
    \frac{1}{\xi_{GL}^2}=\left(\frac{m^* v_s}{\hbar}\right)^2=\frac{1}{R^2}\left(n-\frac{\Phi}{\Phi_0}\right)^2.
\end{equation}
Using the temperature variation of $\xi_{GL}$ in Eq.~\ref{eq:little-parks-xi-var}, we get the following form of fractional depression of $T_c$,
\begin{equation}\label{eq:little-parks-Tc-suppression}
    \frac{\Delta T_c(H)}{T_c}\propto \frac{\xi^2_{GL}(0)}{R^2}\left(n-\frac{\Phi}{\Phi_0}\right)^2
\end{equation}
It is evident from Eq.~\ref{eq:little-parks-Tc-suppression} that the suppression of $T_c$ is maximum when $\left(n-\frac{\Phi}{\Phi_0}\right)=\frac{1}{2}$. On the other hand, when $\left(n-\frac{\Phi}{\Phi_0}\right)=0$, $v_s=0$. As a result, $\abs{\psi}^2=\psi^2_{\infty}$ and $T_c$ is enhanced. For other nonzero values of $v_s$, $T_c$ is less.
\subsection{Linearized GL equation and \texorpdfstring{$H_{c2}$}{}}
If $\psi$ reduces by the application of magnetic field such that $\psi \ll \psi_{\infty}$, the $\abs{\psi}^4$ term in ~\ref{eq:GL-eqn-1} can be dropped and we get the linearized GL equation~\cite{tinkham_introduction_2004}:
\begin{equation}\label{eq:linear-GL-eqn}
    \left(\frac{\bm{\nabla}}{i}-\frac{2\pi\bm{A}}{\Phi_0}\right)^2\psi=-\frac{2m^*\alpha}{\hbar^2}\psi=\frac{\psi}{\xi^2_{GL}(T)}
\end{equation}
If we solve Eq.~\ref{eq:linear-GL-eqn} for a bulk sample with magnetic field along $z$-axis with the choice of vector potential being $A_y=Hx$, we can get the expression of the maxima of field H as,
\begin{equation}
    H_{c2}=\frac{\Phi_0}{2\pi\xi^2_{GL}(T)},
\end{equation}
where the maxima is the upper critical field ( $H_{c2}$ ) introduced before ( Section~\ref{eq:condens_Hc} ). 
\subsection{Abrikosov vortex state}
General solution to the linearized GL equation~\ref{eq:linear-GL-eqn} for a bulk type-II SC was given by Abrikosov~\cite{abrikosov_magnetic_1957-1} as follows:
\begin{equation}
    \psi(x,y)=\sum_{n=-\infty}^{\infty}C_n e^{inky} e^{-\frac{1}{2}\kappa^2(x-\frac{nk}{\kappa^2})^2},
\end{equation}
where the coefficients $k$ and $C_n$ need to be adjusted to minimize the Gibbs free energy subject to the periodicity condition: $C_{n+N}=C_n$. Now, for finding the configuration with minimum energy, Abrikosov coined a new parameter,
\begin{equation}
    \beta_A=\frac{\langle \psi^4 \rangle}{\langle \psi^2 \rangle^2}.
\end{equation}
The configuration with minimum $\beta_A$ will have the minimum energy. It was calculated by Abrikosov~\cite{abrikosov_magnetic_1957-1} that the minimum energy configuration was a square lattice with $\beta_A=1.18$ and lattice constant ( $a_{\square}$ ) as follows:
\begin{equation}
        a_{\square}=\left(\frac{\Phi_0}{B}\right)^{1/2}
\end{equation}
But, later it was found to be incorrect. Kleiner et al.~\cite{kleiner_bulk_1964} showed that it is a triangular ( or hexagonal ) lattice that possesses the minimum energy with $\beta_A=1.16$ and has the nearest neighbor distance ( $a_{\Delta}$ ):  
\begin{equation}
    a_{\Delta}=\left(\frac{4}{3}\right)^{1/4} \times\left(\frac{\Phi_0}{B}\right)^{1/2}=1.075\times\left(\frac{\Phi_0}{B}\right)^{1/2}
\end{equation}
\section{Effect of phase fluctuations: the concept of phase stiffness}\label{sec:phase-fluc-intro}
The superconducting ground state, characterized by the complex order parameter $\psi=\Delta e^{i\phi}$, can be destroyed by two types of excitations: (i) quasi-particle excitations formed due to the breaking of Cooper pairs and (ii) phase fluctuations destroying the global phase coherence. The energy scale associated with the quasi-particle excitations is the pairing energy or superconducting energy gap $\Delta$, while the extent of phase fluctuations is dictated by another energy scale called phase stiffness $J_s$, which is defined by the energy required to apply a phase twist of unit magnitude over a unit length. For conventional bulk superconductors, the phase of the superconducting wave function is rigid, i.e., $J_s \gg \Delta$; hence $\Delta$ is the defining energy scale of the superconducting transition which is well explained by the Bardeen-Cooper-Schrieffer~\cite{bardeen_theory_1957} ( BCS ) and Eliashberg~\cite{eliashberg_interactions_1960} mean-field theories, where phase fluctuations were not considered. However, this hierarchy of $J_s$ and $\Delta$ can be altered in certain situations, which include (i) strongly correlated unconventional superconductors such as underdoped high-$T_c$ cuprates or organic superconductors, and (ii) thin films of extremely disordered conventional superconductors~\cite{sacepe_quantum_2020} such as NbN, InO$_x$, TiN, MoGe, and SrTiO$_3$/LaAlO$_3$ interfacial superconductors. In such situations, the superfluid density becomes very low and the superconducting state can get destroyed by strong phase fluctuations even when the pairing energy $\Delta$ remains finite. Up until now, a comprehensive microscopic theory that can effectively address the entire problem, encompassing both quasi-particle excitations and phase fluctuations on equal footing, remains elusive. However, if we consider only the phase degree of freedom, the phase stiffness $J_s$ can be defined in the framework of GL theory as follows.
\par
When there is no magnetic field, GL equation ( Eq.~\ref{eq:GL-eqn-2} ) for the supercurrent density reduces to:
\begin{equation}\label{eq:GL-eqn-J-no-field}
    \bm{J}=\frac{\hbar e^*}{m^*}\abs{\psi}^2 \bm{\nabla}\phi,
\end{equation}
where we have taken the gauge choice $\bm{A}=0$. If we identify $\abs{\psi}^2$ with $n_s/2$, $n_s$ being the superfluid density in the unit of electrons per unit volume, $\bm{J}$ is given by,
\begin{equation}
    \bm{J}=\frac{n_s}{2}e^*\bm{v_s},
\end{equation}
with $\bm{v_s}$ expressed as,
\begin{equation}
    \bm{v_s}=\frac{\hbar}{m^*}\bm{\nabla}\phi.
\end{equation}
Now, the superfluid kinetic energy density is given by,
\begin{equation}
    \frac{1}{2}\left(\frac{n_s}{2}\right)(m^*v_s^2)=\frac{1}{2}\frac{\hbar^2 n_s}{2m^*}\abs{\bm{\nabla}\phi}^2=\frac{1}{2}\frac{\hbar^2 n_s}{4m}\abs{\bm{\nabla}\phi}^2,
\end{equation}
where effective mass of Cooper pair $m^*$ is replaced by the mass of an electron $m$ such that $m^*=2m$. Now, considering only the phase degree of freedom, the Hamiltonian of the system is expressed as,
\begin{equation}\label{eq:phase-hamiltonian}
    H=\frac{1}{2}\frac{\hbar^2 n_s}{4m}\int \abs{\bm{\nabla}\phi}^2 d^3\bm{r}.
\end{equation}
It is evident from Eq.~\ref{eq:phase-hamiltonian} that the pre-factor in the integral $\frac{\hbar^2n_s}{4m}$ is a measure of the resilience of the superconductor against the phase fluctuations. In order to convert this quantity in the characteristic energy scale, we multiply a suitable length $a$, and define the superfluid stiffness ( $J_s$ ) expressed as,
\begin{equation}
    J_s=\frac{\hbar^2n_sa}{4m}=\frac{\hbar^2a}{4\mu_0\lambda^2_L e^2},
\end{equation}
where the characteristic length scale $a$ for phase fluctuations depends on the relation: $a\approx min(d,\xi)$; $d$ and $\xi$ being the thickness and coherence length respectively. In the second equality, $n_s$ is replaced by London penetration depth $\lambda_L$ by the relation~\ref{eq:london-pen-depth-ns}.
\section{Effect of disorder on superconductivity}\label{sec:effect-of-disorder-intro}
Apart from the search for room-temperature superconductors, another area of research in the field of superconductivity focuses on understanding how external parameters, such as magnetic fields, disorder, dimensionality, or changes in composition, can affect the superconducting state. In this section, we shall specifically discuss the impact of the disorder on the state of superconductivity, particularly in the context of superconducting thin films.
\par
Disorder, arising from impurities, lattice imperfections, or external perturbations, introduces spatial and temporal variations that significantly impact the behavior of superconductors. As the disorder level increases, the transition temperature ( $T_c$ ) gradually decreases, eventually leading to a nonsuperconducting ground state. However, even in the absence of the global superconducting ground state, superconducting correlations continue to considerably influence the electronic properties. These correlations give rise to several intriguing phenomena, such as finite high-frequency superfluid stiffness above $T_c$\cite{crane_survival_2007}, a giant magnetoresistance peak\cite{gantmakher_destruction_1998,steiner_superconductivity_2005,sambandamurthy_experimental_2005,baturina_quantum_2007,nguyen_observation_2009} in strongly disordered superconducting films, the persistence of magnetic flux quantization even after being driven into an insulating state observed in strongly disordered Bi films~\cite{stewart_superconducting_2007}, and the persistence of a pronounced pseudogap~\cite{mondal_phase_2011-1,sacepe_pseudogap_2010,sacepe_localization_2011} at temperatures significantly higher than $T_c$. This interplay between disorder and superconductivity offers an intricate and captivating avenue for condensed matter physics research.
\par
The research on the effect of disorder on superconducting thin films was pioneered by Shalnikov~\cite{shalnikov_superconducting_1938}, followed by Buckel and Hilsch~\cite{buckel_einflus_1954,buckel_elektronenbeugungs-aufnahmen_1954}. The first comprehensive and systematic study of the suppression of $T_c$ with decreasing thickness of the films was done by Strongin et al.~\cite{strongin_superconductive_1968,strongin_destruction_1970}.
\subsection{Beyond Anderson’s hypothesis}
Debate on how the disorder affects superconductivity dates back to the late fifties when Anderson~\cite{anderson_theory_1959,abrikosov_superconducting_1959} predicted that in an s-wave superconductor attractive interaction forming the Cooper pairs would remain unaffected by the presence of non-magnetic impurities. He argued that the Cooper pairs are formed by time-reversed eigenstates, whose density of states should not be strongly perturbed by the disorder. This hypothesis, also known as \enquote{Anderson's theorem}, was also demonstrated in the random potential scattering up to the leading order by Gor'kov~\cite{gorkov_microscopic_1959}. This was then loosely interpreted that $T_c$ would also not be strongly sensitive to disorder. But, subsequent experiments showed that the above is only true in the limit of weak disorder: In the presence of increasingly strong disorder, $T_c$ gets gradually suppressed~\cite{strongin_superconductive_1968,strongin_destruction_1970},  and eventually, the material is driven into a non-superconducting state at a critical disorder. The seminal work by Haviland et al.~\cite{haviland_onset_1989} on the SI transition inspired a lot of experimental research~\cite{hebard_magnetic-field-tuned_1990,yazdani_superconducting-insulating_1995,okuma_anomalous_1998,goldman_superconductorinsulator_1998}, where $T_c$ of a thin superconducting film vanishes at some critical value of the dimensionless sheet conductance: $g=\sigma\hbar/e^2$. In fact, the suppression of $T_c$ with an increase in disorder can happen from two origins, namely Fermionic and Bosonic mechanisms~\cite{larkin_superconductor-insulator_1999,goldman_superconductorinsulator_2003,gantmakher_superconductorinsulator_2010,goldman_superconductor-insulator_2010,goldman_superconductor-insulator_2010-1,lin_superconductivity_2015,sacepe_quantum_2020}.
\subsection{Fermionic mechanism}
In the Fermionic route, suppression of the mean-field transition temperature occurs due to the loss of effective screening with an increase in disorder scattering. Due to less screening, the repulsion among the oppositely charged electrons gets dominant over the attractive pairing interaction, and at a critical disorder, superconductivity is destroyed. There have been several efforts to capture the effect of Coulomb interaction on the suppression of $T_c$ with increasing disorder. Ovchinnikov~\cite{ovchinnikov_fluctuation_1973} calculated the shift in $T_c$ for thick films ( thickness $d\gg l$, where $l$ is the mean free path ) taking into account the electromagnetic fluctuations, whose main contribution was related to the scalar potential or Coulomb interaction. However, interest in the interplay of disorder and electron-electron interaction truly blossomed when Altshuler and Aronov~\cite{altshuler_contribution_1979,altshuler_interaction_1980,altshuler_electronelectron_1985}, in 1979, showed that Coulomb interaction along with the impurity scattering creates a dip in the single-particle DOS at the Fermi level. By applying the diagrammatic technique, Maekawa et al.~\cite{maekawa_localization_1982,maekawa_upper_1983} and Takagi et al.~\cite{takagi_anderson_1982} calculated the lowering of transition temperature with the increasing disorder:
\begin{equation}\label{eq:Tc-suppression-maekawa}
    \ln{\frac{T_c}{T_{c0}}}=-\frac{1}{3}g_l\frac{e^2}{2\pi^2\hbar} R_{\square}\left(ln{\frac{h}{k_B T_c\tau}}\right)^3,
\end{equation}
where $T_{c0}$ is the transition temperature of the sample at the bulk limit, $R_{\square}$ is the sheet resistance of the film, $\tau$ is the mean-free-path time and $g_l$ is a constant dependent on the electron-electron interaction and has the value $1/2$ for the screened Coulomb potential. The above analysis illustrates an abnormal rise in the magnitude of the small momentum transfer component of the Coulomb interaction within the effective amplitude of the electron-electron interaction in the Cooper channel. This escalation of electron repulsion, resulting from the interplay between the Coulomb interaction and disorder, serves as the primary cause for the reduction in $T_c$ within homogeneous films. Finkel'stein~\cite{finkelshtein_metal-insulator_1984,finkelshtein_superconducting_1987,finkelstein_suppression_1994} improved Eq.~\ref{eq:Tc-suppression-maekawa} by employing normalization-group analysis as follows:
\begin{equation}\label{eq:Finkelstein-eqn}
    \frac{T_c}{T_{c0}}=e^\gamma \left(\frac{\frac{1}{\gamma}+\frac{r}{4}-\sqrt{\frac{r}{2}}}{\frac{1}{\gamma}+\frac{r}{4}+\sqrt{\frac{r}{2}}}\right)^{1/\sqrt{2r}},
\end{equation}
where $r=\frac{e^2}{2\pi^2 \hbar} R_{\square}$ and the adjustable parameter $\gamma=\ln \frac{h}{k_B T_{c0}\tau}$, $\tau$ being the transit time. This Eq.~\ref{eq:Finkelstein-eqn} by Finkel'stein is considered a benchmark for the Fermionic route in the case of homogeneous thin films.
\par
There has been another work by Anderson et al.~\cite{anderson_theory_1983} where a possible mechanism of the degradation of $T_c$ has been proposed in the case of high-$T_c$ superconductors taking into account the effect of Coulomb interaction. Based on the scaling theory developed by Abrahams et al.~\cite{abrahams_scaling_1979}, they showed that disorder increases the Coulomb pseudopotential $\mu^*$ or the effective Coulomb repulsion between the elections in the Cooper pair. They came out with the following expression for the variation of $T_c$~\cite{anderson_theory_1983}:
\begin{equation}
    T_c=\frac{\omega_D}{1.45}e^{-\left [\frac{1.04(1+\lambda)}{\lambda-\mu^*(1+0.62\lambda)}\right]},
\end{equation}
where $\omega_D$ is the Debye frequency, and $\lambda$ is the electron-phonon coupling. 
\subsection{Bosonic mechanism}
On the other hand, in the Bosonic mechanism~\cite{gold_dielectric_1986,fisher_presence_1990} disorder scattering diminishes superfluid density ( $n_s$ ) and hence superfluid phase stiffness $J_s$ ( $\sim n_s/m^*$ ). As a result, the phase coherent superconducting state gets vulnerable to phase fluctuations and eventually can get destroyed due to strong phase fluctuations even when the pairing amplitude remains finite. Experimentally, this manifests as a persistence of the superconducting gap in the electronic excitation spectrum, even after the global superconducting state is destroyed. Now, if we warm up the system from the superconducting state, it becomes resistive in DC transport measurements due to thermally excited phase fluctuations way before the superconducting gap $\Delta(T)$ in the quasiparticle density of states decreases toward zero. Thus, in a certain temperature range, the state of superconductivity ( indicated by the gap ) and \enquote{normality} ( DC resistance ) coexist. This state can be qualitatively visualized as being composed of phase incoherent Cooper pairs. This regime is referred to as \enquote{pseudogapped}~\cite{sacepe_pseudogap_2010,sacepe_localization_2011,mondal_phase_2011,dubouchet_collective_2019} state.
\par
The bosonic scenario tends to be realized in superconductors with low electron density due to their weaker shielding and relatively lower rigidity when it comes to phase changes. Consequently, the significance of phase fluctuations is amplified in such cases~\cite{emery_importance_1995,emery_superconductivity_1995,larkin_theory_2005}. Though the Bosonic formalism was developed primarily for disordered homogeneous thin films~\cite{larkin_superconductor-insulator_1999}, granular superconductors are naturally suited for this scenario in the BCS framework. Interestingly, some signatures of the Bosonic mechanism were observed in granular 2D systems~\cite{dynes_two-dimensional_1978,adkins_increased_1980} even before the concepts of SI transition came into the picture.
\subsection{Distingushing features in the Fermionic and Bosonic mechanisms}
Magneto-transport measurements and subsequent scaling analysis have traditionally been employed to differentiate between Fermionic and Bosonic mechanisms~\cite{larkin_superconductor-insulator_1999}. However, relying solely on the success of scaling and the identification of universality classes based on critical exponents may not fully elucidate the microscopic physics underlying the transition or the nature of the insulating state. In fact, the lack of universality in scaling exponents and the large variation of critical resistances across different systems have undermined the confidence in arguments based solely on the scaling analysis in recent years.
\par
However, the distinction between the two scenarios can be conveniently formulated using the complex order parameter, $\psi=\Delta e^{i\phi}$, where $\abs{\psi}$ is the superconducting energy gap $\Delta$ and $\phi$ is the phase of the condensate. When fluctuations are present, the superconducting state is preserved if the correlator~\cite{gantmakher_superconductorinsulator_2010},
\begin{equation}
    G(\bm{r})=\langle \psi(\bm{r})\psi(0)\rangle,
\end{equation}
remains non-vanishing in the limit $\abs{\bm{r}}\rightarrow \infty$, i.e.,  $G(\abs{\bm{r}}\rightarrow \infty) \neq 0$. In the case of the Fermionic route~\cite{valles_superconductivity_1989,valles_electron_1992,hsu_magnetic_1995}, $\Delta$ vanishes at the transition and the phase becomes meaningless automatically. Thus, certain aspects of the BCS theory, such as the complete disappearance of the superconducting gap precisely at the transition temperature, remain intact. On the other hand, $G(\bm{r})$ can vanish due to the phase fluctuations in the Bosonic scenario~\cite{sacepe_pseudogap_2010,sacepe_localization_2011,mondal_phase_2011,dubouchet_collective_2019} even if the modulus of the order parameter remains non-zero, which manifests in the form of a superconducting gap or pseudogap above $T_c$. Recent advancements in local spectroscopy~\cite{sacepe_pseudogap_2010,mondal_phase_2011,sacepe_localization_2011,lemarie_universal_2013,kamlapure_emergence_2013,dubouchet_collective_2019} have provided experimental evidence for the disorder-induced granularity of superconductivity and the existence of preformed Cooper pairs above $T_c$.
\subsection{Fermionic to Bosonic transition}
Recent studies indicate that the aforementioned classification may have been overly simplified. In the vicinity of the critical resistance, the interplay between quantum phase fluctuations, localization effects, and disorder-induced spatial variations in electronic properties disrupts the conventional distinction between amplitude- and phase-driven pathways. This disruption gives rise to novel scenarios in which these two scenarios become intertwined~\cite{sacepe_quantum_2020}. In fact, to explain the absence of universal critical resistance as originally predicted in the Bosonic scenario, Yazdani et al.\cite{yazdani_superconducting-insulating_1995} proposed the possibility of a parallel Fermionic conductance channel at finite temperatures, an idea that is further supported by Gantmakher et al.\cite{gantmakher_superconductorinsulator_2000}.
\par
Experimentally, the significant advancements in recent times can be largely attributed to the utilization of low-temperature scanning tunneling spectroscopy~\cite{sacepe_pseudogap_2010,mondal_phase_2011,sacepe_localization_2011,lemarie_universal_2013,kamlapure_emergence_2013,dubouchet_collective_2019}, scanning squid microscopy~\cite{kremen_imaging_2018}, and high-frequency \sloppy electrodynamic response measurements~\cite{liu_dynamical_2011,chand_phase_2012,mondal_enhancement_2013,cheng_anomalous_2016}. Using these techniques, it has been possible to observe the emergent electronic granularity in homogeneously disordered thin films and the presence of a pseudogap for preformed Cooper pairs. These discoveries have led to the unexpected breakdown of the frequently mentioned Fermionic and Bosonic dichotomy and necessitate a new microscopic description of superconductivity under conditions of strong disorder. It appears to be a common feature in many superconductors such as TiN, NbN and MoGe that the same system can follow the Fermionic route at moderate disorder and crossover to a Bosonic scenario at a stronger disorder~\cite{sacepe_disorder-induced_2008,chand_phase_2012,burdastyh_superconducting_2020,lotnyk_suppression_2017}. It is therefore interesting to investigate whether a system can follow the Fermionic route all the way to the disorder level where the superconducting ground state is completely destroyed.
\section{Thermally activated flux creep}
Anderson and Kim~\cite{anderson_theory_1962,anderson_hard_1964} demonstrated that at finite temperatures, flux lines have the ability to jump from one pinning point to another, surpassing the pinning barrier $U$ due to the driving force of the current and flux-density gradient. This phenomenon, known as flux creep, manifests in two ways: firstly, it gives rise to measurable resistive voltages, and secondly, it leads to gradual changes in trapped magnetic fields. In the absence of current, the flux line has an equal probability of moving in either direction relative to the pinning potential. However, in the presence of a transport current or flux-density gradient, the tilted potential grants priority to downhill creep over uphill movement. The tiny resistivity resulting from this thermally activated flux creep ( TAFF ) is expressed as follows,
\begin{equation}
    \rho_{TAFF}=\rho_{ff} e^{-\frac{U}{k_BT}}
\end{equation}
\section{Effect of ac excitation on vortex lattice}
In a clean superconductor, the interaction between the vortices arranges them in a triangular lattice known as the Abrikosov~\cite{abrikosov_vliyanie_1952,abrikosov_magnetic_1957,abrikosov_magnetic_1957-1,levy_vortices_2000}  vortex lattice ( VL ). However, the inevitable presence of crystalline defects in solids acts as a random pinning potential for the vortices. If these vortices are made to oscillate under the influence of an oscillatory current or magnetic field, their motion is governed by the following competing forces~\cite{van_der_beek_linear_1993}: i) Lorentz force due to the external current density driving the motion, ii) restoring force due to the combined effect of pinning by crystalline defects and repulsion from neighboring vortices, and iii) the dissipative viscous drag of the vortices. In addition, at finite temperature thermal activation can cause the vortices to spontaneously jump over the pinning barrier resulting in thermally activated flux flow ( TAFF ), which produces a small resistance even for external current much below the critical current density ( $J_c$ ). 
\subsection{Gittleman-Rosenblum model}
To model the above problem, Gittleman and Rosenblum~\cite{gittleman_radio-frequency_1966} ( GR ) considered the following simplified picture for a single vortex \!( \!neglecting vortex mass term~\cite{blatter_quantum_1991,simanek_fluxon_1991} and TAFF ), where the displacement $\bm{u}$ of the vortex due to small ac excitation follows the following equation of motion ( force per unit length of the vortex ):
\begin{equation}\label{eq:GR_model}
    \eta \bm{\Dot{u}}+\alpha_L \bm{u}=\Phi_0 \bm{J_{ac}}\times \hat{\bm{n}},
\end{equation}
where $\bm{J_{ac}}$ is the external alternating current density on the vortex containing one flux quantum $\Phi_0$ ( $\hat{\bm{n}}$ being the unit vector along the vortex ), $\eta$ is the viscous drag coefficient on the vortex in the absence of pinning and flux creep, and the restoring pinning force constant $\alpha_L$ is called the Labusch~\cite{labusch_calculation_1969,pan_labusch_2000,laiho_labusch_2005} parameter. Even though this model does not explicitly invoke the interaction between vortices, it nevertheless captures the dynamics of a vortex solid where the pinning acts collectively over a typical length scale over which the vortices maintain their positional order, namely the Larkin length. In this case, the resultant pinning parameters have to be interpreted in a mean-field sense~\cite{van_der_beek_linear_1993}, where they incorporate both the effect of interactions and the pinning potential. As we will show in chapter~\ref{ch:vortex-par-chap}, it can also be applied in certain vortex fluid states as long as the thermally activated motion of the vortices is very slow compared to the excitation frequency. $\alpha_L$ determines the shielding response of the superconductor in the vortex state. Assuming harmonic solution ( $\sim e^{i\omega t}$ ) in the equation of motion ( Eq.~\ref{eq:GR_model} ), we get,
\begin{equation}
    \bm{u}=\Phi_0 \frac{\bm{J_{ac}}\times\bm{\hat{n}}}{\alpha_L+i\omega\eta}
\end{equation}
The resultant $\bm{u}$ modifies the London equation~\cite{tinkham_introduction_2004} in the following way~\cite{van_der_beek_linear_1993,josephson_macroscopic_1966}:
\begin{equation}\label{eq:effective_field_london_eqn}
  \begin{split}
    \bm{A}=-\mu_0 \lambda_L^2 \bm{J_{ac}}+\bm{u}\times\bm{B}
   & =-\mu_0 \left(\lambda_L^2+\frac{\Phi_0 B}{\mu_0 (\alpha_L+i\omega \eta )}\right)\bm{J_{ac}}\\
   & =-\mu_0 \lambda_{eff}^2\bm{J_{ac}}.
    \end{split}
\end{equation}
We see that Eq.~\ref{eq:effective_field_london_eqn} has a form similar to the usual London equation, where the London  penetration depth ( $\lambda_L$ ) is replaced by the effective complex penetration depth ( $\lambda_{eff}$ ),
\begin{equation}\label{eq:eff_lambda_GR}
    \lambda_{eff}=\sqrt{\lambda_L^2+\frac{\Phi_0 B}{\mu_0(\alpha_L+i\omega \eta)}}
\end{equation}
Eq.~\ref{eq:eff_lambda_GR} captures the effect of small ac excitation on vortices in addition to the London penetration depth, where $\lambda_{eff}$ is the effective ac screening length. $\lambda_{eff}$ can be written in the following way,
\begin{equation}\label{eq:eff_lambda_campbell}
    \lambda_{eff}^2=\lambda_L^2+\lambda^2_{vortex}=\lambda_L^2+\lambda_C^2(1+i\omega \tau_0)^{-1}
\end{equation}
Here, $\tau_0=\eta/\alpha_L$ is the vortex relaxation time without any vortex creep and $\lambda_C=\sqrt{B\Phi_0/\mu_0\alpha_L}$ is the Campbell penetration depth which determines how much the magnetic field penetrates the superconductors when you apply small ac excitation in the presence of vortices. We shall discuss about $\lambda_C$ in detail in section~\ref{sec:campbell-penetration-depth}.
\par
Now, looking at Eq.~\ref{eq:eff_lambda_campbell} we observe two frequency regimes: The imaginary or the dissipative part becomes significant at high enough frequency ( flux-flow regime, $\omega\tau_0>1$ ); while as we decrease frequency ( Campbell regime, $\omega\tau_0>1$ ), real part dominates and vortex contribution is given by $\lambda_C$.
\par
$\lambda_{eff}^2$ in Eq.~\ref{eq:eff_lambda_campbell} can also be rewritten in terms of vortex resistivity, $\rho_v$, as:
\begin{equation}\label{eq:eff_lambda_rho_v}
    \lambda_{eff}^2=\lambda_L^2 - i\frac{\rho_v(\omega)}{\mu_0 \omega},
\end{equation}
where $\tau_0=\eta/\alpha_L$ is the vortex relaxation time and $\rho_v$ is the complex vortex resistivity expressed as:
\begin{equation}\label{eq:rho_v_GR}
    \rho_v(\omega)=\rho_{ff} \frac{i\omega \tau_0}{1+ i\omega \tau_0},
\end{equation}
where $\rho_{ff}$ is the dc flux flow resistivity ( $\rho_{ff}$ ) given by the relation~\cite{bardeen_theory_1965,kim_flux-flow_1965}:
\begin{equation}\label{eq:rho_ff_Bardeen-Stephen}
    \rho_{ff}=\frac{B\phi_0}{\eta}=\frac{B}{B_{c2}}\rho_n,
\end{equation}
$\rho_n$ being the normal state resistivity. We shall see in the later sections, how Eq.~\ref{eq:rho_v_GR} modifies in the presence of TAFF. 
\subsection{Campbell penetration depth}\label{sec:campbell-penetration-depth}
Campbell~\cite{campbell_response_1969,campbell_interaction_1971} penetration depth is not a penetration depth in the true sense. It is rather an ac screening response of the superconductor to small amplitude ac excitation in the presence of vortices. When an external ac field is applied, it at first interacts with the vortices on the surface generating an oscillation. This oscillation then propagates towards the centre of the sample through the vortices resulting in a change in screening response. This vortex contribution to the screening response is called Campbell penetration depth, named after Campbell, who first tried to estimate the contribution via a simple calculation which we shall briefly discuss as follows.
\par
Let us consider square vortex lattice for the sake of simplicity in our calculation~\cite{kim_campbell_2021}, where the positions of a certain vortex are given by $x$ and $u$ is the displacement of that vortex due to external ac excitation. In the absence of excitation, the distance between consecutive vortices is given the vortex lattice constant, $a_0=\sqrt{\Phi_0/B_0}$, where $B_0$ is the unperturbed magnetic induction. In the presence of excitation, the inter-vortex distance becomes,
\begin{equation}
\begin{split}
    a(x) & =a_0+u(x+a_0)-u(x)\\
    & =a_0 \left[1+\frac{u(x+a_0)-u(x)}{a_0} \right]\\
    & \approx a_0\left(1+\frac{du}{dx}\right).
\end{split}
\end{equation}
Hence the changed magnetic induction is given by ( assuming one flux quantum per vortex ):
\begin{equation}
    B(x) =\frac{\Phi_0}{a_0 a(x)}=\frac{\Phi_0}{a_0^2(1+\frac{du}{dx})}\approx B_0(1-\frac{du}{dx}).
\end{equation}
Invoking the Maxwell equation, $\mu_0 \bm{J}=\bm{\nabla}\times\bm{B}$, with $\bm{B}=B(x)\bm{\hat{z}}$, we get,
\begin{equation}
    \mu_0 J_y=-\frac{\partial B(x)}{\partial x}=B_0 \frac{d^2u}{dx^2}
\end{equation}
Hence, Lorentz force per unit volume on vortices is given by,
\begin{equation}\label{eq:lorentz-force}
    F_L=JB_0=\frac{B_0^2}{\mu_0}\frac{d^2u}{dx^2}
\end{equation}
Again, the Lorentz force is balanced by the pinning force, which can be written in terms of the Labusch parameter $\alpha_L$. In single vortex approximation, pinning force per unit volume is given by $f_p=-\alpha_L u$ and the areal density of vortices is $n_A=B_0/\Phi_0$. Here, pinning force per unit volume is given by,
\begin{equation}\label{eq:pinning-force}
    F_p=n_Af_p=-\frac{\alpha_L B_0}{\Phi_0}u
\end{equation}
At the steady state,
\begin{equation}
    F_L+F_p=0.
\end{equation}
Now, putting the expressions from Eqs. \ref{eq:lorentz-force} and \ref{eq:pinning-force}, we get,
\begin{equation}
    \frac{B_0 \Phi_0}{\mu_0 \alpha_L}\frac{d^2u}{dx^2}\equiv \lambda_C^2 \frac{d^2u}{dx^2}=u.
\end{equation}
This $\lambda_C=\sqrt{B_0 \Phi_0/\mu_0 \alpha_L}$ is the Campbell penetration depth we discussed in Eq.~\ref{eq:eff_lambda_campbell}.
\subsection{Effect of thermally activated flux flow}
So far, we have not considered the thermal motion of the vortices in GR model which is applicable at higher frequencies. However, at extremely low frequency ( $\omega\tau_0 \ll 1$ ), the thermally activated flux jumps among different metastable states of VL become important which could not be captured in GR model. As it is difficult to map the distribution of random pinning potential barriers ( due to non-uniformity between different metastable state minima ), a typical single activation energy value~\cite{van_der_beek_linear_1993,revenaz_frequency_1994} U ( dependent on both temperature and field ) is usually assumed for the sample under study, in the same spirit that the single $\alpha_L$ value was considered before. It is the factor ($U/k_B T$) that determines the strength of thermally activated motion in the improved models independently proposed by Brandt~\cite{brandt_penetration_1991} and Coffey-Clem~\cite{coffey_unified_1991}. Physically, TAFF relaxes the restoring pinning force over large time scales and is therefore important when measurements are performed at very low frequencies.
\subsubsection{Brandt model}
Brandt~\cite{brandt_penetration_1991} introduced the thermal creep by the thermal relaxation of Labusch parameter, $\alpha_L(t)$. He argued that after a sudden step-like displacement of the vortex lattice, the pinning restoring force should decay exponentially in time ( $t$ ):
\begin{equation}\label{eq:alphaL-Brandt-decay}
    \alpha_L(t)=\alpha_L e^{-t/\tau},
\end{equation}
 where modified vortex relaxation time $\tau \gg \tau_0$, $\tau_0$ being the vortex relaxation time without the thermal creep ( Eq.~\ref{eq:eff_lambda_campbell} ). Brandt suggested the form of $\tau$ as,
\begin{equation}
    \tau=\tau_0 e^{U/k_BT}
\end{equation}
In Fourier space, Eq.~\ref{eq:alphaL-Brandt-decay} becomes~\cite{brandt_penetration_1991,brandt_dynamics_1991,brandt_elastic_1992,brandt_linear_1992,brandt_thin_1994},
\begin{equation}\label{eq:Brandt-modified-alphaL}
    \alpha_L(\omega)=\frac{\alpha_L}{1-i/\omega \tau}.
\end{equation}
Brandt also gave an alternative derivation~\cite{helmut_brandt_flux_1992,brandt_flux-line_1995}\hspace{0.2cm} of the modified expression of $\alpha_L$ ( Eq. \ref{eq:Brandt-modified-alphaL} ) by taking into account the contribution from both elastic velocity due to $\alpha_L$ and plastic velocity due to $\eta$. Effective velocity is given by,
\begin{equation}\label{eq:Brandt-v-eff-alternative}
    v=v_{el}+v_{pl}=\frac{\dot{F}_{eff}}{\alpha_L}+\frac{F_{eff}}{\eta}=\left(\frac{i\omega}{\alpha_L}+\frac{1}{\eta}\right)F_{eff},
\end{equation}
where $F_{eff}=F_{ext}-i\omega \eta$ is the effective driving force density. We have used the simplified relations for a sinusoidal external force of frequency $\omega$: $\dot{F}_{eff}=i\omega F_{eff}$, $\dot{v}=i\omega v$. Now, putting the expressions in Eq.~\ref{eq:Brandt-v-eff-alternative}, we get:
\begin{equation}\label{eq:Brandt-alpha-alternative}
    F_{eff}=F_{ext}-i\omega \eta=\frac{\alpha_L}{1-i/\omega \tau} u, 
\end{equation}
where $\tau=\eta/\alpha$ and $v=i\omega u$. Here, we have 
got the same expression of $\alpha_L$ as in Eq.~\ref{eq:Brandt-modified-alphaL}.

\subsubsection{Coffey-Clem model}\label{sec:Coffey-Clem model-chapter}

Coffey and Clem~\cite{coffey_unified_1991,coffey_theory_1992,coffey_theory_1992-1,coffey_theory_1992-2,coffey_coupled_1992,coffey_magnetic_1991,clem_vortex_1992} addressed the problem of TAFF in a more elaborate manner. To account for the thermally activated flux motion over the pinning barriers, a random Langevin force which depends on the pinning barrier height U is added to the right-hand side of Eq.~\ref{eq:GR_model}. The solution to the above problem is similar to that of a particle undergoing Brownian motion in a periodic potential.
\begin{figure*}[hbt]
	\centering
	\includegraphics[width=16cm]{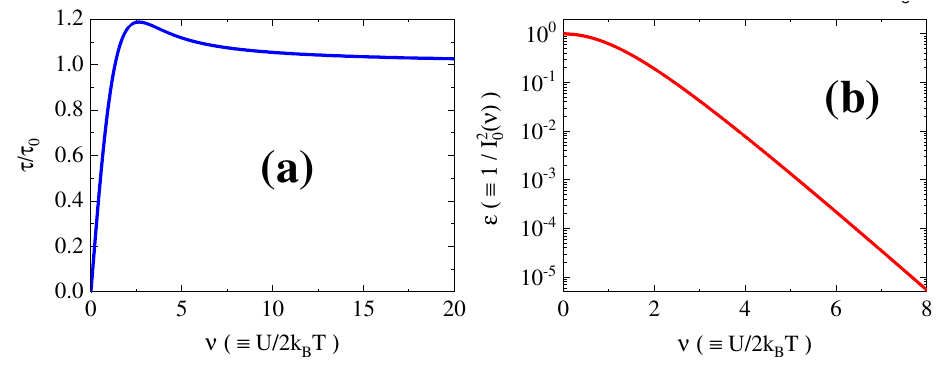}
	\caption[Variation of normalized vortex relaxation time and flux-creep factor as a function of normalized potential barrier]{
    Variation of \textbf{(a)} the normalized vortex relaxation time ($\tau/\tau_0=(I_0^2(\nu)-1)/(I_1(\nu)I_0(\nu))$) and \textbf{(b)} the flux-creep factor ($\epsilon=1/I_0^2(\nu)$) as a function of the normalized potential barrier ($\nu=U/2k_BT$) according to the Coffey-Clem model~\cite{coffey_unified_1991}.}
 \label{fig:tau-epsilon-vs-nu} 
\end{figure*}
In their model, the expression of vortex resistivity ( $\rho_v^{TAFF}$ ) gets modified to the following form~\cite{pompeo_reliable_2008}:
\begin{equation}\label{eq:rho_v_coffey_clem}
    \rho_v^{TAFF}(\omega)=\rho_{ff} \frac{\epsilon+ i\omega \tau}{1+ i\omega \tau}
\end{equation}
where $\tau$ is the vortex relaxation time in the presence of TAFF given by:
\begin{equation}
    \tau=\tau_0  \frac{I_0^2(\nu)-1}{I_1(\nu) I_0(\nu)},
\end{equation}
 and $\epsilon$ is called the flux creep factor ($0<\epsilon<1$) defined by,
 \begin{equation}
     \epsilon=\frac{1}{I_0^2(\nu)}.
 \end{equation}
 Here, $I_p$ is the modified Bessel’s function of the first kind of order $p$, and $\nu=U/2k_B T$; $U$ being the pinning barrier relevant for the thermally activated vortex motion. The variations of $\tau/\tau_0$ and $\epsilon$ with $\nu\equiv U/2k_BT$ are plotted in Fig.~\ref{fig:tau-epsilon-vs-nu}.
 \par
 If one looks at Eq.~\ref{eq:rho_v_coffey_clem} in zero-frequency limit,
 \begin{equation}
      \frac{\rho_v^{TAFF}(\omega \rightarrow 0)}{\rho_{ff}}\approx \epsilon,
 \end{equation}
which implies that the flux-creep factor $\epsilon$ quantifies the extent to which thermal activation facilitates the approach of the electric field, generated by vortex motion, towards the value it would have reached in the absence of pinning. 
 \par
 Hence the Eq.~\ref{eq:eff_lambda_rho_v} is modified to the following form:
 \begin{equation}\label{eq:eff_lambda_Coffey_Clem_simplified}
     \lambda_{eff}^2=\lambda_L^2 - i\frac{\rho_{ff}}{\mu_0 \omega} \frac{\epsilon+ i\omega \tau}{1+ i\omega \tau}.
 \end{equation}
 At $T\rightarrow 0$, $\epsilon\approx 0$ and $\tau\approx \tau_0$, i.e., the effect of thermal creep becomes negligible. Hence Eq.~\ref{eq:eff_lambda_Coffey_Clem_simplified} reduces to Eq.~\ref{eq:eff_lambda_GR}. At high temperatures where $\epsilon\approx 1$, vortices behave as if there are no pinning centers.
\par
Furthermore, the normal fluid skin depth, $\delta_{nf}$ introduces an additional correction such that Eq.~\ref{eq:eff_lambda_GR} is modified as follows:
\begin{equation}\label{eq:eff_lambda_Coffey_Clem}
        \lambda_{eff,CC}^2=\left(\lambda_L^2 - i\frac{\rho_v^{TAFF}}{\mu_0 \omega}\right)\bigg/\left(1+2i \frac{\lambda_L^2}{\delta_{nf}^2}\right),
\end{equation}
where $\delta_{nf}$ follows the phenomenological variation~\cite{coffey_unified_1991} of the form:
\begin{equation}
    \delta_{nf}^2 (t,h)=\frac{(\frac{2\rho_n}{\mu_0 \omega})}{1-f(t,h)}
\end{equation}
which is complementary to the $\lambda_L$ variation considering they come from normal and superconducting electrons respectively in the framework of the two-fluid model:
\begin{equation}
    \lambda_L^2 (t,h)=\frac{\lambda_L^2}{f(t,h)},
\end{equation}
 where $f(t,h)=(1-t^4)(1-h)$ with $t=T/T_c$ and $h=H/H_{c2}$; $T_c$ and $H_{c2}$ being the zero-field superconducting transition temperature and upper critical field respectively.
\par
Consequently, the vortex state is characterized by the minimal set of three parameters: $\alpha_L, \eta$, and $U$.
We shall use this Coffey-Clem equation to fit the experimental in-field penetration depth data in Chapter~\ref{ch:vortex-par-chap}.

\subsection{Universal expression for vortex resistivity}\label{sec:universal_expression-rho}
Pompeo and Silva~\cite{pompeo_reliable_2008} came up with a universal expression for the vortex resistivity $\rho_v$:
\begin{equation}\label{eq:rho_v_universal}
    \rho_v(\omega)=\rho_{ff} \frac{\epsilon_{eff}+ i\omega \tau_{eff}}{1+ i\omega \tau_{eff}}.
\end{equation}
All the models discussed above can be written in terms of Eq.~\ref{eq:rho_v_universal}, where the parameters for various models are written in Table~\ref{table:epsilon-tau-diff-model}.
\begin{table}[hbt]
\caption{\centering The expressions of flux-creep factor $\epsilon_{eff}$ and vortex relaxation time $\tau_{eff}$ for different models.}\label{table:epsilon-tau-diff-model}
\centering
    \begin{tblr}{|c|c|c|c|}
    \hline
        Model & $\displaystyle \epsilon_{eff}$ & $\displaystyle \tau_{eff}$ & Defined parameters\\
        \hline
        Gittleman and Rosenblum~\cite{gittleman_radio-frequency_1966} & $0$ & $\displaystyle\tau_0$ & $\displaystyle\tau_0=\frac{\eta}{\alpha_L}$ \\
        \hline
                Brandt~\cite{brandt_penetration_1991} & $ \displaystyle\frac{\tau_0}{\tau_0 + \tau_B}$ & $\displaystyle \tau_0  \frac{\tau_B}{\tau_0 + \tau_B}$ & $\displaystyle\tau_B=\tau_0 e^{U/k_B T}$ \\
                        \hline
                Coffey and Clem~\cite{coffey_unified_1991} & $\displaystyle\frac{1}{I_0^2(\nu)}$ & $\displaystyle\tau_0\frac{I_0^2(\nu)-1}{I_0(\nu) I_1(\nu)}$ & $\displaystyle\nu=\frac{U}{2k_BT}$ \\
        \hline
    \end{tblr}
\end{table}

	\chapter{Experimental methods}\label{ch:exp-methods}
\section{Sample preparation}
The majority of our samples comprise amorphous Molybdenum Germanium ( \textit{a}-MoGe ) thin films of different thicknesses, grown by pulsed laser deposition ( PLD ) technique. Additionally, a few epitaxial Niobium Nitride ( NbN ) thin films were deposited using the reactive DC magnetron sputtering method. Thermal evaporation was used to deposit silver ( Ag ) on the sample as a contact pad for Corbino geometry in the broadband microwave setup.
\subsection{Pulsed laser deposition}
MoGe thin films were deposited using the pulsed laser deposition~\cite{chrisey_pulsed_1994,lowndes_synthesis_1996} ( PLD ) technique. During the PLD process, an arc-melted Mo$_{70}$Ge$_{30}$ target was ablated by a KrF $248$ $nm$ excimer laser ( \href{https://www.coherent.com/lasers/excimer/compex}{COHERENT COMPex 205} ) while keeping the substrate at room temperature. Both \href{https://www.universitywafer.com/thermal-oxide-deposition-silicon-wafer.html}{surface-oxidized Si} substrates ( with an oxide layer thickness of approximately $200$ $nm$ ) and ($100$) oriented MgO substrates were used based on the experimental requirements. The deposition was performed at a base pressure of $\sim 10^{-7}$ $Torr$, achieved through the use of high vacuum rotary and turbo pumps. For metallic thin films, a relatively high energy density of the incident laser is necessary to maintain the stoichiometry of the amorphous film close to that of the target, compared to oxide thin films. The laser was focused onto a small spot on the target, and laser pulses were bombarded at a repetition rate of $10$ $Hz$, with an effective energy density of around $240$ $mJ/mm^2$ per pulse. The growth rate was $\sim 1$ $nm$ per $100$ pulses. Films with different thicknesses were obtained by varying the number of laser shots.
\begin{figure*}[hbt]
	\centering
	\includegraphics[width=16cm]{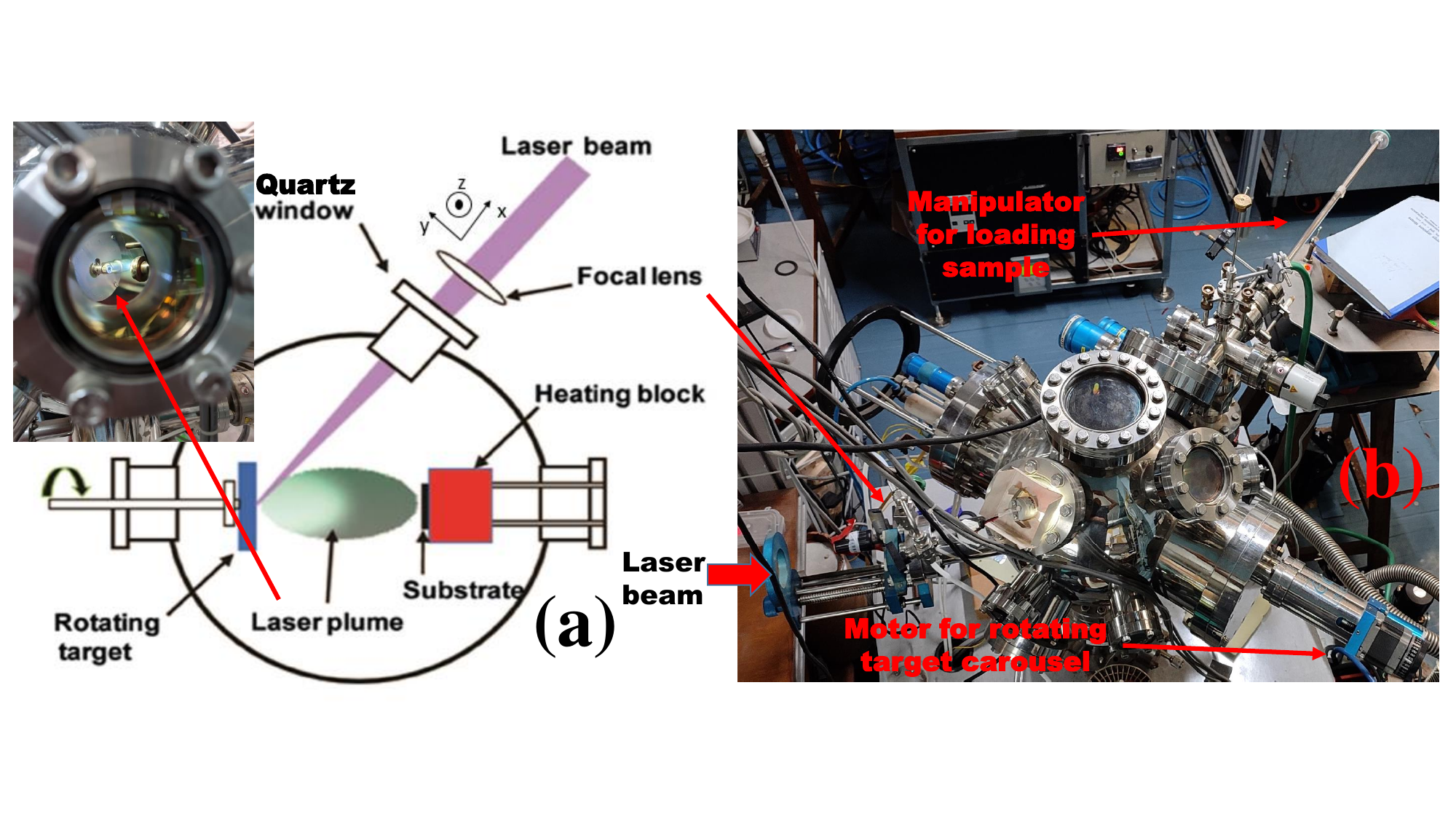}
	\caption[Pulsed laser deposition setup]{
		\label{fig:pld-schematic-pic} 
	\textbf{(a)} Schematic of the pulsed laser deposition (PLD) technique (adapted from Ref.~\citenum{kim_laser_2014}). (inset) A bright plume emanating from the MoGe target ablated by laser pulses and getting deposited on the sample holder placed in front of it (Photo courtesy: Arghya Dutta). \textbf{(b)} Top view of the PLD setup (Photo courtesy: Dr. Subhamita Sengupta).
	}
\end{figure*}
\subsubsection{Optimizing the growth parameters}
The thickness of the thin film is determined by several parameters: (i) the number of laser shots, (ii) laser energy, and (iii) the size of the focused laser spot on the target. The size and position of the laser spot can be controlled by adjusting the position of a converging lens in the laser path. In order to achieve better target ablation, our approach focused on using a smaller laser spot, which corresponds to higher energy density, rather than a larger spot size, as typically used in standard PLD practice. To determine the
spot size, we scanned the size of the spot on thermal paper while changing the position of the lens. Usually, laser ablation is performed by focusing the demagnified image of the aperture onto the target. However, in our setup, we focused on the focal point of the lens to concentrate higher laser intensity and achieve greater ablation on the metal-like surface of the MoGe target. This approach is not commonly used because it can produce side lobes, leading to inhomogeneity in the stoichiometry of the deposited film. Nevertheless, our observations showed that the laser intensity at the side lobe was not sufficient to cause ablation on the MoGe surface. Consequently, we optimized the position of the lens to achieve a smaller spot size.
\begin{figure*}[hbt]
	\centering
	\includegraphics[width=16cm]{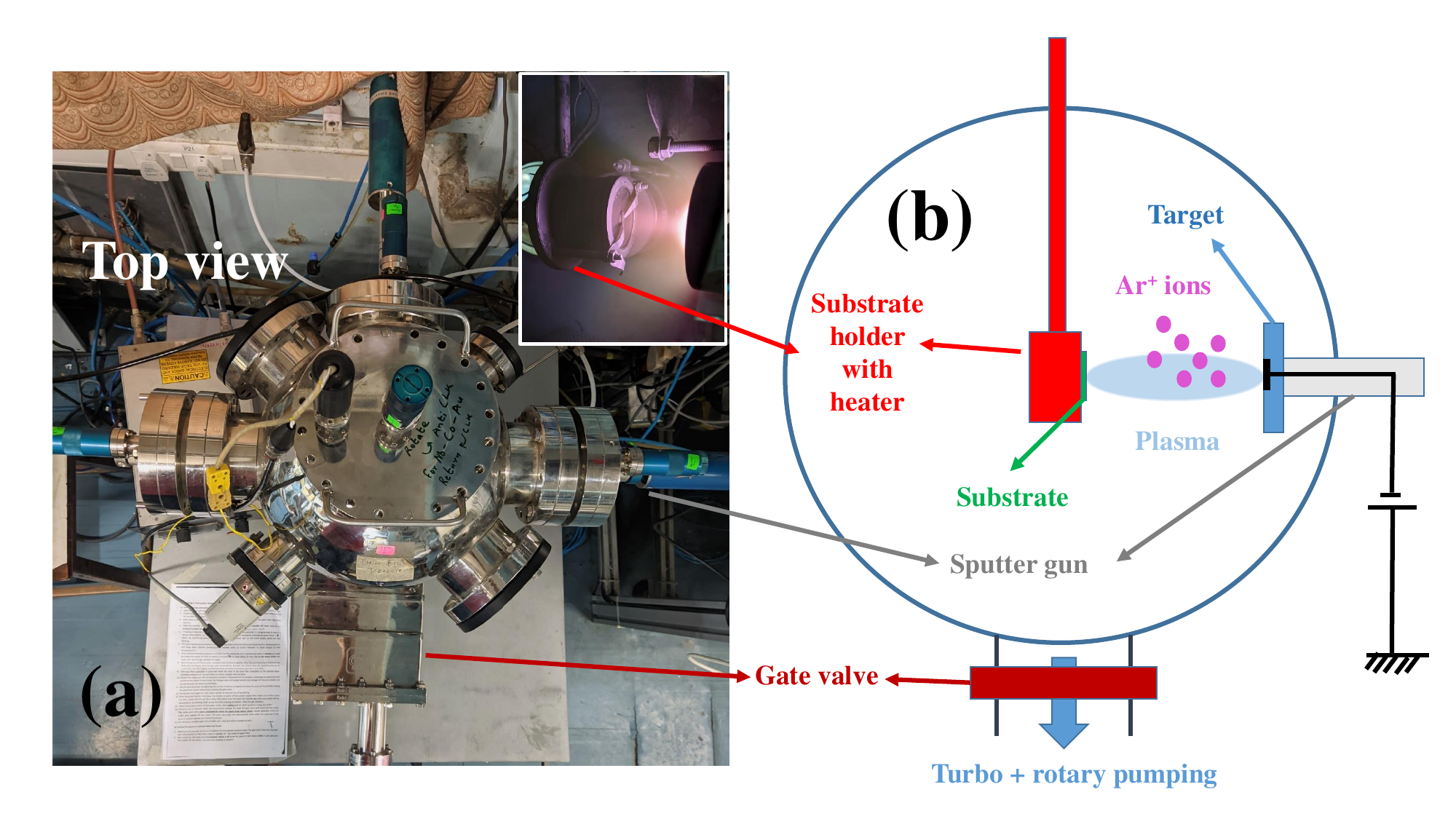}
	\caption[Sputtering setup for thin film deposition]{
		\label{fig:sputtering-schematic-pic} 
	\textbf{(a)} Top view of the Sputtering chamber. (inset) Plasma plume ejecting from the Nb target and getting deposited on the sample holder placed in front of it. (Photo courtesy: Arghya Dutta/ John Jesudasan). \textbf{(b)} Schematic of the sputtering setup. 
	}
\end{figure*}
\subsection{Sputtering}
The NbN thin films were grown using reactive DC magnetron sputtering. The sputtering system, manufactured by \href{https://excelinstruments.biz/}{Excel Instruments}, featured four ports for loading the sputtering gun or viewing inside the chamber. Additionally, it had a top port for mounting the substrate and smaller ports for the sputtering gas inlet, venting, pressure gauge, and other purposes.
\par
To achieve epitaxial NbN films, a high-purity Nb metal ( $99.999\%$ ) from \href{https://www.lesker.com/index.cfm}{Kurt-Lesker} was mounted on the sputtering gun, while single crystal MgO substrates with a lattice constant of $4.212$ \AA \, were utilized. The MgO substrates were cleaned using \href{https://www.cdc.gov/niosh/topics/trichloroethylene/default.html#:~:text=Trichloroethylene%20(CICH%3DCCl2),liver%20damage%2C%20and%20even%20death.}{trichloroethylene} ( TCE ) and securely attached to the sample holder using high-temperature 
\href{https://www.2spi.com/category/silver-paints-pastes/}{silver paste}. Sample patterning was performed using shadow masking.
\par
The chamber was evacuated using a \href{https://www.pfeiffer-vacuum.com/en/products/vacuum-generation/turbopumps/}{Pfeiffer} turbo pump, and the substrate temperature was elevated to $600$ $\degree C$ using an embedded nichrome wire heater. Two mass flow controllers were employed to precisely control the partial pressures of Ar and N\textsubscript{2} gases while maintaining a total pressure of $5\times10^{-3}$ $Torr$ during all depositions. Following the gas introduction, a negative high voltage was applied using an \href{https://www.aplab.com/}{Aplab} high-voltage power supply to initiate plasma discharge. The sputtering power was regulated by adjusting the current passing through the plasma. Pre-sputtering was carried out for approximately $3$ minutes prior to the actual deposition.
\par
The properties of NbN thin films strongly rely on deposition conditions such as sputtering power, Ar/N\textsubscript{2} gas ratio, and deposition time, all of which were systematically varied in this study to obtain the desired film properties.
\begin{figure*}[hbt]
	\centering
	\includegraphics[width=16cm]{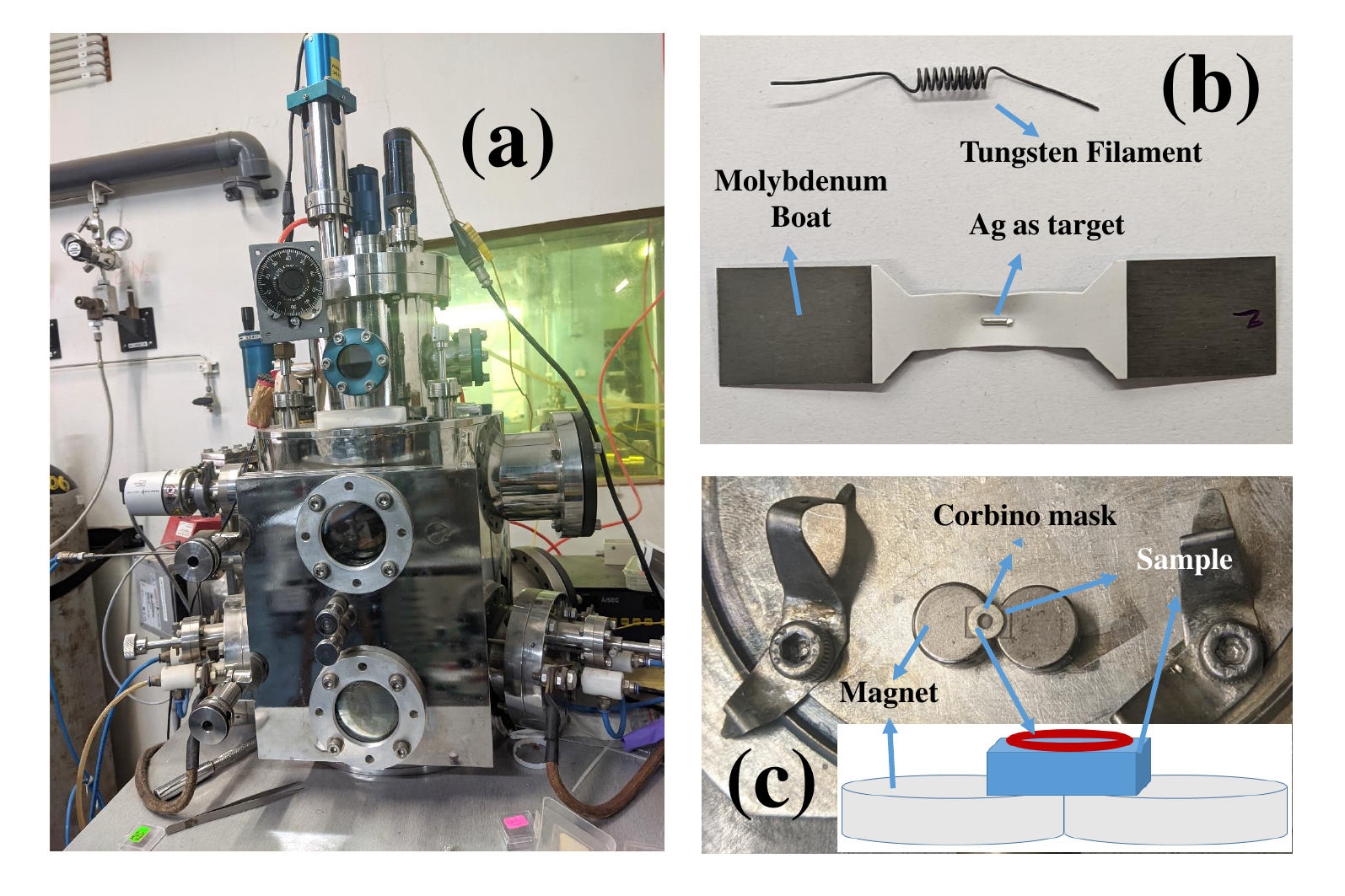}
	\caption[Thermal evaporation setup]{
		\label{fig:evaporation-schematic-pic} 
	\textbf{(a)} Thermal evaporation chamber. \textbf{(b)} Tungsten filament and Molybdenum boat used for thermally evaporating targets such as Silver (Ag). \textbf{(c)} Magnet assembly for the deposition of Corbino-shaped contact pad. (inset) Schematic of the side view of the assembly. (Photo courtesy: Arghya Dutta).
	}
\end{figure*}
\subsection{Thermal evaporation}
Thermal evaporation is a popular technique for depositing thin films onto solid substrates. It involves heating a material in a vacuum chamber until it vaporizes and then allowing the vaporized atoms or molecules to condense onto a substrate, forming a thin film. The material to be evaporated, known as the target material, is heated using resistive heating, and the substrate is maintained at a lower temperature to facilitate condensation and film growth.
\par
In the evaporation chamber, the target is usually kept on a Molybdenum boat or inside a Tungsten filament ( as shown in Fig.~\ref{fig:evaporation-schematic-pic}(b) ) while the sample is placed just above it. For patterning the contact pad in the Corbino form, the sample is kept on top of an assembly of two button magnets such that the annular mask remains attached while the deposition is going on ( Fig.~\ref{fig:evaporation-schematic-pic}(c) ). 
\par
For the deposition of silver ( Ag ), a small piece of Ag wire, with a diameter of $1$ $mm$, is placed inside the filament. Prior to the deposition process, the chamber is subjected to a low-pressure vacuum of approximately $10^{-5}$ $mbar$. During the deposition, a high current of 5 A is applied to the filament, heating it to a red-hot state, causing the Ag to evaporate. The evaporated Ag then gets deposited onto the substrate, which is maintained at a temperature of $150$ $\degree C$.
\begin{figure*}[hbt]
	\centering
	\includegraphics[width=16cm]{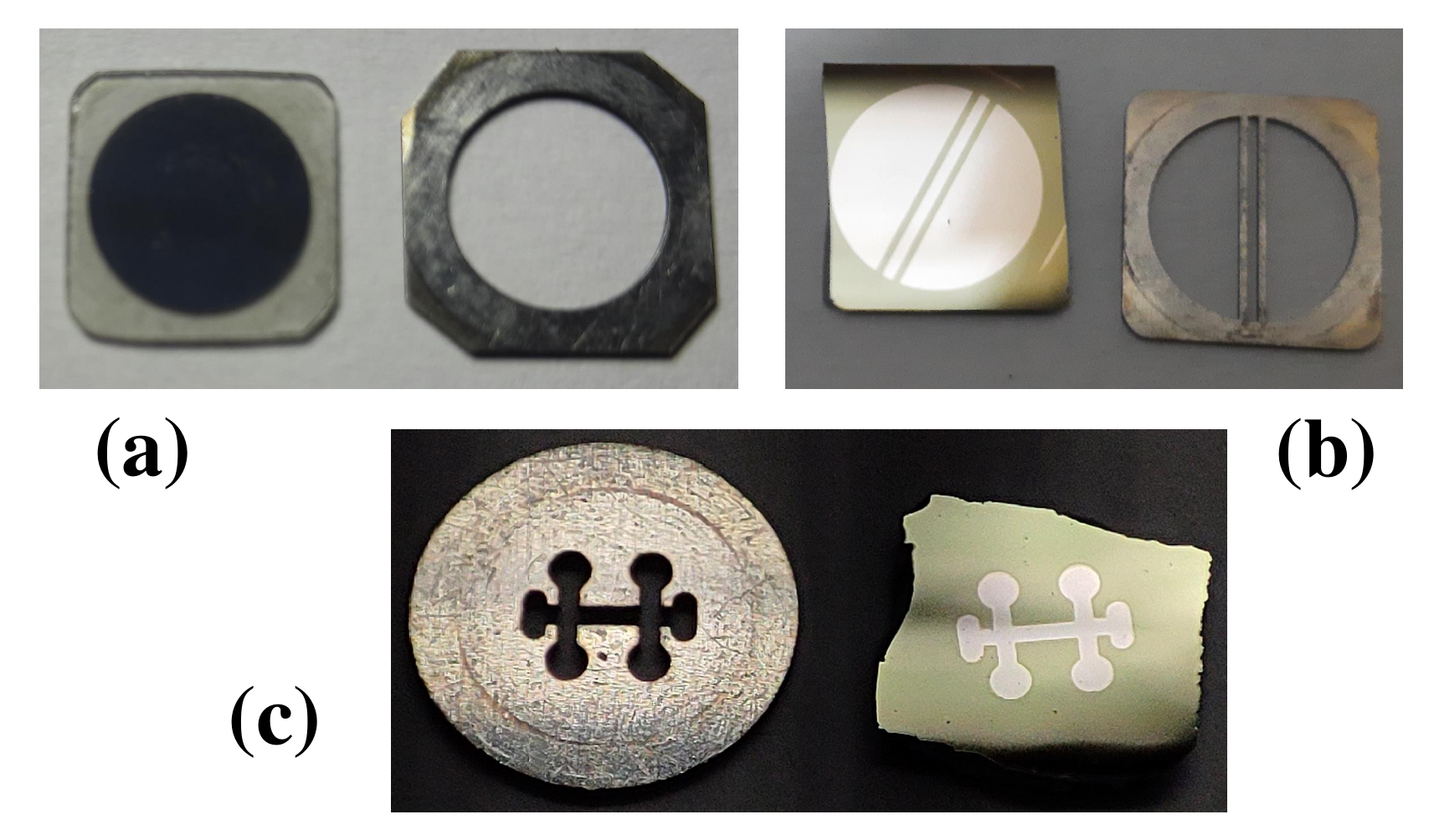}
	\caption[Different masks used for thin film deposition]{
		\label{fig:diff-mask-pic} 
	Deposited thin films and different stainless steel masks used for thin film deposition: \textbf{(a)} $8$-$mm$-diameter \textit{a}-MoGe thin film deposited on MgO substrate with the stainless steel mask, \textbf{(b)} $8$-$mm$-diameter \textit{a}-MoGe thin film grown in surface-oxidized Si substrate with strip lines for measuring the thickness with the mask, and \textbf{(c)} deposited hall-bar-patterned \textit{a}-MoGe thin film with the mask. (Photo courtesy: Dr. Subhamita Sengupta).
	}
\end{figure*}
\subsection{Masks used in deposition}
For two-coil measurements, the sample was deposited in the form of an $8$-$mm$-diameter circle using a shadow stainless steel mask on $10\times10$ $mm^2$ substrate as illustrated in Fig.~\ref{fig:diff-mask-pic}(a), while Fig.~\ref{fig:diff-mask-pic}(b) shows the mask with two additional strip lines in the middle used for measuring thickness on the $8$-$mm$-diameter sample. In the case of transport measurements, hall-bar-shaped geometry was patterned on samples for better sensitivity ( Fig.~\ref{fig:diff-mask-pic}(c) ). For both two-coil and transport measurements, samples were capped with a $2$-$nm$-thick Si layer to prevent surface oxidation.
\par
For the broadband microwave measurements, thin films grown on the substrate without any mask, were placed in the evaporator chamber for silver ( Ag ) deposition using the annular mask to form the Corbino geometry ( Fig.~\ref{fig:evaporation-schematic-pic}(c) ).
\section{Thickness measurements}
The thickness of all films was measured using an Ambios XP2 Stylus profilometer. Here, the metal tip scratches through the sample to find any change in the topography. Thickness measurements were carried out at various positions of the sample and the mean value was taken as the film thickness ( $d$ ). For $d\geq10$ $nm$, the thickness of the film was directly measured using the stylus profilometer whereas for thinner samples it was estimated from the number of laser pulses using two films with $d\geq10$ $nm$ grown before and after the actual run for calibration.
\section{Low-temperature measurements}
\begin{figure*}[hbt]
	\centering
	\includegraphics[width=16cm]{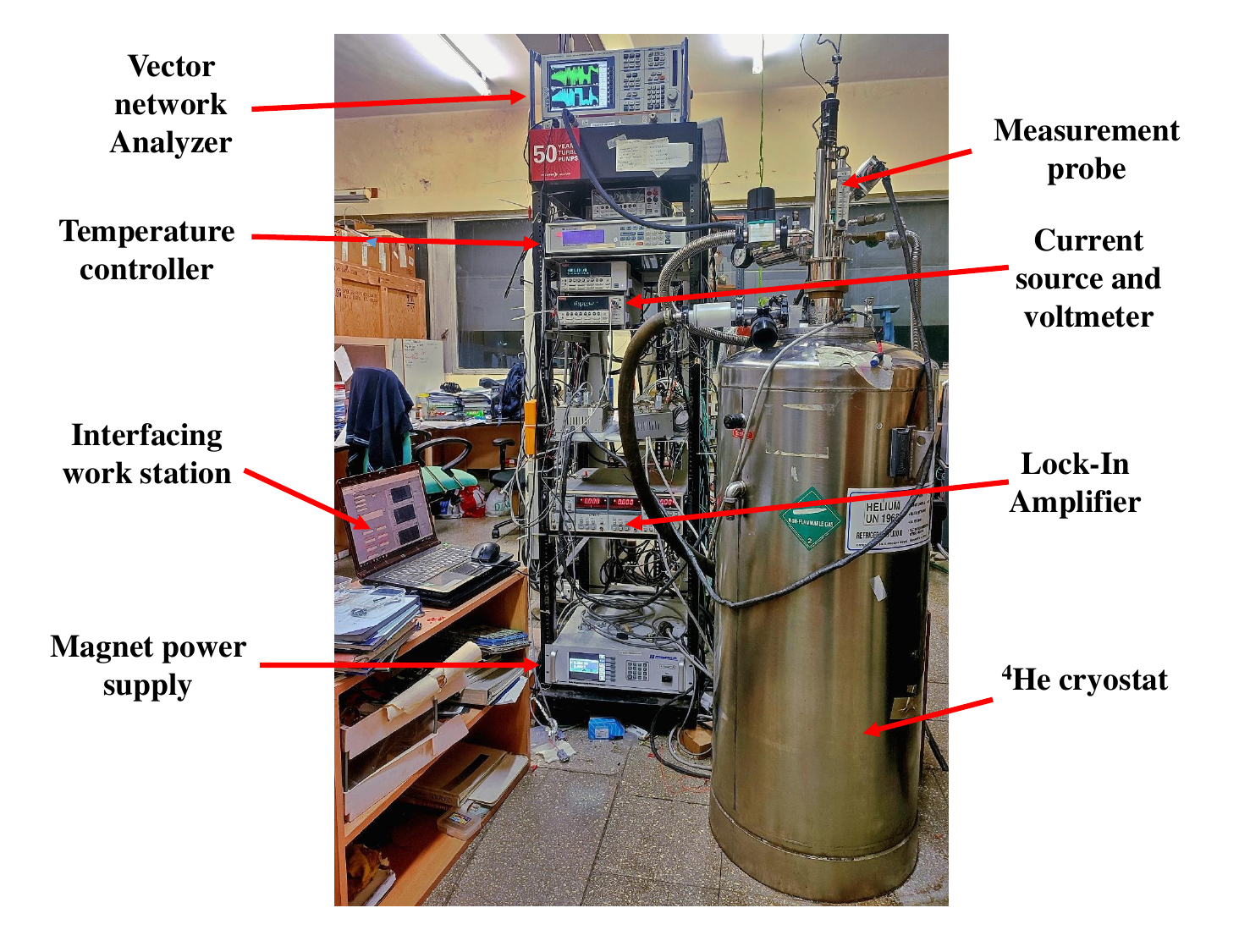}
	\caption[Instrument assembly for low-temperature measurements]{
		\label{fig:measurement-assembly} 
	The Instrument assembly for low-temperature measurements. (Photo courtesy: Dr. Subhamita Sengupta).
	}
\end{figure*}

All our measurements were conducted in our lab using various cryogenic setups to achieve extremely low temperatures. Fig.~\ref{fig:measurement-assembly} shows the measurement assembly comprising the \textsuperscript{4}He cryostat and the measuring instruments. Below, we provide brief descriptions of the cryogenic setups extensively utilized for our experiments.
\begin{figure*}[hbt]
	\centering
	\includegraphics[width=16cm]{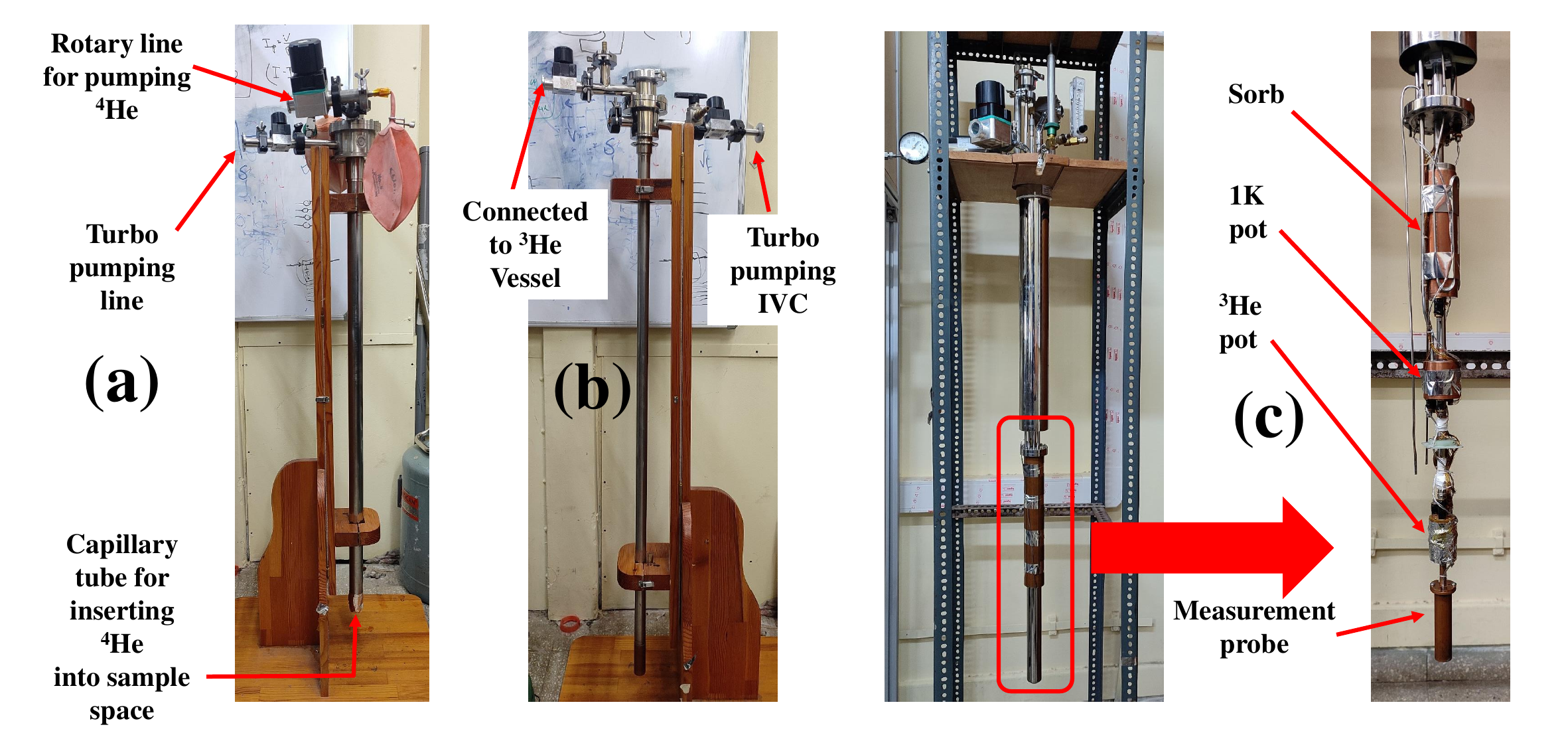}
	\caption[The \textsuperscript{4}He and \textsuperscript{3}He VTI]{
		\label{fig:4He-3He-VTI} 
	\textbf{(a)} \textsuperscript{4}He VTI. \textbf{(c)} Homemade \textsuperscript{3}He VTI design to sit inside the existing \textsuperscript{4}He VTI  \textbf{(c)} Janis \textsuperscript{3}He VTI. (Photo courtesy: Dr. Subhamita Sengupta).
	}
\end{figure*}
\subsection{\texorpdfstring{\textsuperscript{4}}{}He VTI}\label{sec:4He-VTI}
All the in-field two-coil mutual inductance ( Chapter~\ref{ch:vortex-par-chap} ) and broadband microwave measurements above $\sim 1.7$ $K$ were performed using a \textsuperscript{4}He cryostat equipped with a NbTi-Nb\textsubscript{3}Sn superconducting magnet capable of reaching a maximum field of $11$ $T$ ( \href{https://cryomagnetics.com/}{Cryomagnetics Inc.} ). To control the temperature, a custom-made variable temperature insert ( VTI ) built by \href{https://excelinstruments.biz/}{Excel Instruments} was employed. The VTI consists of two concentric thin-walled stainless steel tubes separated by an evacuated annular space. The outer tube is immersed in liquid helium, while the inner tube accommodates the sample insert. Liquid helium is introduced into the sample space through a capillary from the helium bucket. The flow of \textsuperscript{4}He into the sample space is controlled by adjusting the temperature of the capillary with the help of a heater attached to it.
\subsection{Janis \texorpdfstring{\textsuperscript{3}}{}He VTI}\label{sec:janis-3He-VTI}
To conduct measurements at temperatures below $2$ $K$, we utilized a custom-made \textsuperscript{3}He insert provided by \href{https://www.janis.com/}{Janis Research Co}. This insert was specifically designed to fit into our $11$ $T$ NbTi-Nb\textsubscript{3}Sn superconducting magnet cryostat. Zero-field two-coil mutual inductance measurements in Chapter~\ref{ch:phase_fluctuations} were conducted using the \textsuperscript{3}He VTI.
\par
At the start of the cooling procedure, the sorption pump temperature is set to approximately $45$ $K$, and the $1K$ pot is cooled down to around 1.6 K by continuously pumping over liquid \textsuperscript{4}He, through a capillary immersed in the \textsuperscript{4}He liquid bath. The \textsuperscript{3}He, in thermal contact with the 1K pot, is allowed to condense inside the pot. Subsequently, the heater for the sorption pump is deactivated, and it cools down naturally due to the flow of liquid \textsuperscript{4}He in narrow tubes surrounding it. As the sorption pump cools, it adsorbs \textsuperscript{3}He vapor, which further cools the liquid in the pot down to 300 mK. Although the entire setup operates under vacuum, the sample holder maintains thermal contact with the \textsuperscript{3}He pot through a solid cold finger made of copper.
\par
The lowest temperature achieved in this \textsuperscript{3}He VTI was $\sim 330-350$ $mK$. The temperature control of the \textsuperscript{3}He pot is achieved through a combination of regulating the temperature of the sorption pump, which determines the rate of pumping over the liquid \textsuperscript{3}He and applying a small amount of heater power directly at the \textsuperscript{3}He pot.

\subsection{Homemade \texorpdfstring{\textsuperscript{3}}{}He VTI inside existing \texorpdfstring{\textsuperscript{4}}{}He VTI}\label{sec:homemade-3He-VTI}
Some preliminary two-coil measurements were carried out in a custom-made \textsuperscript{3}He VTI manufactured by \href{https://excelinstruments.biz/}{Excel Instruments}, which is designed to sit inside the sample space of our existing \textsuperscript{4}He VTI. The \textsuperscript{3}He VTI consists of two annular stainless steel cylinders, with the lower part made of copper for better heat conduction. The space between these two cylinders, known as the inner vacuum chamber ( IVC ), can be pumped by a turbomolecular pump.
\par
Now, the cooling process is carried out in two steps. In the first step, the sample space in the \textsuperscript{4}He VTI is cooled down to $\sim 1.2$ $K$ using the same procedure mentioned in Section~\ref{sec:4He-VTI}. At the same time, the \textsuperscript{3}He slowly cools via conduction through the copper part, aided by a small amount of \textsuperscript{4}He gas ( exchange gas ) injected into the IVC. Once the system reaches equilibrium at $1.2$ $K$, the \textsuperscript{3}He sample space is isolated by removing the exchange gas from the IVC with the help of the turbo pump.
\par
The second step of cooling involves pumping out the \textsuperscript{3}He gas. Initially, the sample space is pumped by a turbo to $\sim 10^{-6}$ $Torr$. Then, the \textsuperscript{3}He gas is gradually allowed into the sample space from a separate container known as the \textsuperscript{3}He dump vessel. With the temperature at $1.2$ $K$, the system is left for some time to allow all the \textsuperscript{3}He gas to condense into the \textsuperscript{3}He pot. To further decrease the temperature, the condensed \textsuperscript{3}He needs to be pumped similar to the process used for the evaporative cooling of \textsuperscript{4}He in the \textsuperscript{4}He VTI. For this purpose, we utilize a cryopump, which consists of a long pipe filled with activated charcoal. The cryopump, previously kept at room temperature, is then inserted into a helium Dewar to cool the charcoal and start adsorbing all the \textsuperscript{3}He gas in contact with the sample. As a result, the system is further cooled by evaporative cooling. With this homebuilt setup, we were able to achieve the lowest temperature of $700$ $mK$.
\par
Since the sample is in physical contact with the \textsuperscript{3}He in this system, it is necessary to recover the \textsuperscript{3}He from the sample space and return it to the dump before the sample can be taken out at the end of a measurement. For this purpose, the cryopump is used as before to evacuate the \textsuperscript{3}He gas in contact with the sample. The cryostat is then isolated, and the cryopump is taken out of the Dewar to allow the \textsuperscript{3}He gas to desorb and return to the dump. This process is carried out $2-3$ times before the sample is removed from the cryostat.
\par
The main advantage of this system is its greater cooling power since the sample is directly immersed in the liquid \textsuperscript{3}He. Additionally, achieving thermal equilibrium is easier in this dip system compared to a dry vacuum system like the Janis \textsuperscript{3}He system discussed in Section~\ref{sec:janis-3He-VTI}.
\section{Measurement techniques}
\subsection{Two-coil mutual inductance measurements}
\label{sec:two-coil-technique}
For both the works presented Chapters~\ref{ch:phase_fluctuations} and \ref{ch:vortex-par-chap}, we have primarily used a two-coil mutual inductance setup. In this section, I shall provide a detailed discussion of this setup. Magnetic field decays exponentially in a superconductor over a characteristic length known as the penetration depth ( $\lambda$ ). Using the two-coil mutual inductance setup, we measure the absolute value of $\lambda$ from the screening response of the superconductor.
\begin{figure*}[hbt]
	\centering
	\includegraphics[width=16cm]{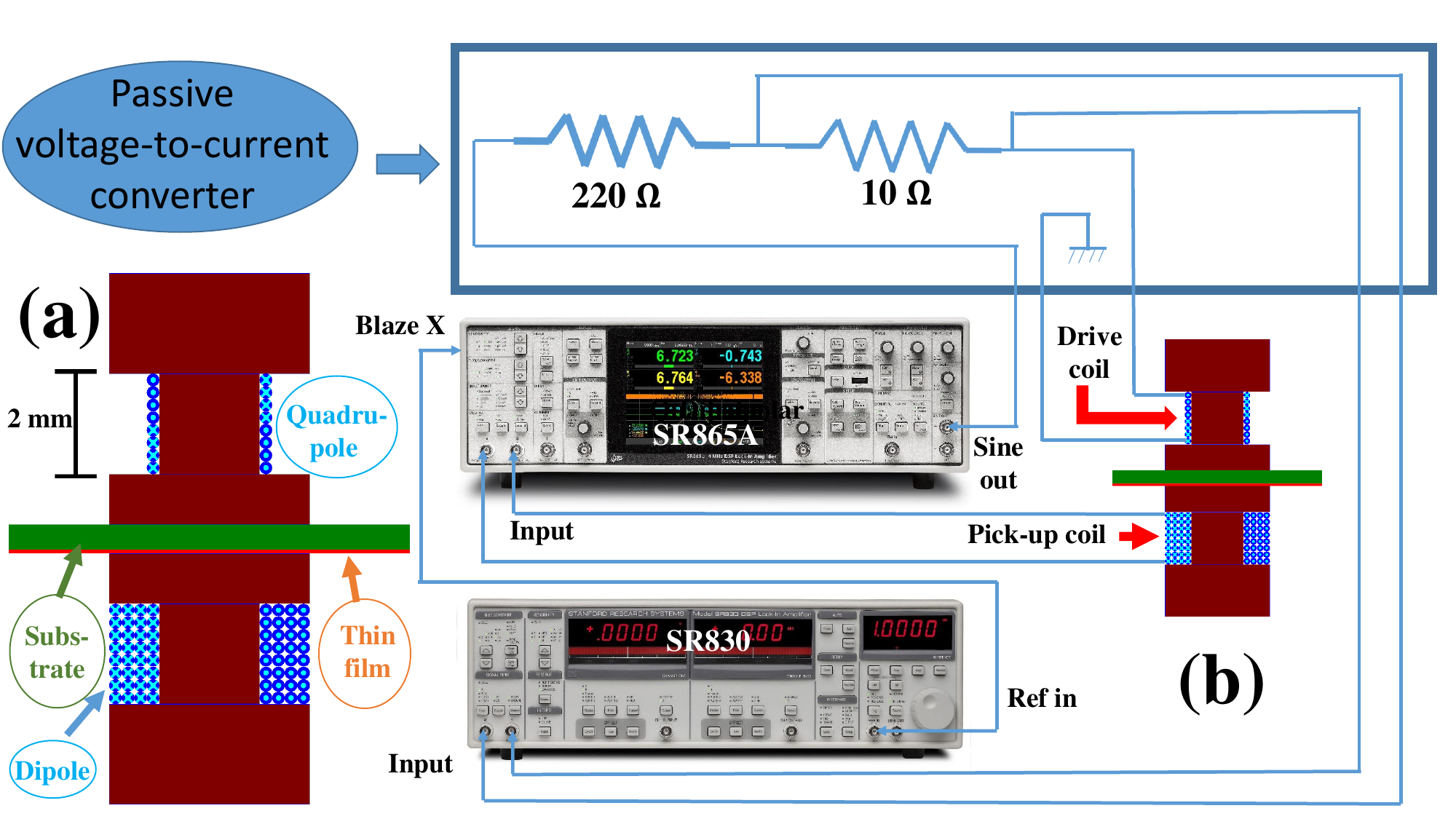}
	\caption[Schematic of the two-coil circuit]{
		\label{fig:two-coil-circuit} 
	\textbf{(a)} Schematic of the two-coil assembly: quadrupole as drive coil (top) and dipole as pick-up coil (bottom) with sample sandwiched in between. The coil wire diameter ($50$ $\mu m$) is drawn bigger than the actual for clarity. \textbf{(b)} Schematic of the electronic circuit for the two-coil mutual inductance measurement. The \href{https://www.thinksrs.com/products/sr865a.html}{SR865A} Lock-in amplifier sends a voltage $V$ at a frequency of $kHz$ to the $V$-$I$ converter, connected in series with the primary drive coil, and measures the pick-up voltage in the secondary coil. Simultaneously, the \href{https://www.thinksrs.com/products/sr810830.htm}{SR830} Lock-in amplifier, locked at the same frequency, measures the voltage drop across a $10$ $\Omega$ resistance that is in series with the drive coil.
	}
\end{figure*}

\begin{figure*}[hbt]
	\centering
	\includegraphics[width=16cm]{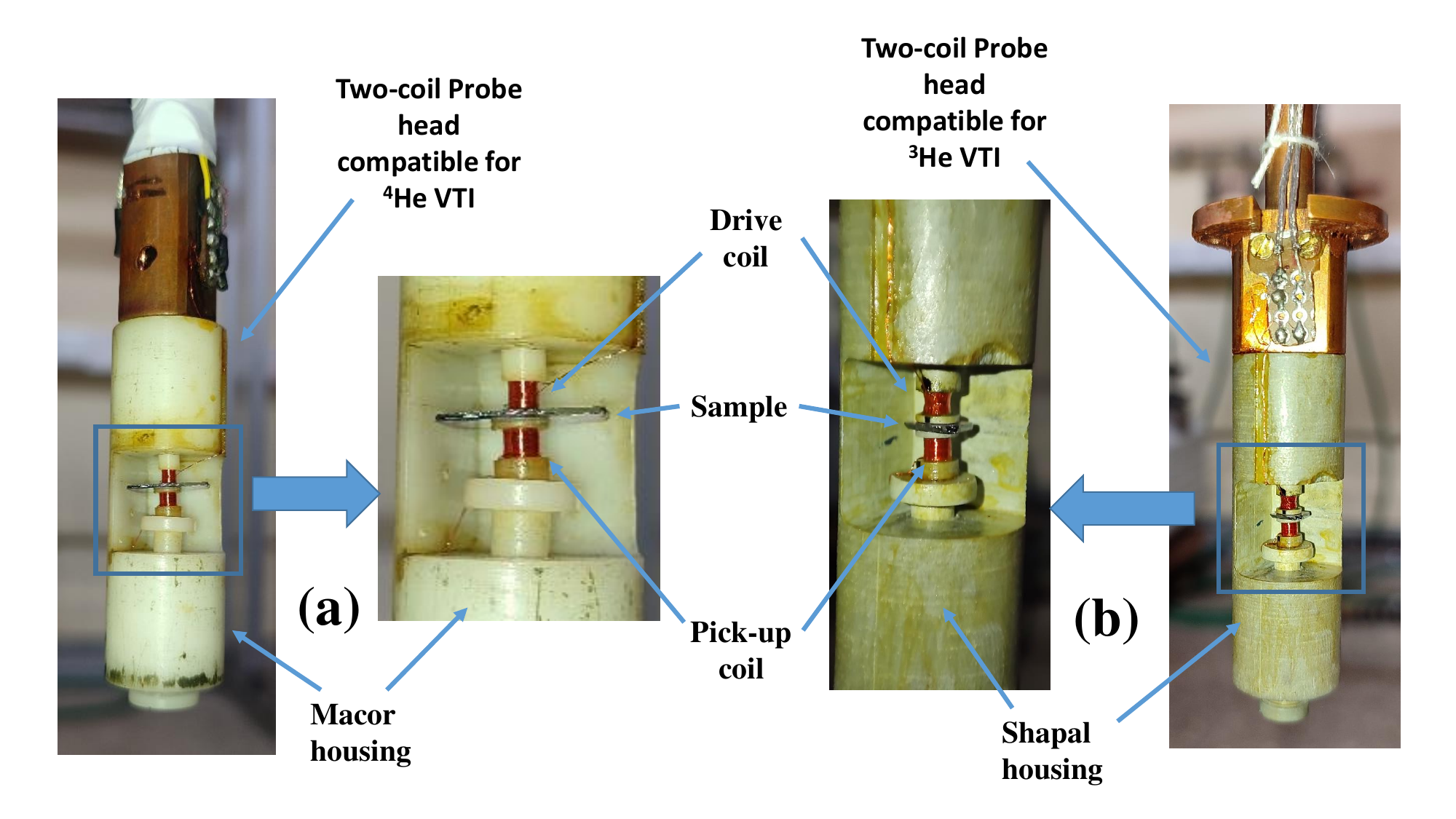}
	\caption[The two-coil mutual inductance probe head]{
		\label{fig:two-coil-probe-head} 
	\textbf{(a)} The two-coil mutual inductance probe head made of Macor compatible with \textsuperscript{4}He VTI. \textbf{(b)} The two-coil mutual inductance probe head machined from Shapal compatible with \textsuperscript{3}He VTI. (Photo courtesy: Dr. Subhamita Sengupta).
	}
\end{figure*}
\subsubsection{Instrument and measurement details}
We sandwich the superconductor between the primary quadrupolar coil and the secondary dipolar coil. In the case of \textsuperscript{4}He VTI, Both the coils are wound on $2$-$mm$-diameter bobbins made out of \href{https://www.dupont.com/delrin.html#:~:text=Delrin%C2%AE%20acetal%20homopolymer%20(Polyoxymethylene,%C2%B0C)%20and%20good%20colorability.}{Delrin} ( Fig.~\ref{fig:two-coil-probe-head}(a) ). The entire coil assembly is kept inside a \href{https://www.morgantechnicalceramics.com/en-gb/products/macor-machinable-glass-ceramic/macor-properties/#:~:text=MACOR%C2%AE%20composition,machining%20can%20be%20an%20irritant.}{Macor} housing surrounded by a heater-can made of copper ( Cu ) in a helium ( He ) gas environment, which provides excellent thermal equilibrium. However, achieving thermal equilibrium in \textsuperscript{3}He VTI is slightly difficult, where the whole coil assembly is kept within $\sim10^{-6}$ vacuum. To overcome the challenge, both the coil bobbins and the outer housing are made of \href{https://precision-ceramics.com/uk/materials/shapal/}{Shapal}, which offers higher thermal conductivity compared to Macor ( Fig.~\ref{fig:two-coil-probe-head}(b) ).
\par
The quadrupolar primary coil comprises 15 clockwise turns wound in the one half of the coil and 15 anticlockwise turns in the other half of the coil, while the dipole secondary coil has 120 turns wound in 4 layers. 50 $\micro m$ diameter copper wire is used for both coils. The complex mutual inductance ( $M=M'+iM''$ ) between the two coils is measured by passing a small ac excitation current ( $I_{ac}$ ) through the primary coil and measuring the resulting in-phase and out-of-phase components of the induced voltage, $V^{in}$ and $V^{out}$, in the secondary using a lock-in amplifier,
\begin{equation}\label{eq:mutual-inductance-formula}
    M'(M'')=\frac{V^{out} (V^{in})}{\omega I_{ac}},
\end{equation}
where $\omega$ is the excitation frequency. The schematic of the electronic circuit used in the measurement is shown in Fig.~\ref{fig:two-coil-circuit}. For the voltage-to-current ( $V-I$ ) converter employed to send the ac excitation current, we used a simple circuit with resistances in series instead of the usual $V-I$ converter comprising OPAMPs. The reason for this choice was that we conducted preliminary measurements at high frequencies ( $MHz$ ) using the \href{https://www.thinksrs.com/products/sr865a.html}{SR865A} Lock-in amplifier, where the common OPAMPs ceased to work effectively. To address this issue, we attempted this passive $V-I$ converter consisting only of resistances in series and observed that in the low-temperature range, the coil resistance changed negligibly, resulting in very little current variation. As a result, we were able to maintain a practically constant current source in the low-temperature range ( below $10$ $K$ ). We have been using this simple $V-I$ converter for low frequencies ( $kHz$ ) as well.
\begin{figure*}[hbt]
	\centering
	\includegraphics[width=16cm]{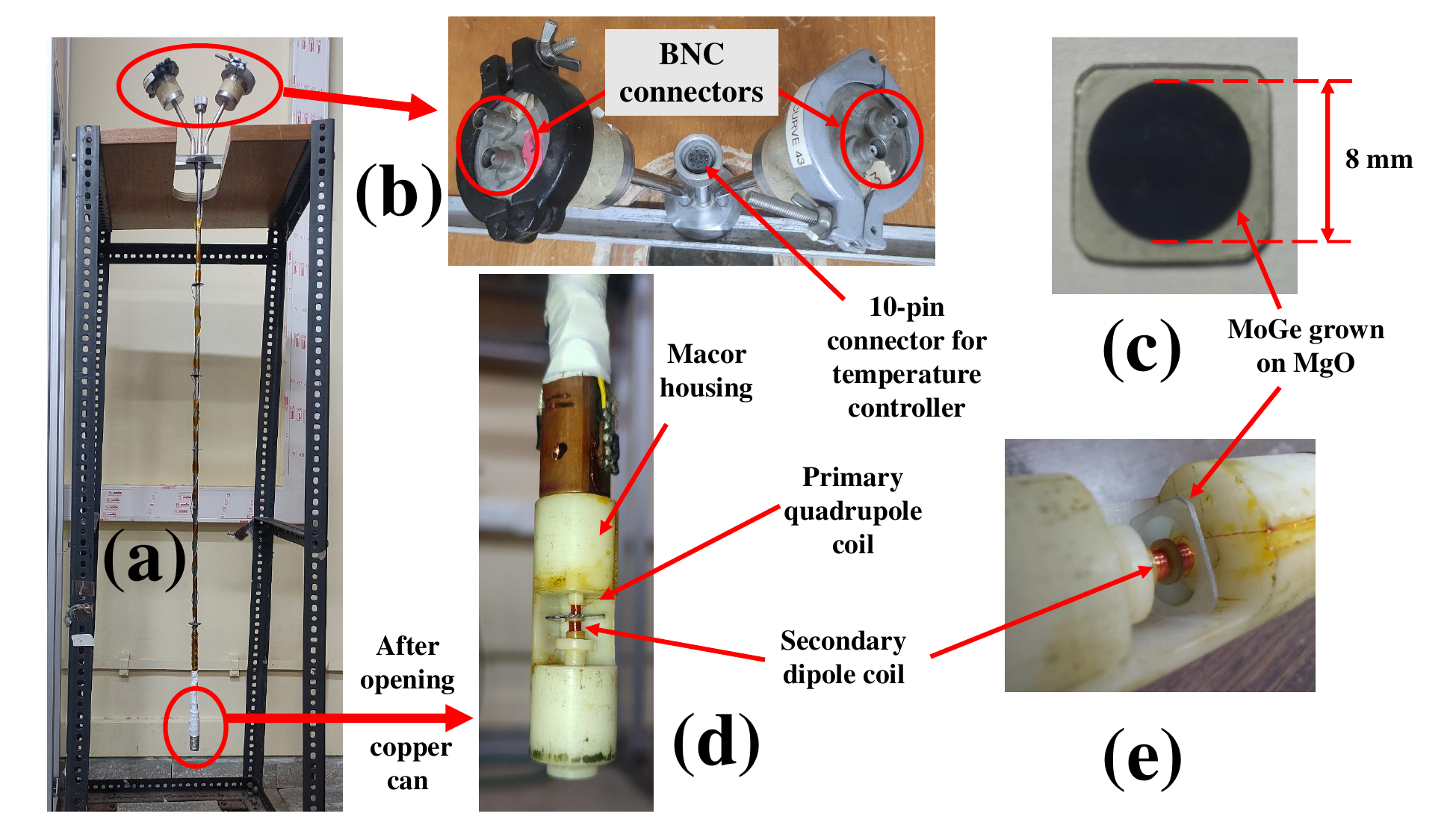}
	\caption[The two-coil mutual inductance setup]{
		\label{fig:two-coil-insert} 
	\textbf{(a)} The two-coil mutual inductance insert compatible with \textsuperscript{4}He VTI. \textbf{(b)} The connector assembly on the upper part of the insert consists of BNC connectors used to connect the drive and pick-up coils to the lock-in amplifiers, along with a $10$-pin connector for temperature measurements. \textbf{(c)} $8$-$mm$-diameter \textit{a}-MoGe film grown on ($100$) oriented MgO substrate. \textbf{(d)} The two-coil probe, enclosed within a macor housing, is displayed. \textbf{(e)} The two-coil assembly: quadrupole as drive coil (top) and dipole as pick-up coil (bottom) with sample sandwiched in between. \textbf{(f)} The side view of the two-coil assembly shows the MoGe film, deposited on MgO, positioned facing the secondary coil. (Photo courtesy: Dr. Subhamita Sengupta and Somak Basistha).
	}
\end{figure*}
\par
The use of a quadrupolar coil as the primary ensures fast radial decay of the ac magnetic field on the film which minimizes edge effects on the measurements ( see Section~\ref{sec:two-coil-quad-dipole-compare} ). The degree of shielding between the primary and the secondary is governed by $\lambda$ of the superconductor, while skin depth ( $\delta$ ) captures the loss in the system, which eventually gets reflected in $M'$ and $M''$ respectively. One can represent both $\lambda$ and $\delta$ in a single complex form known as complex penetration depth ( $\lambda_{eff}$ ) such that,
\begin{equation}
    \lambda_{eff}^{-2}=\lambda^{-2}+i\delta^{-2}.
\end{equation}
However, extracting $\lambda_{eff}^{-2}$ from M is not straightforward because $M'$ and $M''$ both depend on $\lambda$ and $\delta$. For this, we numerically solve the coupled Maxwell and London equations for the geometry of the coils and samples using finite element analysis and create a lookup table $M=M'+iM''$ for a range of values of $\lambda$ and $\delta$. $\lambda_{eff}$ is then extracted by comparing the experimentally measured value of M with the corresponding value in the lookup table. In Section~\ref{sec:two-coil-analysis}, we shall give an overview of how we extract $\lambda$ and $\delta$ from experimental $M$.
\par
Though this method is primarily applied to zero-field measurements, it can also be applied in the presence of a magnetic field.
\subsubsection{Analysis}\label{sec:two-coil-analysis}
In this section, we shall discuss the general scheme to formulate the lookup table between $(M', M'')$ and $(\lambda, \delta)$. We have followed a numerical scheme suggested by Turneaure et al.~\cite{turneaure_numerical_1996,turneaure_numerical_1998}, which takes into account the finite radius of the films. 
\par
We start with the mutual inductance expression between the two coils given by,
\begin{equation}\label{eq:M-expn-A-twocoil}
    M=\frac{1}{I_d}\sum_{j=1}^{n_p}\oint \bm{A}_j^p \cdot d\bm{l},
\end{equation}
where $n_p$ is the number of loops in the pickup coil, $\bm{A}_j^p$ is the total vector potential on the $j^{th}$ turn of the pickup coil. $\bm{A}_j^p$ consists of two contributions: (i) $\bm{A}_j^{d\rightarrow p}$ due to the drive current $I_d$ ( denoted by $I_{ac}$ in Eq.~\ref{eq:mutual-inductance-formula} ) and (ii) $\bm{A}_j^{film\rightarrow p}$ due to the induced current on the superconducting film generated by excitation magnetic field due to $I_d$ in drive coil. Now, utilizing the azimuthal symmetry of the coaxial coil-sample assembly in our setup, Eq.~\ref{eq:M-expn-A-twocoil} simplifies as follows,
\begin{equation}\label{eq:M-expn-A-twocoil-simpl}
\begin{split}
    M
    & =\frac{2\pi}{I_d}\sum_{j=1}^{n_p}r^p_jA^p_j\\
    & =\frac{2\pi}{I_d}\sum_{j=1}^{n_p}r^p_j\left[A^{d\rightarrow p}_j+A^{film \rightarrow p}_j\right],
\end{split}
\end{equation}
where $A_j^p$ is the $\phi$-component of the total vector potential at the radius of $j^{th}$ loop of the pickup coil, $r^p_j$. 
\par
Now, the vector potential due to the current in the drive coil is a standard problem of vector potential between two concentric rings and can be expressed as~\cite{jackson_classical_2012},
\begin{equation}\label{eq:A-drive-to-pickup}
    A^{d\rightarrow p}_j=\frac{\mu_0 I_d}{4\pi}\sum_{i}^{n_d}\frac{4r_i^d}{m_{ij}}\left[\frac{(2-k_{ij}^2)K(k_{ij})-2E(k_{ij})}{k^2_{ij}}\right],
\end{equation}
with
\begin{equation}
    m_{ij}=\sqrt{(r_i^d+r_j^p)^2+D^2_{ij}},
\end{equation}
and, $E(k_{ij})$ and $K(k_{ij})$ are the complete elliptic integrals with argument,
\begin{equation}
    k^2_{ij}=\frac{4r^d_ir_j^p}{(r_i^d+r_j^p)^2+D^2_{ij}},
\end{equation}
where $r^d_j$ is the radius of the $i^{th}$ loop in the drive coil and $D_{ij}$ is the distance between $i^{th}$ drive loop and $j^{th}$ pickup loop. The summation in Eq.~\ref{eq:A-drive-to-pickup} runs on the loops of the drive coil, $n_d$ being the number of drive loops.
\par
In order to determine the vector potential due to the screening currents in the film, we need to know the current density across the film, $\bm{J}_{film}$. Now, in the local limit, vector potential in the film is related to the current density  according to London's equation as follows,  
\begin{equation}\label{eq:J-film-london-eqn}
    \bm{J}^{film}(\bm{r})=-\frac{\bm{A}^{film}(\bm{r})}{\mu_0 \lambda^2}.
\end{equation}
Also, the vector potential in the film can be expressed in the following self-consistent manner,
\begin{equation}\label{eq:J-film-A-film-self-consistent}
    \bm{A}^{film}(\bm{r})=\bm{A}^{d}(\bm{r})+\frac{\mu_0}{4\pi}\int d^3\bm{r'}\frac{\bm{J}^{film}(\bm{r'})}{\abs{\bm{r-r'}}},
\end{equation}
where $\bm{A}^{d}(\bm{r})$ is the contribution at $\bm{r}$ caused by the drive coil and $r'$ is the film coordinate.
\begin{equation}\label{eq:A-drive-total}
    \bm{A}^{d}(\bm{r})=-\mu_0\lambda^2 \bm{J}^{film}(\bm{r})-\frac{\mu_0}{4\pi}\int d^3\bm{r'}\frac{\bm{J}^{film}(\bm{r'})}{\abs{\bm{r-r'}}}.
\end{equation}
Now, for simplification in the calculation, the $z$-integral is replaced by a multiplicative factor,
\begin{equation}
    d_{eff}=\lambda \tanh{\left(\frac{d}{\lambda}\right)},
\end{equation}
which is a good approximation to the exact solution for a stripline geometry. Also, after doing the $\phi$-integration, we express Eq.~\ref{eq:A-drive-total} in terms of single $\rho$-integration,
\begin{equation}\label{eq:A-drive-total-rho-int}
\begin{split}
    A^{d}(\rho)=-\mu_0\lambda^2 J^{film}(\rho) 
    & -\frac{\mu_0}{2\pi}d_{eff}\int d\rho'J^{film}(\rho')\frac{\rho+\rho'}{\rho}\\
    & \times \left[\left(1-\frac{k^2}{2}\right)K(k)-E(k)\right],
\end{split}
\end{equation}
where $E(k)$ and $K(k)$ are the complete elliptic integrals, with the argument $k$ defined as,
\begin{equation}
    k^2=\frac{4\rho\rho'}{(\rho+\rho')^2}.
\end{equation}
To numerically solve Eq.~\ref{eq:A-drive-total-rho-int}, we divide the circular film into N number of concentric annular rings. Thus, the integration in Eq.~\ref{eq:A-drive-total-rho-int} gets converted to a matrix equation,
\begin{equation}\label{eq:A-drive-total-rho-int-matrix-form}
\begin{split}
    A^{d}_i(\rho_i)=-\mu_0\lambda^2 J^{film}_i(\rho_i) 
    & -\frac{\mu_0}{2\pi}d_{eff}\sum_j \Delta\rho_j J^{film}_j\frac{\rho_i+\rho_j}{\rho_i}\\
    & \times \left[\left(1-\frac{k^2}{2}\right)K(k)-E(k)\right],
\end{split}
\end{equation}
where $\rho_i$ is the radius of the $i^{th}$ annular ring. Now, Eq.~\ref{eq:A-drive-total-rho-int-matrix-form} can be expressed as,
\begin{equation}\label{eq:matrix-eqn}
    a_{ij}c_j=b_i,
\end{equation}
where
\begin{equation}
\begin{split}
    a_{ij} & =\left[\frac{\lambda^2}{d_{eff}}+\frac{\Delta R}{2\pi}\left(\ln{\frac{16\pi\rho_j}{\Delta R}-2}\right)\right]\delta_{ij}\\
    & +\frac{\Delta R}{2\pi}\frac{\rho_i+\rho_j}{\rho_i}\left[\left(1-\frac{k^2}{2}\right)K(k)-E(k)\right](1-\delta_{ij}),
\end{split}
\end{equation}
\begin{equation}
    b_i=\frac{A^{d}_i}{\mu_0},
\end{equation}
\begin{equation}
    c_i=-d_{eff}J^{film}_i,
\end{equation}
where the first term in $a_{ij}$ corresponds to the self-term of the loop from the self-mutual inductance calculated by Gilchrist and Brandt~\cite{gilchrist_screening_1996}.
\par
Now, by numerically solving Eq.~\ref{eq:matrix-eqn}, we can calculate the current density in $i^{th}$ annular ring, which can be used to calculate $A^{film\rightarrow p}$. Hence, for a given $\lambda_{\omega}^{2}=\lambda^{-2}+i\delta^{-2}$, $M$ can be determined from Eq.~\ref{eq:M-expn-A-twocoil-simpl}.
\par
We generally take a $100\times100$ sets of ($\lambda,\delta$) and calculate the corresponding $100\times100$ matrix of ($M', M''$) values.
\begin{figure*}[hbt]
	\centering
	\includegraphics[width=16cm]{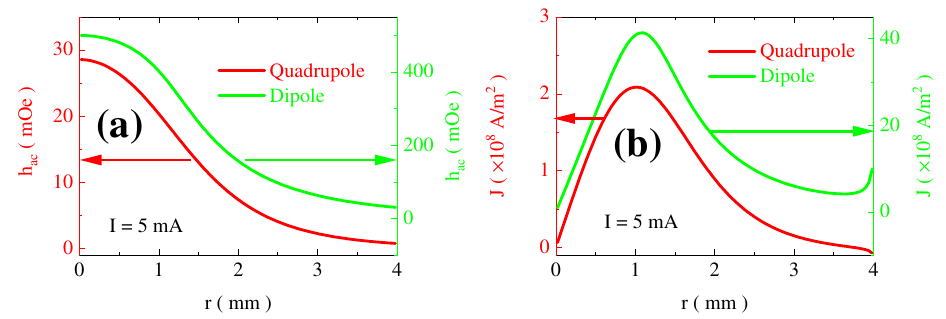}
	\caption[Comparison of simulated radial distribution of field and induced current density due to quadrupole and dipole coils]{
		\label{fig:two-coil-H-J-vs-r-dipole-quad} 
	Comparison of simulated radial distribution of \textbf{(a)} magnetic field and \textbf{(b)} induced current density due to quadrupolar and dipolar coils. In the simulation, the following parameters were considered: thickness $d=21$ $nm$, $\lambda=570$ $nm$, and $\delta=\infty$.
	}
\end{figure*}
\subsubsection{Advantage of using quadrupole as drive coil}\label{sec:two-coil-quad-dipole-compare}
Despite producing a weaker signal compared to the dipolar coil, we opted to use a quadrupolar coil as the drive coil. In Fig.~\ref{fig:two-coil-H-J-vs-r-dipole-quad}, we present a comparison of the simulated field and induced current density in the film for our setup geometry using both the quadrupolar and dipolar coils. It is evident that the field corresponding to the quadrupolar coil diminishes significantly at the boundary of the film ( radius $r=4$ $mm$ ), while a substantial field is still present in the case of the dipolar coil ( Fig.~\ref{fig:two-coil-H-J-vs-r-dipole-quad}(a) ). Additionally, the induced current density on the superconducting film due to the quadrupole nearly becomes zero at the film's edge, whereas that induced by the dipole exhibits a finite value with a sharp peak at the edge ( Fig.~\ref{fig:two-coil-H-J-vs-r-dipole-quad}(b) ). Consequently, using the quadrupolar coil renders the edge effects negligible, simplifying our calculation of the penetration depth.
\subsection{Broadband microwave spectroscopy using Corbino geometry}
Resonant-cavity techniques are commonly used for highly sensitive and stable conductivity measurements at microwave frequencies. However, these techniques are limited to discrete frequencies determined by the resonant modes of the cavity, providing information only at specific frequencies. In contrast, broadband microwave spectroscopy serves as a complementary technique, allowing for the measurement of the electrodynamic response of materials across a wide frequency range, typically spanning from $MHz$ to $GHz$. This enables researchers to obtain continuous information on the behavior of materials as a function of frequency, providing a more comprehensive understanding of their electromagnetic properties. By combining resonant-cavity techniques with broadband microwave spectroscopy, researchers can gain insights into the behavior of materials at different frequencies, leading to a more detailed characterization of their electromagnetic properties and potential applications.
\begin{figure*}[hbt]
	\centering
	\includegraphics[width=16cm]{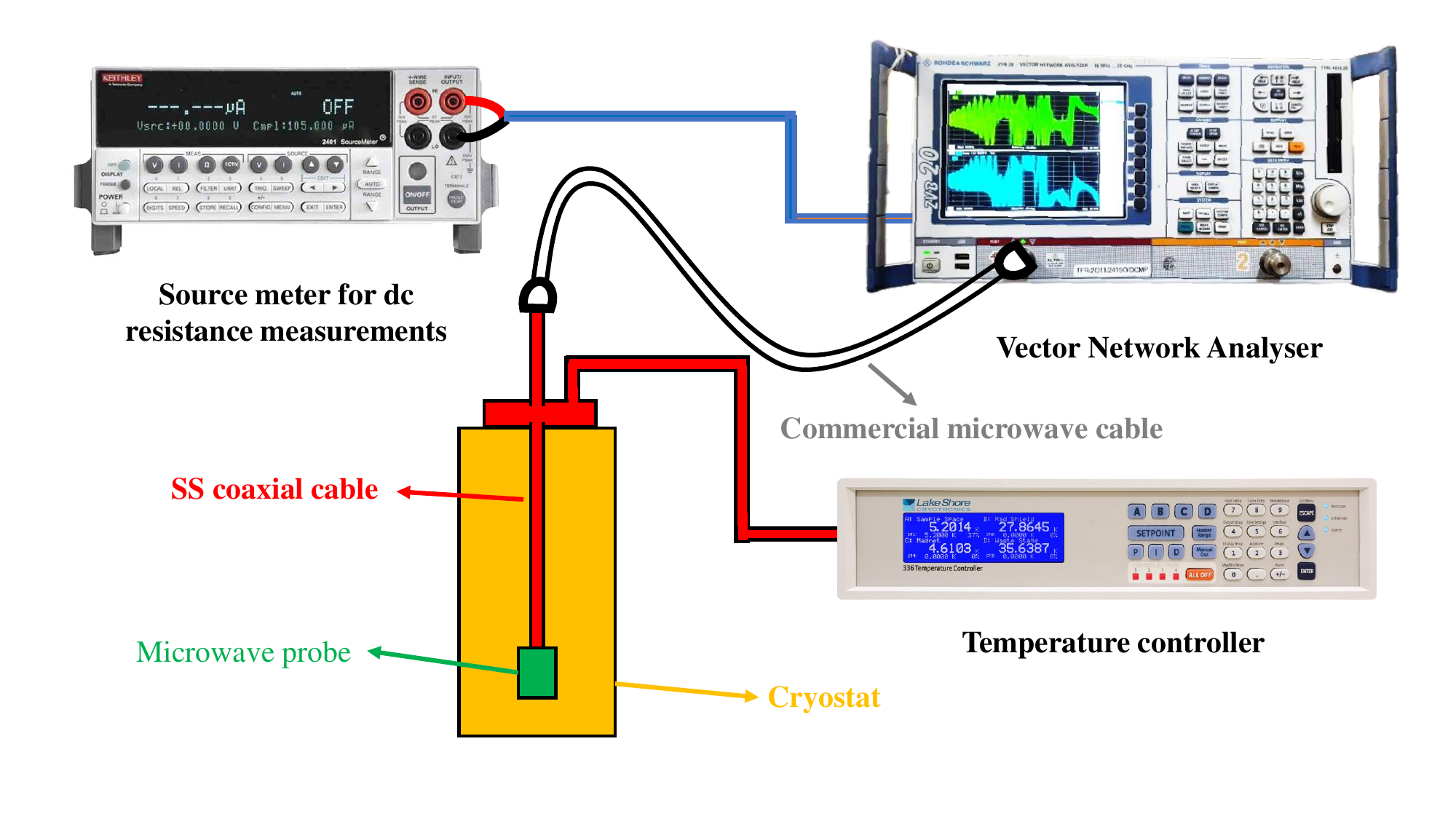}
	\caption[Schematic of the broadband microwave circuit]{
		\label{fig:broadband-microwave-circuit} 
	Schematic of the electronic circuit of the broadband microwave setup. The microwave signal at a frequency of $GHz$ range is transmitted through both the commercial microwave cable and the SS-coaxial cable from the vector network analyzer (VNA), enabling the measurement of the complex reflection coefficient. The Lakeshore controller is utilized to monitor the temperature, while simultaneous two-probe DC resistance measurement is conducted using  the source meter connected via the VNA.
	}
\end{figure*}
\subsubsection{Instrument and measurement details}
\begin{figure*}[hbt]
	\centering
	\includegraphics[width=16cm]{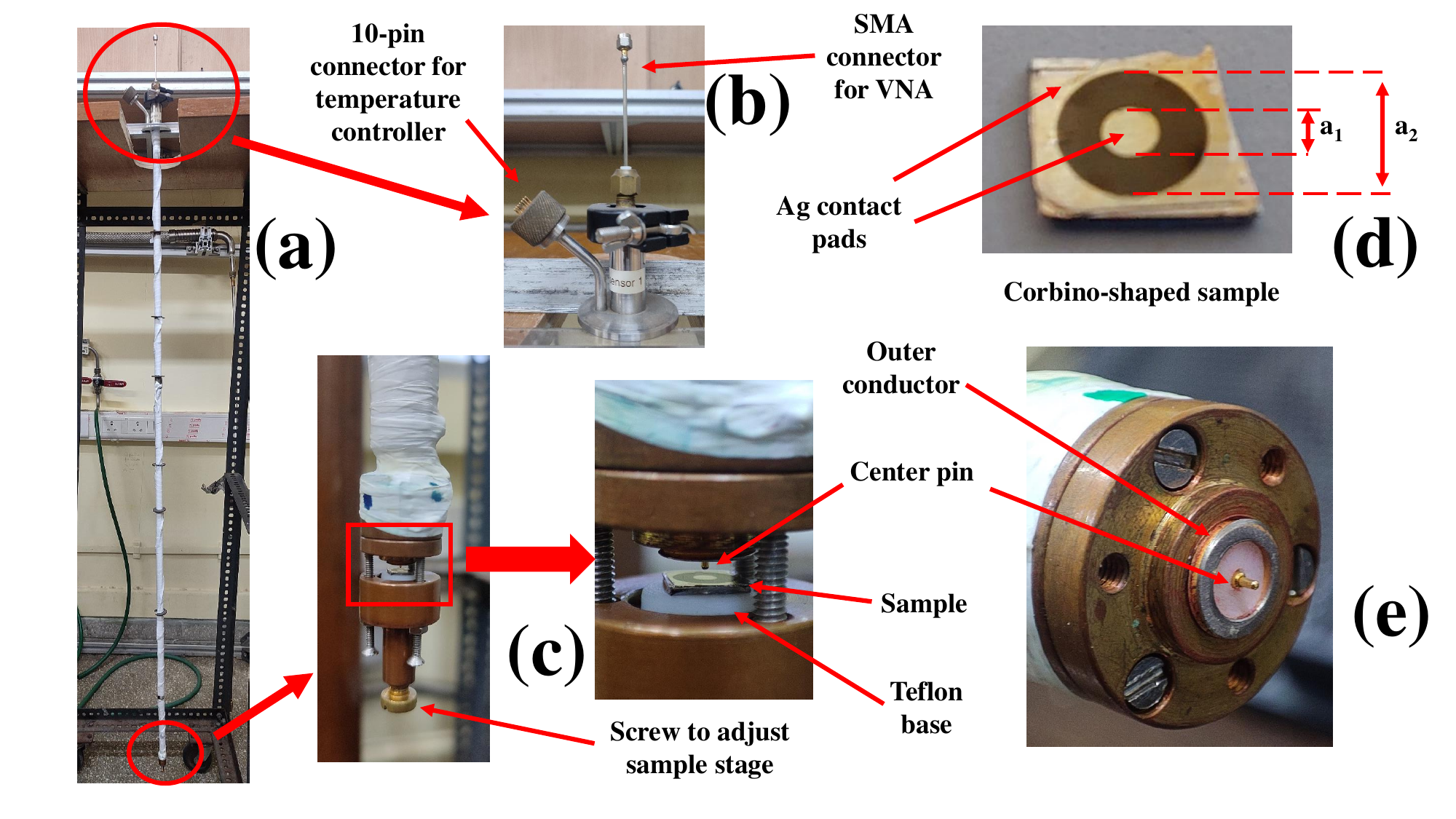}
	\caption[The broadband microwave insert]{
		\label{fig:broadband-microwave-insert} 
	\textbf{(a)} The broadband microwave insert compatible with \textsuperscript{4}He VTI. \textbf{(b)} Connector assembly on the upper part of the insert comprising SMA connector to send microwave signal from VNA and $10$-pin connector to measure the temperature. \textbf{(c)} The microwave probe, which is connected to the lower part of the SS coaxial cable, is displayed. It is enclosed within a housing made of gold-plated copper. At the center of the SMA connector, a spring-loaded pin is attached to ensure good electrical contact with the sample, while the sample is placed on a Teflon base. \textbf{(d)} Corbino-shaped sample with silver deposited as a contact pad. \textbf{(e)} Probe head without the sample. (Photo courtesy: Dr. Subhamita Sengupta and Somak Basistha).
	}
\end{figure*}

Our laboratory has a custom-built setup for broadband microwave spectroscopy, following the method proposed by J. C. Booth~\cite{booth_broadband_1994}.\! We utilize a vector network analyzer\! ( VNA ) from \href{https://www.rohde-schwarz.com/manual/zvb/}{Rhode and Schwarz}, which is capable of generating microwave signals in the frequency range of $10$ $MHz$ to $20$ $GHz$ and measuring the complex reflection coefficient. The VNA is connected to a flexible microwave cable with a characteristic impedance of 50 $\Omega$. For low-temperature measurements, we extend the setup with a $50$ $cm$ long $0.181"$ SS-coaxial cable, connected via an SMA connector to the commercially available microwave cable. The sample under investigation is positioned at the lower end of the SS-coaxial cable, acting as a terminator and reflecting back the microwave signal toward the VNA. A zoomed-in illustration of the lower end of the SS-coaxial cable, attached to the microwave probe, is shown in Fig.~\ref{fig:broadband-microwave-insert}.
\par
The microwave probe in our setup features a female SMA connector, with a spring-loaded \href{https://www.kita-mfg.com/}{Kita} pin made of beryllium copper inserted at its center. This design ensures reliable electrical contact even at low temperatures, accommodating the thermal contraction of the cable and connector assembly. The sample is positioned on a Teflon base, which can be adjusted in height using spring and screw arrangements from below. The housing of the probe is constructed with gold-plated copper for improved performance.
\par
In order to fit the sample within the probe, the sample dimensions are maintained at approximately $5$ $mm \times 5$ $mm$. To ensure better electrical contact with the SMA connector and center pin, a silver ( Ag ) contact pad with a thickness of approximately $200$ $nm$ is deposited on the thin film in a Corbino geometry. This is achieved by thermally evaporating Ag using an annular mask with inner and outer diameters of $1.5$ $mm$ and $4$ $mm$, respectively. The Corbino geometry allows the self-field generated by the current flowing through the film to be parallel to the surface of the film, eliminating edge effects. Temperature is monitored using \href{https://www.lakeshore.com/products/categories/overview/temperature-products/cryogenic-temperature-controllers/model-336-cryogenic-temperature-controller}{Lakeshore 336 controller}, and at each temperature, the frequency is swept across the entire available range, and the corresponding complex reflection coefficient is measured using a vector network analyzer ( VNA ). Additionally, the two-probe DC electrical resistance is measured at the same temperature using a \href{https://www.tek.com/en/datasheet/series-2400-sourcemeter-instruments}{Keithley 2400 source meter} via the bias port on the back side of the VNA.
\par
If $S_{meas}$ is the measured complex reflection coefficient, then sample impedance, $Z_L$ is given as,
\begin{equation}\label{eq:Z_L-expn-microwave}
    Z_L=Z_0 \left(\frac{1+S_{meas}}{1-S_{meas}}\right).
\end{equation}
Here, $Z_0$ is the characteristic impedance of the transmission line which is $50 \Omega$ in our case. Now, the complex conductivity, $\sigma$ is obtained as,
\begin{equation}\label{eq:sigma-expn-microwave}
    \sigma=\frac{\ln{(\frac{a_2}{a_1}})}{2\pi d Z_L},
\end{equation}
where $a_2$ and $a_1$ are outer and inner radius of the sample and $d$ is the film thickness.
\par
We can then find out the complex penetration depth $\lambda_{\omega}$ using the relation,
\begin{equation}\label{eq:lambda-expn-microwave}
    \lambda^{-2}_\omega=i\mu_0 \omega \sigma.
\end{equation}
\subsubsection{Error correction}
In microwave measurements, ensuring accurate measurement of the reflection coefficient becomes particularly challenging at lower temperatures and higher frequencies due to signal attenuation along transmission cables and unwanted reflections from improper joints. To overcome these systematic errors, precise calibration of the probe is essential to extract only the sample contribution. The calibration process involves establishing a reference plane, which, in this case, is the plane of the Corbino probe. Any contribution from the other side of this plane is considered the sample contribution, as it is experimentally impossible to separate the contact resistance between the Corbino probe and the sample from the sample impedance.
\par
There are three main types of errors expressed by a model. The first one is Directivity ( $E_D$ ), which arises from reflected signals due to improper soldering or joints. The second error is Reflection tracking ( $E_R$ ), which occurs due to the damping of the signal along the coaxial cable. The third error is Source mismatch ( $E_S$ ), which happens when the reflected signal from the sample gets reflected again from improper junctions and adds to the incoming signal. To mitigate these error contributions, precise calibration is conducted using three known standards: the short standard, the open standard, and the load standard. The reflection coefficient for a short terminator is $-1$, while for an open terminator, it is $1$. As for the load, its reflection coefficient value depends on its impedance, as specified by Eq.~\ref{eq:Z_L-expn-microwave}.
There are \href{https://www.rohde-schwarz.com/in/products/test-and-measurement/manual-calibration-and-verification-kits/rs-zv-z2xx-network-analyzer-calibration-kits_63493-418432.html}{commercially available calibrators} for room temperature, but they are not suitable for low-temperature calibration. Therefore, alternative methods need to be employed for low-temperature calibration~\cite{booth_broadband_1994,stutzman_broadband_2000,scheffler_broadband_2005,kitano_broadband_2008}:
\subsubsection{Short standard}
A thick Al/Au film or the superconducting spectra of an ordered thick superconductor at a temperature far below $T_c$ can be used as a short standard.
\subsubsection{Open standard}
The sample stage is left empty, and the Teflon base of the sample stage with a thickness of $1.5$ $mm$ acts as the open standard.
\subsubsection{Load standard}
Metallic NiCr films have proven to be reliable load standards due to their temperature-independent resistivity over a wide temperature range, spanning from a few Kelvin to room temperature. This characteristic makes them excellent candidates for load calibration purposes, with typical resistance values falling within the range of $20-30$ $\Omega$. Traditionally, these NiCr films are thermally evaporated onto a glass substrate to serve as calibration standards.
\par
However, in our recent experiments, we adopted an alternative approach. Instead of employing three distinct choices for load standards, we utilized a superconducting short standard along with two load standards. These load standards were based on the spectra of the actual sample above its critical temperature ( $T_c$ ) at two different temperatures. The advantage of this approach was that it enabled us to measure two of the three calibrators during the same thermal cycle as the actual sample. By doing so, we could reduce potential errors associated with variations in heating and cooling cycles within the experimental setup. This improvement in the calibration process contributes to more accurate and reliable measurements.
\par
Below are the expressions for the three error terms concerning the three calibration standards:
\begin{equation}
    E_D(\omega)=\frac{M_1(M_2-M_3)A_2A_3+M_2(M_3-M_1)A_3A_1+M_3(M_1-M_2)A_1A_2}{(M_1-M_2)A_1A_2+(M_2-M_3)A_2A_3+(M_3-M_1)A_3A_1},
\end{equation}
\begin{equation}
    E_R(\omega)=\frac{(M_1-M_2)(M_2-M_3)(M_3-M_1)(A_1-A_2)(A_2-A_3)(A_3-A_1)}{[(M_1-M_2)A_1A_2+(M_2-M_3)A_2A_3+(M_3-M_1)A_3A_1]^2}
\end{equation}
\begin{equation}
    E_S(\omega)=\frac{M_1(A_2-A_3)+M_2(A_3-A_1)+M_3(A_1-A_2)}{(M_1-M_2)A_1A_2+(M_2-M_3)A_2A_3+(M_3-M_1)A_3A_1}
\end{equation}
The expressions provided above utilize $M_i$ to represent the measured reflection coefficient and $A_i$ to represent the actual reflection coefficient for the $i^{th}$ calibrator. Once the error terms are established as a function of frequency at different temperatures, the corrected reflection coefficient, denoted as $S_{corr}$, can be computed for each frequency and temperature using the given formula,
\begin{equation}
    S_{corr}=\frac{S_{meas}-E_D}{E_R+E_S(S_{meas}-E_D)}.
\end{equation}
Subsequently, the obtained $S_{corr}$ can be utilized in Eqs.~\ref{eq:Z_L-expn-microwave}, \ref{eq:sigma-expn-microwave} and \ref{eq:lambda-expn-microwave} to derive the electrodynamic properties of the sample in the frequency domain.
\subsection{Scanning tunneling microscopy/ spectroscopy (STM/S)}
All Scanning Tunneling Spectroscopy/Microscopy ( STS/M ) measurements on the \textit{a}-MoGe thin films were conducted using a custom-built \textsuperscript{3}He STM capable of operating at temperatures as low as $350$ $mK$ and equipped with a superconducting magnet of $90$ $kOe$. The STM is connected to a sample preparation chamber, which includes a dc magnetron sputter gun, Ar ion milling, and a recently installed Ferrovac GmbH ion pump. Data acquisition and basic analysis are performed using a Rev 9 controller from RHK Tech., which serves both functions. A Pt-Ir tip ( $300$ $\mu m$ diameter ) is utilized for scanning the sample, and it is mounted on a molybdenum sample holder. The tip is attached to a piezoelectric tube that can be controlled in $\pm X$, $\pm Y$, and $Z$ directions. The piezo-tube is positioned on a coarse positioner, controlled by the sixth channel of Attocube controller ANC300, operating at $30$ $V$ and $250$ $Hz$ mode, utilizing the raising part of the pulse. Vibration isolation is achieved by floating the cryostat on four air-cushion legs, further damped by a piezo-controlled active vibration isolator from Newport Smart Table.
\begin{figure*}[hbt]
	\centering
	\includegraphics[width=16cm]{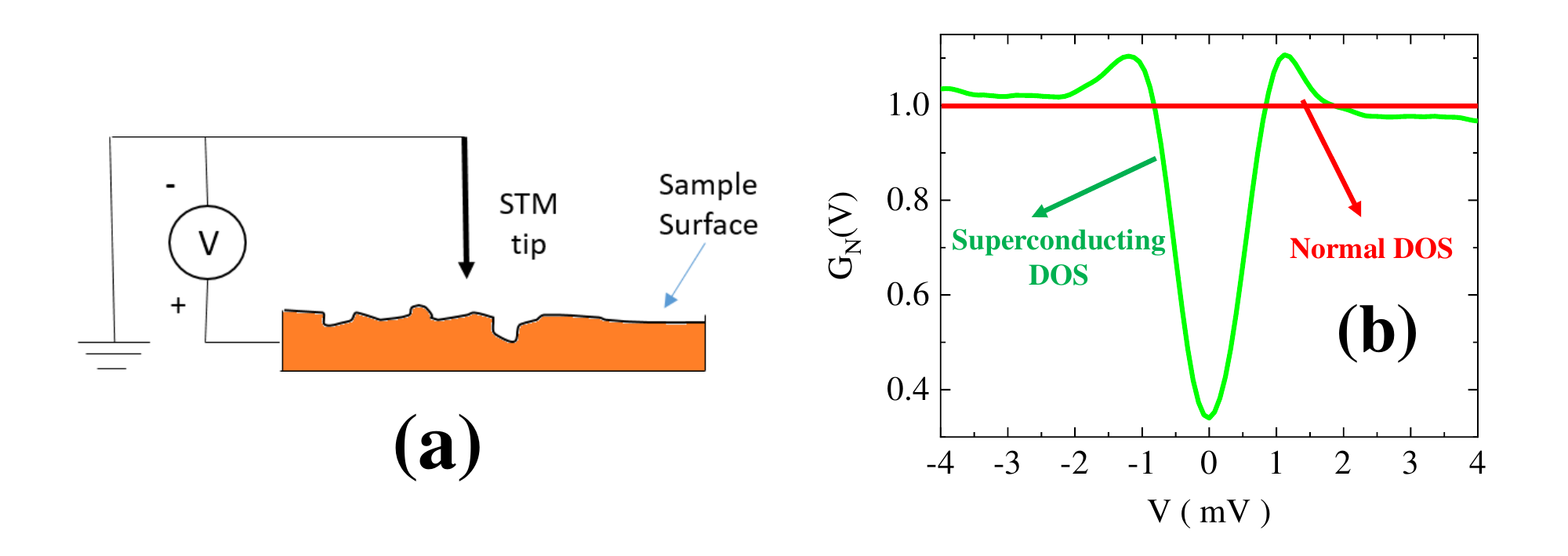}
	\caption[Schematic of scanning tunneling microscope and typical conductance spectrum]{
		\label{fig:STM-schematic-spectra} 
	\textbf{(a)} Schematic of scanning tunneling microscope. \textbf{(b)} Representative normalized tunneling conductance $G_N(V)$ vs $V$ spectrum.
	}
\end{figure*}
\par
The operation of STS/M is based on the phenomenon of tunneling, as explained in section ~\ref{sec:tunneling}. In terms of the instrument, the tunneling current falls within the range of pico-nano-amperes, which is then pre-amplified by a factor of $10^9$ and converted to digital signals using an analog-to-digital converter for acquisition by the computer. The feedback loop, depending on the set current for the measurement, controls the distance between the tip and the surface of the sample to maintain a constant current mode of operation, enabling scanning over rough surfaces. Specific parameters for different purposes are determined through trial and error, such as setting the bias at $10$ $mV$ and the set current at $50$ $pA$ for topography imaging, for instance.
\subsubsection{Bias spectroscopy}
In addition to topography data, an interesting type of data that can be obtained using STS/M is bias spectra, specifically the differential conductance ( $dI/dV$ or $G_{ns}$ ) as a function of bias. To acquire this data, we apply an alternating voltage of $dV = 150 \mu V$ and $\sim 2.677$ $kHz$ to our bias voltage and momentarily pause the feedback loop to record the $dI$ using the built-in lock-in amplifier. This allows us to vary the bias from $-V$ ( e.g., $-6$ $mV$ ) to $+V$ ( e.g., $+6$ $mV$ ), and obtain the $dI/dV$ vs. $V$ spectra. To normalize the data, we divide the full spectra by $G_{ns}(V \gg \Delta)$ so that $G_{ns}$ and $G_{nn}$ ( normalized differential conductance ) become unity at high bias.
\par
Using\! the\! $G_{ns}(V)$ vs $V$ data,\! one\! can extract\! information such as\! the\! BCS gap ( $\Delta$ ) by\! fitting\! the $G_{ns}(V)$ curve using Eq.~\ref{eq:tunneling_conductance}.\! However, the\! BCS density of states\! (\! DOS\! )\! is\! slightly\! modified\! to incorporate\! a $\Gamma$ factor\!, accounting for non-thermal sources of broadening. The modified form of the equation is:
\begin{equation}
\frac{N_s(E)}{N(0)}=
\begin{cases}
Re\left(\frac{\abs{E}+i\Gamma}{\sqrt{(\abs{E}+i\Gamma)^2}-\Delta^2}\right)& (E>\Delta)\\
0 & (E<\Delta)
\end{cases}
\end{equation}
An experimental $G_{ns}(V)$ obtained from a $20$-$nm$-thick \textit{a}-MoGe film is shown in Figure, and it can be fitted using equation 5 to obtain $\Delta=1.3$ $meV$ and $\Gamma=0.2$ $meV$ at $450$ $mK$. It is possible to perform the same experiment and fitting for different temperatures to extract $\Delta(T)$ values, which can be further fitted using the universal BCS curve.
\begin{figure*}[hbt]
	\centering
	\includegraphics[width=16cm]{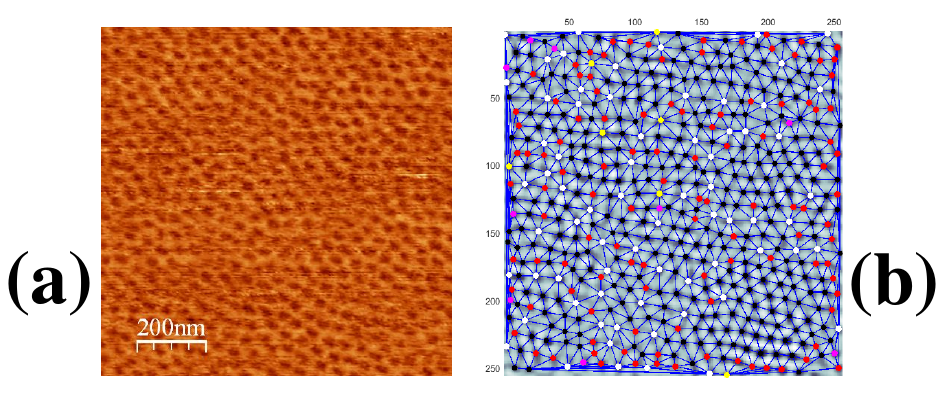}
	\caption[Typical vortex image and Delaunay triangulation]{
		\label{fig:vortex-image-delaunay} 
	\textbf{(a)} Representative vortex image obtained using scanning tunneling spectroscopy ($450$ $mK$, $10$ $kOe$, $6$-$nm$-thick \textit{a}-MoGe film). \textbf{(b)} Representative Delaunay triangulated vortex image ($450$ $mK$, $55$ $kOe$, $21$-$nm$-thick \textit{a}-MoGe film). Color codes are used to differentiate vortices based on the number of nearest neighbors or coordination numbers: Black - $6$, Red - $5$, White - $7$, Magenta - $4$, and Yellow - $8$.
	}
\end{figure*}

\subsubsection{Vortex imaging}
By utilizing the fact that a vortex core behaves as a normal metallic region, where the superconducting gap and coherence peaks are suppressed ( as shown in Fig.~\ref{fig:STM-schematic-spectra}(b) ), we set our bias voltage ( $V_{peak}$ ) at the location of the coherence peak and record the differential conductance across the surface. This results in a $G(V=V_{peak})$ map, as depicted in Fig.~\ref{fig:vortex-image-delaunay}, where vortices are visualized as minima due to their lower value of $G(V=V_{peak})$ compared to the superconducting regions.
\subsubsection{Delaunay Triangulation}
Once the positions of the vortices are determined, all the nearest neighbor points are joined by the Delaunay triangulation method such that the circumcircle of any triangle made of three nearest neighbor points consists of no other point. Topological defects are then identified from the lattice points having coordination numbers larger or smaller than 6 ( Colour code: black - 6, red - 5, white - 7, magenta - 4, and yellow - 8 ).
\begin{figure*}[hbt]
	\centering
	\includegraphics[width=16cm]{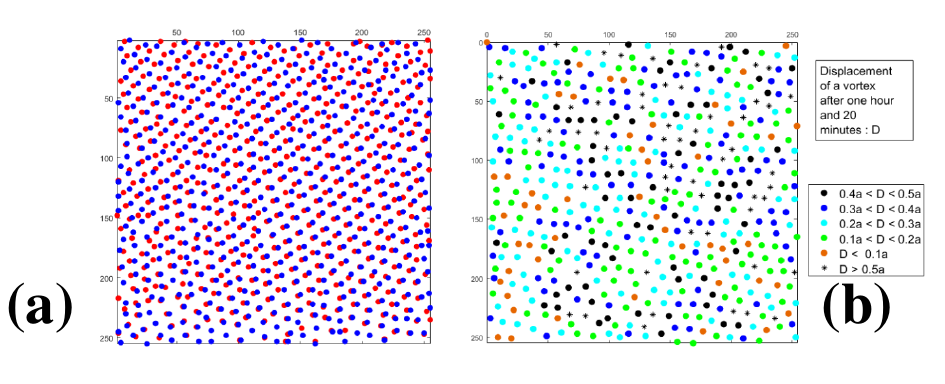}
	\caption[Tracking the movement of vortices by STM]{
		\label{fig:vortex-movement-tracking} 
	\textbf{(a)} Shift in vortex positions tracked        through two representative consecutive vortex images at the same position: Red – initial position, Blue – after 1 hour and 20 minutes ($450$ $mK$, $55$ $kOe$, $21$-$nm$-thick \textit{a}-MoGe film). \textbf{(b)} Classification of Vortices according to their shifted positions after one hour and 20 minutes: color codes mentioned in the plot.
	}
\end{figure*}
\subsubsection{Vortex Movement Tracking}
To track how much the vortices are moving with time, we scanned successive vortex images at the intervals of 1 hour and 20 minutes and compared the positions of the vortices in those images to see how much they moved. Each vortex was then classified according to its displacement\hspace{0.1cm} being a fraction of\hspace{0.1cm} the theoretical vortex lattice constant (\! $a_{\Delta}$ )\! given by:
\begin{equation}
    a_{\Delta}=\left(\frac{4}{3}\right)^{1/4} \times\left(\frac{\Phi_0}{B}\right)^{1/2}=1.075\times\left(\frac{\Phi_0}{B}\right)^{1/2}
\end{equation}

\subsection{Magneto-transport setup}
\begin{figure*}[hbt]
	\centering
	\includegraphics[width=16cm]{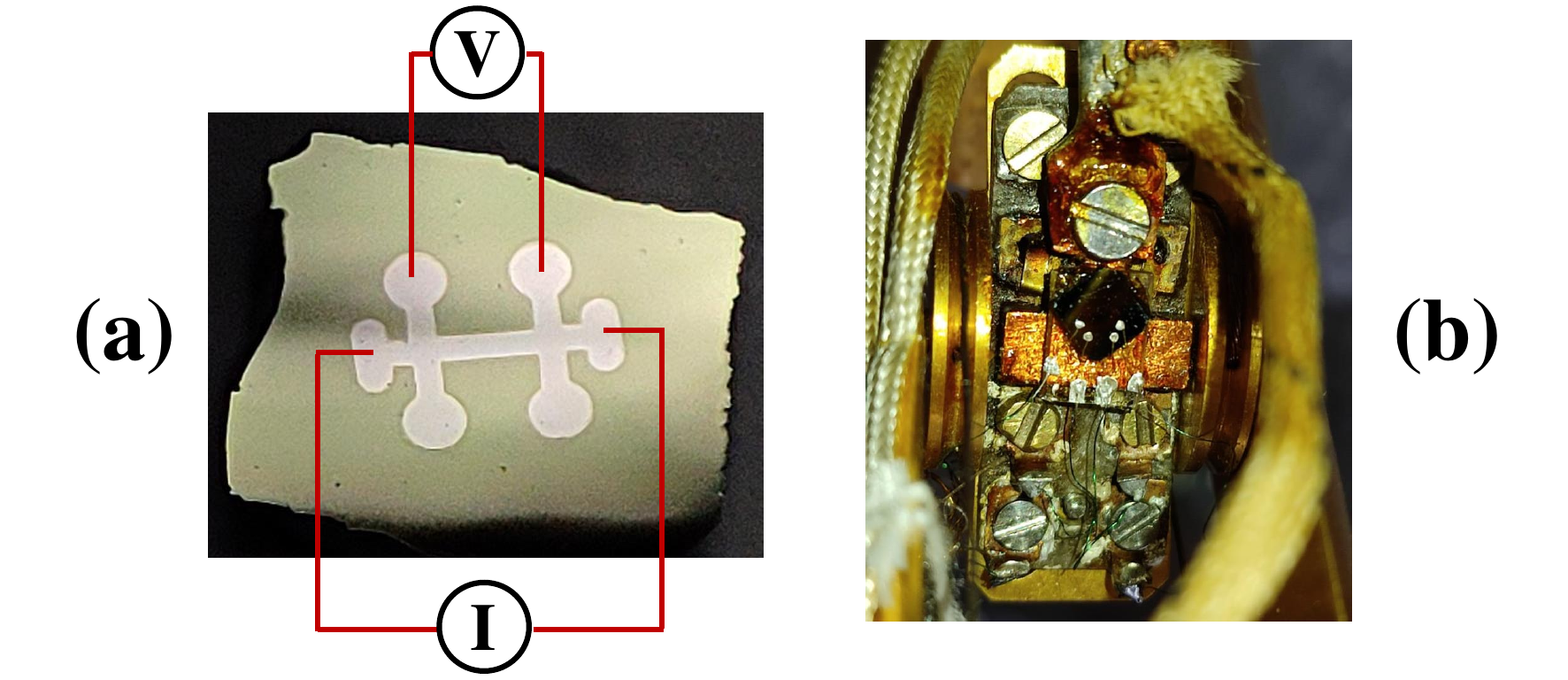}
	\caption[Hall-bar geometry for four-probe magneto-transport setup]{
		\label{fig:4probe-hallbar} 
	\textbf{(a)} Schematic of Hall-bar geometry with four-probe contacts for current (I) and voltage (V) measurements. \textbf{(b)} Sample with four-probe contacts placed in the measurement setup. (Photo courtesy: Dr. Subhamita Sengupta).
	}
\end{figure*}

Transport measurements were done by a standard four-probe technique using a current source and nanovoltmeter. To improve the sensitivity, samples were patterned in the shape of a hall bar ( Fig.~\ref{fig:4probe-hallbar} ). The bridge between the two voltage probes was $0.3$ $mm$ wide ( $w$ ) and $1.3$ $mm$ long ( $L$ ).
\vspace{1 cm}
\vspace{0.5 cm}
\par
The protocol for making contacts for the magneto-resistance measurements for MoGe samples is described briefly. To avoid excessive heating in the case of direct soldering at the sample surface, contacts are made in two steps. The sample is attached to a slightly bigger glass slide. At first, thin gold wires are attached to the voltage and current contact pads using \href{https://www.epotek.com/docs/en/Datasheet/E4110.pdf}{two-component epoxy from EPOTEK}. The other end of the gold wire is glued by epoxy on a small glass slide glued to the sample. This glass end is then connected to the PCB board via a thin copper wire.

\begin{figure*}[hbt]
	\centering
	\includegraphics[width=16cm]{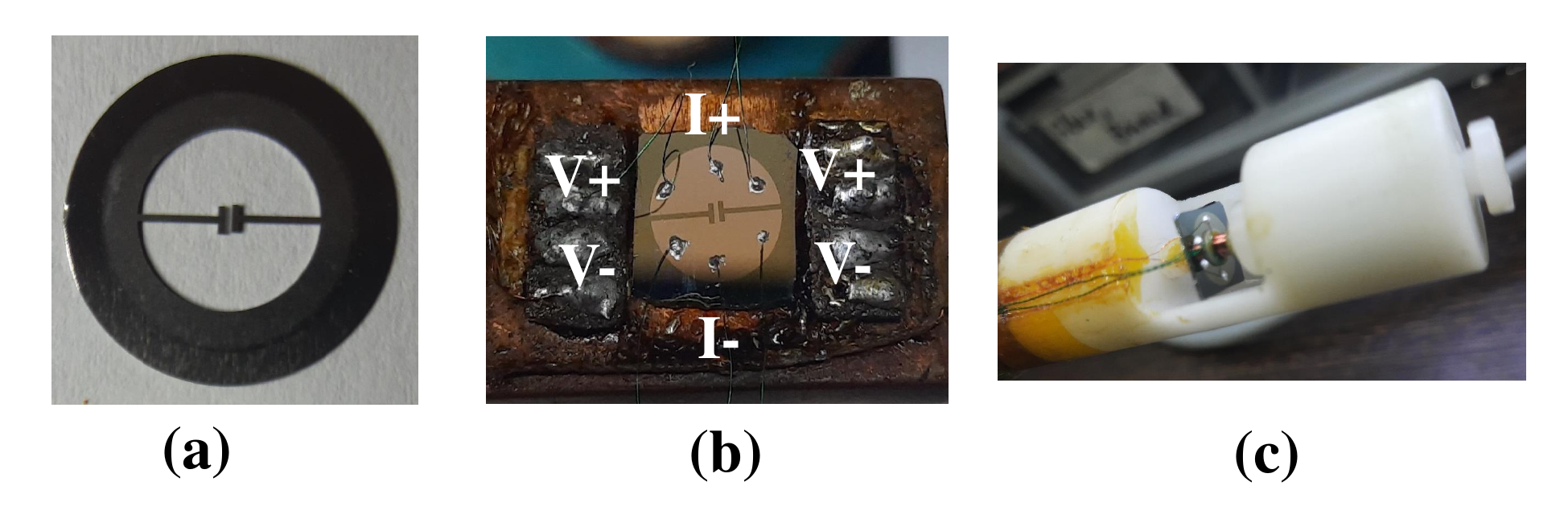}
	\caption[Setup for simultaneous two-coil mutual inductance and magneto-transport measurements]{
		\label{fig:two-coil-4probe-hallbar} 
	\textbf{(a)} Stainless steel mask for depositing thin films for simultaneous two-coil mutual inductance and magneto-transport measurements. \textbf{(b)} Contacts for current (I) and voltage (V) measurements. \textbf{(c)} Sample with four-probe contacts placed inside the two-coil measurement setup.
	}
\end{figure*}
\vspace{1 cm}
\subsection{Setup for simultaneous two-coil mutual inductance and magneto-transport measurements}
In this setup, we have made additional contacts to measure magneto-resistance alongside the two-coil mutual inductance measurements. To create contacts for the resistance measurements, we deposited the thin film using a shadow mask, as shown in Fig.~\ref{fig:two-coil-4probe-hallbar}(a), in a manner that allows the current to flow through a small bridge at the center of the $8$-$mm$-diameter sample. The sample dimensions are kept the same as our two-coil setup, except for the small bridge, aiming to closely mimic the two-coil environments.
\subsection{SQUID-VSM setup}
In Chapter~\ref{ch:vortex-par-chap}, we conducted characterization of $20$-$nm$-thick \textit{a}-MoGe films grown on both (100) oriented MgO and surface-oxidized Si substrates using a commercial Superconducting Quantum Interference Device-Vibrating Sample Magnetometer\! ( SQUID-VSM ) setup. The SQUID-VSM combines two crucial components: a SQUID, an extremely sensitive magnetometer that operates based on the quantum interference of superconducting currents, and a vibrating sample magnetometer, which enables precise control of the sample's magnetic orientation.
\par
The SQUID measures the magnetic flux generated by the sample, while the vibrating sample setup applies a small oscillating magnetic field to the sample, facilitating the measurement of its magnetic response as a function of field strength and temperature. This remarkable technique excels at detecting even the tiniest magnetic signals, rendering it invaluable for studying superconductors, magnetic materials, and other systems with distinctive magnetic properties. The SQUID-VSM measurements yield valuable insights into various aspects of magnetic behavior, including the determination of critical temperatures, characterization of superconducting and magnetic phase transitions, analysis of hysteresis phenomena, and quantification of magnetic moments.
\par
\begin{figure*}[hbt]
	\centering
	\includegraphics[width=16cm]{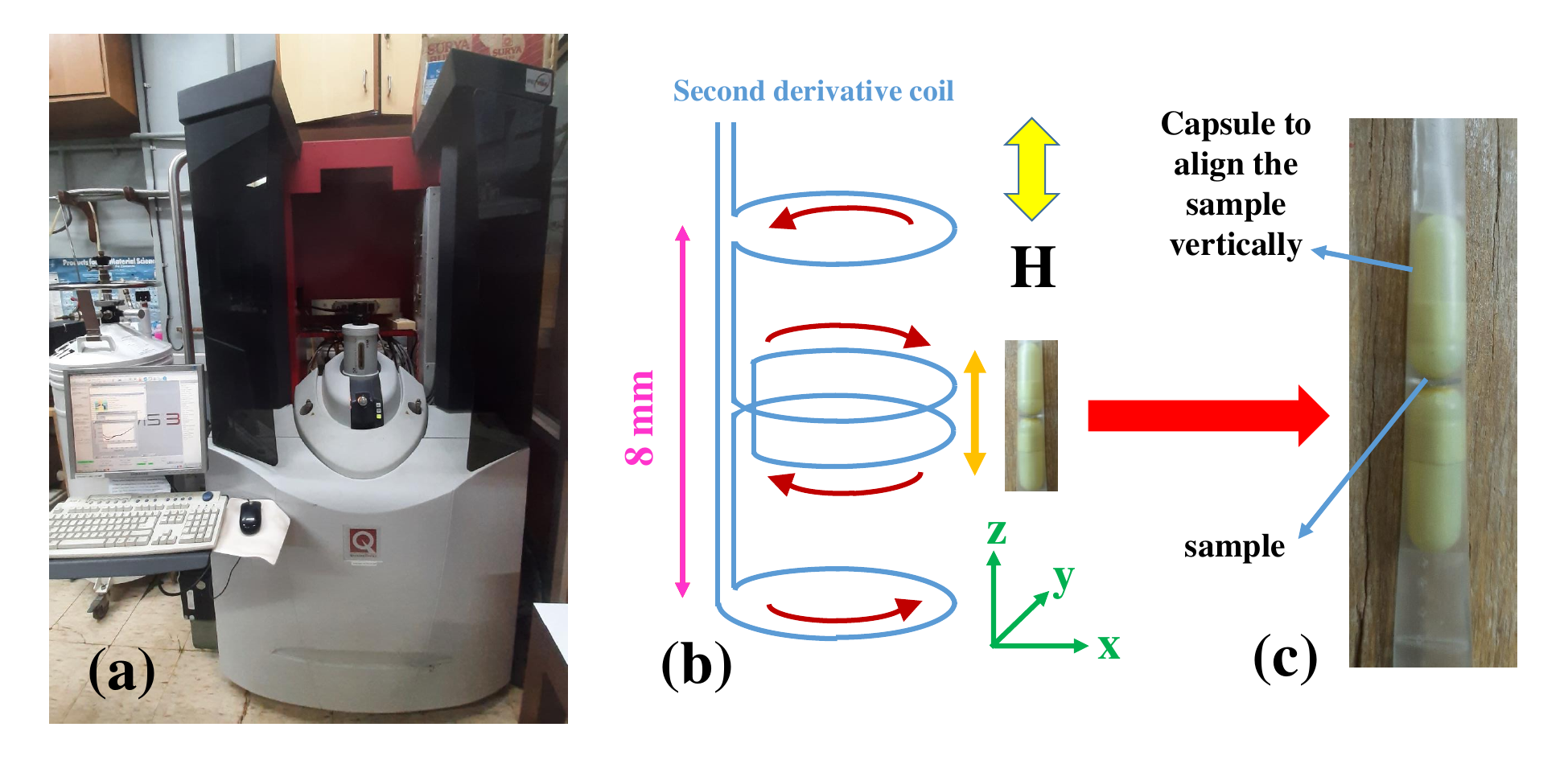}
	\caption[SQUID-VSM setup]{
		\label{fig:squid-vsm} 
	\textbf{(a)} The assembly of the SQUID-VSM setup. \textbf{(b)} The schematic illustrates the pickup coil geometry in the SQUID-VSM systems used, with the direction of current indicated by brown arrows. The vertical orange arrows approximate the length and direction of sample movement during measurement. (Adapted from Ref.~\citenum{boekelheide_artifacts_2016}). \textbf{(c)} Sample is aligned perpendicular to the applied magnetic field by sandwiching between two plastic capsules.
	}
\end{figure*}
For our experiments, we employed a \href{https://www.qdusa.com/products/mpms3.html}{SQUID-VSM system manufactured by Quantum Design}. This high-quality instrument allows measuring samples within a temperature range of $300$ $K$ to $2$ $K$ while subjecting them to magnetic fields of up to $70$ $kOe$. To ensure that the sample is aligned perpendicular to the external magnetic field, it is sandwiched between two empty plastic capsules ( Fig.~\ref{fig:squid-vsm}(c) ) inside a straw. The straw, containing the sample, is securely attached to a carbon-fiber rod, which is then inserted into a helium flow cryostat. Finally, the sample assembly is positioned between two pick-up coils to facilitate the magnetic measurements. During the measurements, we applied a constant rate of vibration to the sample ( with a peak amplitude of $3$ $mm$ and a frequency of $40$ $Hz$ ), allowing us to capture the resulting change in flux using the SQUID, and thus enabling the determination of the sample's magnetic moment.

	\chapter{Our model system: amorphous Molybdenum Germanium thin films}\label{ch:MoGe-sample}
\chaptermark{Our model system: \textit{a}-MoGe thin films}
\section{Sample details}
Our samples consist of amorphous Molybdenum Germanium ( \textit{a}-MoGe ) thin films of different thicknesses. The amorphous phase can be considered the most disordered state of a solid. Similar to their crystalline counterparts, lattice vibrations or phonons also occur in amorphous solids. An amorphous solid can be thought of as the limit where the unit cell becomes infinitely large~\cite{thorpe_phonons_1976}, and the Brillouin zone reduces to just the point $\bm{k} = 0$. Notably, the electron-phonon coupling is found to be stronger for amorphous superconductors, with the superconducting gap ( $\Delta(0)$ ) far exceeding the Bardeen-Cooper-Schrieffer ( BCS ) ratio ( $\Delta(0)/k_BT_c\sim 1.76$ ), and in some cases, they exhibit $T_c$ values higher than their crystalline counterparts~\cite{bergmann_amorphous_1976,tsuei_amorphous_1981}. Thanks to their exceptionally short electronic mean free paths, these amorphous materials serve as excellent model systems for exploring the impact of disorder on superconductivity~\cite{collver_superconductivity_1973,bergmann_amorphous_1976,finkelshtein_superconducting_1987,finkelstein_suppression_1994,stewart_superconducting_2007,spathis_nernst_2008,liu_dynamical_2011,proslier_atomic_2011,shammass_superconducting_2012,ivry_universal_2014}. The emergence of novel vortex phases in these amorphous superconductors has become an area of growing interest~\cite{berghuis_dislocation-mediated_1990,yeh_high-frequency_1997,okuma_vortex_2001,guillamon_direct_2009,roy_melting_2019}. Additionally, the low pinning strength of amorphous superconductors suggests a highly soft vortex lattice, making them ideal candidates for investigating the Berezinskii-Kosterlitz-Thouless-Halperin-Nelson-Young ( BKTHNY ) phase transition~\cite{ryzhov_berezinskiikosterlitzthouless_2017}.
\par
The \textit{a}-MoGe samples were grown by pulsed laser deposition ( PLD ) technique, on surface-oxidized Si and ($100$) oriented MgO substrates. Zero-field superconducting transition temperatures ( $T_c$ ) varied from $7.5$ $K$ to $0.7$ $K$ for thicknesses $40$ $nm$ to $1.8$ $nm$. During the PLD run, an arc-melted Mo$_{70}$Ge$_{30}$ target was ablated using a $248$ $nm$ excimer laser keeping the substrate at room temperature. Laser pulses of comparatively high energy density ( $\sim 240$ $mJ/mm^2$ per pulse ) were bombarded to maintain the stoichiometry of the amorphous film close to the stoichiometry of the target. The growth rate was $\sim 1$ $nm/ 100$ pulse.
\par
The disorder in thin amorphous MoGe films is tuned by changing thickness or dimensionality. We can quantify disorder by Ioffe Regel parameter, $k_Fl$ as,
\begin{equation}
    k_Fl=\frac{\hbar (3\pi^2)^{2/3}}{[n_H(25K)]^{1/3}\rho(295K)e^2},
\end{equation}
where $k_F$ is the Fermi wave vector, $l$ is the mean free path, $n_H(25K)$ is the Hall carrier density at $25$ $K$, $\rho(295K)$ is the resistivity at $295$ $K$ and $e$ is the electron charge.
\begin{figure*}[hbt]
	\centering
	\includegraphics[width=16cm]{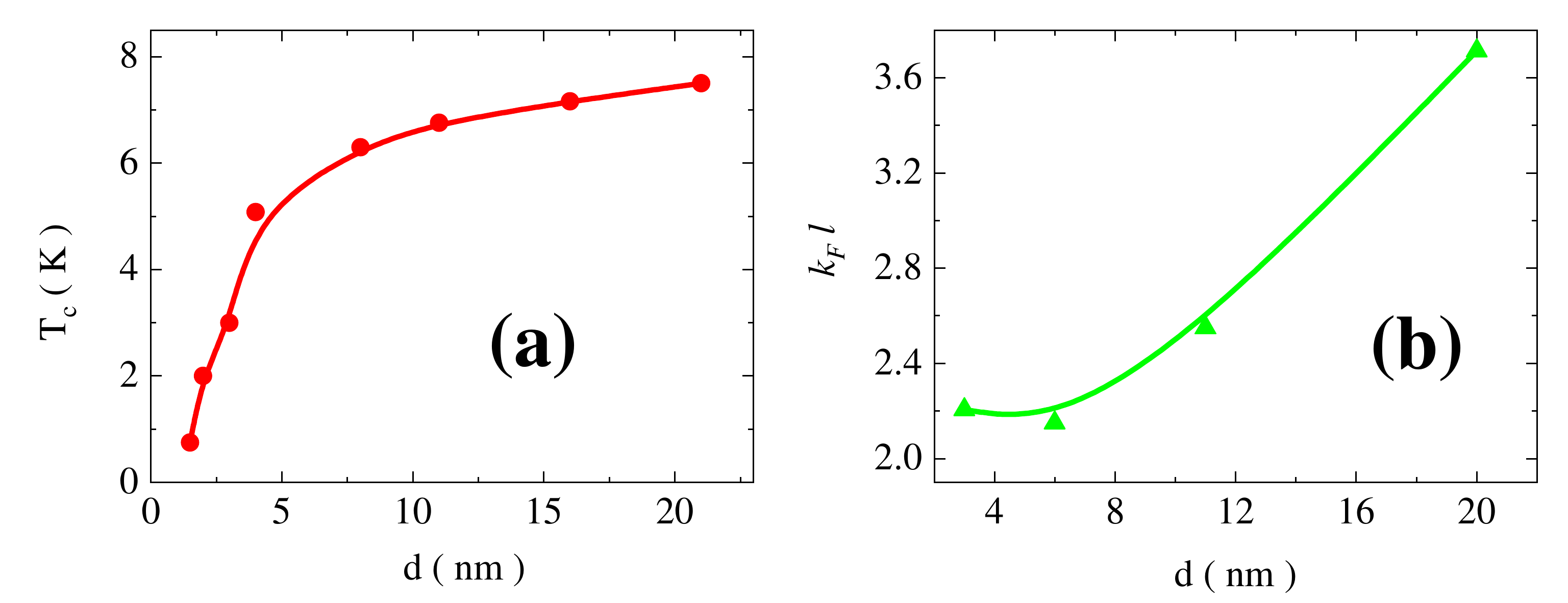}
	\caption[Variation of $T_c$ and $k_Fl$ for \textit{a}-MoGe thin films of different thicknesses.]{
		\label{fig:kfl-MoGe-data} 
	Variation of \textbf{(a)} $T_c$ (connected red circles) and \textbf{(b)} $k_Fl$ (connected green triangles) for \textit{a}-MoGe thin films of different thicknesses.
	}
\end{figure*}
\subsection{Target preparation}
\begin{figure*}[hbt]
	\centering
	\includegraphics[width=16cm]{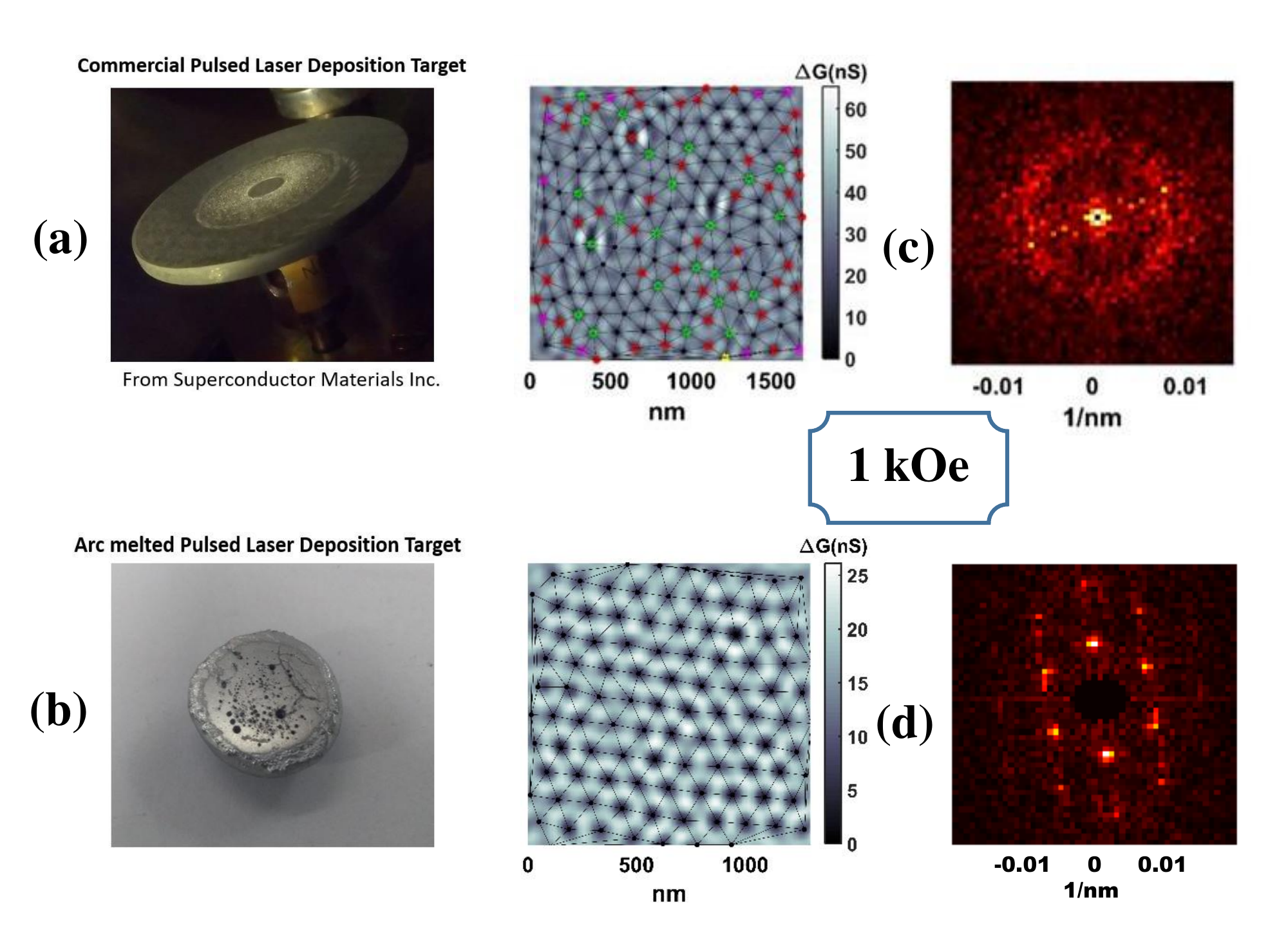}
	\caption[Commercial and arc-melted MoGe target]{
		\label{fig:commercial-arcmelted-target} 
	\textbf{(a)} Commercial MoGe target purchased from \href{https://www.scm-inc.com/}{Superconductor Materials Inc.} and \textbf{(b)} arc-melted MoGe target prepared in Prof. Arumugam Thamizhavel's lab (picture courtesy: Dr. Indranil Roy). \textbf{(c)}-\textbf{(d)} (left) Representative vortex images (conductance map $\Delta G$ recorded at a fixed bias, $V=1.52$ $mV$ for $1$ $kOe$ at 2 K) of $20$-$nm$-thick MoGe films prepared from the commercial and arc-melted target respectively. The black dots represent vortices connected by black lines to illustrate the Delaunay triangulation. The topological defects are indicated by the red, green, magenta, and yellow points, corresponding to 5, 7, 4, and 8-fold coordination, respectively. While topological defects are present in the former image, there is no defect in the visible area for the latter. (right) 2D Fourier transforms (FT) of the corresponding images on the left, with a scale bar in $nm^{-1}$. FT image for the film made from the commercial target diffuses to a ring, while that from the arc-melted target shows six spots. Vortex images were taken by Dr. Indranil Roy.
	}
\end{figure*}
\subsubsection{Commercial target}
We started with commercial MoGe targets purchased from \href{https://www.scm-inc.com/}{Superconductor Materials Inc.}. But the problem with commercial ones is that there are impurities including oxygen residue as observed in EDAX data. Commercial targets are typically prepared using press-and-sinter powder metallurgy~\cite{mpif_videos_conventional_2018}, where powdered metal components are mixed and compacted to a desired shape before being subjected to sintering or high-temperature heating. During the powdering procedure of the metal components, oxides may inadvertently mix in when the molten metal is passed through high-pressure water, resulting in the formation of powder via the water atomization process. The presence of higher oxygen residue or impurities does not significantly affect the $T_c$, but it appears to enhance pinning, resulting in stronger pinning forces. As we are dealing with extremely low pinned samples, it is preferable to have the least amount of impurities.
\subsubsection{Home-made target}
So, we turned to a homemade target which was prepared in \href{https://www.tifr.res.in/crystalgrowth/}{Prof. Thamizhavel’s lab} using a tetra-arc furnace. At first, a commercially available Molybdenum rod was taken along with small pieces of Germanium. The components were taken in stoichiometric amounts ( $Mo: Ge \equiv 3:1$ ) and placed on a copper hearth. This hearth is water-cooled and plays the role of an anode in the arc-melting procedure employed to make the MoGe target buttons. The hearth along with the starting materials are loaded onto a chamber in the tetra-arc furnace. The furnace is equipped with a combination of high vacuum rotary and turbo pumps, which bring down the pressure inside the chamber to $\sim 10^{-6}$ $mbar$. The chamber was flushed three times with gaseous Argon to drive out any impurities that might have been left. The arcs, made of Tungsten, acting as cathodes, are ignited by sending a current of $\sim 50$ $A$. When the temperature of the chamber rises to $\sim 3000$ $K$, the materials kept on the hearth start melting. The melting was performed in an Ar atmosphere with pressure slightly less than $1$ $atm$. The chamber is heated for about 10 minutes and then rapidly cooled to achieve an amorphous material. The end product takes a dome shape, with a concave upper surface. The alloy obtained is then flipped upside down, placed on the hearth, and re-melted again. The process is carried out a total of four times to ensure homogeneity. Two such targets were made of a diameter of around $18$ $mm$ and a thickness of around $3.4 - 3.5$ $mm$. Subsequently, the MoGe was sliced by wire cutting and then the top surface was polished with fine emery paper to obtain a flat, smooth ablating surface which helps to ascertain a fixed plume direction during the laser shots. No oxygen residue was found in EDAX for films synthesized from a homegrown MoGe target.
\par
It is evident from Fig.~\ref{fig:commercial-arcmelted-target} that the MoGe thin films produced from the commercial target exhibit stronger pinning, manifesting in the form of topological defects in the vortex image. In contrast, the films made from the arc-melted target show no defects in the visible area of the vortex image, implying significantly lower pinning.

\section{Hexatic phase in \textit{a}-MoGe}
Melting of 3D VL can happen by increasing temperature or magnetic field, similar to the melting of any crystalline solid via a first-order phase transition through the “Lindemann~\cite{lindemann_uber_1910} criterion” where the lattice vibration amplitude exceeds a certain fraction of the lattice constant. However, for 2D solid melting can happen through an alternate route via two continuous phase transitions mediated by topological defects as predicted by Berezinskii-Kosterlitz-Thouless-Halperin-Nelson-Young ( BKTHNY )~\cite{benfatto_berezinskii-kosterlitz-thouless_2013,halperin_theory_1978,young_melting_1979,ryzhov_berezinskiikosterlitzthouless_2017} theory. In fact, this two-step melting was observed for 2D vortex lattice of weakly pinned \textit{a}-MoGe thin film~\cite{roy_melting_2019}. In the first step of melting, thermally excited free dislocations proliferate in the VL, forming an intermediate state between crystalline solid and liquid, which is called Hexatic fluid when the solid ( e.g., VL ) has hexagonal symmetry. Then in the second step, dislocations dissociate into isolated disclinations producing isotropic liquid. Hexatic vortex fluid ( HVF ) differs from regular fluid in that it has short-range positional order like in isotropic vortex liquid ( IVL ), but retains the quasi-long-range orientational order of vortex solid ( VS ). In the HVF state, vortices at low temperatures move extremely slowly, and hence no stark change in pinning properties was observed in the VS-HVF transition during the magneto-transport measurements~\cite{dutta_collective_2020} in \textit{a}-MoGe thin films. Moreover, VL was found to be extremely sensitive to small electromagnetic perturbations in the HVF state in \textit{a}-MoGe~\cite{dutta_extreme_2019}.
\begin{figure*}[hbt]
	\centering
	\includegraphics[width=16cm]{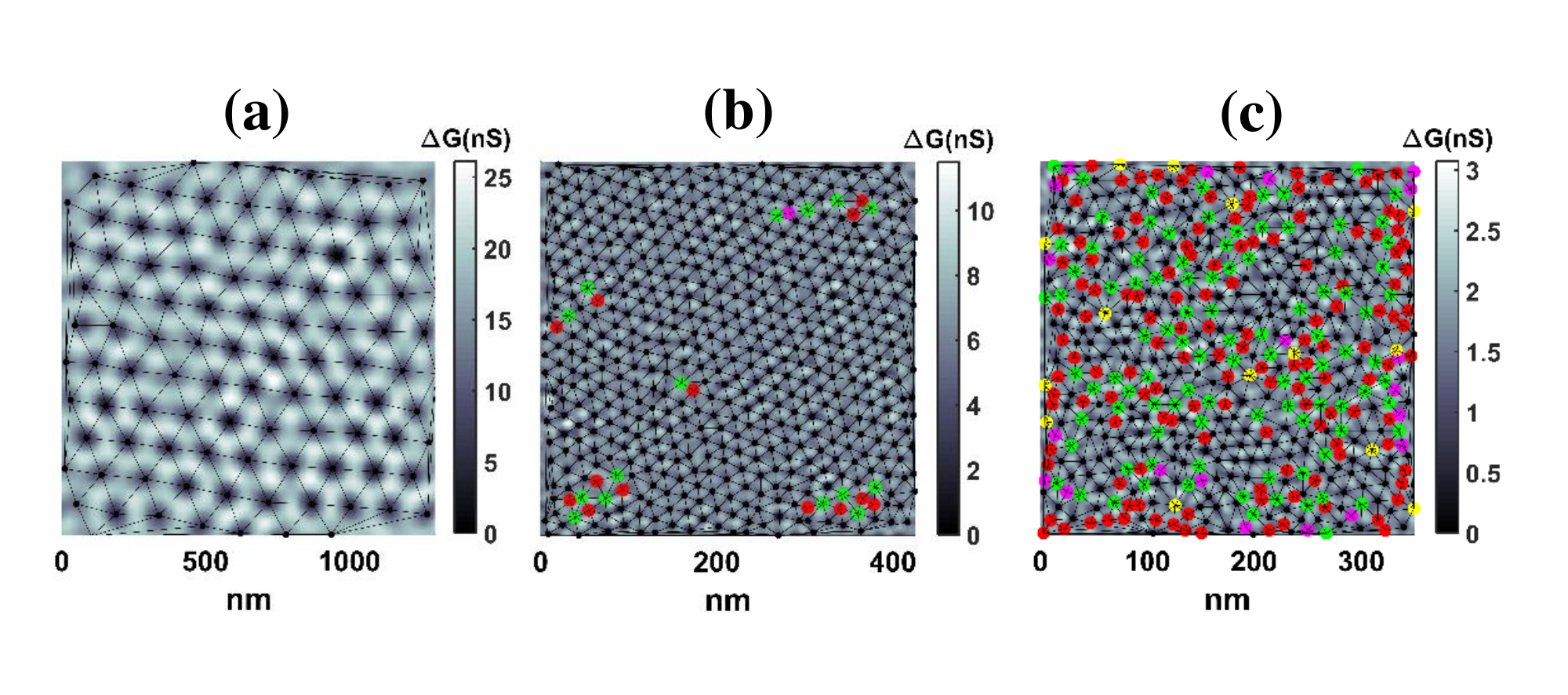}
	\caption[Representative STM images of vortices in vortex solid (VS), hexatic vortex fluid (HVF), and isotropic vortex liquid (IVL) in a $20$-$nm$-thick \textit{a}-MoGe film]{
		\label{fig:VS-HVF-IVL} 
	Representative STM images of vortices in \textbf{(a)} vortex solid (VS), \textbf{(b)} hexatic vortex fluid (HVF), and \textbf{(c)} isotropic vortex liquid (IVL) in $20$-$nm$-thick \textit{a}-MoGe film (adapted from Ref.~\citenum{dutta_collective_2020}).
	}
\end{figure*}
\par
We note that although the TAFF characteristics in the I-V curves undergo nontrivial modifications across the transition from vortex solid to hexatic vortex fluid~\cite{roy_melting_2019}, the flux flow regime of the VL is not significantly altered. This is because the movement of the dislocations, which provides the small mobility of vortices in the TAFF regime, is much slower compared to the vortex velocity in the flux flow regime. Therefore, we can apply the general arguments of collective pinning developed by Larkin and Ovchinnikov ( LO ) to understand the extrapolated $J_c$ obtained from the I-V curve in the flux flow regime.
\begin{figure*}[hbt]
	\centering
	\includegraphics[width=8cm]{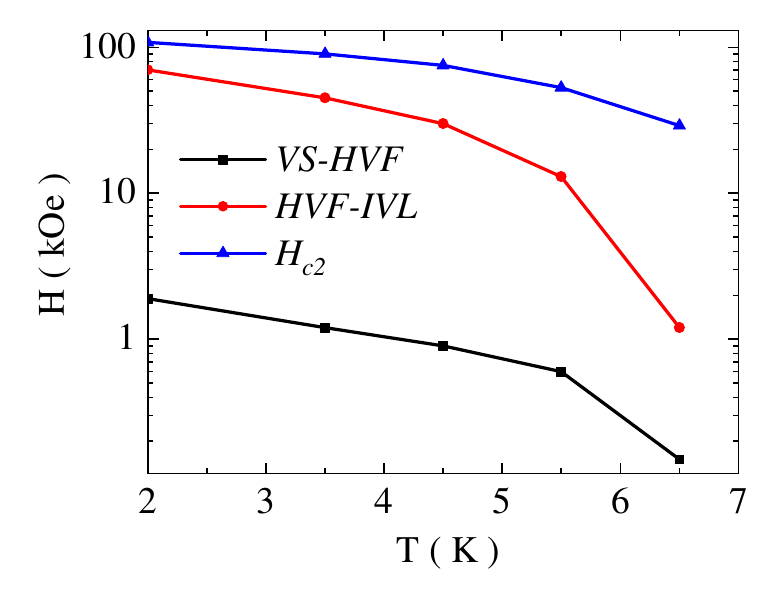}
	\caption[Phase space boundaries in a 20-nm-thick \textit{a}-MoGe film]{
		\label{fig:phase-boundary-20nm-hexatic-paper} 
	Phase space boundaries in a 20-nm-thick \textit{a}-MoGe fim: VS-HVF (connected black squares), HVF-IVL (connected red circles) and $H_{c2}$ (connected blue triangles) respectively (VS: Vortex solid, HVF: Hexatic vortex fluid and IVL: Isotropic vortex liquid) (adapted from Ref.~\citenum{roy_melting_2019}).
	}
\end{figure*}
\par
In 2-D, the LO theory incorporates two fundamental elements: (i) the existence of Larkin domains with short-range order in a VL lacking long-range order and (ii) a finite shear modulus $C_{66}$ of the vortex lattice. In the hexatic state, the characteristic length scale $R_c$ represents the typical separation between two dislocations. On the other hand, in the hexatic state, the modulus of the orientational stiffness, known as the Frank constant $\kappa$, assumes the same role as $C_{66}a_{\Delta}^2$ ( where $a_{\Delta}$ is the vortex lattice constant ) does in a vortex solid~\cite{keim_franks_2007}. Formally, $\kappa$ is the coefficient of the term $\frac{1}{2}\abs{\nabla \theta}^2$ in the free energy density, where $\theta$ represents the bond angle~\cite{halperin_theory_1978}. Therefore, by establishing these two formal associations, we can employ the LO arguments to comprehend the magnetic field variation of $J_c$, which we shall discuss in Section~\ref{sec:peak-effect-Jc-rhoTAFF}.
\begin{figure*}[hbt]
	\centering
	\includegraphics[width=16cm]{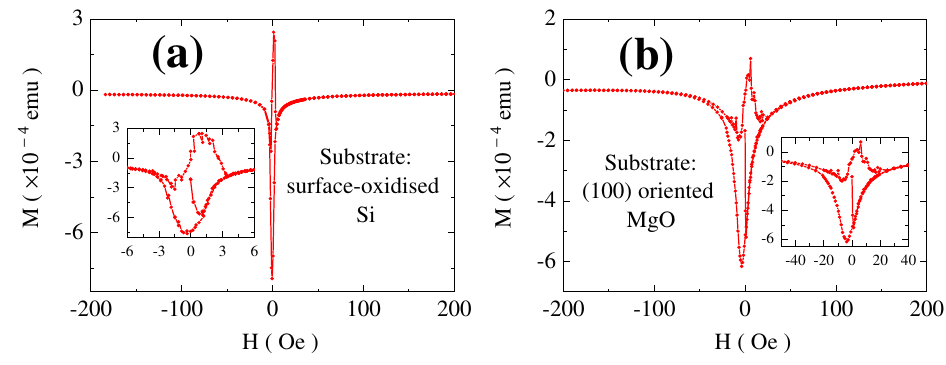}
	\caption[M-H loop in $20$-$nm$-thick \textit{a}-MoGe films]{
		\label{fig:squid-MH-loop-MoGe-on-Si-MoGe} 
	M-H loop for $20$-$nm$-thick \textit{a}-MoGe films grown on \textbf{(a)} surface-oxidized Si and \textbf{(b)} ($100$) oriented MgO.
	}
\end{figure*}
\section{Weakly pinned vortex lattice in \textit{a}-MoGe}

\subsection{Extremely small M-H loop}
The weakly pinned nature of the $20$-$nm$-thick \textit{a}-MoGe thin film is reflected in the very small M-H loop measured using the SQUID-VSM technique ( Fig.~\ref{fig:squid-MH-loop-MoGe-on-Si-MoGe} ). We have done the magnetization measurements on both surface-oxidized Si ( Fig.~\ref{fig:squid-MH-loop-MoGe-on-Si-MoGe}(a) ) and ($100$) oriented MgO ( Fig.~\ref{fig:squid-MH-loop-MoGe-on-Si-MoGe}(b) ). It is observed that the MH-loop is bigger in the case of the MgO substrate than the Si substrate, suggesting a slightly stronger pin in the former. 
\subsection{ZFC and FC data}
\begin{figure*}[hbt]
	\centering
	\includegraphics[width=16cm]{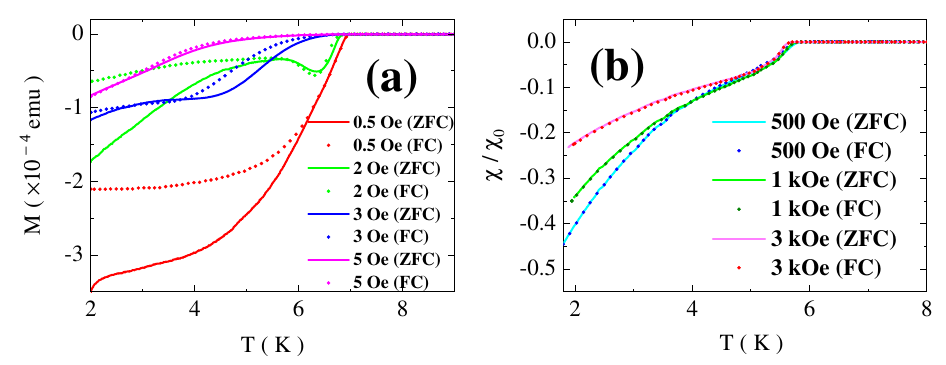}
	\caption[ZFC and FC data in a 20-nm-thick \textit{a}-MoGe film]{
		\label{fig:ZFC-FC-data-20nm-MoGe } 
	\textbf{(a)} Temperature variation of Magnetization ($M$) in the zero-field-cooled (ZFC) and field-cooled (FC) states at $0.5$ $Oe$, $2$ $Oe$, $3$ $Oe$, and $5$ $Oe$ respectively measured on a $20$-$nm$-thick \textit{a}-MoGe film using SQUID-VSM technique. There is a significant gap between the ZFC and FC states at $0.5$ $Oe$, which gradually decreases as the magnetic field is increased. Eventually, the ZFC and FC curves nearly converge for fields at or above $5$ $Oe$. \textbf{(b)} The normalized real part of the ac susceptibility, $\chi$, as a function of temperature for the FC and ZFC states at $500$ $Oe$, $1$ $kOe$ and $3$ $kOe$ ($\chi_0$ is the zero-field susceptibility at $1.5$ $K$) measured using the two-coil mutual inductance setup on a $20$-$nm$-thick \textit{a}-MoGe film (adapted from Ref.~\citenum{roy_melting_2019}). The FC and ZFC curves at each field overlap with each other showing the very weakly pinned nature of the VL. The density of points for the FC curves has been reduced for clarity.
	}
\end{figure*}
The extremely small M-H loop in Fig.~\ref{fig:squid-MH-loop-MoGe-on-Si-MoGe}(a) is corroborated by the field-cooled ( FC ) and zero-field-cooled ( ZFC ) data ( Fig.~\ref{fig:ZFC-FC-data-20nm-MoGe }(a) ) measured on the same sample grown on surface-oxidized Si. In the presence of pinning, the vortex lattice ( VL ) can exhibit multiple metastable states with varying degrees of order. As pinning increases, the susceptibility in the ZFC and FC states begins to display distinct responses, representing two limiting cases. It is observed in Fig.~\ref{fig:ZFC-FC-data-20nm-MoGe }(a) that with increasing field, the difference between ZFC and FC diminishes and they nearly become identical above $5$ $Oe$, signifying extremely weak pinning.
\par
Weak pinning nature was also evident in previous ac susceptibility measurements, where the FC and ZFC data on similar samples~\cite{roy_melting_2019} showed a negligible difference for fields above 500 Oe, indicating very weak pinning in the sample ( Fig.~\ref{fig:ZFC-FC-data-20nm-MoGe }(b) ).
\section{Signatures of Peak effect in \textit{a}-MoGe}
\subsection{Peak in \texorpdfstring{$J_c$}{} and  \texorpdfstring{$\rho_{TAFF}$}{}}\label{sec:peak-effect-Jc-rhoTAFF}
One of the most fascinating findings of our magneto-transport measurements on $20$-$nm$-thick \textit{a}-MoGe films, is the presence of shallow minima in $\rho_{TAFF}$, which is associated with the \enquote{peak effect} in critical current density $J_c$. It is a common feature in many weakly pinned type-II superconductors, where at the boundary between a Bragg glass and a vortex glass~\cite{henderson_metastability_1996}, a dip in $\rho$ is accompanied by a corresponding peak in $J_c$ for $J \gtrsim J_c$. However, in a Bragg glass and the vortex glass, where $\rho_{TAFF}$ is extremely small, the peak effect is typically not detected in measurements with $J \ll Jc$. In contrast, in our study, we observe the peak effect at the boundary between two vortex fluid states, where $\rho_{TAFF}$ has a finite value. Therefore, our first objective is to investigate the internal consistency between the observed variations in $J_c$ and $\rho_{TAFF}$.
\begin{figure*}[hbt]
	\centering
	\includegraphics[width=10cm]{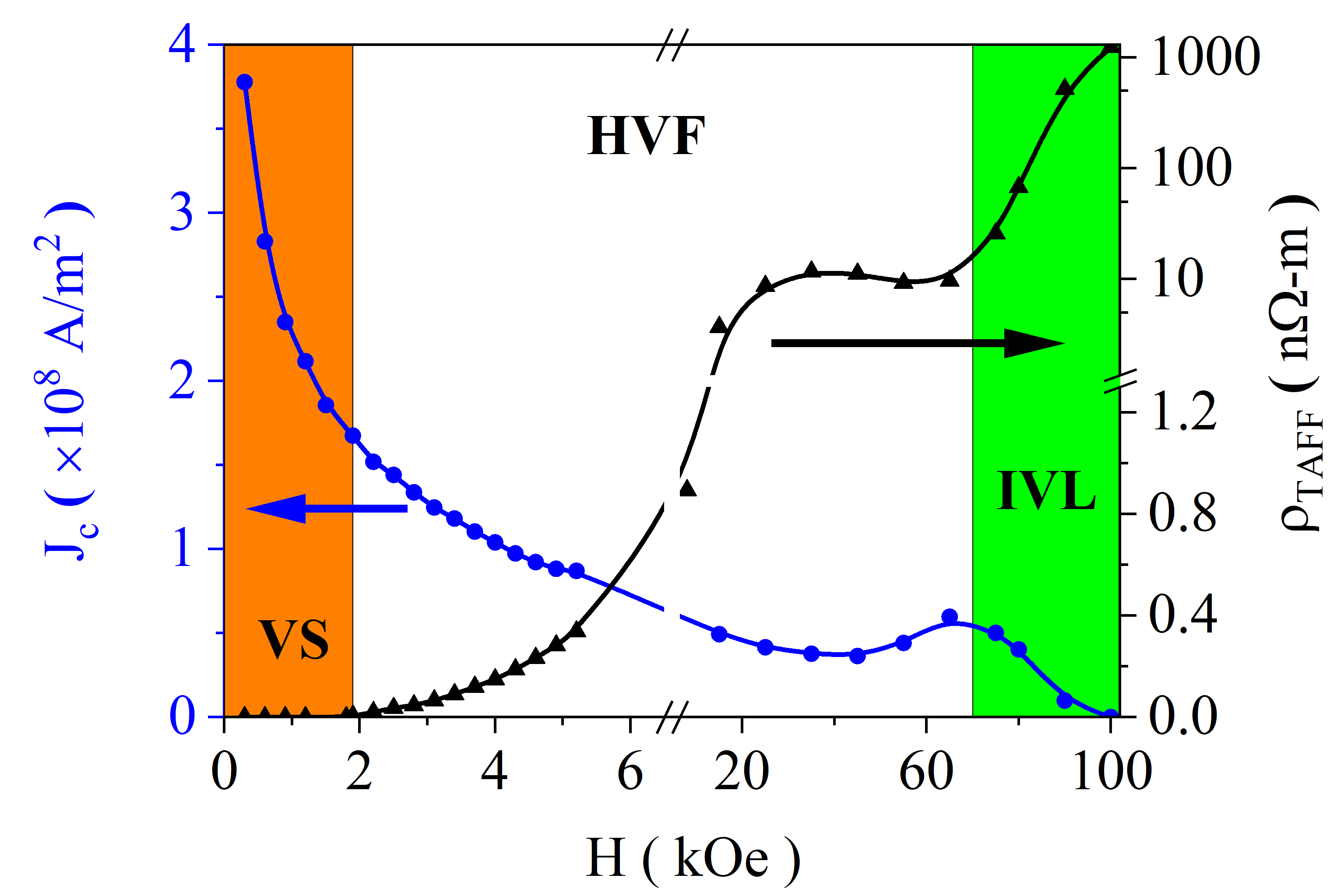}
	\caption[Peak effect in $J_c$ and $\rho_{TAFF}$ in a 20-nm-thick \textit{a}-MoGe film]{
		\label{fig:Jc-rhoTAFF-vs-H} 
	Field variation of $J_c$ (connected blue circles) and $\rho_{TAFF}$ (connected black triangles) at $2$ $K$ for a $20$-$nm$-thick MoGe film. Here, three distinct phases of vortex lattice: vortex solid (VS), hexatic vortex fluid (HVF), and isotropic vortex liquid (IVL) are shown in orange, white, and green color area respectively  (Adapted from Ref.~\citenum{dutta_collective_2020}).
	}
\end{figure*}
\par
As depicted in Fig.~\ref{fig:Jc-rhoTAFF-vs-H}, a shallow minimum is observed in $\rho_{TAFF}$ at the boundary between the hexatic vortex fluid and the isotropic vortex liquid. We demonstrate that this phenomenon arises due to the non-monotonic variation of the flux flow critical current density, $J_c$, which can be explained by the theory of weak collective pinning. The correlation between the peak in $J_c$ and the minima in $\rho_{TAFF}$ is evident; both occur at the same magnetic field. Therefore, the reduction in $\rho_{TAFF}$ reflects the increase in $J_c$, where enhanced collective pinning influences the vortex lattice and reduces thermally activated flux motion in the TAFF region, resulting in an effective decrease in linear resistivity.
\par
We will now elucidate the origin of the maximum in the flux flow critical current, $J_c$. In the case of a well-ordered vortex lattice with an area of $R_c^2$, the pinning force exerted on the vortex lattice can be expressed as $(\langle f^2 \rangle N)^{1/2}$, where $f$ represents the elementary pinning force and $N$ denotes the total number of random pinning centers within this area. Thus, the volume pinning force density is given by, 
\begin{equation}
    F_p=\frac{(\langle f^2 \rangle N)^{1/2}}{R_c^2 d}=\frac{(\langle f^2 \rangle n_A)^{1/2}}{d R_c},
\end{equation}
where $n_A$ is the areal density of pinning centers and $d$ is the thickness of the film. Equating $F_p$ with the Lorentz force density we obtain the flux flow critical current density,
\begin{equation}\label{eq:JcB-MoGe-chap}
    J_cB=\frac{\left(\langle f^2 \rangle n_A\right)^{1/2}}{d R_c}.
\end{equation}
Depending on the nature of pinning, the variation of $f$ follows the general form~\cite{kramer_fundamental_1978}: 
\begin{equation}
    \langle f^2 \rangle n_A\propto b^n (1-b)^2,
\end{equation}
where $b=B/B_{c2}$. Here, $n=1$ corresponds to pinning due to dislocation loops~\cite{pande_firstorder_1976}  and $n=3$ due to impurities and vacancies~\cite{campbell_flux_1972} respectively. We have observed that $n=1$ correctly describes the pinning in a similar $20$-$nm$-thick MoGe sample~\cite{dutta_collective_2020}. Thus from Eq.~\ref{eq:JcB-MoGe-chap} we obtain,
\begin{equation}\label{eq:JcB-B-dep-MoGe-chap}
    J_cB\propto\frac{b^{1/2}(1-b)}{R_c}\Rightarrow J_c\propto\frac{b^{-1/2}(1-b)}{R_c}.
\end{equation}
We can now gain a qualitative understanding of the variation of $J_c$ based on Eq.~\ref{eq:JcB-B-dep-MoGe-chap}. As the magnetic field increases within the hexatic vortex fluid, the density of dislocations gradually rises, resulting in a decrease in $R_c$. Near the phase transition between the hexatic fluid and the isotropic liquid, the rapid decline in $R_c$ can offset the slow decrease in $J_c$ from the numerator, leading to an increase in $J_c$. Generally, this argument remains valid until $R_c$ reaches a limiting value around $R_c \sim a_{\Delta}$, where $a_{\Delta}$ represents the vortex lattice constant. Beyond this limiting value of $R_c$, $J_c$ will decrease due to the reduction of $(\langle f^2 \rangle n_A)^{1/2}$.
\begin{figure*}[hbt]
	\centering
	\includegraphics[width=8cm]{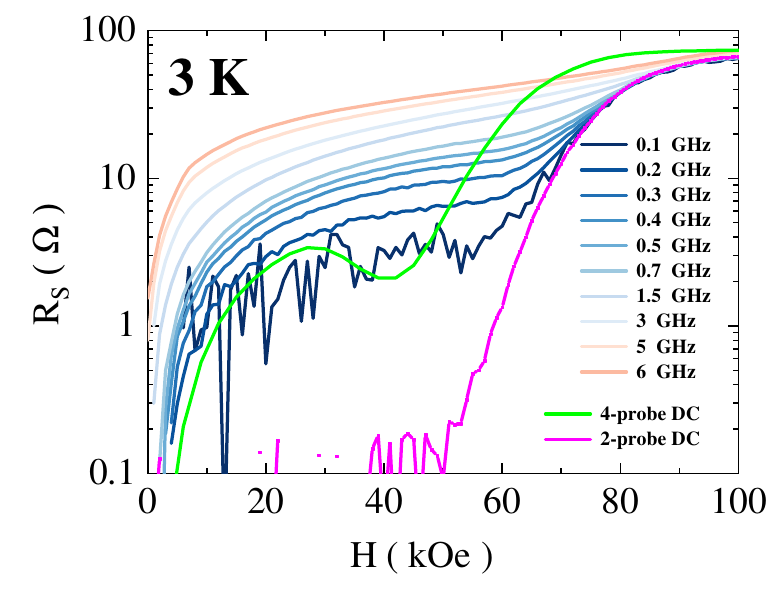}
	\caption[A hint of peak effect in broadband microwave measurements in amorphous $20$-$nm$-thick MoGe film]{
		\label{fig:broadband-microwave-peak-effect} 
	A hint of peak effect in the field dependence of surface resistance $R_s$ for different frequencies at $3$ $K$ in amorphous $20$-$nm$-thick MoGe film using broadband microwave spectroscopy. DC 2-probe measurement was done simultaneously using a current source meter, but not sensitive enough to capture the effect (shown as the pink line in the same plot). For comparison, 4-probe DC resistance, measured separately using hall-bar geometry, is also plotted in the same plot (green line). 
	}
\end{figure*}
\vspace{-0.5 cm}
\subsection{Hint of peak-like feature in broadband microwave spectroscopy}
Figure~\ref{fig:broadband-microwave-peak-effect} illustrates the surface resistance $R_s$ plotted against the magnetic field at $3$ $K$ for various frequencies obtained through broadband microwave spectroscopy using Corbino geometry. Notably, a subtle plateau-like feature resembling the peak effect is observed at $GHz$ frequencies. Bhangale and co-workers~\cite{bhangale_peak_2001,bhangale_peak_2002,bhangale_peak-effect_2002,banerjee_peak_2002} demonstrated the occurrence of the peak effect in the temperature or magnetic field dependence of microwave surface resistivity during the order-disorder transition of the flux-line lattice in DyBa$_2$Cu$_3$O$_{7-\delta}$ and YBa$_2$Cu$_3$O$_{7-\delta}$. They proposed a plausible explanation for this effect within the framework of collective pinning, attributing it to the non-monotonic behavior of the Labusch parameter associated with the peak in the critical current $J_c$. This explanation is also applicable to the HVF-IVL transition in \textit{a}-MoGe. However, at $3$ $K$, the peak feature is not very prominent, which further diminishes at higher frequencies. For comparison, the 4-probe DC surface resistance, measured on a similar \textit{a}-MoGe film with a thickness of $20$ nm, is plotted on the same graph. Additionally, the 2-probe DC surface resistance, measured simultaneously with microwave spectroscopy, is included, although it is not sensitive enough to capture the peak effect. 
\begin{figure*}[hbt]
	\centering
	\includegraphics[width=16cm]{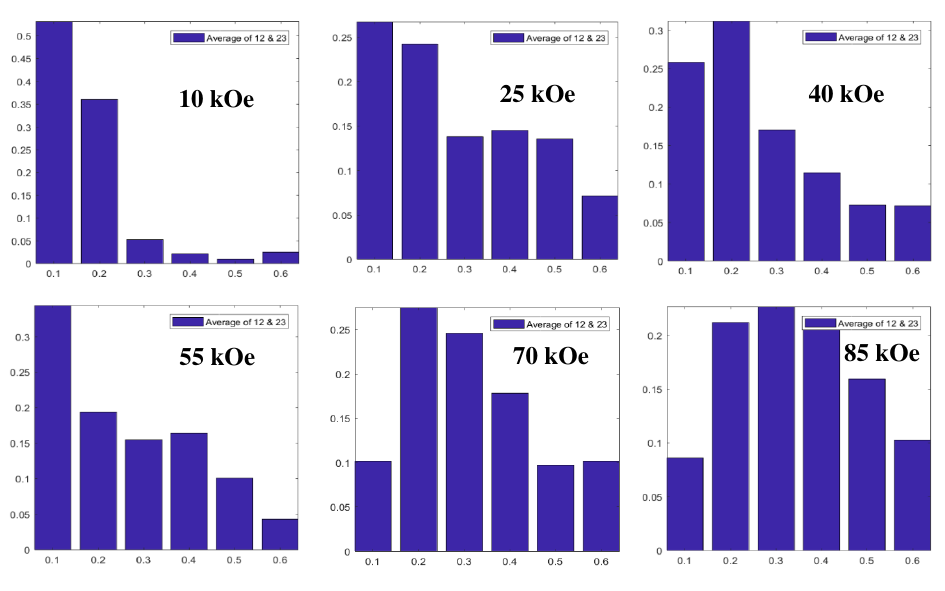}
	\caption[Histogram of the number of vortices with different displacements as a fraction of vortex lattice constant at $2$ $K$ for different fields in amorphous $20$-$nm$-thick MoGe film]{
		\label{fig:vortex-movement-histogram-diff-H} 
	Histogram of the number of vortices with different displacements as a fraction of vortex lattice constant ($a_{\Delta}$) at $2$ $K$ for different fields in an amorphous $20$-$nm$-thick MoGe film. Bars in each histogram are normalized by the total number of vortices so that the sum of all the pillars is unity.
	}
\end{figure*}
\vspace{-0.5 cm}
\subsection{Peak effect in the movement of vortices using scanning tunneling spectroscopy}
In order to examine the dynamics of vortices in the field region exhibiting the peak effect, we monitored their motion using a scanning tunneling microscope at intervals of one hour and 20 minutes. Fig.~\ref{fig:vortex-movement-histogram-diff-H} illustrates a histogram presenting the number of vortices exhibiting various displacements, expressed as a fraction of the vortex lattice constant $a_{\Delta}$, after one hour and 20 minutes. This analysis was conducted through three consecutive scans at a specific temperature ( $2$ $K$ in Fig.~\ref{fig:vortex-movement-histogram-diff-H} ) across various magnetic fields.
\par
Upon examining the first distribution at $10$ $kOe$, it is evident that the average displacement of the vortices is minimal. This is because, at low fields, the influence of vortex-vortex interaction is insignificant compared to the strength of pinning. As the field is increased, the movement of vortices intensifies due to enhanced vortex-vortex repulsion as more vortices come into proximity with each other ( resulting in a decrease in the vortex lattice constant ), as observed at $25$ $kOe$.
\par
With further increases in the field, the motion of vortices appears to be impeded, indicating that pinning forces become increasingly dominant over vortex-vortex interaction. This can be explained by the fact that as the field approaches $H_{c2}$, the energy associated with vortex-vortex interaction decreases rapidly, following $(H - H_{c2})^2$, while the energy related to vortex-impurity interaction varies as $(H - H_{c2})$~\cite{pippard_possible_1969}. Consequently, the former diminishes more rapidly than the latter~\cite{ganguli_real_2016}. As a result, vortices tend to pass through the pinning centers as vortex-impurity interaction starts to prevail over vortex-vortex interaction.
\par
Finally, at $55$ $kOe$, the pinning potential becomes exceptionally strong, leading to the majority of the vortices becoming pinned and resulting in a highly rigid vortex lattice. Beyond $55$ $kOe$, the average displacement of the vortices begins to increase again, indicating that the vortex lattice is approaching the threshold of melting.
\begin{figure*}[hbt]
	\centering
	\includegraphics[width=8cm]{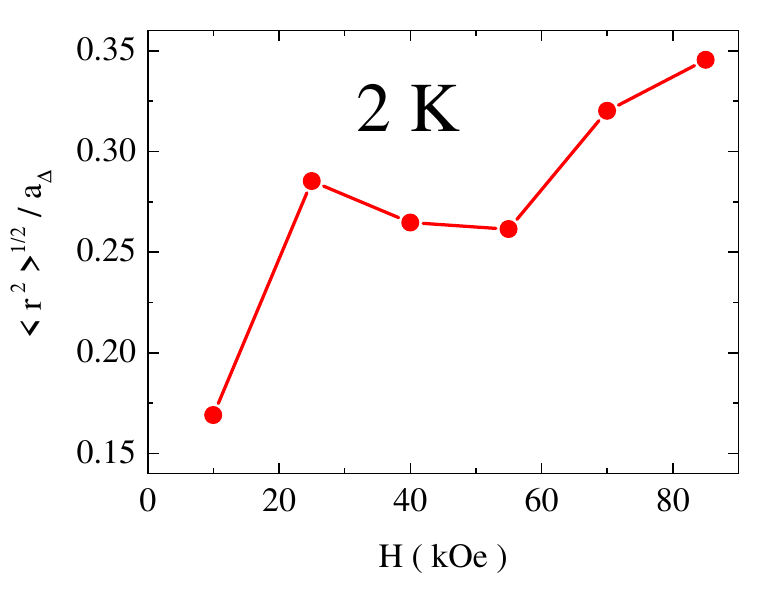}
	\caption[RMS displacement of vortices as a fraction of vortex lattice constant at $2$ $K$ for different fields in amorphous $20$-$nm$-thick MoGe film]{
		\label{fig:vortex-avg-displacement-2K} 
	RMS displacement of vortices (in the unit of inter-vortex distance, $a_{\Delta}$)  at $2$ $K$ for different fields in amorphous $20$-$nm$-thick MoGe film, calculated from Fig.~\ref{fig:vortex-movement-histogram-diff-H}.
	}
\end{figure*}
\par
To consolidate the findings from the histograms presented in Fig.\ref{fig:vortex-movement-histogram-diff-H}, we have plotted the root-mean-square ( RMS ) displacement across all the vortices in Fig.\ref{fig:vortex-avg-displacement-2K} as a function of the magnetic field. As anticipated, this plot exhibits a dip at $55$ $kOe$. This phenomenon serves as a microscopic manifestation of the \enquote{Peak Effect}.
\begin{figure*}[hbt]
	\centering
	\includegraphics[width=16cm]{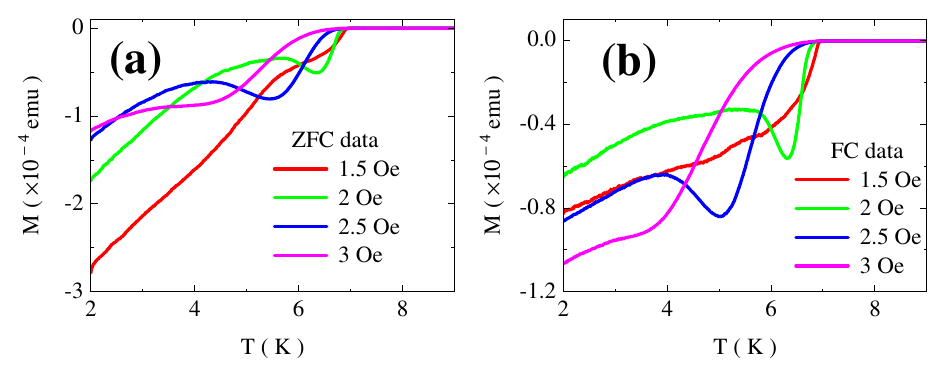}
	\caption[Signature of peak effect at extremely low fields in SQUID-VSM measurements in amorphous $20$-$nm$-thick MoGe film]{
		\label{fig:squid-vsm-peak-effect} 
	Signature of peak effect in both \textbf{(a)} zero-field-cooled (ZFC) and \textbf{(b)} field-cooled (FC) states at extremely low fields in SQUID-VSM measurements in an amorphous $20$-$nm$-thick MoGe film. 
	}
\end{figure*}
\subsection{Unusual peak-like feature in low-field magnetization data}
In Fig.~\ref{fig:squid-vsm-peak-effect}, we present an unusual peak-like signature observed around $H_{c1}$ in a $20$-$nm$-thick \textit{a}-MoGe film grown on an oxidized Si substrate. As depicted in Fig.~\ref{fig:squid-vsm-peak-effect}, both the zero-field-cooled ( ZFC ) and field-cooled ( FC ) states exhibit a dip in the magnetization data. However, this characteristic is only sustained within a narrow range of magnetic field values ( $2$ - $2.5$ $Oe$ ) before leveling off above $3$ $Oe$. We do not yet have a complete understanding of the origin of this \enquote{peak effect} and needs to be explored more.
\section{\textit{a}-MoGe grown on anodic alumina template}
This section delves into the impact of periodic pinning centers on the vortex lattice in \textit{a}-MoGe. Instead of using surface-oxidized Si or MgO as substrates, we have deposited the \textit{a}-MoGe thin film on the background of a nanoporous anodic alumina membrane ( AAM ) template, which acts as an array of pinning centers inside the superconductor. As demonstrated in Section~\ref{sec:Little-Parks-intro}, the presence of a physical hole or void in the superconductor results in oscillatory behavior of the order parameter and related thermodynamic quantities with the magnetic field, known as the Little-Parks effect~\cite{little_observation_1962,parks_fluxoid_1964}. Furthermore, it has been revealed~\cite{thakur_vortex_2009,kumar_origin_2015,roy_dynamic_2017} that in an array of voids, or an antidot array, the cancellation of circulating supercurrents around each antidot can lead to a Little-Parks-like quantum interference ( QI ) effect.
\par
In an antidot array, each vortex experiences two distinct forces: the confining potential of the antidot, which remains nearly constant with the magnetic field, and the field-dependent confining potential arising from the repulsive interaction with surrounding vortices. At matching fields, where the number of vortices in each antidot is an integer, the vortices become tightly confined within a cage formed by neighboring vortices. This phenomenon, known as the vortex matching effect ( VME ), gives rise to a pronounced oscillatory response of the superconductor to the magnetic field. The VME manifests as oscillations in magnetoresistance\hspace{0.2cm} close to the superconducting\hspace{0.2cm} transition temperature ( $T_c$ ), exhibiting a period corresponding to the first matching field ( $H_m$ ). In addition, there are noticeable periodic variations with the magnetic field in all dynamic quantities influenced by the motion of flux lines under external driving, including the magnetic shielding response and the critical current.
\begin{figure*}[hbt]
	\centering
	\includegraphics[width=16cm]{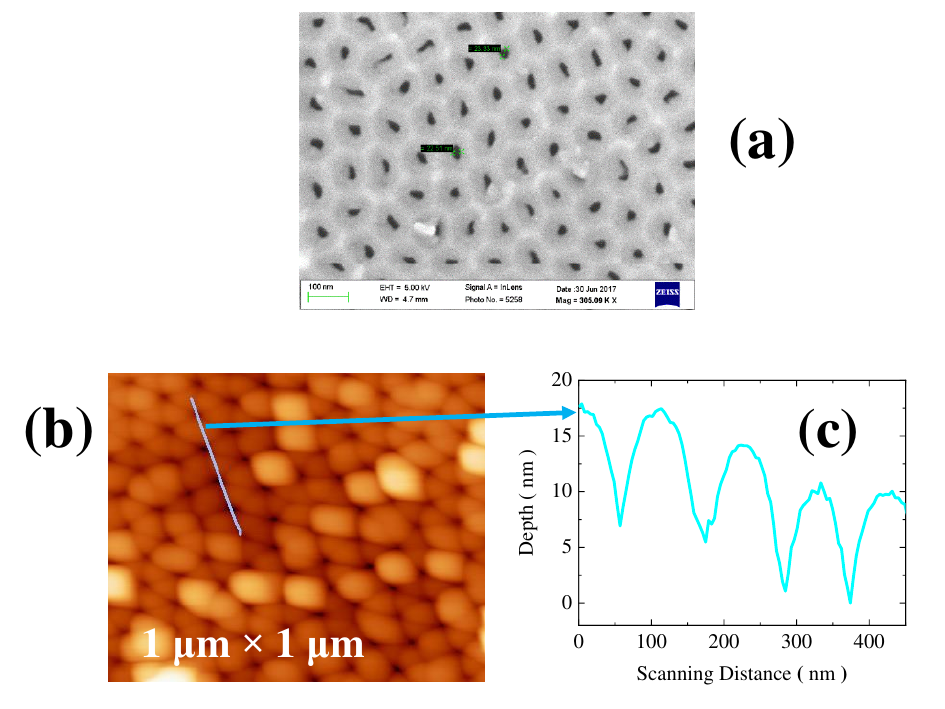}
	\caption[SEM and AFM images of a MoGe sample grown on anodic alumina template]{
		\label{fig:little-parks-MoGe-12000shots-AFM-SEM} 
	$(a)$ Scanning electron microscopy (SEM) images of a representative MoGe sample grown on an AAM template (film thickness $= 55$ $nm$ and $T_c=6.1$ $K$). $(b)$ AFM scan of the same sample. $(c)$ Depth profile of the sample surface along the line shown in $(b)$, passing through multiple pores on the template. AFM scan was performed with help from Mr. Vivas Bagwe.
	}
\end{figure*}
\par
The samples utilized in this study comprise $3$-$mm$-diameter MoGe films grown on a nanoporous anodic alumina membrane \hspace{-0.1cm}(\hspace{-0.1cm} AAM )\hspace{-0.1cm} using\hspace{-0.1cm} pulsed\hspace{-0.1cm} laser\hspace{-0.1cm} deposition\hspace{-0.1cm} ( PLD ). The AAM template, with a thickness of $50$ $\mu m$, pore diameter of approximately $35$ $nm$, and inter-hole distance of around $100$ $nm$, is affixed onto a Si substrate with GE Varnis prior to deposition. In Fig.~\ref{fig:little-parks-MoGe-12000shots-AFM-SEM}(a), a scanning electron micrograph ( SEM ) of a representative sample ( $55$ $nm$ MoGe grown on AAM ) is presented. Fig.~\ref{fig:little-parks-MoGe-12000shots-AFM-SEM}(b) depicts the surface scan of the sample using an atomic force microscope ( AFM ). The depth profile, shown in Fig.~\ref{fig:little-parks-MoGe-12000shots-AFM-SEM}(c), corresponds to a representative line traversing multiple pores seen in Fig.~\ref{fig:little-parks-MoGe-12000shots-AFM-SEM}(b). The depth profile reveals that the pores are filled up to $10$ $nm$ after deposition of around $50$ - $60$ $nm$ of MoGe.
\begin{figure*}[hbt]
	\centering
	\includegraphics[width=16cm]{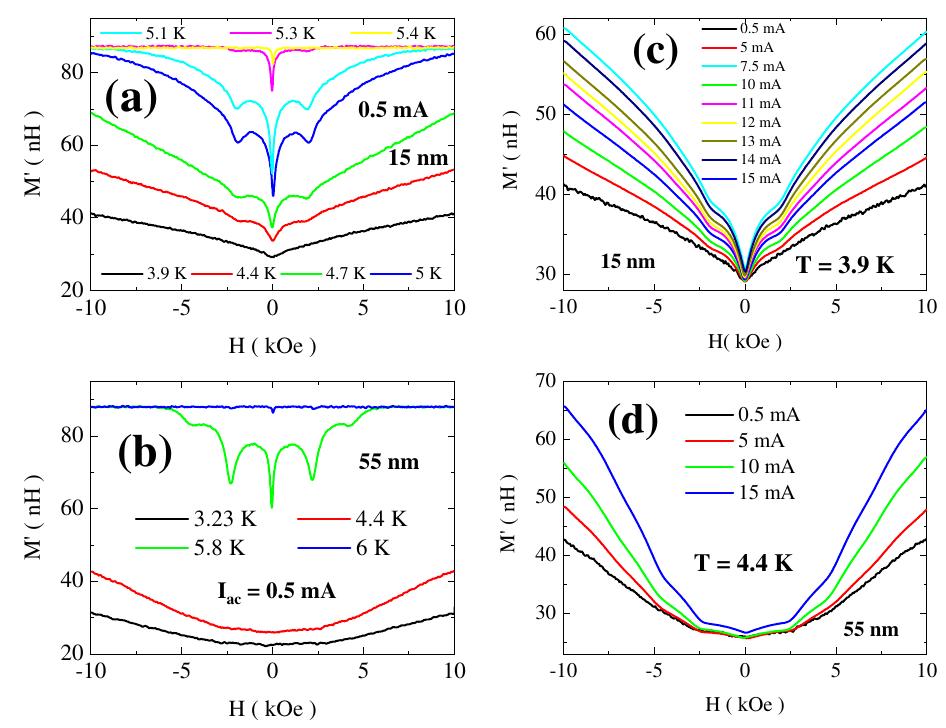}
	\caption[Vortex matching effect in amorphous MoGe film]{
		\label{fig:little-parks-MoGe} 
	\textbf{(a)} - \textbf{(b)} Variation of screening response $M'$ as a function of magnetic field $H$ at different temperatures with $I_{ac} \sim 0.5$ $mA$ in MoGe deposited on AAM template for thicknesses around $15$ $nm$ and $55$ $nm$ respectively (with $T_c$ around $5.3$ $K$ and $6.1$ $K$ respectively). VMEs manifest as minima in $M'$ at matching fields. \textbf{(c)} - \textbf{(d)} $M'-H$ for different oscillation amplitudes $I_{ac}$ at $3.9$ $K$ and $4.4$ $K$ for $15$-$nm$- and $55$-$nm$-thick samples respectively.
	}
\end{figure*}
\par
We\hspace{0.1cm} begin by examining\hspace{0.1cm} the field\hspace{0.1cm} variation\hspace{0.1cm} of the screening\hspace{0.1cm} response, represented by ( $M'-H$ plots ), at different temperatures using a small excitation current ( ac amplitude $I_{ac}$ = $0.5$ $mA$ ), for two distinct samples: (i) a $15$-$nm$-thick MoGe film deposited on AAM, as shown in Fig.~\ref{fig:little-parks-MoGe}(a), with a $T_c$ of approximately $5.3$ $K$; and (ii) a $55$-$nm$-thick MoGe film deposited on AAM, as shown in Fig.~\ref{fig:little-parks-MoGe}(b), with a $T_c$ of around $6.1$ $K$. For both samples, the matching effect is very prominent around near $T_c$. Here, only one matching field is found at around $2$ $kOe$, consistent with the expected matching field, $H_M=2\Phi_0/\sqrt{3}a^2$, $\Phi_0$ represents the flux quantum and $a$ denotes the average lattice constant of the antidot array. This is probably because at high fields, vortices start moving, causing an increasing $M'$ response. As a result, oscillations are suppressed by this background response.
\par
The diminishing effect of decreasing temperature on the matching effect can be explained as follows. The magnitude of the oscillation observed in the screening response is determined by the ratio~\cite{roy_dynamic_2017,kumar_two-coil_2013} $F_I/F$, where (i) $F_I$ represents the repulsive force between vortices, which reduces the compressibility of the vortex lattice when each vortex is surrounded by neighboring vortices, and (ii) $F$ is the restoring force on a vortex due to the confining potential created by the surrounding superconductor. For the antidot array~\cite{kumar_two-coil_2013}, the ratio $F_I/F\sim \xi_{GL}/a$. As the temperature decreases, $\xi_{GL}$ follows the usual GL relation, $\xi_{GL}\propto (1-T/T_c)^{-1/2}$, resulting in less pronounced oscillations. From Fig.~\ref{fig:little-parks-MoGe}(a) and (b), it is evident that oscillation nearly vanishes at $3.9$ $K$ and $4.4$ $K$ respectively.
\par
We now explore the effect of VME on the excitation amplitude $I_{ac}$. In Fig.~\ref{fig:little-parks-MoGe}(c)-(d), we look at the magnetic field evolution of $M'$ for different $I_{ac}$ at the lowest temperature where the oscillations start to diminish. It is evident that the oscillations get enhanced as $I_{ac}$ increase. This is consistent with the transition of the vortex lattice from a Mott-like to a metal-like state, previously observed~\cite{roy_dynamic_2017} in a superconducting NbN film comprising a periodic array of holes.

	\chapter{The role of phase fluctuations in ultrathin \textit{a}-MoGe films: Crossover from Fermionic to Bosonic regime}
\chaptermark{Fermionic to Bosonic crossover in ultrathin \textit{a}-MoGe}
\label{ch:phase_fluctuations}
\section{Introduction}
In the case of conventional superconductors, superconductivity is well explained by the celebrated Bardeen-Cooper-Schrieffer~\cite{bardeen_theory_1957} ( BCS ) theory, which states that the superconducting state is primarily dependent on two phenomena: (i) weak attractive interaction between a pair of electrons with opposite momenta and opposite spins forming a spin-zero Boson-like object called Cooper pairs and (ii) condensation of all such Cooper pairs into a phase-coherent macroscopic quantum state. The superconducting state is given by $\Psi=\Delta e^{i\varphi}$. But things get changed when the disorder is introduced.
\par
Disorder, stemming from impurities, lattice imperfections, or external perturbations, introduces spatial and temporal variations that can have a substantial impact on the behavior of superconductors. With an increase in disorder level, the transition temperature ( $T_c$ ) gradually decreases, eventually leading to a nonsuperconducting ground state. However, even after the destruction of the global superconducting ground state, superconducting correlations continue to significantly influence the electronic properties. These correlations manifest through several intriguing phenomena: finite high-frequency superfluid stiffness above $T_c$~\cite{crane_survival_2007}, giant magnetoresistance peak~\cite{gantmakher_destruction_1998,steiner_superconductivity_2005,sambandamurthy_experimental_2005,baturina_quantum_2007,nguyen_observation_2009} in strongly disordered superconducting films, the persistence of magnetic flux quantization even after being driven into an insulating state observed in strongly disordered Bi films\cite{stewart_superconducting_2007}, and the persistence of pronounced pseudogap~\cite{mondal_phase_2011-1,sacepe_pseudogap_2010,sacepe_localization_2011} at temperatures significantly higher than $T_c$. This interplay between disorder and superconductivity presents an intricate and captivating avenue of study in condensed matter physics. In Section~\ref{sec:effect-of-disorder-intro}, a detailed discussion has been provided regarding the influence of the disorder on superconductivity. I shall however touch upon a few points in this chapter for the sake of maintaining continuity.
\section{Theoretical motivation}
Debate on how the disorder affects superconductivity dates back to the late fifties when Anderson~\cite{anderson_theory_1959} predicted that, in an s-wave superconductor, attractive interaction forming the Cooper pairs would remain unaffected by the presence of non-magnetic impurities. This was then interpreted that $T_c$ would also not be strongly sensitive to disorder. But, subsequent experiments showed that the above is only true in the limit of weak disorder: In the presence of increasingly strong disorder, $T_c$ gets gradually suppressed~\cite{strongin_destruction_1970,goldman_superconductorinsulator_1998}, and eventually, the material is driven into a non-superconducting state at a critical disorder. In fact, the suppression of $T_c$ with an increase in disorder can happen from two origins, namely Fermionic and Bosonic mechanisms~\cite{larkin_superconductor-insulator_1999,goldman_superconductorinsulator_2003,gantmakher_superconductorinsulator_2010,goldman_superconductor-insulator_2010,goldman_superconductor-insulator_2010-1,lin_superconductivity_2015}.
\subsection{Fermionic and Bosonic mechanism}
In the Fermionic route~\cite{anderson_theory_1983,finkelshtein_superconducting_1987,finkelstein_suppression_1994,maekawa_localization_1982,maekawa_upper_1983}, the repulsion among the oppositely charged electrons becomes dominant over the attractive pairing interaction due to the loss of effective screening with an increase in disorder, and thus the mean-field transition temperature is suppressed,  and at a critical disorder superconductivity is destroyed. On the other hand, in the Bosonic mechanism~\cite{gold_dielectric_1986,fisher_presence_1990}, disorder scattering diminishes superfluid density\! ( $n_s$ ) and hence superfluid phase stiffness $J_s$  ( $\sim n_s/m^*$ ). As a result, the phase coherent superconducting state gets vulnerable to phase fluctuations~\cite{emery_importance_1995,emery_superconductivity_1995} and eventually can get destroyed due to strong phase fluctuations even when the pairing amplitude remains finite~\cite{ghosal_role_1998,ghosal_inhomogeneous_2001,bouadim_single-_2011}. Experimentally, this manifests as a persistence of the superconducting gap in the electronic excitation spectrum, known as the pseudogap, even after the global superconducting state is destroyed~\cite{sacepe_pseudogap_2010,sacepe_localization_2011,mondal_phase_2011,dubouchet_collective_2019}. However, recent studies indicate that these two mechanisms need not be exclusive: The interplay between quantum phase fluctuations, localization effects, and disorder-induced spatial variations in electronic properties disrupts the conventional distinction between amplitude- and phase-driven pathways. The same system can follow the Fermionic route at moderate disorder and crossover to a Bosonic scenario at a stronger disorder~\cite{sacepe_disorder-induced_2008,chand_phase_2012,burdastyh_superconducting_2020}. It is therefore interesting to investigate whether a system can follow the Fermionic route all the way to the disorder level where the superconducting ground state is completely destroyed.
\subsection{Reason for choosing \textit{a}-MoGe thin film: popular candidate for Fermionic route}
Several amorphous films of conventional superconductors including our sample \textit{a}-MoGe have long been observed to follow the Fermionic route towards the destruction of superconductivity~\cite{graybeal_localization_1984,goldman_superconductor-insulator_2010,gantmakher_superconductorinsulator_2010,lin_superconductivity_2015}. However, the above evidence was primarily through the transport measurements, where the variation of sheet resistance ( $R_s$ ) with $T_c$ is consistent with the Fermionic theory proposed by Finkel'stein~\cite{finkelshtein_superconducting_1987,finkelstein_suppression_1994}. Nevertheless, the presence of quantum phase fluctuations at very low thicknesses has also been recognized~\cite{yazdani_superconducting-insulating_1995}. Our motivation was to see if \textit{a}-MoGe still shows Fermionic behaviour even if we increase the disorder or if there is a transition to Bosonic regime at stronger disorder.
\section{Samples used in this work}
The \textit{a}-MoGe thin films used in this study were grown on oxidized silicon substrates using pulsed laser deposition technique ablating a homegrown Mo\textsubscript{70}Ge\textsubscript{30} arc-melted target. Film thickness ( $d$ ) was varied between $20$ $nm$ to $1.8$ $nm$ with $T_c$ varying between $7$ $K$ to $1.8$ $K$ respectively. For $d > 10$ $ nm$, the thickness of the film was directly measured using a stylus profilometer whereas for thinner samples it was estimated from the number of laser pulses using two films with $d > 10$ $nm$ grown before and after the actual run for calibration.
\section{Experimental data}
In this work, we have explored \textit{a}-MoGe films of varying thickness using a combination of low-temperature penetration depth ( $\lambda$ ), scanning tunneling spectroscopy ( STS ), and magneto-transport measurements. Magneto-transport and penetration depth measurements were performed in \textsuperscript{3}He cryostats operating down to $300$ $mK$. STS measurements were performed using a home-built scanning tunneling microscope  ( STM ) operating down to $450$ $mK$ and in magnetic fields up to $90$ $kOe$.
\begin{figure*}[hbt]
	\centering
	\includegraphics[width=16cm]{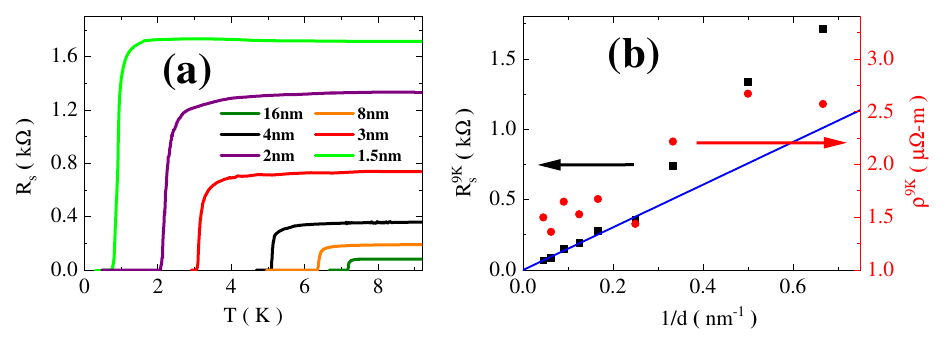}
	\caption[Sheet resistance $R_s$ vs $T$ for \textit{a}-MoGe films with different thicknesses, and variation of normal state sheet resistance and resistivity as a function of inverse thickness]{
    \textbf{(a)} $R_s$ vs $T$ for \textit{a}-MoGe films with 
    different thicknesses. \textbf{(b)} Variation of $R_s^{9K}$ (black squares) and $\rho_N^{9K}$ (red circles) as a function of $1/d$, where $d$ is the thickness of the film. The blue line shows the linear $R_s^{9K}-1/d$ variation for thickness down to 4 nm.}
 \label{fig:Rs-rho-diff-d} 
\end{figure*}
\subsection{Transport data}
At first, we present the results from transport measurements on samples of varying thicknesses. We have done all the measurements on standard hall-bar geometry designed by shadowing masks for better sensitivity. Fig.~\ref{fig:Rs-rho-diff-d}(a) shows the sheet resistance, $R_s$, as a function of temperature for samples with different thicknesses. $R_s$ is defined as follows:
\begin{equation}\label{eq:Rs-rho-R-relation}
    R_s=\frac{\rho}{d}=R\frac{w}{l},
\end{equation}
where $\rho$ is the resistivity, $d$ is the thickness, $R$ is the resistance, $w$ is the width, and $l$ is the length of the sample. As $R$ is proportional to the length and inversely proportional to the width of the sample geometry, it is evident from Eq.~\ref{eq:Rs-rho-R-relation} that $R_s$ is independent of the area of the film ( i.e., $l\times w$ ), but not the thickness of the film. Hence, it is an ideal parameter to characterize resistive behavior and quantify disorder in amorphous 2D thin films of varying thicknesses~\cite{goldman_superconductorinsulator_1998}. In our sample \textit{a}-MoGe, similar to other amorphous superconducting thin films, thickness is used as the tuning parameter for modulating the level of disorder~\cite{graybeal_localization_1984,strongin_destruction_1970,raffy_localization_1983}. In the case of amorphous superconductors such as \textit{a}-MoGe thin films, defects in the substrate affect the films more as we go to lower thickness, whereas the effect of the substrate is screened for thicker samples. Hence, the inverse of thickness has been conventionally taken as a measure of disorder in amorphous superconducting thin films~\cite{strongin_destruction_1970}. In Fig.~\ref{fig:Rs-rho-diff-d}(b), variation of normal state sheet resistance ( taken as the sheet resistance at $9$ $K$, $R_s^{9K}$ ) and normal state resistivity ( taken as the resistivity at $9$ $K$, $\rho_N^{9K}$ ) are plotted as a function of the inverse of the thickness ( $1/d$ ). For $d > 4$ $nm$, $R_s^{9K}$  varies linearly with $1/d$, showing that the increase in the sheet resistance is primarily a geometric effect ( Fig.~\ref{fig:Rs-rho-diff-d}(b) ). Below this thickness, $R_s^{9K}$  follows an upward trend and the corresponding resistivity ( $\rho^{9K}$ ) shows an increase from approximately $1.5$ $\mu \Omega-m$ to $2.6$ $\mu \Omega-m$.
\par
\begin{figure*}[hbt]
	\centering
	\includegraphics[width=16cm]{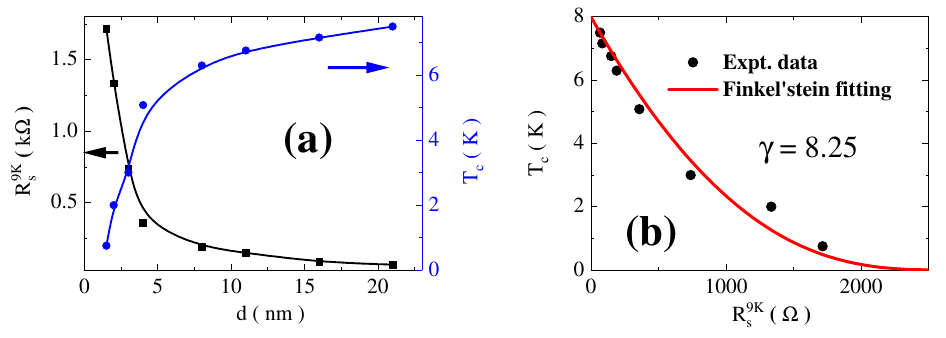}
	\caption[Normal state $R_s$ and $T_c$ s as a function of film thickness and the Finkel’stein fit]{
		\label{fig:Rs-Tc-Finkelstein-fit} 
		\textbf{(a)} $R_s^{9K}$ (black squares) and $T_c$ (blue circles) as a function of film thickness ($d$); the solid lines are guides to the eye. \textbf{(b)} $T_c$ vs $R_s^{9K}$ (black circles) for \textit{a}-MoGe thin films of different thicknesses with fit to Finkel’stein model (red line).
	}
\end{figure*}
We define $T_c$ as the temperature where the resistance becomes $< 0.05\%$ of the normal state value. Fig.~\ref{fig:Rs-Tc-Finkelstein-fit}(a) shows the variation of the normal state sheet resistance ( taken as the sheet resistance at 9 K, $R_s^{9K}$ ) and $T_c$ as a function of film thickness. With decreasing thickness, $T_c$ decreases whereas $R_s$ increases, but remains \! well below the quantum resistance, $h/(4e^2) = 6.45$ $k\Omega$ ( where $e$ is the electron charge and $h$ is the Planck constant ).
\par
To compare with the previous results in the literature, we have plotted the $T_c$ as a function of the corresponding normal state values of $R_s$ at $9$ $K$ for all the samples which fit well with the Finkel’stein~\cite{finkelshtein_superconducting_1987,finkelstein_suppression_1994} model which considered the suppression of $T_c$ according to Fermionic route,
\begin{equation}
    \frac{T_c}{T_{c0}}=e^\gamma \left[\frac{\frac{1}{\gamma}+\frac{r}{4}-\sqrt{\frac{r}{2}}}{\frac{1}{\gamma}+\frac{r}{4}+\sqrt{\frac{r}{2}}}\right]^{{1}/{\sqrt{2r}}},
\end{equation}
where $T_{c0}$ is the $T_c$ of the bulk sample, $r=\frac{e^2}{2\pi^2 \hbar} R_s^{9K}$ and the adjustable parameter $\gamma=\ln \frac{h}{k_B T_{c0}\tau}$, $\tau$ being the transit time. Our data fit with almost the same value as in Refs.~\citenum{finkelshtein_superconducting_1987} and \citenum{graybeal_localization_1984}. But this result alone does not validate that it is following the Fermionic route. Moreover, it is not known whether the Bosonic route will also have a similar variation of $T_c$ since it has not been explored in detail yet. To investigate it further, we go to STS and penetration depth measurements.

\subsection{Scanning tunneling spectroscopy data}\label{sec:stm-data-chap4}
\begin{figure*}[hbt]
	\centering
	\includegraphics[width=16cm]{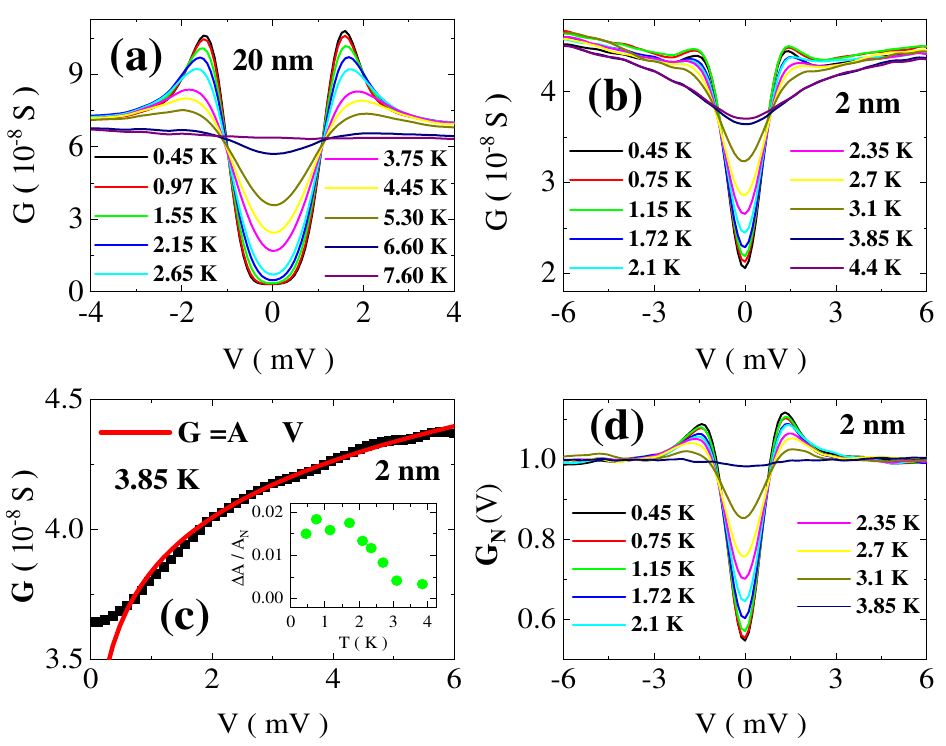}
	\caption[Representative tunneling $G(V)$ vs $V$ spectra and normalization for Altshuler-Aronov-type electron-electron interactions]{
		\label{fig:spectra-Altshuler-Aronov} 
		\textbf{(a), (b)} Representative tunneling spectra $G(V)$ vs $V$ for $20$-$nm$- and $2$-$nm$-thick films respectively at different temperatures using scanning tunneling microscope. \textbf{(c)} The fit of the spectra for the $2$-$nm$-thick sample at $3.85$ $K$ to $G(V)=A \times V^{\alpha}$ expected from Altshuler-Aronov-type electron-electron interactions, where $A=3.84\times10^{-8}$ and $\alpha=0.076$; the deviation at very low bias is due to thermal smearing at finite temperature. (inset) The fractional change of total spectral weight with respect to normal state. \textbf{(d)} The spectra for $2$ $nm$ after normalizing with the spectra at $4.4$ $K$.
	}
\end{figure*}

We next concentrate on the STS measurements. The tunneling conductance ( $G(V)\equiv \frac{dI}{dV}\big\vert_V$ vs $V$ ) spectra were recorded over a $32\times32$ grid over $200$ $nm \times 200$ $ nm$ area at each temperature. Fig.~\ref{fig:spectra-Altshuler-Aronov}(a) and (b) show the average spectra at different temperatures for representative samples ( here $20$-$nm$- and $2$-$nm$-thick \textit{a}-MoGe films respectively ). At low temperatures, the spectra for all samples have the characteristic features of a superconductor: a depression in $G(V)$ at low bias corresponding to the superconducting energy gap, $\Delta$, and the presence of coherence peaks at the gap edge. In addition, for the $2$-$nm$-thick sample, we observe a broad V-shaped, nearly temperature-independent background. This feature, also observed in other disordered superconductors~\cite{chockalingam_tunneling_2009,lemarie_universal_2013,carbillet_spectroscopic_2020},  is attributed to the Altshuler-Aronov type electron-electron interactions in disordered metals~\cite{rollbuhler_coulomb_2001}, which is discussed in Section~\ref{sec:altshuler-aronov}. 

\subsubsection{The Altshuler-Aronov Anomaly and Normalization of the tunneling spectra}\label{sec:altshuler-aronov}
Apart from the low bias feature associated with the superconductivity, the $G(V)-V$ tunneling spectra contain a broad $V$-shaped nearly temperature-independent background that extends up to high bias. With an increase in temperature at some temperature, the low bias feature disappears; above this temperature, the spectra overlap with each other but the $V$-shape remains. For the $d = 2$ $nm$ sample, this temperature is $3.85$ $K$ ( Fig.~\ref{fig:spectra-Altshuler-Aronov}(b) ). We observe that the $V$-shape can be fitted with a functional form $G(V)=A V^{\alpha}$, consistent with the predicted dependence from extensions of Altshuler-Aronov-like interactions~\cite{rollbuhler_coulomb_2001} beyond the perturbative limit ( Fig.~\ref{fig:spectra-Altshuler-Aronov}(c) ). In order to remove the $V$-shaped background and keep only the feature related to superconductivity, we divide all the spectra with the spectrum at $4.4$ $K$ ( Fig.~\ref{fig:spectra-Altshuler-Aronov}(d) ). One consistency check of this procedure is to verify that the normalized spectra, $G_N (V)$, conserves the total spectral weight, i.e., $\lim\limits_{V \gg \Delta/e} \int_{-V}^V G_N(V')dV'$ between the normal and the superconducting state. Defining the quantity,
\begin{equation}
    A(T)= \int_{-6mV}^{6mV} G_N(V',T)dV',
\end{equation}
we plot in the inset of Fig.~\ref{fig:spectra-Altshuler-Aronov}(c),
\begin{equation}
    \frac{\Delta A}{A}=\frac{A(T)-A_N}{A_N},
\end{equation}
where $A_N$ is the value of the integral in the normal state, i.e., for $G_N (V)=1$. We observe that $\Delta A/A<0.02$ which is zero within the limits of resolution of our measurement.
\begin{figure*}[hbt]
	\centering
	\includegraphics[width=15.5cm]{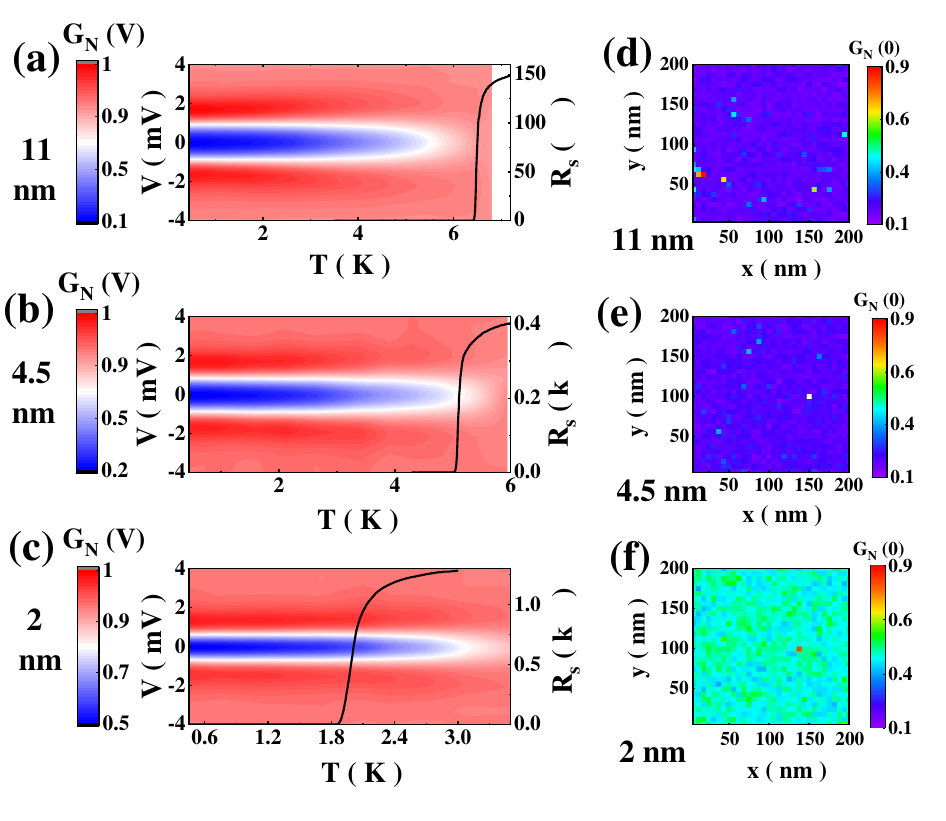}
	\caption[Normalized tunneling spectra and zero-bias conductance maps for $11$-, $4.5$- and $2$-$nm$-thick films]{
		\label{fig:spectra-diff-d} 
		\textbf{(a)-(c)} Intensity plot of normalized spectra $G_N(V)$ as a function of bias voltage ($V$) and temperature for three films with thickness $11$ $nm$, $4.5$ $nm$, $2$ $nm$ respectively. \textbf{(d)-(f)} Spatial maps of zero-bias conductance (ZBC), $G_N(0)$, at $450$ $mK$ for $11$, $4.5$, and $2$ $nm$ respectively.
	}
\end{figure*}
\par
To extract the superconducting contribution alone, we calculated the normalized spectra, $G_N(V)$ vs $V$, by dividing it with the spectra obtained at high temperatures where the low bias feature associated with superconducting pairing disappears. In Fig.~\ref{fig:spectra-diff-d}(a)-(c), the similar temperature dependence of $G_N (V)$ spectra at different thicknesses are represented as intensity plots along with $R_s$ measured on the same samples ( hall-bar-patterned after taking out of STM ). We can see that for thicknesses $11$ $nm$ or higher, the superconducting gap vanishes very close to the transition from the transport measurements consistent with the expectation from Bardeen-Cooper-Schrieffer theory~\cite{tinkham_introduction_2004}. But for the sample with $4.5$ $nm$ thickness, we can see a small hint of a pseudogap, where a soft gap in the tunneling spectra extends approximately $0.5$ $K$ above $T_c$. For the $2$-$nm$-thick sample, pseudogap exists until almost double the $T_c$ ( $\sim 1.8 K$ ). We define the pseudogap vanishing temperature, $T_{\Delta}$, as the temperature where $G_N(0)\approx 0.95$. We observe that even though the coherence peaks are greatly suppressed compared to the BCS estimate, pseudogaps continue to exist up to a temperature much above $T_c \sim 1.8$ $K$ till about $3.1$ $K$. This is consistent with theories where the pseudogap arises from the destruction of global phase coherence state~\cite{bouadim_single-_2011}, but inconsistent with theories that ascribe the pseudogap to parity gap~\cite{feigelman_eigenfunction_2007}. It is also pertinent to note that we observe significant spectral weight inside the gap, which is similar to earlier observations in strongly disordered in situ grown~\cite{kamlapure_emergence_2013,lemarie_universal_2013} NbN ( different from ex situ grown~\cite{sacepe_pseudogap_2010,sacepe_localization_2011} TiN and InO\textsubscript{x} which do not show states inside the gap ), even though its origin is not completely understood~\cite{bucheli_pseudo-gap_2015,dentelski_tunneling_2018}.
\par
\begin{figure*}[hbt]
	\centering
	\includegraphics[width=16cm]{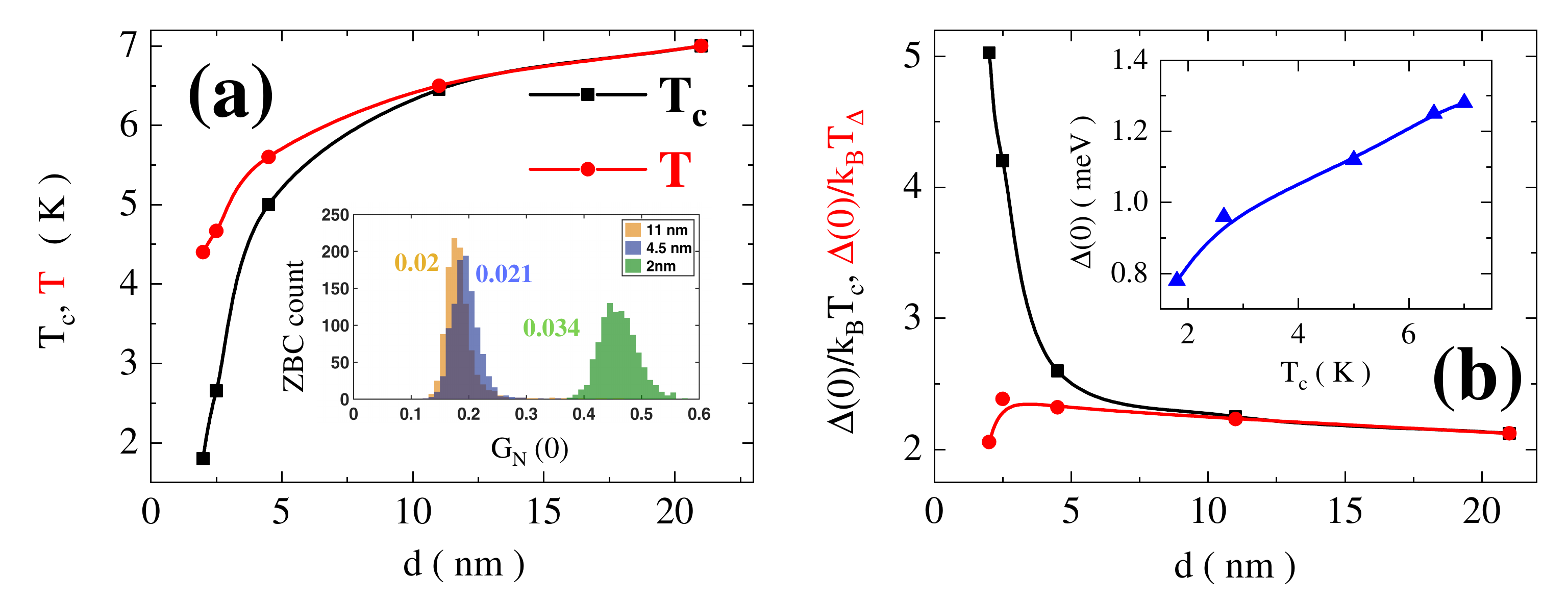}
	\caption[$\Delta(0)/k_B T_c$ and $\Delta(0)$ as a function of $T_c$, and variation of $T_c$ and gap-vanishing temperature $T_{\Delta}$ with film thickness]{
		\label{fig:Delta-Tc-d} 
		\textbf{(a)} Variation of $T_c$ and $T_{\Delta}$ (connected black squares and red circles respectively) with respect to film thickness $d$. Here, $T_c$ is the transition temperature where the resistance becomes $< 0.05\%$ of the normal state value, while $T_{\Delta}$ is the gap-vanishing temperature where $G_N(0)\approx 0.95$. (inset) Histogram of the ZBC for samples of different thicknesses; the root mean square width of each ZBC distribution is written next to each histogram. \textbf{(b)} $\Delta(0)/k_B T_c$ and $\Delta(0)/k_B T_\Delta$ (connected black squares and red circles respectively) as a function of film thickness $d$. (inset) Variation of $\Delta(0)$ with $T_c$ (connected blue triangles).
	}
\end{figure*}

The zero bias conductance ( $G_N(0)$ ) maps obtained at $450$ $mK$ ( Fig.~\ref{fig:spectra-diff-d}(d)-(f) ) reveal that the superconducting state becomes progressively inhomogeneous as we move to lower thicknesses. This nanoscale inhomogeneity~\cite{sacepe_disorder-induced_2008,mondal_phase_2011,lemarie_universal_2013,kamlapure_emergence_2013,zhao_disorder-induced_2019} in the thinnest sample ( $d=2$ $nm$ ) is also evident in the histogram in the inset of Fig.~\ref{fig:Delta-Tc-d}(a), where the root mean square width of the distribution of $G_N(0)$ values increases significantly when $d=2$ $nm$. It should be noted that the surface morphology remains homogeneous down to the thinnest sample.
 \begin{figure*}[hbt!]
	\centering
	\includegraphics[width=14cm]{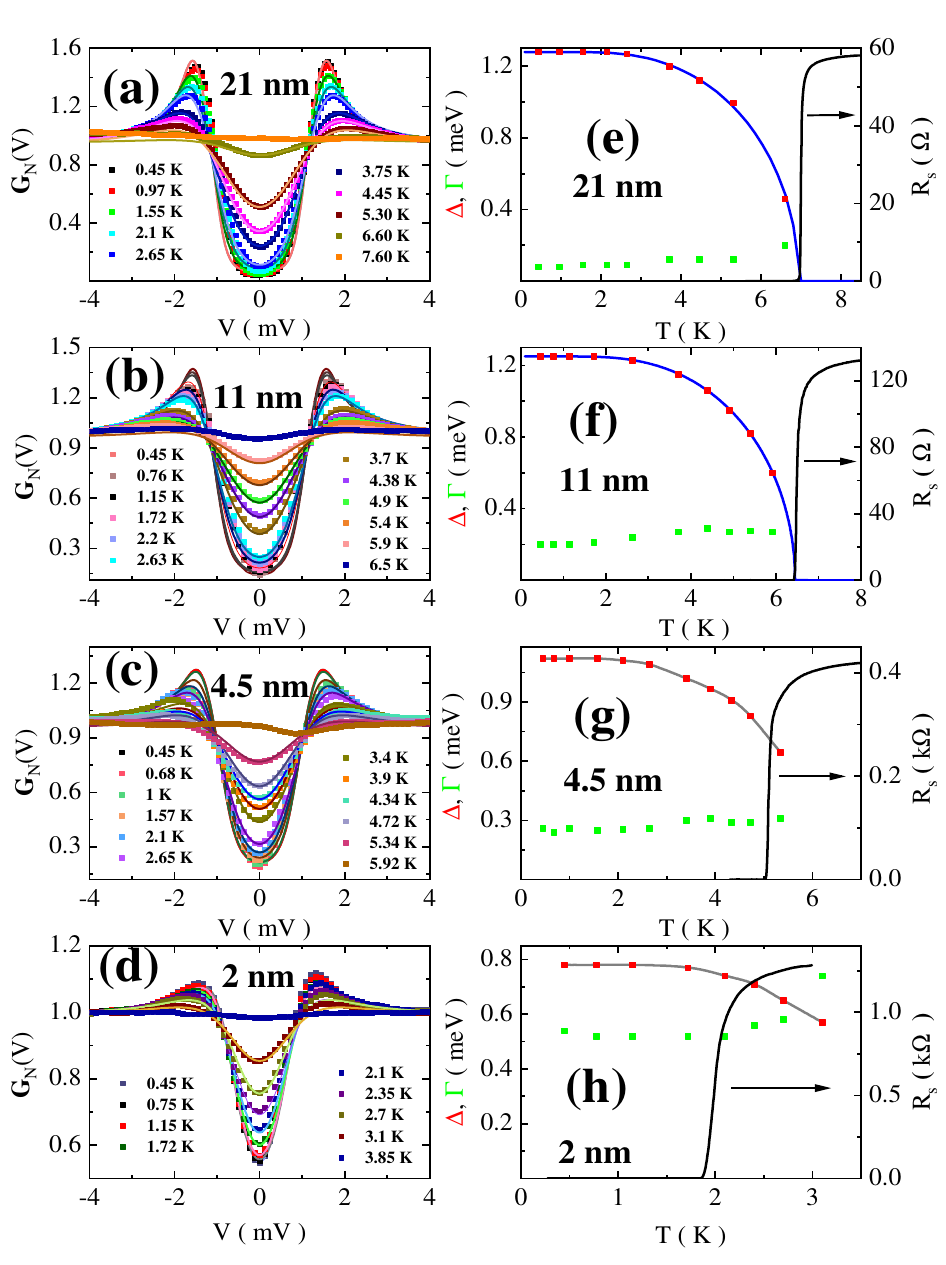}
	\caption[Normalized tunneling spectra for different thicknesses and temperature variation of $\Delta$ and $\Gamma$ extracted from the spectra]{
		\label{fig:spectra-diff-d-Delta-Gamma} 
		\textbf{(a)-(d)} The normalized $G_N(V)-V$ spectra along with the fits with the BCS+ $\Gamma$ model for 4 samples with different thicknesses: $21$, $11$, $4.5$, and $2$ $nm$ respectively. \textbf{(e)-(h)} Temperature variation of the corresponding $\Delta$ and $\Gamma$ extracted from the fits; the temperature variation of the sheet resistances are shown in the same panels. The blue lines in \textbf{(e)} and \textbf{(f)} show the BCS fit to the temperature variation of $\Delta$. The solid grey lines in \textbf{(g)} and \textbf{(h)} are guides to the eye; for these two thicknesses pseudogaps are observed above $T_c$. 
	}
\end{figure*}
\subsubsection{Fitting of the normalized tunneling spectra and temperature variation of superconducting energy gap}

The tunneling conductance between a normal-metal tip and a superconductor is theoretically described by the equation~\cite{tinkham_introduction_2004},
\begin{equation}
    G(V)\propto \frac{1}{R_N}\int_{-\infty}^{\infty}N_s(E)\frac{\partial f(E+eV)}{\partial E} dE,
\end{equation}
where
\begin{equation}
    N_s(E)=Re\left(\frac{E+i\Gamma}{\sqrt{\left(E+i\Gamma\right)^2-\Delta^2}}\right)
\end{equation}
is the single-particle density of states in the superconductor. When $\Gamma=0$, this expression reduces to the usual BCS expression ( Eq.~\ref{eq:tunneling_conductance} ). $\Gamma$ is a phenomenological parameter that is incorporated in the density of states to account for possible broadening in the presence of disorder~\cite{dynes_direct_1978} ( see Section~\ref{sec:dynes-par} for detailed discussion on $\Gamma$ ). Fig.~\ref{fig:spectra-diff-d-Delta-Gamma}(a)-(d) show the fit of the normalized $G_N(V)-V$ spectra for $21$-,$11$-,$4.5$-, and $2$-$nm$-thick films respectively, while the temperature variation of $\Delta$ and $\Gamma$ for the corresponding samples are shown in \ref{fig:spectra-diff-d-Delta-Gamma}(e)-(h). The temperature variations of $R_s$ are also shown in the right panels.
\par
$T_{\Delta}$ has been plotted in Fig.~\ref{fig:Delta-Tc-d}(a) along with the $T_c$ from transport measurements. It is observed\hspace{0.1cm} that they\hspace{0.1cm} start\hspace{0.1cm} to deviate\hspace{0.1cm} below $11$ $nm$. After\hspace{0.1cm} extracting\hspace{0.1cm} $\Delta$ at $450$ $mK$ ( denoted by $\Delta(0)$ ) using BCS + $\Gamma$ model, we observe that $\Delta(0)/(k_B T_c)$ increases rapidly for samples below $5$ $nm$ thickness ( Fig.~\ref{fig:Delta-Tc-d}(b) ) reaching a value of $5$ at $2$ $nm$. \textcolor{black}{In contrast, the ratio $\Delta(0)/(k_B T_{\Delta})$ hovers between $2$ and $2.5$, as expected for a strong-coupling type-II superconductor}. The emergence of a pseudogap between $T_c$ and $T_{\Delta}$, coupled with the anomalously large value of $\Delta(0)/(k_B T_c)$ signals a departure from the BCS ( or Fermionic ) scenario~\cite{sacepe_disorder-induced_2008,mondal_phase_2011,lotnyk_suppression_2017} for \textit{a}-MoGe films with a thickness below $5$ $nm$.
\par
To summarize, the STM data reveal three key findings: Firstly, the gap $\Delta(T)$ for the two thinnest films appears to extrapolate to zero at a temperature $T_{\Delta}$ well above the onset of dc resistivity at $T_c$. Secondly, a consistent low-temperature gap ratio is observed across all four films, regardless of thickness ( both thick and thin ), $\Delta/(k_B T_{\Delta})\approx 2$. Thirdly, there is evidence pointing to a mechanism responsible for broadening the peak in the density of states for quasiparticles.
\color{black}
\subsection{Penetration depth data}
\begin{figure*}[hbt]
	\centering
	\includegraphics[width=16cm]{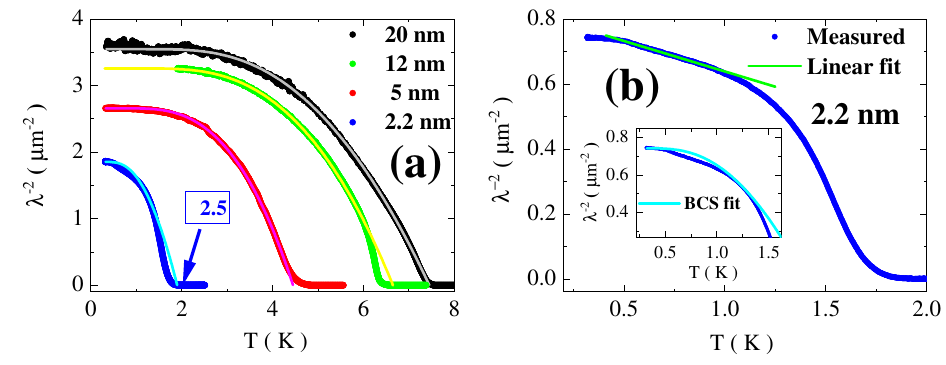}
	\caption[ $\lambda^{-2}$ as a function of temperature for films with different thicknesses]{
		\label{fig:lambda-vs-T-diff-d} 
		\textbf{(a)} $\lambda^{-2}$ as a function of temperature for films with different thicknesses ($d$). For $2.2$ $nm$, $\lambda^{-2}$  is multiplied by $2.5$ for clarity. Solid lines represent the temperature variation expected from the dirty-limit BCS theory. \textbf{(b)} Temperature variation of $\lambda^{-2}$ for $2.2$-$nm$-thick sample (blue circles); the green line is fit to the linear T region. (inset) $\lambda^{-2}$ vs $T$ for $2.2$ $nm$ zoomed at low-temperature region along with the dirty-limit BCS fit (cyan line).
	}
\end{figure*}

We now examine the penetration depth ( $\lambda$ ) data. The temperature dependence of $\lambda^{-2}$ for different thicknesses ( $d$ ) is shown in Fig.~\ref{fig:lambda-vs-T-diff-d}(a). As evident from Fig.~\ref{fig:lambda-vs-T-diff-d}(a), for larger thicknesses ( $d>5$ $nm$ ) $\lambda^{-2}$  vs $T$ follows the dirty limit BCS equation~\cite{tinkham_introduction_2004} ( \ref{eq:lambda-bcs-dirty} ):
\begin{equation}\label{eq:lambda-vs-T-dirty-BCS-limit-chap4}
    \frac{\lambda^{-2}(T)}{\lambda^{-2}(0)}=\frac{\Delta(T)}{\Delta(0)} \tanh \left [\frac{\Delta(T)}{2k_B T}\right],
\end{equation}
where $\Delta(0)$ is constrained within $10\%$ of the value obtained from tunneling measurements at $450$ $mK$ and $\Delta(T)/\Delta(0)$ is assumed to have BCS temperature dependence for a weak-coupling s-wave superconductor ( Eq.~\ref{eq:gap_var_BCS} ). But as the thickness is decreased, $\lambda^{-2}$ deviates from the BCS behavior as evident in the $2.2$ $nm$ sample data ( zoomed in the inset of Fig.~\ref{fig:lambda-vs-T-diff-d}(b) ). As seen from Fig.~\ref{fig:lambda-vs-T-diff-d}(b), $\lambda^{-2}$  vs $T$ for the most disordered sample ( $2.2$ $nm$ ) slowly varies at low temperature and crosses over to a linear variation before decreasing rapidly close to transition.
\par
\begin{figure*}[hbt]
	\centering
	\includegraphics[width=16cm]{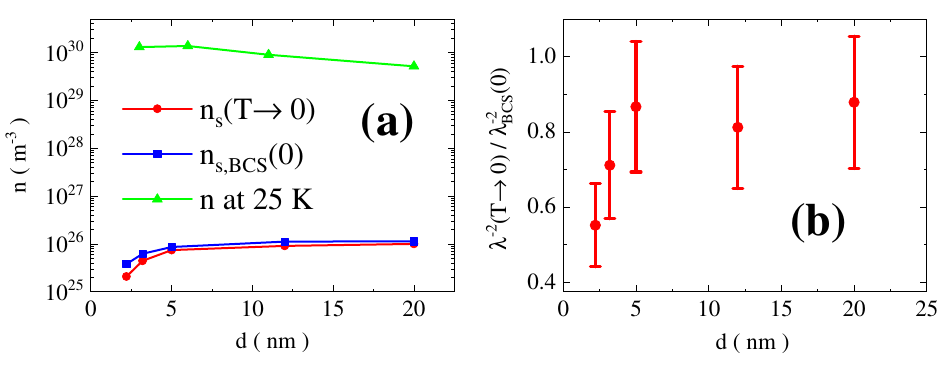}
	\caption[Comparison of $\lambda^{-2}(T\rightarrow0)$ with BCS estimate]{
		\label{fig:lambda-ns-diff-d} 
		\textbf{(a)} Comparison of the thickness dependence of $\lambda_{expt}^{-2}(T\rightarrow0)$ and $\lambda_{BCS}^{-2}(0)$ converted to superfluid density (connected red circles and blue squares respectively). For comparison total carrier density obtained by Hall measurements at $25$ $K$ has been added (connected green triangles). \textbf{(b)} $\lambda_{expt}^{-2}(T\rightarrow0)/\lambda_{BCS}^{-2} (0)$ for different thicknesses, with error bars set at $20\%$ of the corresponding $y$-axis values.
	}
\end{figure*}

At low temperatures, $\lambda^{-2}$ saturates towards a constant value for all samples. We now analyze the thickness variation of $\lambda^{-2} (T\rightarrow 0)$. With a decrease in thickness, $\lambda^{-2}(T\rightarrow0)$ progressively decreases by more than an order of magnitude. Within BCS theory, with an increase in disorder, superfluid density, $n_s$ ( $\equiv m^*/\mu_0 e^2 \lambda^2$, where $m^*$ is the effective mass and $\mu_0$ is the vacuum permeability ) gets suppressed from the electronic carrier density, $n$, due to an increase in electron scattering. In the dirty limit, this is captured by the relation,
\begin{equation}\label{eq:lambda0-dirty-BCS-chap4}
   \lambda_{BCS}^{-2}(0)=\frac{\pi \mu_0 \Delta(T\rightarrow0)}{\hbar \rho_N},
\end{equation}
where $\hbar$ is the reduced Planck constant, and $\rho_N$ is the resistivity in the normal state ( at $10$ $K$ ). This suppression is evident in Fig.~\ref{fig:lambda-ns-diff-d}(a), where the experimental $n_s (T\rightarrow0)$ and $n_{s,BCS}(0)$ are observed to be suppressed by four orders of magnitude from the total carrier density in the normal state ( at $25$ $K$ ) measured from Hall measurements~\cite{basistha_growth_2020}. Moreover, we plot $\frac{\lambda^{-2}(T\rightarrow0)}{\lambda_{BCS}^{-2}(0)}$ for different thicknesses to compare the experimental saturated zero temperature value with the dirty BCS estimate ( Fig.~\ref{fig:lambda-ns-diff-d}(b) ). Here, we observe that the ratio is nearly 1 within the error bar for higher thicknesses, but measured $\lambda^{-2}(T\rightarrow0)$ falls significantly below the disorder-suppressed BCS value below $d \sim 5$ $nm$, reaching a value $\sim 0.55$ for $2.2$-$nm$-thick sample. To explain this additional suppression in the $2.2$ $nm$ data, one has to invoke the effect phase fluctuation in the sample, which was not considered in the dirty BCS fit.
\section{Effect of phase fluctuations}
Since the suppression of $\lambda^{-2}(0)$ and the linear-$T$ variation are consistent with the expectations from quantum and classical longitudinal phase fluctuations~\cite{ebner_superfluid_1983,roddick_effect_1995}, we now attempt a comparison of these features with theory~\cite{de_palo_effective_1999,benfatto_low-energy_2004}. For the $2.2$-$nm$-thick sample, we estimate $\xi\sim 8$ $nm$ from $H_{c2}$, such that it is in the 2D limit. Now, the resilience of a superconductor against phase fluctuations is given by the superfluid stiffness ( $J_s$ ), which for a two-dimensional superconductor ( $d < \xi$ ) is expressed as,
\begin{equation}\label{eq:Js-expn-chap4}
    J_s=\frac{\hbar^2 n_s d}{4m^*}.
\end{equation}
A rough estimate of the suppression of $n_s$ due to quantum phase fluctuations can be obtained by considering two energy scales~\cite{benfatto_phase_2001}: The Coulomb energy, $E_c=\frac{e^2}{2 \epsilon_0 \epsilon_B \xi}$ and $J_s(T\rightarrow0)$, where $\epsilon_0$  is the vacuum permeability and $\epsilon_B$ is the background dielectric constant. Using the carrier density measured from Hall effect measurements $n=4.63\times 10^{29}$ $m^{-3}$ and the plasma frequency~\cite{tashiro_unusual_2008} $\Omega_p=1.625\times 10^{16}$ $Hz$, we estimate $\epsilon_B=\frac{e^2 n}{\epsilon_0 m \Omega_p^2} \sim 5.6$. Here, the suppression of the superfluid density due to quantum phase fluctuations is given by,
\begin{equation}\label{eq:ns-phase-fluc-corr}
    \frac{n_s}{n_s^0 }=e^{-\langle(\Delta\theta)^2\rangle/4},
\end{equation}
 where
 \begin{equation}
     \langle(\Delta \theta)^2\rangle\approx \frac{1}{5\pi}\sqrt{\frac{E_c}{J_s(0)}},
 \end{equation}
and $n_s^0$ is the bare superfluid density in the absence of phase fluctuations. We obtain $\frac{n_s}{n_s^0} \sim 0.78$. This value would get further reduced in a disordered system if the local superfluid density is spatially inhomogeneous~\cite{barabash_conductivity_2000,seibold_superfluid_2012,maccari_broadening_2017,maccari_disordered_2019}. Though it is difficult to quantify this effect in our film, from the large spatial variation of $G_N(0)$ in the $2$-$nm$-thick films ( Fig.~\ref{fig:spectra-diff-d}(f) - (h) ), we believe that this effect could be substantial. This additional suppression of the superfluid density due to inhomogeneity reflects in the emergence of a finite-frequency absorption that is expected to occur at relatively low energies, i.e., below the $2\Delta$ threshold for quasiparticle absorption in dirty superconductors~\cite{cea_optical_2014,seibold_application_2017}. This effect has been observed in NbN and InO\textsubscript{x} films~\cite{crane_fluctuations_2007,cheng_anomalous_2016}. Such low energy dissipative mode has a feedback effect on quantum phase fluctuations such that the effect of quantum corrections gets reduced and the quantum to classical cross-over shifts to a lower temperature~\cite{de_palo_effective_1999}. Considering these uncertainties at the moment we can only state that the observed value of $\frac{\lambda^{-2}(300mK)}{\lambda^{-2}_{BCS}(0)}\sim 0.55$ falls in the correct ballpark.
\par
\begin{figure*}[hbt]
	\centering
	\includegraphics[width=8cm]{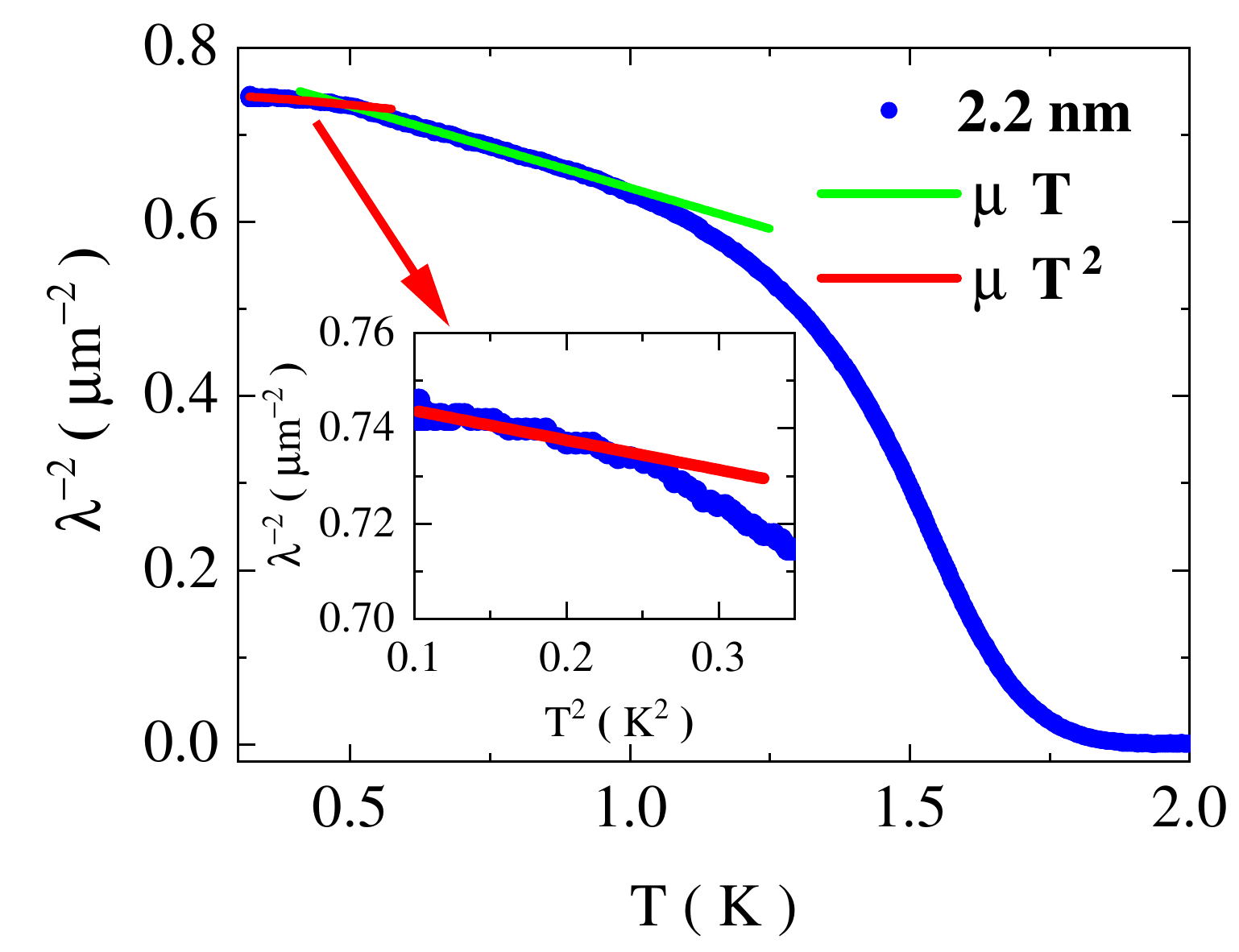}
	\caption[Temperature variation of $\lambda^{-2}$ for $2.2$-$nm$-thick sample highlighting quadratic ($T^2$) dependence at very low temperatures]{
		\label{fig:lambda-vs-T-2p2nm-quadratic-linear-fits} 
		Temperature variation of $\lambda^{-2}$ for $2.2$-$nm$-thick sample highlighting quadratic ($T^2$) dependence at very low temperatures which becomes linear ($T$) at around $0.5$ $K$. (inset) Expanded view of $T^2$ dependence of $\lambda^{-2}$ at low temperatures.
	}
\end{figure*}
Upon further scrutiny of the $\lambda^{-2}$ variation in the low-temperature range, particularly below the linear regime, we observe a shift towards quadratic ( $T^2$ ) behavior in the $2.2$-$nm$-thick sample ( see Fig.~\ref{fig:lambda-vs-T-2p2nm-quadratic-linear-fits} ). This $T^2$ trend at lower temperatures aligns with findings documented in various reports in the literature~\cite{pond_penetration_1991,annett_superconducting_1992,hirschfeld_effect_1993,bonn_microwave_1993,zuev_search_2006,zuev_crossover_2008}. In our disordered \textit{a}-MoGe thin film, the rationale behind this quadratic variation can be elucidated by considering the role of dissipation. Theoretical predictions~\cite{lee_localized_1993} and experimental observations in $d$-wave superconductors suggest the presence of low-energy dissipation, as evidenced by high-frequency conductivity measurements\cite{lee_-b_1996,corson_nodal_2000}. Recent studies on amorphous InO\textsubscript{x}~\cite{crane_fluctuations_2007} and strongly disordered NbN~\cite{mondal_phase_2011} films indicate that low-energy dissipation may also manifest in disordered s-wave superconductors. Although the precise origin of this dissipation remains unclear, quantum phase fluctuations contribute to the $T^2$ temperature depletion for the superfluid density ( $n_s\propto 1/\lambda^2$ ) in the presence of dissipation~\cite{benfatto_phase_2001,mondal_phase_2011}. This is expressed as $n_s/n_{s0}=1-DT^2$ at low temperatures, where $D$ is directly proportional to the dissipation. For the $2.2$-$nm$-thick sample with a critical temperature of approximately $1.8$ K, the $T^2$ variation becomes distinctly evident below $500$ $mK$, while a linear variation predominates beyond this temperature threshold.
\color{black}
\section{Comparison of \texorpdfstring{$T_c$}{} and \texorpdfstring{$T_\Delta$}{}}\label{sec:Tc-T_Delta-comparison-phase-fluc}
In exploring the impact of phase fluctuations, we turn our attention to the comparison between the critical temperature ( $T_c$ ) and the superconducting gap \enquote{vanishing} temperature ( $T_\Delta$ ). We have already observed in Fig.~\ref{fig:Delta-Tc-d}(a) that $T_c$ and $T_\Delta$ are almost identical for higher thicknesses and only start to deviate below $11$ $nm$. Interestingly, across all film thicknesses, both thick and thin, we consistently observe a low-temperature gap ratio of $\Delta(0)/(k_B T_{\Delta})\approx 2$ ( Fig.~\ref{fig:Delta-Tc-d}(b) ), implying that $T_\Delta$ corresponds to the expected mean-field transition.
\begin{table}[hbt]
\caption{Comparison of $T_c$, $T_{\Delta}$ and $\Gamma$ (in $K$) for two thinnest samples of thickness $d$ (in $nm$) as curated from Fig~\ref{fig:spectra-diff-d-Delta-Gamma} and Fig.~\ref{fig:Delta-Tc-d}.}\label{table:Tc-T*-thinnest-samples}
\centering
    \begin{tblr}{|c|c|c|c|c|c|}
    \hline
        $d$ & $T_c$ & $T_{\Delta}$ & $(T_c^{Bulk}-T_{\Delta})$ & $\Gamma$ & $\Gamma/(T_c^{Bulk}-T_{\Delta})$ \\
        \hline
        $2$ & $1.8$ & $4.4$ & $2.6$ & $6.1$ & $2.3$ \\
        \hline
        $4.5$ & $5$ & $5.6$ & $1.4$ & $2.9$ & $2.1$ \\
        \hline
    \end{tblr}
\end{table}
\par
Notably, the difference between the bulk critical temperature ( $T_c^{Bulk}$ ) of the thickest film ( $7$ $K$ for $21$ $nm$) and the measured $T_\Delta$ of thinner films corresponds closely to half of the tunneling \enquote{pair-breaking} parameter $\Gamma$, as evident in Table~\ref{table:Tc-T*-thinnest-samples}. This correlation, while not conclusive, hints at a consistent pairing interaction in all films, capable of yielding a $T_c$ of $7$ $K$. However, the expected \enquote{mean-field} transition is suppressed by pair-breaking mechanisms such as electron-electron scattering or thermal phase fluctuation~\cite{lee_electron-electron_1989}. This aligns with the idea that the phenomenological broadening parameter $\Gamma$ embodies the influence of the pair-breaking mechanism within the system.
\par
Collectively, these findings strongly suggest that $T_\Delta$ essentially represents the \enquote{mean-field} superconducting transition temperature ( $T_{\Delta}\approx T_{c0}$ ), diminished in thinner films due to Fermionic effects. The presence of Bosonic effects, potentially linked to phase fluctuations, further reduces the observed $T_c$ below $T_{\Delta}$.
\section{Interpretation of penetration depth data from \enquote{Mean-field} perspective}\label{sec:penetration-depth-from-STM-mean-filed}
In this section, we shall attempt to compare and interpret the measured penetration depth data from the \enquote{mean-field} viewpoint in light of scanning tunneling spectroscopy ( STS ) data. It is revealed in the STS data ( Section~\ref{sec:stm-data-chap4} ) that the \enquote{gap}, which signifies the suppression of the low-energy quasi-particle density of states, disappears at $T_\Delta$. As discussed in Section~\ref{sec:Tc-T_Delta-comparison-phase-fluc}, $T_\Delta$ is essentially the expected \enquote{mean-field} superconducting transition temperature. Mean-field theories, which do not consider phase fluctuations, suggest that the superfluid density also disappears at $T_{\Delta}$. These theories encompass BCS~\cite{bardeen_theory_1957}, the Abrikosov-Gor'kov theory~\cite{abrikosov_problem_1961}, and the Eliashberg theory~\cite{eliashberg_interactions_1960}. The Abrikosov-Gor'kov~\cite{abrikosov_problem_1961} theory incorporates elastic pair-breaking, revealing that pair-breaking smears the BCS peak in the quasiparticle density of states. On the other hand, the Eliashberg theory~\cite{eliashberg_interactions_1960}, which includes inelastic electron-phonon scattering, indicates a low-energy suppression in the density of states, but not a true gap, reminiscent of the Dynes phenomenological formula~\cite{dynes_direct_1978}.
\begin{figure*}[hbt]
	\centering
	\includegraphics[width=8cm]{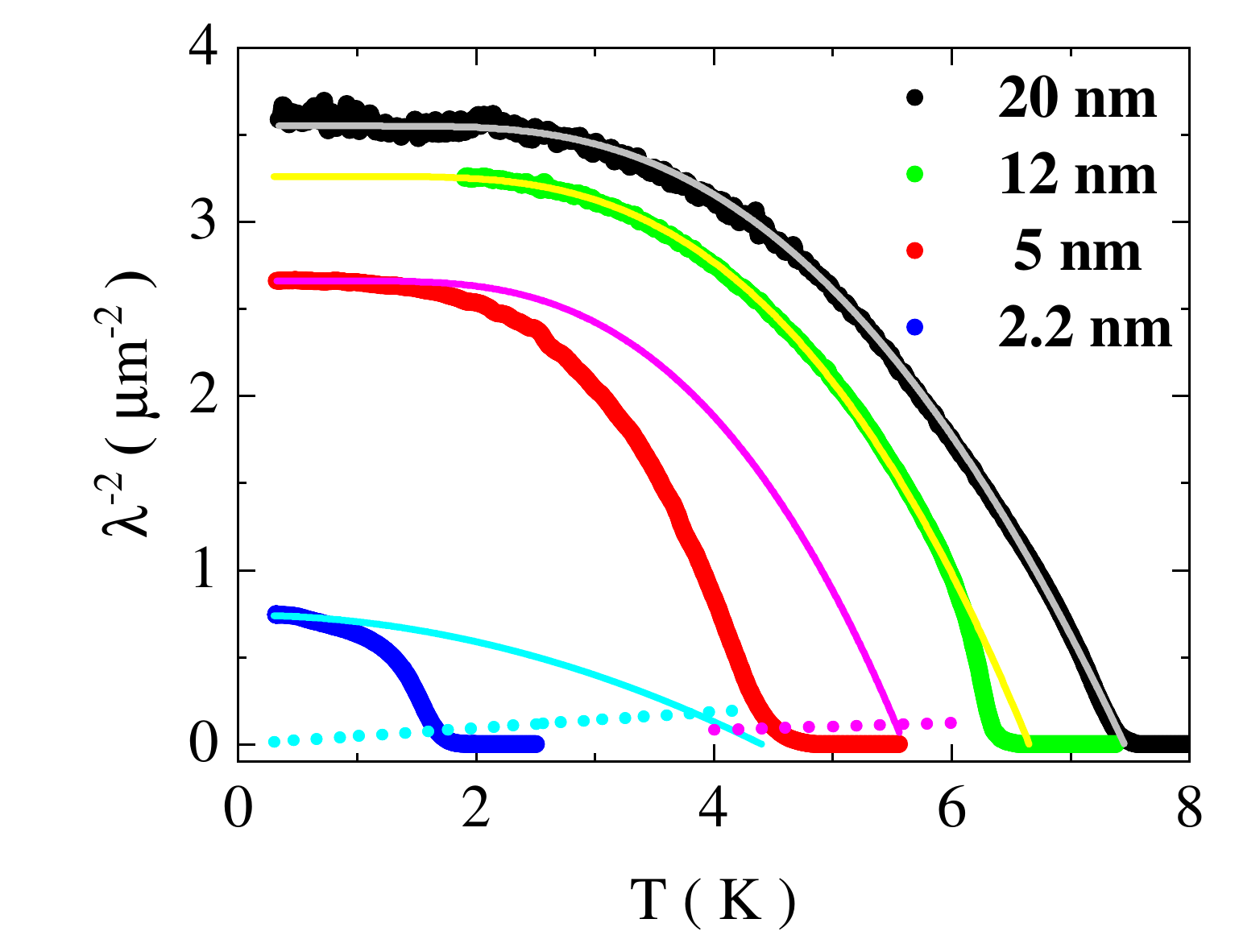}
	\caption[ $\lambda^{-2}$ as a function of temperature for films with different thicknesses with BCS variation taking $T_{Delta}$ as transition]{
		\label{fig:lambda-vs-T-diff-d-T_Delta-fits} 
		$\lambda^{-2}$ as a function of temperature for films with different thicknesses. Solid lines represent the temperature variation expected from the dirty-limit BCS theory with respect to $T_{\Delta}$. For $2.2$ $nm$ quadratic fit has been used (cyan line). For $2.2$ and $5$ $nm$, BKT lines are drawn (cyan and magenta dots respectively).
	}
\end{figure*}
\par
Now, in Fig.~\ref{fig:lambda-vs-T-diff-d-T_Delta-fits}, we juxtapose the temperature dependence of the measured superfluid density ( $\propto 1/\lambda^2(T)$ ) with a \enquote{mean-field} superfluid density that diminishes at $T_\Delta$. For the thick samples, we apply the dirty-BCS temperature dependence of $\lambda^{-2}(T)$ ( Eq.~\ref{eq:lambda-vs-T-dirty-BCS-limit-chap4} ) to fit the data. However, in the case of the thinner samples ( $d<5$ $nm$ ), where $\Delta(T)$ deviates from the BCS equation ( refer to Fig.~\ref{fig:spectra-diff-d-Delta-Gamma} ), the use of the BCS functional form is not justified. Instead, we adopt a mean-field approach for the $2.2$-$nm$-thick sample, assuming a simple quadratic form for superfluid density:
\begin{equation}\label{eq:lambda-T-quadratic-T_Delta}
    \frac{\lambda^{-2}(T)}{\lambda^{-2}(0)}\bigg\vert_{mean field}=1-(T/T_\Delta)^2.
\end{equation}
This Eq.~\ref{eq:lambda-T-quadratic-T_Delta} is consistent with Fig.~\ref{fig:lambda-vs-T-2p2nm-quadratic-linear-fits}, where the quadratic temperature ( $T^2$ ) dependence is observed at low temperatures before transitioning to linearity in the presence of thermal phase fluctuations.
\par
The fits in Fig.\ref{fig:lambda-vs-T-diff-d-T_Delta-fits} are distinct from those in Fig.\ref{fig:lambda-vs-T-diff-d} in that we now fix the transition at $T_\Delta$ instead of finding the best-fit $T_c$. This approach provides a \enquote{mean-field} perspective on the temperature variation of superfluid density. Fig.~\ref{fig:lambda-vs-T-diff-d-T_Delta-fits} highlights the temperatures between $T_c$ and $T_{\Delta}$, known as the \enquote{pseudogap} region, where presumably some Bosonic mechanism, such as thermal phase fluctuations, triggers the onset of DC resistance.
\par
To further emphasize the anomalously strong phase fluctuations, we have incorporated the universal BKT~\cite{berezinskii_destruction_1972,kosterlitz_ordering_1973,kosterlitz_critical_1974} line into Fig.~\ref{fig:lambda-vs-T-diff-d-T_Delta-fits} for the two thinnest films, given by,
\begin{equation}
\begin{split}
    J_s &=\frac{2}{\pi}T\\
    \Rightarrow \frac{1}{\lambda^2}&=\frac{8\mu_0e^2}{\pi\hbar^2 d}\times T,
\end{split}
\end{equation}
where $J_s$ is the superfluid stiffness ( defined in Eq.~\ref{eq:Js-expn-chap4} ) and $d$ is the thickness of the thin film. The intersection of the BKT line with the \enquote{mean-field} $\lambda^{-2}(T)$ visually indicates the region where conventional phase fluctuation theory predicts the disappearance of superfluid density~\cite{nelson_universal_1977,minnhagen_two-dimensional_1987}. It is observed in Fig.~\ref{fig:lambda-vs-T-diff-d-T_Delta-fits} that the superfluid density for both $2.2$ and $5$ $nm$ falls well before the possible intersection point mentioned above. Consequently, it becomes evident that phase fluctuations suppress superfluid density much more vigorously than anticipated. Furthermore, the spatial inhomogeneity in the local superfluid density within the thinnest films could amplify the overall suppression of the superfluid density. This results in a reduction of the critical temperature ( $T_c$ ) in the thinnest films, where both superfluid density and resistivity vanish, well below the anticipated occurrence of the familiar Kosterlitz-Thouless-Berezinskii transition.
\color{black}
\section{Comparison of energy scales}\label{sec:Js-Delta-energy-compare-phase-fluc}
At last, to reconcile the penetration depth measurements with the emergence of the pseudogap, we discuss the energy scales from the above two measurements: superfluid stiffness $J_s(0)$ from the penetration depth measurements and superconducting gap or pairing energy $\Delta(0)$ from STS measurements. $\Delta(0)$ is the energy scale important for the superconducting transition in conventional superconductors according to the BCS theory where the superconducting state is resilient to phase fluctuations. On the contrary, for superconductors vulnerable to phase fluctuations, the energy scale corresponding to $J_s(0)$, which is the measure of the resilience of a superconductor against phase fluctuations, plays an important role. Thus, in general, the superconducting $T_c$ is determined by the lower of these two energy scales~\cite{emery_importance_1995}. When $\Delta(0) \ll Js(0)$, the superconducting transition temperature is given by~\cite{benfatto_low-energy_2004}, $T_c  \sim \Delta(0)/(A k_B)$, where $A\sim 2$ for \textit{a}-MoGe. On the other hand, when $Js(0)\ll \Delta(0)$, thermal phase fluctuations play a dominant role, where the superconducting state gets destroyed due to thermal phase fluctuations at $T_c \sim J_s (0)/(Bk_B )$ ( B is a constant of the order of unity ) even if the pairing amplitude remains finite up to higher temperatures.
\par
 \begin{figure*}[hbt]
	\centering
	\includegraphics[width=8cm]{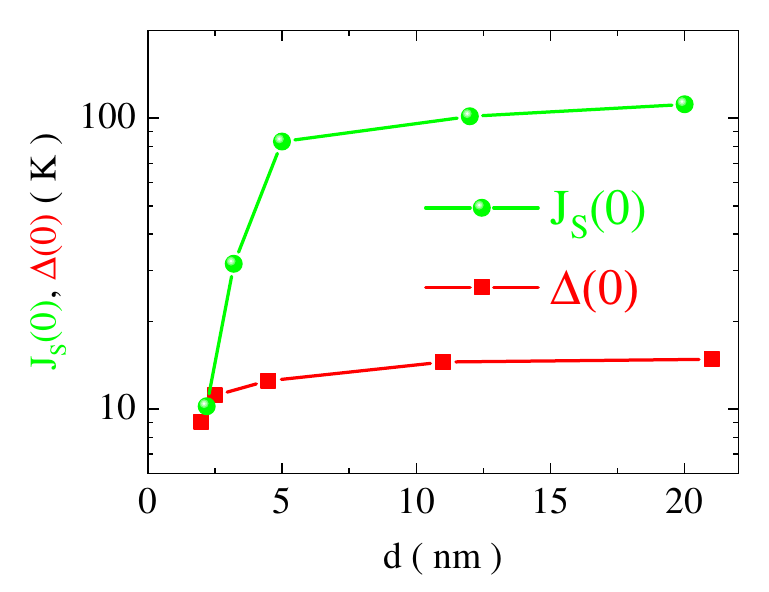}
	\caption[Comparison of $J_s$ and $\Delta$ as a function of thickness $d$.]{
		\label{fig:Js-Delta} 
		Comparison of Superfluid stiffness $J_s(0)$ and pairing energy $\Delta(0)$ in the limit $T\rightarrow0$ as a function of thickness $d$.
	}
\end{figure*}
It is observed in Fig.~\ref{fig:Js-Delta} that $J_s(0)$ is an order of magnitude larger than $\Delta(0)$ for larger thicknesses ( $d>5$ $nm$ ). But $J_s(0)$ drops sharply below $5$ $nm$ and becomes comparable to $\Delta(0)$ around $d\sim 2$ $nm$. Thus, around this thickness, we expect phase fluctuation to dominate the superconducting properties. Indeed, in the corresponding STS intensity maps ( Fig.~\ref{fig:spectra-diff-d}(d)-(e) ), a soft gap appears in STS for $d\sim 5$ $nm$, while pseudogap persists up to almost double of $T_c$ for $d\sim 2$ $nm$. The observation of the pseudogap suggests that at this thickness the pairing amplitude remains finite even when the global phase coherent state has been destroyed by phase fluctuations. In addition, since the film is in the 2D limit, the measured temperature dependence of the superfluid stiffness near $T_c$ turns out to be consistent with a Berezinskii-Kosterlitz-Thouless~\cite{kosterlitz_ordering_1973,kosterlitz_critical_1974,minnhagen_two-dimensional_1987} ( BKT ) jump smeared by disorder-induced inhomogeneity~\cite{mondal_role_2011,yong_robustness_2013} ( see Appendix~\ref{ap:BKT-fit-2p2nm} ).

\section{Summary and conclusion}
In summary, we have shown that the suppression of superconductivity in \textit{a}-MoGe with decreasing film thickness has two regimes. At moderate disorder, $T_c$ decreases but $\Delta/k_BT_c$ shows only a small increase ( Fig.~\ref{fig:Delta-Tc-d}(a) ) and $R_s^{9K}$ vs $T_c$ ( Fig.~\ref{fig:Rs-Tc-Finkelstein-fit}(b) ) follows the Finkel’stein model both consistent with the Fermionic scenario. But, at stronger disorder, the system moves from Fermionic to Bosonic regime, where the pairing amplitude remains finite and manifests in the form of pseudogap even after the superconducting state is destroyed by phase fluctuations. This evolution of the superconducting state is consistent with earlier observations on disordered NbN, TiN, and InO\textsubscript{x}~\cite{mondal_phase_2011,sacepe_pseudogap_2010,sacepe_localization_2011}, where the eventual destruction of superconductivity through the Bosonic route is well established. 
\par
In this context, we would like to note that some early tunneling measurements performed at low temperatures on thin superconducting amorphous films ( Bi, Pb ) appeared consistent with the Fermionic scenario~\cite{valles_superconductivity_1989,valles_electron_1992}. However, these measurements were performed on planar tunnel junctions where the influence of the normal-metal electrode on the superconductor through the relatively low resistance tunnel barrier and the possibility of chemical mixing at the interface are difficult to completely rule out. Also, it should be approached with caution to draw the conclusion that the disappearance of the gap in the insulating regime implies the absence of Cooper pairs, because vanishing of the gap in superconducting tunneling can also arise from other factors, such as pair breaking caused by disorder~\cite{meyer_gap_2001}. 
\par
On the other hand, the observation of the Bosonic scenario in \textit{a}-MoGe, which was widely believed to follow the Fermionic route, raises the important question on whether a disordered superconductor can follow the Fermionic route all the way to the destruction of superconductivity or whether all superconductors become Bosonic at sufficiently strong disorder. We conclude with the following conjecture, which needs to be addressed in future theoretical and experimental studies.
\begin{conjecture*}
    In the limit of extreme disorder, the suppression of the superconductivity in all superconductors will follow the Bosonic mechanism, regardless of whether it follows the Fermionic or Bosonic route at moderate disorder.
\end{conjecture*}


	\chapter{Study of vortex dynamics in \textit{a}-MoGe using low-frequency two-coil mutual inductance measurements: extracting vortex parameters}\label{ch:vortex-par-chap}
\chaptermark{Study of vortex dynamics in \textit{a}-MoGe}


\section{Introduction}

Understanding the dynamics of vortices in type-II superconductors is of paramount importance both from a fundamental standpoint and for the practical application of these materials~\cite{eley_challenges_2021,fomin_perspective_2022,naibert_imaging_2021,chaves_enhancing_2021,li_vortex_2021,sang_pressure_2021}. In type-II superconductors, when we apply a magnetic field larger than the lower critical field $H_{c1}$, quantized flux lines ( vortices ) penetrate the sample. In a clean superconductor, the interaction between the vortices arranges them in a triangular lattice known as the Abrikosov~\cite{abrikosov_vliyanie_1952,abrikosov_magnetic_1957,abrikosov_magnetic_1957-1,levy_vortices_2000}  vortex lattice ( VL ). However, the inevitable presence of crystalline defects in solids acts as a random pinning potential for the vortices. 
\par
If these vortices are made to oscillate under the influence of an oscillatory current or magnetic field, their motion is governed by the following competing forces~\cite{van_der_beek_linear_1993}: i) Lorentz force due to the external current density driving the motion, ii) restoring force due to the combined effect of pinning by crystalline defects and repulsion from neighboring vortices, and iii) the dissipative viscous drag of the vortices. In addition, at finite temperature thermal activation can cause the vortices to spontaneously jump over the pinning barrier resulting in thermally activated flux flow ( TAFF ), which produces a small resistance even for external current much below the critical current density ( $J_c$ ).
\par
In this work, we have studied the dynamics of vortices in the presence of small ac excitation ( $0.05$ $mA$ and $30$ $kHz$ for our two-coil~\cite{turneaure_numerical_1996,turneaure_numerical_1998,kamlapure_measurement_2010,kumar_two-coil_2013,mondal_phase_2013,roy_dynamic_2017,gupta_superfluid_2020} measurements ) in weakly-pinned $20$-$nm$-thick \textit{a}-MoGe film, where the ac magnetic shielding response is primarily governed by the dynamics of vortices. The vortex contribution can be formally accounted for through an effective penetration depth ( the Campbell~\cite{campbell_response_1969,campbell_interaction_1971} penetration depth or $\lambda_C$ ) which operates in addition to the London~\cite{london_electromagnetic_1935} penetration depth ( $\lambda_L$ ). To analyze the experimental temperature-dependent variation of the penetration depth data, we utilized mean-field models to describe the motion of vortices under small ac excitation.

\section{Theoretical background: Coffey-Clem Model}

To interpret the extracted penetration depth data from the two-coil mutual inductance measurements, we have made use of a mean-field model proposed by Coffey and \sloppy Clem~\cite{coffey_unified_1991,coffey_coupled_1992,coffey_magnetic_1991,coffey_theory_1992,coffey_theory_1992-1,coffey_theory_1992-2,clem_vortex_1992} for the motion of vortices in response to small ac excitation. This model was discussed in detail in section~\ref{sec:Coffey-Clem model-chapter}. For continuation in this chapter, I shall briefly state the relevant part needed for our analysis. According to this model, the equation of motion of a single vortex ( force per unit length ) is given by:

\begin{equation}
    \label{eq:Coffey_Clem_model}
    \eta \bm{\Dot{u}}+\alpha_L \bm{u}=\Phi_0 \bm{J_{ac}}\times \hat{\bm{n}}+\bm{F(U,T)},
\end{equation}
where $\bm{J_{ac}}$ is the external alternating current density on the vortex containing one flux quantum $\Phi_0$ ( $\bm{\hat{n}}$ being the unit vector along the vortex ), $\eta$ is the viscous drag coefficient on the vortex, $\alpha_L$ is the restoring pinning force constant called the Labusch parameter~\cite{labusch_calculation_1969,laiho_labusch_2005,pan_labusch_2000}, and $\bm{F(U,T)}$ is a random Langevin ( thermal ) force depending on the pinning barrier $U$ for the thermally activated flux flow ( TAFF ) of the vortices. Coffey and Clem~\cite{coffey_unified_1991,coffey_coupled_1992,coffey_magnetic_1991,coffey_theory_1992,coffey_theory_1992-1,coffey_theory_1992-2,clem_vortex_1992} ( CC ) solved Eq.~\ref{eq:Coffey_Clem_model} and derived the following equation for ac penetration depth:
\begin{equation}\label{eq:Coffey_Clem_eqn}
    \lambda_{ac}^2=\left(\lambda^2_L-i \frac{\rho_v^{TAFF}}{\mu_0\omega}\right)\bigg/\left(1+2i\frac{\lambda^2_L}{\delta^2_{nf}}\right),
\end{equation}
where normal fluid correction is given by the normal fluid skin depth ( $\delta_{nf}$ ) and the  vortex contribution is expressed in the vortex resistivity ( $\rho^{TAFF}_v$ ) term given by,
\begin{equation}
    \rho_v^{TAFF}=\rho_{ff} \frac{\epsilon+i\omega\tau}{1+i\omega\tau},
\end{equation}
where $\tau$ is the vortex relaxation time in the presence of TAFF expressed as,
\begin{equation}
    \tau=\frac{\eta}{\alpha_L}\frac{I_0^2 (\nu)}{I_1 (\nu) I_0 (\nu)},
\end{equation}
and $\epsilon$ is the vortex creep parameter given by,
\begin{equation}
    \epsilon=\frac{1}{I_0^2(\nu)}.
\end{equation}
Here, $I_p$ is the modified Bessel’s function of the first kind of order p. The effect of TAFF enters in the argument of $I_p$ through the parameter $\nu$ defined as,
\begin{equation}
    \nu=\frac{U}{2k_B T}.
\end{equation}

\section{Sample used in this work}

Our sample consists of a $20$-$nm$-thick ( $d$ ) weakly-pinned amorphous Molybdenum Germanium ( \textit{a}-MoGe ) thin film, grown on ($100$) oriented MgO substrate by pulsed laser deposition ( PLD ) technique, ablating an arc-melted Mo\textsubscript{70}Ge\textsubscript{30} target. The zero field superconducting transition \hspace{0.1cm}temperature ( $T_c$ ) for the \hspace{0.1cm}$20$-$nm$-thick samples was $\sim 7$ $K$ ( similar to the samples in ref.~\citenum{roy_melting_2019,dutta_collective_2020} ).
\begin{figure*}[hbt]
	\centering
	\includegraphics[width=16cm]{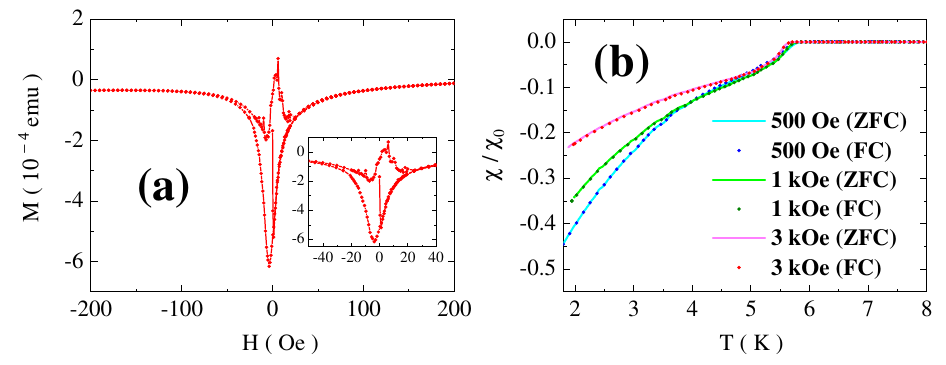}
	\caption[$M-H$ loop and comparison of ZFC and FC data for weakly-pinned $20$-$nm$-thick \textit{a}-MoGe film]{
		\label{fig:MH-loop-ZFC-Fc} 
	\textbf{(a)} $M-H$ loop for a $20$-$nm$-thick \textit{a}-MoGe film grown on ($100$) oriented MgO substrate measured using the SQUID-VSM technique. The loop has been enlarged in the inset. \textbf{(b)} Normalized real part of the ac susceptibility, $\chi$, as a function of temperature for the FC and ZFC state at $500$ $Oe$, $1$ $kOe$ and $3$ $kOe$ ($\chi_0$ is the zero-field susceptibility at $1.5$ $K$) on a similar sample (adapted from Ref.~\citenum{roy_melting_2019}). The FC and ZFC curves at each field overlap with each other showing the very weakly pinned nature of the VL. The density of points for the FC curves has been reduced for clarity.
	}
\end{figure*}
\par
The weakly pinned nature of the $20$-$nm$-thick \textit{a}-MoGe thin film is reflected in the very small M-H loop measured using the SQUID-VSM technique ( Fig.~\ref{fig:MH-loop-ZFC-Fc}(a) ). Weak pinning nature was also evident in the previous ac susceptibility measurements ( Fig.~\ref{fig:MH-loop-ZFC-Fc}(b) ), where the field-cooled ( FC ) and zero-field-cooled ( ZFC ) data on similar samples~\cite{roy_melting_2019} showed a negligible difference for fields above 500 Oe.
\begin{figure*}[hbt]
	\centering
	\includegraphics[width=16cm]{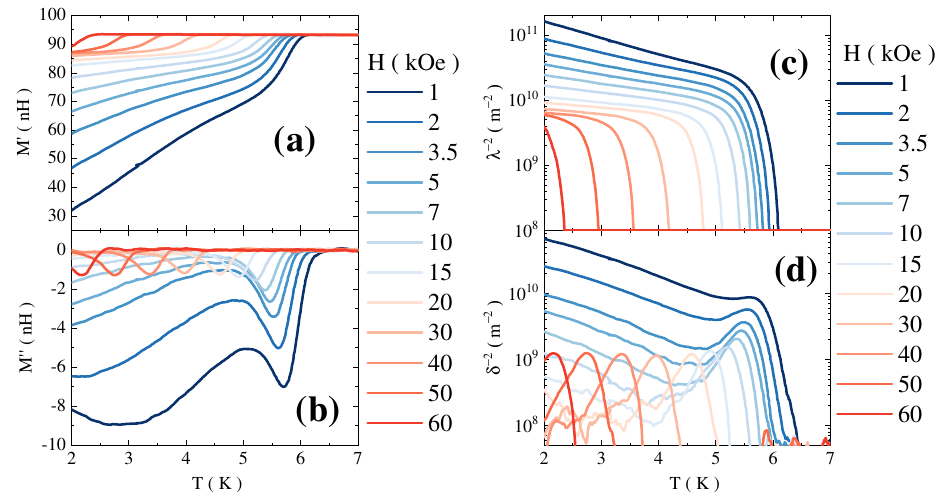}
	\caption[Temperature variation of experimentally measured $M'$ and $M''$ at different fields with the corresponding $\lambda^{-2}$ and $\delta^{-2}$]{
		\label{fig:M-lambda-delta-diff-H} 
		\textbf{(a), (b)} Temperature variation of experimentally measured $M'$ and $M''$ respectively at the different fields ($1 - 60$ $kOe$);
		\textbf{(c), (d)}~Corresponding $\lambda^{-2}$ and $\delta^{-2}$ as a function of temperature respectively extracted using finite element analysis described in Section~\ref{sec:two-coil-technique}.
	}
\end{figure*}
\section{Experimental data}
In Fig.~\ref{fig:M-lambda-delta-diff-H}(a) and (b), we plot the temperature variation of $M'$ and $M''$ for different fields measured using the two-coil technique. From the data, the inverse square of the penetration depth $\lambda^{-2}$ and the skin depth $\delta^{-2}$ data were extracted using the process mentioned in Section~\ref{sec:two-coil-technique} and shown in Fig.~\ref{fig:M-lambda-delta-diff-H}(c) and (d) respectively. The ac penetration depth can be broken down in terms of $\lambda$ and $\delta$ as,
\begin{equation}\label{eq:lambda_eff_lambda_delta_chap4}
    \lambda^{-2}_{ac}=\lambda^{-2}_{eff}=\lambda^{-2}+i\delta^{-2}.
\end{equation}
Eq.~\ref{eq:lambda_eff_lambda_delta_chap4} can also be expressed in terms of the effective electrical conductivity of the film as follows,
\begin{equation}
    \sigma_{eff}=\frac{1}{i\mu_0\omega\lambda^2_{eff}}=\frac{1}{\mu_0\omega\delta^2}+\frac{1}{i\mu_0\omega\lambda^2}.
\end{equation}
Alternatively, the effective impedance of the film is,
\begin{equation}\label{eq:rho_eff-lambda-vortex-chap-5}
    \rho_{eff}=i\mu_0\omega\lambda^2_{eff}=i\mu_0\omega(\lambda^2_{L}+\lambda_{vortex}^2),
\end{equation}
\par
It is evident that the electrical impedances of both the superfluid, denoted by $i\mu_0\omega\lambda_L^2$, and the vortices, represented by $i\mu_0\omega\lambda_{vortex}^2$, are in series ( also see Eq.~\ref{eq:eff_lambda_campbell} ). Consequently, their impedances add, in contrast to their conductivities. It should be noted that Eq.~\ref{eq:rho_eff-lambda-vortex-chap-5} is just the simplified version of Eq.~\ref{eq:Coffey_Clem_eqn} barring the normal-fluid correction.
\par
Now, to gain a sense of the values, let us tabulate the experimental details for our sample: film surface area = $\pi \times 4^2 = 50.3$ $mm^2$, film thickness $d = 20$ $nm$, normal state resistivity $\rho_n = R_n d = 1.55$ $\mu\Omega-m$, normal state sheet resistance $R_n = 77.5$ $\Omega$, zero-field superconducting transition temperature $T_c = 7.05$ $K$, zero-temperature upper critical field $H_{c2}(0) = 11.5$ $T = 115$ $kOe$, GL coherence length $\xi(0) = 5.35$ $nm$, inverse square of the London penetration depth $\lambda^{-2}_L\equiv\lambda^{-2}(T=0, B=0) = 3.55$ $\mu m^{-2}$, Pearl penetration depth $\lambda_P \equiv  2\lambda^2_L/d \approx 28$ $\mu m$. For vortex-creating dc perpendicular fields ( $B$ ) from $1$ $kOe$ to $60$ $kOe$, the intervortex distance ( $a_\Delta=1.075(\Phi_0/B)^{1/2}$ ) decreases from $155$ $nm$ to $20$ $nm$.
\par
For the dc fields studied herein, the impedance of the superfluid is negligible, being less than $10\%$ of the impedance of vortices. Hence, the statement above that the experimental parameters in Fig.~\ref{fig:M-lambda-delta-diff-H} refer only to vortices. So, the dominant contribution to $\sigma_{eff}$ comes from the vortices, i.e., $\sigma_{eff} \approx \sigma_{vortex}$.
\color{black}
\par
Now, coming back to Fig.~\ref{fig:M-lambda-delta-diff-H}, we first concentrate on $\lambda^{-2}$. We observe that $\lambda^{-2}(T\rightarrow 0)$ decreases as a function of field signifying an increase in the Campbell penetration depth. With an increase in temperature, $\lambda^{-2}$ smoothly starts decreasing at low temperatures before it drops rapidly eventually going below the resolution limit.
\begin{figure*}[hbt]
	\centering
	\includegraphics[width=8cm]{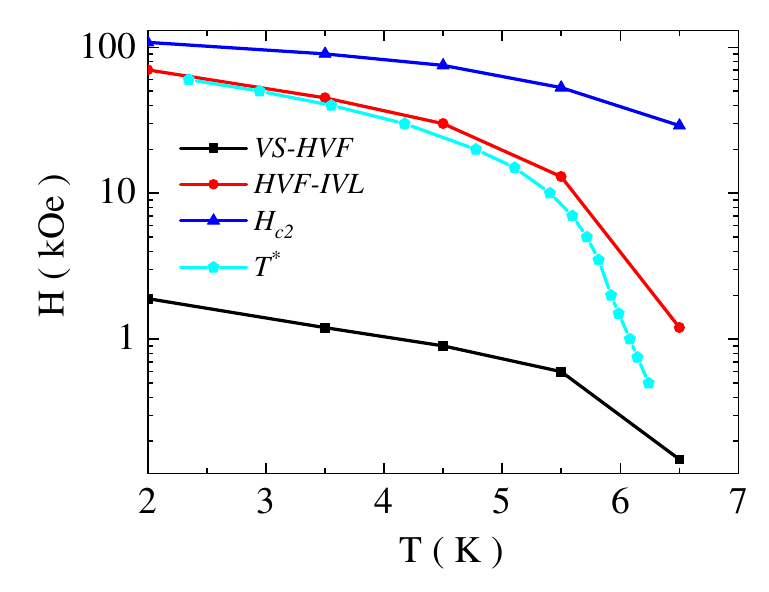}
	\caption[“$\lambda^{-2}$-vanishing” temperature points ($T^*$) in the $H-T$ phase boundary]{
		\label{fig:H-T-phase-boundary} 
		“$\lambda^{-2}$-vanishing” temperature points ($T^*$) were noted for all the fields from Fig.~\ref{fig:M-lambda-delta-diff-H},  where $\lambda^{-2} (T>T^*)$ drops below the resolution limit $10^8$ $m^{-2}$ and plotted (Cyan connected pentagons) in the H-T parameter space along with earlier phase space boundaries~\cite{roy_melting_2019}: VS-to-HVF (connected black squares), HVF-to-IVL (connected red circles) and $H_{c2}$ (connected blue triangles) respectively (VS: Vortex solid, HVF: Hexatic vortex fluid and IVL: Isotropic vortex liquid).
	}
\end{figure*}
\par
From earlier measurements~\cite{roy_melting_2019}, we know that the vortex state in a $20$-$nm$-thick \textit{a}-MoGe film undergoes two transitions: From a vortex solid ( VS ) at low fields and temperatures to a hexatic vortex fluid ( HVF ), and then from a hexatic vortex fluid ( HVF ) to an isotropic vortex liquid ( IVL ). In Fig.~\ref{fig:H-T-phase-boundary}, we plot the temperature $T^*$ for every field at which $\lambda^{-2}(T>T^*)$ drops below our resolution limit of $10^8$ $m^{-2}$. In the same graph, we plot the locus of the transitions from VS to HVF and HVF to IVL obtained earlier~\cite{roy_melting_2019} from the scanning tunneling spectroscopy and magneto-transport measurements. At low temperatures, $T^* (H)$ is just below the HVF-IVL boundary~\cite{roy_melting_2019} showing that the screening response vanishes in the IVL. However, we do not see any signature of the VS-HVF transition in the temperature variation of $\lambda^{-2}$. This is due to the fact that at low temperatures the mobility of the vortices in the HVF is extremely slow~\cite{dutta_collective_2020}, and hence the ac pinning response in HVF is practically indistinguishable from the vortex solid. This is also consistent with earlier magneto-transport measurements where no sharp change in pinning properties was observed across the VS-HVF transition at low temperatures. In fact, it was shown earlier~\cite{dutta_collective_2020} that the Larkin-Ovchinnikov~\cite{larkin_pinning_1979} collective pinning model, originally developed to explain vortex pinning in an imperfect vortex solid is largely applicable in the HVF state as well, at temperatures well below $T_c$. At higher temperatures, $T^*$ bends towards the VS-HVF boundary. This can be understood from the fact that the mobility of the vortices increases with increasing temperatures due to thermal fluctuations and now, even in the HVF, the vortices experience very little restoring force due to pinning, thereby destroying the shielding response.
\par
Coming to $\delta^{-2}$ ( Fig.~\ref{fig:M-lambda-delta-diff-H}(d) ), we observe that for fields above $10$ $kOe$, $\delta^{-2}$ shows a single dissipative peak close to $T^*$ as expected from fluctuations close to a transition. However, at low fields, in addition to this dissipative peak, we observe an increase in $\delta^{-2}$ with a decrease in temperatures. This increase in $\delta^{-2}$ at low temperatures indicates the presence of additional dissipative modes at low temperatures which is not accounted for by the theoretical framework used here.
\begin{figure*}[hbt]
	\centering
	\includegraphics[width=16cm]{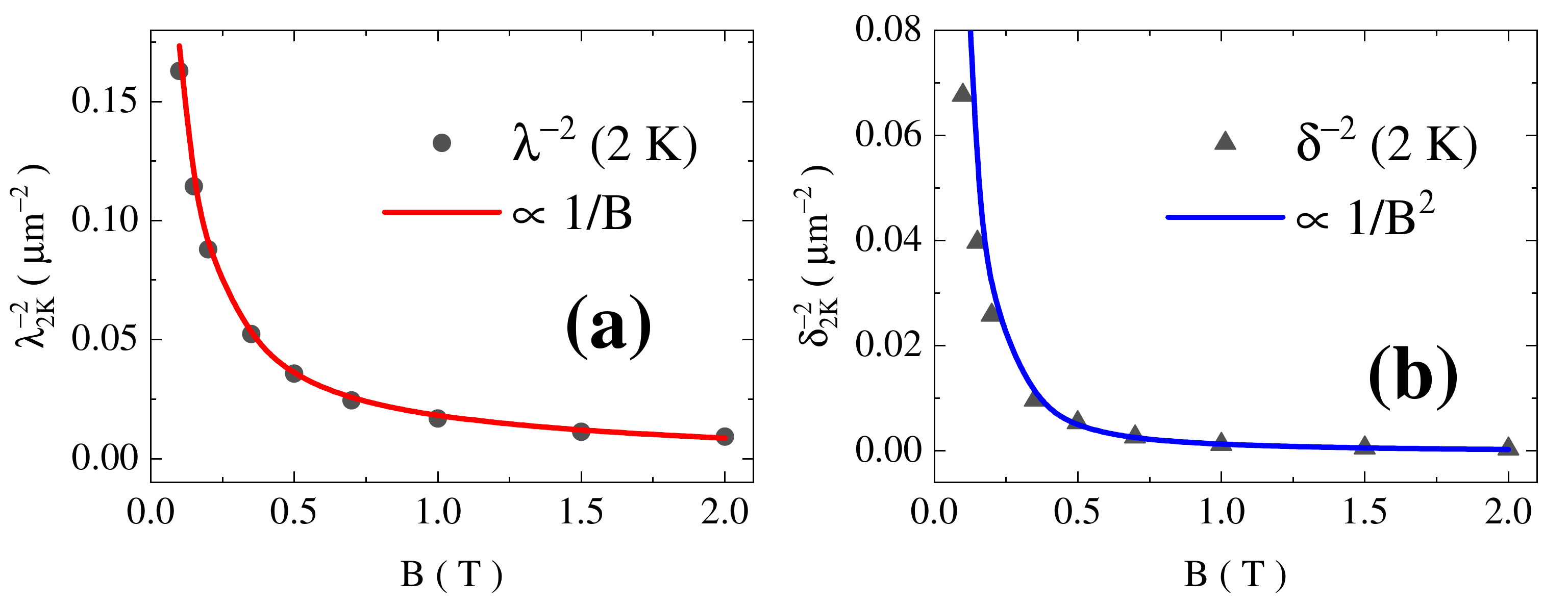}
	\caption[“Magnetic field variation of $\lambda^{-2}$ and $\delta^{-2}$ at $2$ $K$]{
		\label{fig:lambda-delta-vs-B-with-fits} 
		\textbf{(a)}, \textbf{(b)} Magnetic field variation of $\lambda^{-2}$ and $\delta^{-2}$ values at $2$ $K$ (black circles and black triangles respectively) extracted from Fig.~\ref{fig:M-lambda-delta-diff-H}(c) and (d) respectively. They are observed to follow: $\lambda^{-2}_{2K}=k_{\lambda}(1/B)$ (red line) and $\delta^{-2}_{2K}=k_{\delta}(1/B^2)$ (blue line), where the proportionality constants are $k_{\lambda}=1.7\times10^{10}$ $T/m^2$ and $k_{\delta}=1.1\times10^9$ $T^2/m^2$ respectively. The magnetic field (B) values are in Tesla (T).
	}
\end{figure*}
\par
Now, We shall explore the magnetic field variation of low-temperature values of $\lambda^{-2}$ and $\delta^{-2}$. Theoretically, we expect both of them to follow $1/B$ variation as the vortex impedance is proportional to the density of vortices, hence to the applied field:
\begin{equation}
    \rho_{vortex}=\frac{1}{\sigma_{vortex}}=\left(\frac{1}{\mu_0\omega\delta^2}+\frac{1}{i\mu_0\omega\lambda^2}\right)^{-1}\propto B
\end{equation}
Fig.~\ref{fig:lambda-delta-vs-B-with-fits}(a) shows that for $H < 20$ $kOe$ $\lambda^{-2}\propto1/B$ for $2$ $K$ as expected. On the contrary, $\delta^{-2}$ should be vanishingly small at low T, but it is not, and it is proportional to $\propto1/B^2$ as shown in Fig.~\ref{fig:lambda-delta-vs-B-with-fits}(b), not as expected. This anomaly should be investigated in future studies.
\color{black}
\begin{figure*}[hbt]
	\centering
	\includegraphics[width=8cm]{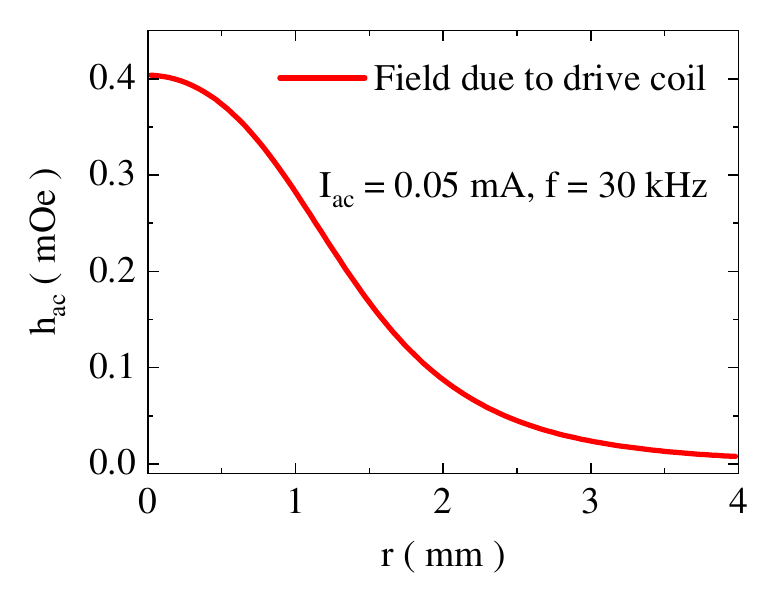}
	\caption[Simulated radial distribution of field on the sample due to primary quadrupole coil]{
		\label{fig:h_ac-vs-radius-quadrupole-coil} 
	Simulated radial distribution of the magnetic field on the sample due to ac excitation ($I_{ac}=0.05$ $mA$, $f=30$ $kHz$) at the primary quadrupole coil.
	}
\end{figure*}
\section{Applicability of Coffey-Clem equation: linearity of ac excitation amplitude}\label{sec:Applicability_of_Coffey-Clem_equation}
\sectionmark{Applicability of Coffey-Clem equation}

In order to quantitatively analyse the data, we fit the data with Eq.~\ref{eq:Coffey_Clem_eqn}. First, we note that in order for the harmonic approximation ( Eq.~\ref{eq:Coffey_Clem_model} ) to hold, the ac excitation needs to satisfy the following condition~\cite{van_der_beek_linear_1993}:
\begin{equation}
    h_{ac}\ll \mu_0 J_c \lambda_C=(\mu_0 J_c B r_f)^{1/2}\equiv h_p.
\end{equation}
Using previously published data~\cite{dutta_collective_2020} on magneto-transport measurements on a similar film, $J_c B \sim 2 \times 10^7 - 3 \times 10^8$ $N/m^3$ within our field of interest ( $1 - 60$ $kOe$ ) and $r_f \sim \xi_{GL} \sim 5.5$ $nm$, where $\xi_{GL}$ is the Ginzburg-Landau ( GL ) coherence length. Using the above values we estimate\hspace{0.1cm} $h_p \sim 3-14$ $Oe$ which\hspace{0.1cm} is several\hspace{0.1cm} orders\hspace{0.1cm} of magnitudes\hspace{0.1cm} larger than\hspace{0.1cm} our $h_{ac}$ ( Fig.~\ref{fig:h_ac-vs-radius-quadrupole-coil} ). Furthermore, from the normal state resistivity of the film, $\rho_N \sim 1.55$ $\mu \Omega-m$  we estimate $\delta_{nf}(0) \sim 3.5$ $mm$. Since $\lambda_L(0) \sim 587$ $nm$, we obtain,
  \begin{equation}
      \frac{\lambda_L^2}{\delta^2_{nf}} \sim 5 \times 10^{-8}.
  \end{equation}
  We can therefore neglect this term in Eq.~\ref{eq:Coffey_Clem_eqn} and set the denominator to unity. The simplified equation now reads as follows:
\begin{equation}\label{eq:Coffey_Clem_eqn_simplified}
    \lambda_{ac}^{-2}=\lambda^{-2}+i\delta^{-2}=\left( \lambda_L^2-i\frac{1}{\mu_0\omega}\frac{B\Phi_0}{\eta}\frac{\epsilon+i\omega\tau}{1+i\omega\tau} \right)^{-1}
\end{equation}
\section{Temperature dependence of vortex parameters}

We have done the subsequent analysis on $\lambda^{-2}$ data only. Now, we attempt to fit the temperature variation of the real part of experimental $\lambda_{ac}^{-2}$ data with the Eq.~\ref{eq:Coffey_Clem_eqn_simplified}, where we also have to make some assumptions on the temperature dependence of the vortex parameters: $\eta$, $\alpha_L$, and $U$. In the following section, I shall discuss some of the existing forms of temperature variations in literature and then state the form used in our subsequent analysis.

\begin{figure*}[hbt]
	\centering
	\includegraphics[width=8cm]{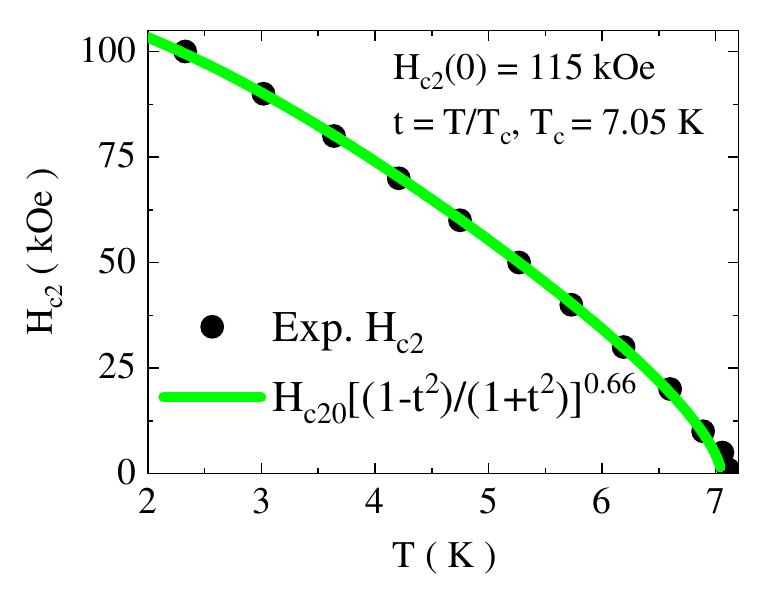}
	\caption[$H_{c2}$ as a function of temperature]{
		\label{fig:Hc2-vs-T} 
	Temperature dependence of $H_{c2}$ ($kOe$) adapted from ref.~\citenum{roy_melting_2019} (black circles), defined by the criterion: $\rho(H_{c2})\sim 0.95\rho_n$, along with the fit: $H_{c2}(0) [(1-t^2)/(1+t^2 )]^{0.66}$ (green line), with $H_{c2}(0)=115$ $kOe$, $t=T/T_c$, $T_c=7.05$ $K$. Temperature variation of $\eta$ was taken to be the same as this form of $H_{c2}$.
	}
\end{figure*}

\subsection{\texorpdfstring{$T$}{} variation of \texorpdfstring{$\eta$}{}}

The temperature dependence of $\eta$ is assumed to follow the variation of $H_{c2}$ according to the Bardeen-Stephen formula~\cite{bardeen_theory_1965,kim_flux-flow_1965}:
\begin{equation}
    \eta=\frac{\mu_0 \Phi_0 H_{c2}}{\rho_n}.
\end{equation}
$H_{c2}$ was earlier~\cite{roy_melting_2019} measured from isotherms in magneto-transport experiments and defined from the criterion, $\rho(H_{c2}) \sim 0.95\rho_n$. As shown in Fig.~\ref{fig:Hc2-vs-T}, $H_{c2}$ follows the empirical variation~\cite{roy_melting_2019}: $H_{c2} \propto [(1-t^2)/(1+t^2 )]^{0.66}$, where $t=T/T_c$, $T_c$ being the zero-field superconducting transition temperature. Therefore, variation of $\eta$ is taken as:
\begin{equation}
    \eta=\eta_0 \left(\frac{1-t^2}{1+t^2}\right)^{0.66},
\end{equation}
where $\eta_0=\frac{\mu_0 \Phi_0 H_{c2}(0)}{\rho_n}$. Our sample ( amorphous MoGe thin film ) being an s-wave superconductor, $\eta_0$ is assumed to be field-independent~\cite{ghosh_field_1997}. For our sample, $\eta_0\approx 1.55\times10^{-8}$ $Ns/m^2$ ( using $B_{c2}\equiv\mu_0H_{c2}(0)=11.5$ $T$ and $\rho_n=1.5-1.55$ $\mu\Omega-m$ ).
\subsection{\texorpdfstring{$T$}{} variation of \texorpdfstring{$U$}{}}
\subsubsection{By scaling approach}
There have been various temperature variations of $U$ suggested in the literature. While Palstra et al.~\cite{palstra_thermally_1988,palstra_critical_1989} used a temperature-independent $U$ to analyse the arrhenius plots in YBa\textsubscript{2}Cu\textsubscript{3}O\textsubscript{7} and Bi\textsubscript{2.2}Sr\textsubscript{2}Ca\textsubscript{0.8}Cu\textsubscript{2}O\textsubscript{${8+\delta}$}, there were other predictions for temperature variations for $U$. Let us start with the scaling approach adopted by Yeshurun and Malozemoff~\cite{yeshurun_giant_1988} ( YM ). They took cues from the Anderson-Kim model~\cite{anderson_theory_1962,anderson_hard_1964} and predicted the temperature dependence for different scenarios. I shall discuss the possible forms for the temperature dependence of $U$ following their arguments. It is to be noted that only the temperature dependence terms are included in the following forms, where the constant coefficients are omitted for simplification. We shall make use of the well-known empirical temperature variation of $H_c$ and $\xi$ given by~\cite{tinkham_introduction_2004,rose-innes_introduction_1994},
\begin{equation}\label{eq:Hc-T-var-empirical}
    H_c(t)\sim (1-t^2),
\end{equation}
and
\begin{equation}\label{eq:coherence-length-xi-T-var-empirical}
    \xi(t)\sim \left(\frac{1+t^2}{1-t^2}\right)^{1/2},
\end{equation}
where $t=T/T_c$, $T_c$ being the superconducting transition temperature.
\par
YM started with the form of $U$ proposed by Anderson and Kim~\cite{anderson_theory_1962,anderson_hard_1964},
\begin{equation}\label{eq:U0-anderson-kim-form}
    U \sim H_c^2\xi^3,
\end{equation}
where $H_c$ is the thermodynamic critical field and $\xi$ is the coherence length. Using the empirical temperature variation of $H_c$ and $\xi$, we get the form,
\begin{equation}\label{eq:U-anderson-kim-T-dep}
    U(t)\sim (1-t^2)^{1/2} (1+t^2)^{1/2}.
\end{equation}
Eq.~\ref{eq:U-anderson-kim-T-dep} can be approximated in the vicinity of transition temperature ($T \rightarrow T_c \equiv t\rightarrow1$) in the following form~\cite{yeshurun_giant_1988}:
\begin{equation}
    U(t)\sim (1-t)^{1/2}.
\end{equation}
\par
Another form of $U$ was suggested by YM for high fields, specifically when more and more flux lines penetrate the sample. When the vortex lattice constant $a_0=1.075(\Phi_0/B)^{1/2}$ becomes significantly smaller than the penetration depth $\lambda$, especially in the case of high-$\kappa$ superconductors, we can expect a crossover in the pinning behavior due to the collective effects. This transition was indeed observed at $H_{cross}=0.2H_{c2}$ in the critical current measurements on $YBa_2Cu_30_7$\cite{kes_thermally_1989}, which corresponds to $a_0\approx6\xi$\cite{yeshurun_giant_1988}. However, it should be noted that $H_{cross}$ may vary depending on the system. Beyond this threshold field $H_{cross}$, $U$ is expected to be constrained by $a_0$ in two dimensions, while $\xi$ remains the minimum characteristic length in the third dimension. Accordingly, YM proposed the following form:
\begin{equation}
    U\sim H_c^2 a_0^2 \xi.
\end{equation}
Inserting the previous Temperature dependencies of $H_c$ and $\xi$ and, using $a_0^2\sim 1/H$, we get the form\cite{kim_flux-creep_1991,zeldov_optical_1989}:
\begin{equation}\label{eq:U-Yeshurun-a0-T-dep}
    U(t,H)\sim\frac{(1-t^2)^{3/2} (1+t^2)^{1/2}}{H}\sim \frac{(1-t^2)(1-t^4)^{1/2}}{H}.
\end{equation}
At $t\rightarrow1$, Eq.~\ref{eq:U-Yeshurun-a0-T-dep} becomes:
\begin{equation}\label{eq:U-Yeshurun-a0-T-dep-simplified}
    U(t,H)\sim \frac{(1-t)^{3/2}}{H}.
\end{equation}
The $U$ expression in Eq.~\ref{eq:U-Yeshurun-a0-T-dep-simplified} has been used extensively in the literature~\cite{yeshurun_giant_1988,tinkham_resistive_1988,hettinger_flux_1989,zeldov_optical_1989}. YM~\cite{yeshurun_giant_1988} was able to qualitatively explain the \enquote{quasi de Almeida - Thouless}~\cite{almeida_stability_1978,muller_flux_1987,carolan_superconducting_1987,tuominen_time-dependent_1988,morgenstern_numerical_1987} or the irreversibility line, $(1-t)\propto H^{2/3}$ by substituting the form of $U$ from Eq.~\ref{eq:U-Yeshurun-a0-T-dep-simplified} into the the critical current expression.
\subsubsection{Using collective pinning model}\label{sec:U-dep-collective-pinning}
The functional dependence of $U$ can also be estimated in the framework of the collective pinning model. It is to be noted that the $20$-$nm$-thick MoGe film has been earlier shown to be in 2D limit following the condition~\cite{kes_two-dimensional_1983}: $J_c^{1/2}\ll 0.2 B^{1/4} d^{-1}$ from the critical current ( $J_c$ ) measurements on a similar sample~\cite{dutta_collective_2020} ( $d$ is the thickness of the thin film ). It is also expected because the thickness of the thin film ( $\sim 20$ $nm$ ) is much less than the characteristic bending length of the vortices~\cite{yazdani_competition_1994}( few $\mu m$ ). Therefore, in the 2D scenario, the characteristic length along the vortex direction, previously represented by the coherence length in 3D, is now substituted by the film thickness $d$. 
\par
Now, according to the 2D collective pinning theory, the activation barrier $U$ governing the thermal creep rate can be estimated by the shearing elastic energy of the vortex lattice due to the small displacement ( $u$ ) of a single vortex as~\cite{eley_accelerated_2018,blatter_vortices_1994},
\begin{equation}
    U \sim C_{66} \left(\frac{\xi}{R_c}\right)^2 V_c,
\end{equation}
where the displacement $u$ has been taken as the core dimension, $\xi$, and $C_{66}$, $R_c$ and $V_c$ are the shear modulus, transverse Larkin length, and collective pinning volume of the VL respectively. In the case of the 2D collective pinning model, $V_c\sim R_c^2 d$ and hence $U\sim C_{66} \xi^2 d$. Using the well known variation of $C_{66}$ from the literature~\cite{larkin_pinning_1979,brandt_shear_1976,brandt_elastic_1986,brandt_flux-line_1995}, $C_{66} \sim \mu_0 H_c^2 h(1-h)^2$, the temperature variation of $U$ becomes,
\begin{equation}
    U(t)\sim H_c^2(t) \xi^2(t).
\end{equation}
Now, Using the known empirical dependence of $H_c$ and $\xi$~\cite{tinkham_introduction_2004,rose-innes_introduction_1994} ( Eqs.~\ref{eq:Hc-T-var-empirical} and \ref{eq:coherence-length-xi-T-var-empirical} ), one obtains,
\begin{equation}\label{eq:U0-T-dep-collective-pinning}
    U(t,H)=U_0(H)(1-t^2)(1+t^2).
\end{equation}
As an illustration of the magnitude of $U$, consider a $3$-$nm$-thick NbN film, which was reported to be measured at $U_0 \approx 20$ $K$~\cite{kundu_effect_2019} using transport.
\subsection{\texorpdfstring{$T$}{} variation of \texorpdfstring{$\alpha_L$}{}}
The restoring force term $\alpha_L \bm{u}$ in Eq.~\ref{eq:Coffey_Clem_model} can be represented as the derivative of pinning potential $U$,
which means that the spring constant $\alpha_L$ can be expressed as $\frac{\partial^2 U}{\partial \bm{u}^2}$, i.e., $\alpha_L$ depends on the curvature of the pinning potential at the bottom. Though the exact shape and curvature of $U$ are difficult to predict, an estimate of $\alpha_L$ can be found from the energy consideration of the core of the vortex when subjected to the small displacement considered in Eq.~\ref{eq:Coffey_Clem_model}. If we equate the total condensation energy in the displaced core with the stored elastic energy: $\mu_0 H_c^2 (\pi \xi^2 d)=\frac{1}{2} \alpha_L \xi^2 d$, we get the temperature variation of $\alpha_L$ as~\cite{wu_pinning_1990,hebard_pair-breaking_1989}:
\begin{equation}\label{eq:alphaL-T-variation}
    \alpha_L(t)\sim H_c^2(t)\sim (1-t^2)^2.
\end{equation}
A further modification was suggested by Belk et al.~\cite{belk_frequency_1996} taking into consideration the effect of quasi-particle scattering due to pinning centers formed by point defects~\cite{van_der_beek_dislocation-mediated_1991,thuneberg_elementary_1989}:
\begin{equation}
    \alpha_L(t)\sim \frac{(1-t^2)^2}{(1+t)^4}.
\end{equation}
Values of $\alpha_L$~\cite{golosovsky_high-frequency_1996,golosovsky_vortex_1994} were reported to be approximately $3 \times 10^5$ $N/m^2$ for YBa\textsubscript{2}Cu\textsubscript{3}O\textsubscript{7-x}, measured using the parallel plate resonator ( PPR ) technique.
\subsection{Correction to \texorpdfstring{$U$}{} and \texorpdfstring{$\alpha_L$}{} due to thermal fluctuations}

A significant modification of the functional forms was suggested by Feigel’man et al.~\cite{feigelman_thermal_1990} and Koshelev et al.~\cite{koshelev_frequency_1991}  considering the effect of thermal fluctuations. They suggested that the smearing of the pinning potential due to thermal fluctuations would result in the exponential decay of $\alpha_L$ and $U$. Experimentally, such exponential decay was indeed observed for YBa$_2$Cu$_3$O$_7$ and Bi$_2$Sr$_2$CaCu$_2$O$_8$ thin films~\cite{golosovsky_high-frequency_1996,golosovsky_vortex_1994,hanaguri_rf_1994} for temperatures lower than $0.8 T_c$. Since the two-dimensional vortex state in thin \textit{a}-MoGe films is extremely sensitive to small perturbations~\cite{dutta_extreme_2019}, it is expected that the vortex lattice would also be susceptible to thermal fluctuations.
\section{Fitting with experimental \texorpdfstring{$\lambda^{-2}$}{} data}
\begin{figure*}[hbt]
	\centering
	\includegraphics[width=16cm]{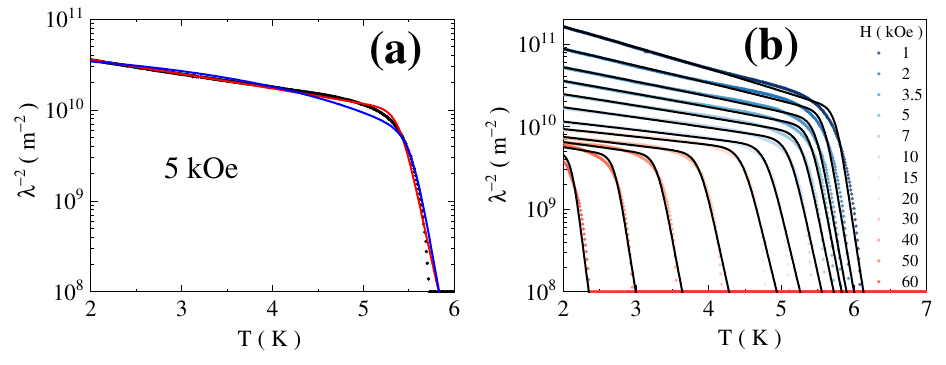}
	\caption[Temperature variation of $\lambda^{-2}$ with fit from Coffey-Clem equation]{
		\label{fig:lambda-vs-T-fit-Coffey-Clem} 
		\textbf{(a)} Experimental $\lambda^{-2}$ vs $T$ data at a representative field (5 kOe) (black circles), along with the fit to Eq.~\ref{eq:Coffey_Clem_eqn_simplified} using (i) $\alpha_L=\alpha_{L0} e^{-T/T_0}, U=U_0 e^{-T/T_0}$ (red line) and (ii) $\alpha_L=\alpha_{L0} (1-t^2)^2, U=U_0 (1-t^2)(1+t^2)$ (blue line).  \textbf{(b)} $\lambda^{-2}$ vs $T$ at different fields along with the fit to Eq.~\ref{eq:Coffey_Clem_eqn_simplified} with the best-fit parameters following the method discussed in Section~\ref{sec:best-fit-error-bar}.
	}
\end{figure*}

Our attempts to fit the data with the above temperature dependence for $U$ and $\alpha_L$ estimated from the 2D collective pinning model ( Eqs.~\ref{eq:U0-T-dep-collective-pinning} and \ref{eq:alphaL-T-variation} respectively ) resulted in poor fit, particularly at low magnetic fields. On the other hand, a much better fit over the entire field range was obtained by assuming temperature dependence of the form:
\begin{equation}\label{eq:alphaL-exp-T-dep}
    \alpha_L \sim \alpha_{L0} e^{-\left(\frac{T}{T_0}\right)},
\end{equation}
\begin{equation}\label{eq:U-exp-T-dep}
    U \sim U_{0} e^{-\left(\frac{T}{T_0}\right)}.
\end{equation}
Here, $T_0$ is the characteristic temperature dependent on the effect of thermal fluctuations. For comparison, fits using both the temperature variations on a representative field $5$ $kOe$ are shown in Fig.~\ref{fig:lambda-vs-T-fit-Coffey-Clem}(a). We also tried other combinations of temperature variations of $U$ and $\alpha_L$, but the fitting was worse than the above two combinations ( see Appendix~\ref{sec:fitting-diff-model} for further details ). However, from Fig.~\ref{fig:lambda-vs-T-fit-Coffey-Clem}(a), it is clearly evident that the dominant effect on temperature dependence comes from the smearing of pinning potential due to thermal fluctuations. Nevertheless, a small deviation is observed at higher temperatures close to the knee region above which $\lambda^{-2}$ drops rapidly. This is however unsurprising since this is close to the boundary where the harmonic approximation assumed in Eq.~\ref{eq:Coffey_Clem_model} will start to break down. For $50$ and $60$ $kOe$, the temperature range of data is too small to perform a reliable best fit, but the qualitative variation is captured by using extrapolated parameters from lower fields.
\par
In Fig.~\ref{fig:lambda-vs-T-fit-Coffey-Clem}(b), we have shown the fits ( black lines ) to the experimental $\lambda^{-2}$ data according to the real part of Eq.~\ref{eq:Coffey_Clem_eqn_simplified} where we have taken the exponential dependence of $\alpha_L$ and $U$, where the fitting parameters are $\alpha_{L0}$, $U_0$ and $T_0$. Further details on the fitting procedure and finding the best-fit procedure are shown in Appendix~\ref{sec:fitting-diff-model} and \ref{sec:best-fit-error-bar}.
\section{Field dependence of vortex lattice parameters}
\begin{figure*}[hbt]
	\centering
	\includegraphics[width=16cm]{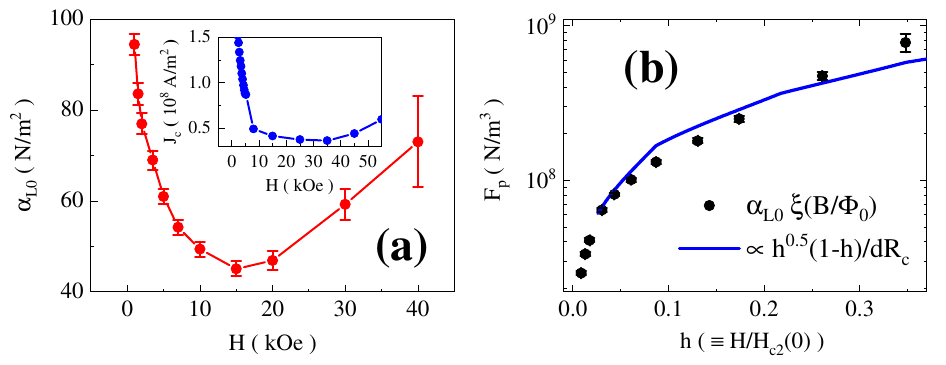}
	\caption[Field dependence of $\alpha_{L0}$]{
		\label{fig:alphaL0-field-dependence} 
		\textbf{(a)} Field dependence of $\alpha_{L0}$ (connected red circles). (inset) Critical current density ($J_c$) as a function of field (blue connected circles) measured on a similar sample from ref.~\citenum{dutta_collective_2020}. \textbf{(b)} Pinning force density $F_p ( =\alpha_{L0} \xi B/\Phi_0 )$ vs normalized field $h ( =H/H_{c2}(0)  )$ (black circles) along with the theoretical field variations proportional to $h^{0.5} (1-h)/dR_c$ (blue line) (semi-log scale). Field variation of $R_c$ is taken from ref.~\citenum{dutta_collective_2020} and $d = 20$ $nm$ is the film thickness. Error bars for the parameters were determined following the protocol explained in Section~\ref{sec:best-fit-error-bar}.
	}
\end{figure*}

As we assumed the temperature dependence of the parameters in Eq.~\ref{eq:Coffey_Clem_eqn_simplified} as explained above, the field dependence of the vortex parameters was reflected in the fitted coefficients $U_0$, $\alpha_{L0}$, and $T_0$ as shown in Fig.~\ref{fig:U0-field-dependence}, \ref{fig:alphaL0-field-dependence} and \ref{fig:T0-field-dependence} respectively. In the following sections, I shall discuss some existing field variations from the literature for perspective, before discussing our data.
\subsection{Field variation of \texorpdfstring{$U_0$}{}}
In literature, there have been several examples of power-law decay in the field variation of $U$ such as $H^{-\beta}$, $\beta$ depending on the materials and the choice of the temperature dependence model. If indeed $U$ is following the Eq.~\ref{eq:U-Yeshurun-a0-T-dep} or \ref{eq:U-Yeshurun-a0-T-dep-simplified}, $\beta$ is supposed to be $1$. Zeldov et al.~\cite{zeldov_optical_1989} found $\beta \approx 0.95$ while Kim et al.~\cite{kim_flux-creep_1991} got $\beta \approx 0.73$ in YBa$_2$Cu$_3$O$_{7-\delta}$ considering Eq.~\ref{eq:U-Yeshurun-a0-T-dep} as the temperature dependence. Palstra et al.~\cite{palstra_thermally_1988} found that $\beta\approx\frac{1}{2}$ for $H_{\parallel}$ and $\beta\approx\frac{1}{6}$ for $H_{\perp}$ considering the temperature-independent $U$ for Bi$_{2.2}$Sr$_2$Ca$_{0.8}$Cu$_2$O$_{8+\delta}$ ( $H_{\parallel}$ and $H_\perp$ are the field parallel and perpendicular to $a-b$ plane ).
\begin{figure*}[hbt]
	\centering
	\includegraphics[width=8cm]{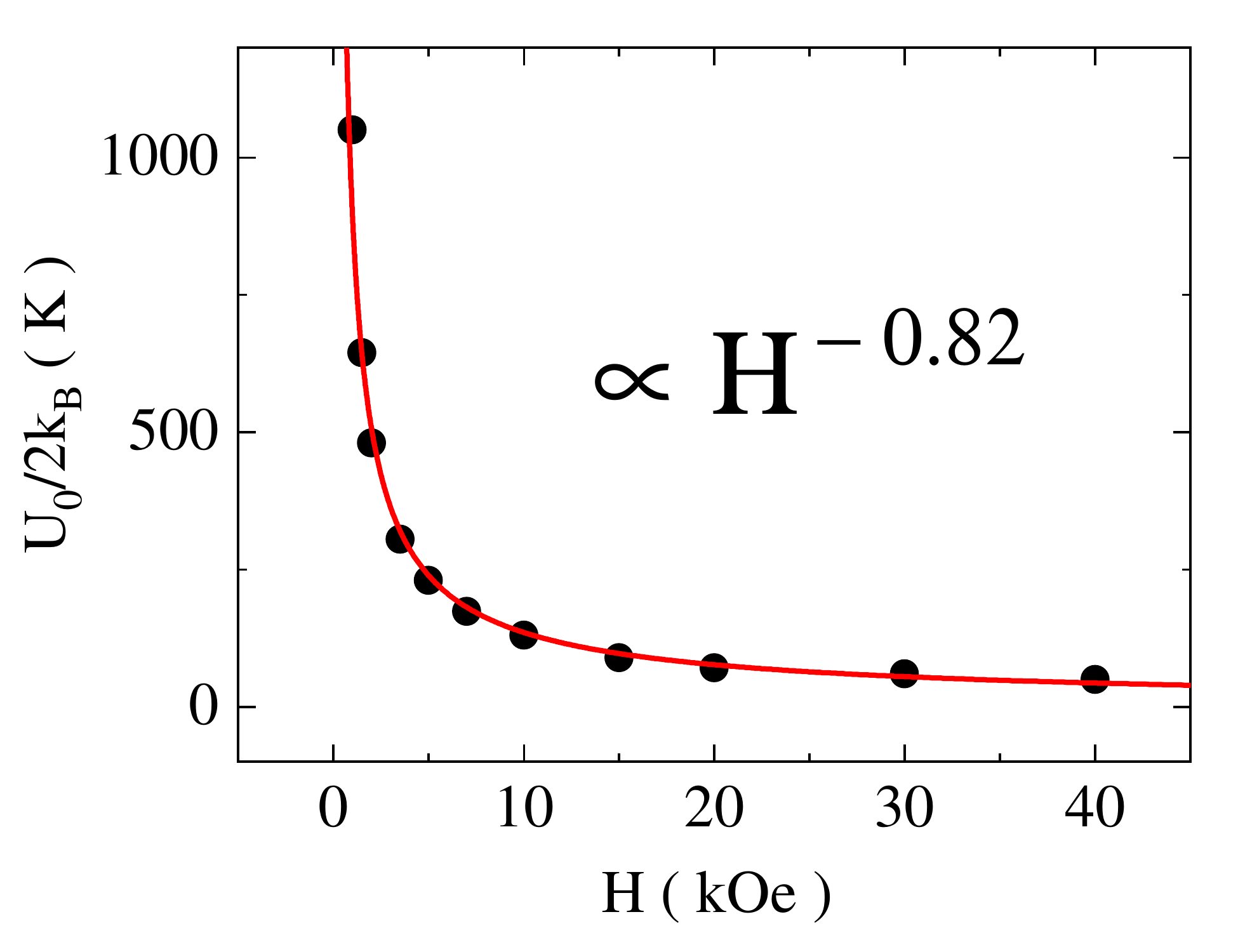}
	\caption[Field dependence of $U_0/2k_B$]{
		\label{fig:U0-field-dependence} 
		$U_0/2k_B$ vs $H$ (black circles) which follows power law decay (red line): $U_0\propto H^{-0.82}$
	}
\end{figure*}
\par
On the other hand, logarithmic dependence was suggested by Feigel'man et al.~\cite{feigelman_pinning_1990} arising from the interaction of dislocations in the vortex lattice. However, after the formation of a free dislocation pair, the potential barrier for moving the dislocations is quite less than the energy to form the pair. So, the defining contribution comes from the formation of dislocation pairs given by~\cite{ephron_observation_1996,white_collective_1993,brunner_thermally_1991}: $U_0(H)\approx U_0'\ln{(H_0'/H)}$, where $U_0'=\frac{\Phi_0^2 d}{64 \mu_0 \pi^2 \lambda^2}$ and $H_0'\approx H_{c2}$. The derived expression is an approximation, where some factors of the order of unity were neglected in the argument.
\par 
However, as seen in Fig.~\ref{fig:U0-field-dependence}, $U_0$ follows the power decay law: $H^{-0.82}$ which is consistent with the earlier studies~\cite{graybeal_observation_1986} on \textit{a}-MoGe, though our exponent is larger than the earlier value of 0.66, calculated from magneto-transport measurements.

\subsection{Field variation of \texorpdfstring{$\alpha_{L0}$}{}}
Fig.~\ref{fig:alphaL0-field-dependence}(a) demonstrates the non-monotonic behavior of $\alpha_{L0}$. Initially, it decreases with the field and reaches a shallow minimum at $15$ $kOe$ before starting to increase again. This minimum closely resembles the pattern observed in the field-dependent variation of $J_c$ measured on similar samples~\cite{dutta_collective_2020}. The similarity can be understood based on the theory of collective pinning. Within Larkin-Ovchinnikov~\cite{larkin_pinning_1979} theory of collective pinning, the pinning force on a vortex ( per unit length ) for a displacement $u$ is given by,
\begin{equation}\label{eq:pinning_force_indiv}
    F_p^{\Phi_0}=\alpha_{L0}u\approx \frac{\Phi_0}{B} \frac{\left(\langle f_{\alpha_{L0}}^2 \rangle n_A\right)^{1/2}}{d R_c}
\end{equation}
where $f_{\alpha_{L0}}$ is the elementary restoring pinning force and $n_A$ is the areal density of pinning centres. $F_p^{\Phi_0}$ is governed by two counteracting effects: the variation of $f_{\alpha_{L0}}$ and $R_c$ with magnetic field. With increase in field, $f_{\alpha_{L0}}$ and $R_c$ control the behaviour of $\alpha_{L0}$ making it non-monotonic with field. Depending on the nature of pinning, the variation of $f_{\alpha_{L0}}$ follows the general form~\cite{kramer_fundamental_1978}: 
\begin{equation}
    \langle f_{\alpha_{L0}}^2 \rangle n_A\propto Ch^n (1-h)^2,
\end{equation}
which suggests that,
\begin{equation}
    F_p^{\Phi_0} \propto \frac{h^{\frac{n}{2}- 1} (1-h)}{dR_c},
\end{equation}
where $n=1$ corresponds to pinning due to dislocation loops~\cite{pande_firstorder_1976}  and $n=3$ due to impurities and vacancies~\cite{campbell_flux_1972} respectively. We have earlier observed that $n=1$ correctly describes the pinning in a similar sample~\cite{dutta_collective_2020}. Thus, at low fields the initial decrease in $\alpha_{L0}$ comes from the dominant numerator because of a decrease in elementary pinning force with an increasing field. However, above a certain field, the decrease in $R_c$~\cite{dutta_collective_2020} with increasing fields starts to dominate, thus increasing $\alpha_{L0}$ ( Fig.~\ref{fig:alphaL0-field-dependence} ). Similar field variation in $J_c$ from the magneto-resistance measurements \hspace{0.1cm}was observed on\hspace{0.1cm} a similar sample~\cite{dutta_collective_2020} ( inset of Fig.~\ref{fig:alphaL0-field-dependence}(a) ) as a precursor of the peak effect observed in the sample. We could not capture the peak at higher fields since the shielding response goes below our resolution limit. To further analyse the data quantitatively, we express Eq.~\ref{eq:pinning_force_indiv} in terms of pinning force density ( $F_p$ ):
\begin{equation}\label{eq:pinning_force_bulk}
    F_p\approx \alpha_{L0} \xi \left(\frac{B}{\Phi_0}\right)\propto \frac{h^{n/2}(1-h)}{d R_c},
\end{equation}
where displacement of a vortex is assumed to be of the order of coherence length $\xi$. Using the field variation of $R_c$ obtained from STS images in Ref.~\citenum{dutta_collective_2020}, in Fig.~\ref{fig:alphaL0-field-dependence}(b), we plot the variation of $F_p$ as a function of the normalized field $h$ ( $\equiv H/H_{c2}(0)$ ) from Eq.~\ref{eq:pinning_force_bulk} where the proportionality constant is taken as an adjustable parameter. The qualitative agreement is indeed very good when we consider that $R_c$ has been measured on a different sample.

\begin{figure*}[hbt]
	\centering
	\includegraphics[width=8cm]{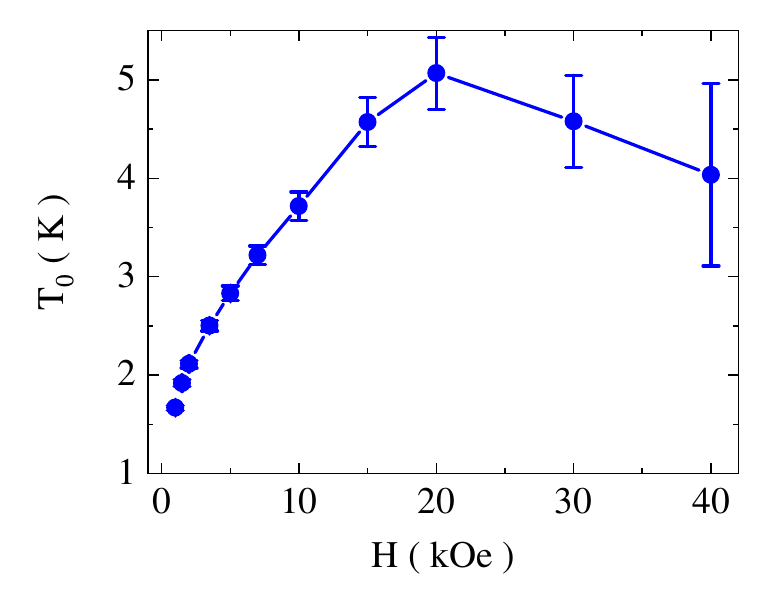}
	\caption[Field dependence of $T_0$]{
		\label{fig:T0-field-dependence} 
		Field variation of the characteristic temperature $T_0$ (Used in the the temperature dependence: $\alpha_L( U )= \alpha_{L0} ( U_0 ) e^{-T/T_0}$ (connected blue circles).
	}
\end{figure*}

\subsection{Field variation of \texorpdfstring{$T_0$}{}}

Finally, in Fig.~\ref{fig:T0-field-dependence}, we observe that $T_0$ increases from $1.5 - 5$ $K$ up to $20$ $kOe$, and then exhibits a gentle decrease in the same range of fields where we observe the increase in $\alpha_{L0}$. Even though we do not have a model to explain this variation at the moment, we believe that the initial increase is related to the increase in rigidity of the VL with the increasing field as the vortex lattice is squeezed. The decrease in the high field is more difficult to understand. However, in this range, the error bar on the extracted value of $T_0$ is large ( due to the small range of temperature considered for the fit ) and it is difficult at the moment to assess the significance of this decrease. The variation of $T_0$ with the field needs to be explored further in the future.

\section{Inconsistency with \texorpdfstring{$\delta^{-2}$}{}}\label{sec:delta_inconsistency}

\begin{figure*}[hbt]
	\centering
	\includegraphics[width=9cm]{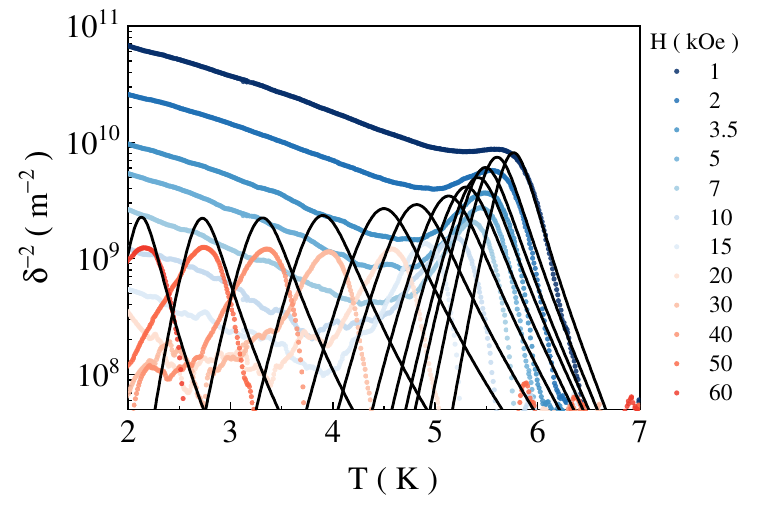}
	\caption[Experimental $\delta^{-2}$ vs $T$ with fit to the imaginary part of the Coffey-Clem equation with best-fit parameters from $\lambda^{-2}$ vs $T$ fits]{
		\label{fig:delta-vs-T-fit} 
		Experimental $\delta^{-2}$ vs $T$ data with fit to the imaginary part of Eq.~\ref{eq:Coffey_Clem_eqn_simplified} with best-fit parameters used in Fig.~\ref{fig:lambda-vs-T-fit-Coffey-Clem}.
	}
\end{figure*}

In Fig.~\ref{fig:delta-vs-T-fit}, we have plotted the corresponding imaginary part of Eq.~\ref{eq:Coffey_Clem_eqn_simplified} with the same parameters used for fitting $\lambda^{-2}$. At high fields, the simulated curve qualitatively captures the dissipation peak observed close to $T^*$. However, at low fields, we observe an additional increase at low temperatures which is not captured within this model. This increase signals some additional mode of dissipation present in the system. One possibility is that at low fields the oscillatory field excites additional internal modes within a Larkin domain in the soft lattice that is not accounted for in the mean-field description considered in Eq.~\ref{eq:Coffey_Clem_model}.
\section{Discussion}

This work was inspired by two influential papers~\cite{liu_microwave_2013, misra_measurements_2013} published around a decade ago. One of the fascinating results in the phenomenology of superconductors is that the superfluid density is proportional to the inverse square of the London penetration depth, which is a measurable quantity. Using the Ginzburg-Landau theory~\cite{Ginzburg:1950sr,landau_theory_1965}, it is possible to connect the superfluid density to an energy scale called the superfluid stiffness, which measures the ability of the superconductor to withstand phase fluctuations. This means that by measuring the penetration depth, it is possible to accurately determine the superfluid stiffness, providing an elegant solution.
\begin{figure*}[hbt]
	\centering
	\includegraphics[width=8cm]{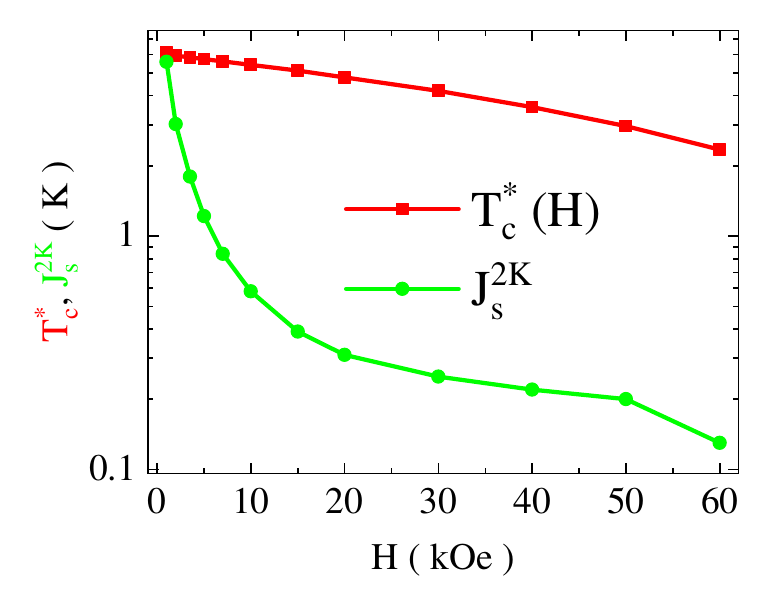}
	\caption[Comparison of superfluid stiffness $J_s^{2K}$ directly calculated from $\lambda^{-2}$ at $2$ $K$ and the in-field transition temperature $T_c^*$.]{
		\label{fig:Js-from-lambda-Tc-compare-vor-par} 
		(Connected green circles) Field variation of superfluid stiffness $J_s^{2K}$ directly calculated from in-field $\lambda^{-2}$ values at $2$ $K$. For comparison, the in-field transition temperatures ($T_c^{*}$) are also plotted (connected red squares), i.e., where $\lambda^{-2}$ falls below the resolution limit ($10^8$ $m^{-2}$) in Fig.~\ref{fig:M-lambda-delta-diff-H}(c).
	}
\end{figure*}
\par
However, this approach has limitations and should not be used in the presence of a magnetic field, when vortices nucleate within the
system. In such cases, the measured penetration depth will be influenced by the motion of pinned vortices, which is not related to the superfluid density or superfluid stiffness. In fact, A. M. Campbell \cite{campbell_response_1969, campbell_interaction_1971} had already discovered this phenomenon in the 1960s while studying the shielding response of pinned vortices. Nonetheless, the above-mentioned papers~\cite{liu_microwave_2013,misra_measurements_2013} disregarded this and treated the vortex lattice in a coarse-grained manner, such that the impact of vortex motion on bulk measurements get nullified across a sizable sample. This allowed for the extension of the Ginzburg-Landau relation in the presence of a magnetic field and the direct estimation of the superfluid stiffness from the ac penetration depth. Consequently, it was concluded that the superfluid stiffness becomes zero at a specific critical magnetic field. To check the consistency of this argument, we have directly calculated $J_s^{2K}$ using the well-known formula: $J_s=\frac{\hbar^2 d}{4 \mu_0 e^2 \lambda^2}$, where $d$ is the film thickness, and $\lambda^{-2}_{2K}$ is taken from Fig.~\ref{fig:M-lambda-delta-diff-H}(c). In Fig.~\ref{fig:Js-from-lambda-Tc-compare-vor-par}, we present a comparison of this calculated $J_s^{2K}$ with the in-field transition temperature ( $T_c^*$ ) obtained from Fig.~\ref{fig:M-lambda-delta-diff-H}. The plot clearly indicates that with an increase in the magnetic field, $J^{2K}_s$ decreases rapidly while $T_c^*$ slowly decreases but remains at least an order of magnitude greater than the former. However, as discussed in Section~\ref{sec:Js-Delta-energy-compare-phase-fluc}, $J_s$ should either be much greater than $T_c$ or of the same order, depending on the relative importance of phase fluctuations. But, it can never be an order of magnitude less than $T_c$, as observed in Fig.~\ref{fig:Js-from-lambda-Tc-compare-vor-par}. This discrepancy implies that it was not correct, in the first place, to directly connect the superfluid stiffness with the penetration depth data in the presence of the magnetic field.
\par
The confusion arises from the fact that we typically measure the penetration depth using a finite-frequency screening response for practical reasons. However, in a zero magnetic field, the measurement is equivalent to what would be measured with a dc field, namely the London~\cite{london_electromagnetic_1935} penetration depth. In contrast, the situation is entirely different in the presence of a magnetic field. The dc field almost entirely penetrates the sample in the form of vortices. So, if we do a dc magnetization ( $M$ ) measurement, measured $M$ will be zero, indicating an almost infinite coarse-averaged penetration depth. However, this is not what we actually measure. The measured penetration depth corresponds to the ac screening response, which results from the fact that the vortices are unable to respond to small changes in the magnetic field due to pinning. This quantity, known as the Campbell~\cite{campbell_response_1969,campbell_interaction_1971} penetration depth, is not necessarily linked to the superfluid stiffness in the same way as the London penetration depth, and therefore cannot be used interchangeably. While it is possible to calculate an energy scale using the zero-field formula for superfluid stiffness, doing so would not be physically meaningful.
\par
The purpose of this work is to support our point of view by analyzing in-field penetration depth data in detail and demonstrating that the variation in penetration depth with a magnetic field can be explained quantitatively by the electrodynamic response of pinned vortices along with the London penetration depth.

\section{Conclusion}
In this work, we have demonstrated a method of extracting vortex parameters from penetration depth measurements using the low-frequency mutual inductance technique. We fitted the measured $\lambda^{-2}$  vs $T$ plots at different fields with a model developed by Coffey and Clem, where the vortex parameters were determined as fitted parameters. The accuracy of the extracted parameter values depends on their correct temperature dependence models. Though the CC model has been extensively used ( with some approximations ) to extract vortex parameters from the microwave vortex resistivity measurements~\cite{golosovsky_high-frequency_1996,pompeo_reliable_2008,wu_frequency_1995}, it has not been used very much for low-frequency ac susceptibility measurements barring a few~\cite{pasquini_linear_1997,pasquini_dynamic_1999}. Our data fitted well for exponential temperature variations of both the Labusch parameter and activation potential barrier, suggesting that the dominant effect of temperature comes from the smearing of pinning potential due to thermal fluctuations. However, using the CC model we could not fully capture the variation of the skin depth signifying the loss in the system which suggests that there might be additional modes of dissipation present in the system beyond the model presented here. One limitation of our approach is that we have to pre-assume the temperature dependence of some parameters in our analysis. If the chosen temperature dependence is not accurate for a material, the shape of the fitted curve does not reproduce the experimental data accurately. If one knows the precise temperature dependence of the vortex parameters for the concerned material, the estimate of the final result only gets better.

	\chapter{Summary and outlook}\label{ch:summary}
\section{Summary}
In this thesis, we have focused on the evolution of superconductivity in amorphous Molybdenum Germanium ( \textit{a}-MoGe ) thin films and the extraction of vortex parameters using the low-frequency mutual inductance technique. Our research has provided valuable insights into the behavior of superconductors and their electromagnetic response, contributing to the understanding of superconductivity and its underlying mechanisms.
\par
Firstly, we investigated the effect of decreasing film thickness on the suppression of superconductivity in \textit{a}-MoGe thin films. We observed two distinct regimes: a Fermionic regime at moderate disorder and a Bosonic regime at stronger disorder. In the Fermionic regime, the critical temperature ( $T_c$ ) decreased with decreasing film thickness, while the ratio of the superconducting gap ( $\Delta$ ) to the temperature ( $k_BT_c$ ) showed only a small increase. The behavior of the sheet resistance at the normal state as a function of $T_c$ followed the Finkel’stein model, consistent with the Fermionic scenario. However, in the Bosonic regime, the system transitioned to a phase-fluctuation-dominated state. Even after the superconducting state was destroyed by phase fluctuations, the finite pairing amplitude manifested as a pseudogap. Our observations aligned with earlier studies on disordered superconductors, suggesting the universality of the Bosonic route in the eventual destruction of superconductivity.
\par
Secondly, we presented a method for extracting vortex parameters using the low-frequency mutual inductance technique. By analyzing the temperature dependence of the measured penetration depth data, we successfully determined vortex parameters such as the Labusch parameter, vortex lattice drag coefficient, and pinning potential barrier. Our analysis revealed the dominant effect of thermal fluctuations on the vortex behavior. However, the Coffey-Clem ( CC ) model used in our analysis had limitations in fully capturing the variation of the skin depth, indicating the presence of additional dissipation modes in the system. Nonetheless, our method provided valuable insights into the behavior of vortices in \textit{a}-MoGe thin films and their interaction with thermal fluctuations.
\par
In conclusion, our research sheds light on the evolution of superconductivity in \textit{a}-MoGe thin films and the characteristics of vortices. This thesis aims to enhance our understanding\hspace{0.1cm} of the\hspace{0.1cm} suppression\hspace{0.1cm} of superconductivity \hspace{0.1cm}with decreasing\hspace{0.1cm} film thickness ( or increasing disorder ) in MoGe thin films through various measurements, with a particular emphasis on tunneling measurements of the superconducting gap and two-coil measurements of superfluid density, conducted on films of varying thicknesses. While extensive two-coil measurements on the same material already exist~\cite{turneaure_effect_2000}, our study introduces tunneling ( STM ) data to enhance the interpretation of superfluid density. \textcolor{black}{The work reveals that the effect of phase fluctuations is much stronger than expected, i.e., they suppress the critical temperature ( $T_c$ ) of the thinnest films ( where superfluid density and resistivity vanish ) well below the predicted occurrence of the familiar Kosterlitz-Thouless-Berezinskii transition. Additionally, spatial inhomogeneity of the local superfluid density in the thinnest films might accentuate the suppression of the total superfluid density.} We have observed the transition from the Fermionic to the Bosonic regime in the suppression of superconductivity, highlighting the importance of disorder and phase fluctuations in determining the behavior of superconductors. Furthermore, our method of extracting vortex parameters using the low-frequency mutual inductance technique has provided a means to investigate the dynamics of vortices in superconducting systems.

\section{Future problems}
\begin{enumerate}
    \item It would be interesting to explore the universality of the Bosonic mechanism in other disordered superconductors and investigate exceptions to the Fermionic route of superconductivity destruction.

    \item Refining the analysis methods for extracting vortex parameters and investigating additional dissipation modes will greatly enhance our understanding of superconducting behavior. It is evident that the simple Bardeen-Stephen model falls short of fully capturing the flux flow viscosity, necessitating modifications to account for additional dissipation mechanisms. One such mechanism is the contribution of shear viscosity arising from the relative motion of vortices. By incorporating these additional factors into our analysis, we can gain a more comprehensive picture of the intricate dynamics and interactions of vortices in superconducting systems.

    \item A hint of a peak-like feature in the field variation of microwave surface resistance in the GHz range was observed during the transition from the Hexatic fluid to the isotropic fluid state. This finding aligns with the results obtained from magneto-transport, scanning tunneling microscopy ( STM ), and penetration depth measurements, thereby emphasizing the need for a detailed exploration of this phenomenon. By conducting further investigations, we can gain a deeper understanding of the underlying mechanisms and dynamics associated with this peak-like feature.

    \item An unusual peak-like feature was observed in the extremely weakly pinned \textit{a}-MoGe sample at a very low field near $H_{c1}$. Despite extensive investigation, a complete understanding of this phenomenon remains elusive, necessitating further exploration in future studies. By delving deeper into the nature and origin of this peak-like feature, we can unravel the underlying mechanisms and gain valuable insights into the unique behavior exhibited by the system under these specific conditions.

    \item We can explore the magnetization ( $M$ ) data of MoGe samples with varying thicknesses to systematically understand the $M-H$ loops and, consequently, the underlying pinning strengths. By analyzing the $M-H$ curves, we can obtain valuable insights into the behavior and response of the samples to external magnetic fields. This systematic investigation will provide a comprehensive understanding of the pinning mechanisms and their influence on the superconducting properties of the MoGe samples. Such knowledge is crucial for optimizing the design and performance of superconducting materials and devices.

\end{enumerate}

	
	\appendix
	
	\begin{appendices}
		
\chapter{BKT transition in \texorpdfstring{$2.2$}{}-\texorpdfstring{$nm$}{}-thick MoGe film}\label{ap:BKT-fit-2p2nm}
Even for our thinnest film ( $\sim 2.2$ $nm$ ), the low-temperature stiffness $J_s(T \rightarrow 0)\approx 10$ K ( Fig.~\ref{fig:Js-Delta} ), calculated from Eq.~\ref{eq:Js-expn-chap4}, is still large enough that the Berezinskii-Kosterlitz-Thouless~\cite{berezinskii_destruction_1972,kosterlitz_ordering_1973,kosterlitz_critical_1974} ( BKT ) temperature $T_{BKT}$ is very close\hspace{0.1cm} to the BCS\hspace{0.1cm} mean-field\hspace{0.1cm} temperature ( $T_{BCS}$ ). In addition, the sample inhomogeneity emerging at strong disorder is expected to smear out the BKT signatures, for example, the universal jump of the superfluid stiffness~\cite{maccari_broadening_2017}, as already observed experimentally in thin films~\cite{mondal_role_2011,yong_robustness_2013}, of NbN. To quantify these effects, we analyzed the temperature dependence of the superfluid stiffness in the thinnest film within the same scheme described in Ref.~\citenum{mondal_role_2011,yong_robustness_2013}.
\par
 \begin{figure*}[hbt]
	\centering
	\includegraphics[width=8cm]{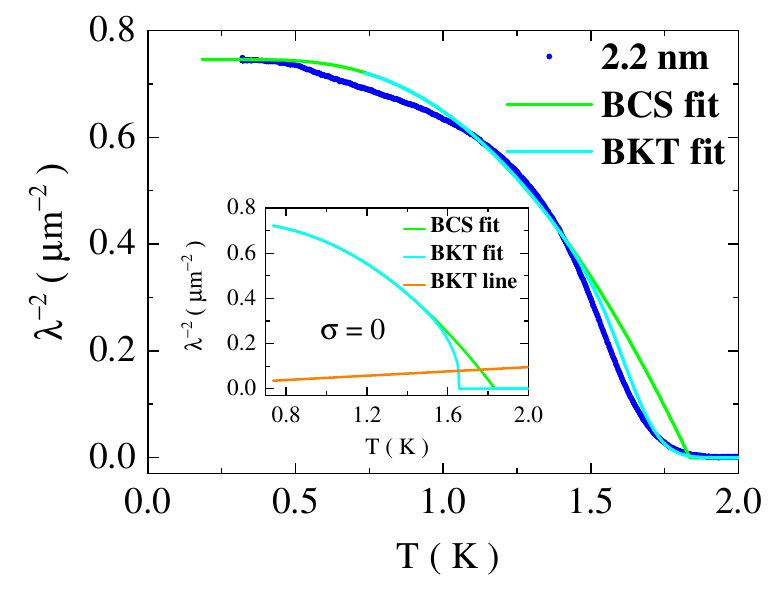}
	\caption[$\lambda^{-2}$ vs $T$ for $2.2$-$nm$-thick sample with dirty BCS and BKT fits]{
		\label{fig:lambda-BKT-2p2nm} 
		Temperature variation of $\lambda^{-2}$ (blue circles) for the $2.2$-$nm$-thick sample along with the fits from dirty-limit BCS formula (green line) and BKT theory (cyan line); The inset shows the corresponding BCS (green line) and BKT (cyan line) variation for the hypothetical case of zero disorder, i.e., $\sigma=0$, where $\sigma$ is the standard deviation in the Gaussian distribution~\ref{eq:gaussian-distr}. The universal BKT line~\cite{nelson_universal_1977,minnhagen_two-dimensional_1987} is shown in orange.
	}
\end{figure*}
Both quasiparticle excitations and phase fluctuations ( longitudinal and transverse ) contribute to the depletion of $J_s$ towards zero. The usual approach used to analyze the BKT effect is to include all the other excitations except vortices in a phenomenological ansatz $J_s^{BCS}$ which serves as starting point for the RG equations. Here we extract $J_s^{BCS}$ from a phenomenological fit of the superfluid stiffness with a dirty-limit BCS-like expression,
\begin{equation}\label{eq:Js-BCS-estimate}
    \frac{J_s^{BCS}(T)}{J_s^{BCS}(0)}=\frac{\Delta(T)}{\Delta(0)} \tanh{\frac{\Delta(T)}{2k_BT}},
\end{equation}
by using $\Delta(0)/T_{BCS}$ as a free parameter. The resulting $J_s^{BCS}(T)$ here acts as a mean-field ansatz for the renormalization group ( RG ) analysis~\cite{berezinskii_destruction_1972,kosterlitz_ordering_1973,kosterlitz_critical_1974,minnhagen_two-dimensional_1987,benfatto_berezinskii-kosterlitz-thouless_2013}.
\par
However, to account for the sample inhomogeneity observed by STM we will follow the same procedure suggested in Ref.~\citenum{mondal_role_2011,yong_robustness_2013}. We will assume that the local BCS stiffness $J_{s,BCS}^{i}$ is distributed according to a given probability density $P(J_s^i)$, and that the local BCS transition temperature $T_{BCS}^i$ scale accordingly. Then by solving numerically the RG equations, we can compute the local stiffness $J_s^i(T)$. The overall superfluid stiffness is then computed phenomenologically as an average value $J_s^{av}$:
\begin{equation}\label{eq:J-avg}
    J_s^{av}(T)=\sum_i P(J_s^i)J_s^i(T).
\end{equation}
For the sake of concreteness, we will use for $P(J_s^i)$ a Gaussian distribution centered around $J_s^0$:
\begin{equation}\label{eq:gaussian-distr}
    P(J_s^i)=\frac{1}{\sqrt{2\pi}\sigma}e^{-(J_s^i-J_s^0)^2/2\sigma^2}.
\end{equation}
When all the stiffness $J_s^i(T)$ are different from zero, as is the case at low temperatures, the average stiffness will be centered around the center of the Gaussian distribution (Eq.~\ref{eq:gaussian-distr}) so that it will coincide with $J_s^0(T)$. However, by approaching $T_{BKT}$ defined by the average $J_s^0(T)$, not all the patches make the transition at the same temperature so that the BKT jump is rounded and $J_s^{av}$ remains finite above the average $T_{BKT}$.  This leads to a rather symmetric smearing of the superfluid-density jump with respect to the abrupt downturn observed for the clean case, which is progressively more pronounced for increasing $\sigma/J_s^0(T=0)$. It is worth noting that such a phenomenological approach accounts rather well for experiments in thin films on NbN~\cite{mondal_role_2011,yong_robustness_2013}, and it has been recently validated theoretically by Monte Carlo simulations within an inhomogeneous 2D XY model in the presence of correlated disorder~\cite{maccari_broadening_2017}.
\par
In Fig.~\ref{fig:lambda-BKT-2p2nm}, we show the result of the above fitting procedure for the $2.2$-$nm$-thick \textit{a}-MoGe film. The curve labeled BCS represents the BCS fit of the average stiffness based on Eq.~\ref{eq:Js-BCS-estimate} above. From the fit, we obtain the values $T_{BCS}=1.84$ $K$ and $\Delta(0)/(k_B T_{BCS})=1.9$, that is consistent with the STM estimate. As mentioned above, the BCS fit cannot capture the linear depletion at low temperatures, but it captures rather well the overall suppression due to quasiparticle excitations up to a temperature $T \sim 1.5$ $K$. Here, the BKT curve starts to deviate from the BCS one due to the effect of bound vortex-antivortex pairs, as is evident in the inset where we show the homogeneous case ($\sigma=0$). The abrupt jump is however smeared out in the presence of inhomogeneity, as shown in the main panel, where the BKT line corresponds to the average procedure encoded in Eq.~\ref{eq:J-avg} for $\sigma/J_s^0=0.05$. The overall fit reproduces reasonably well the experimental trend, with a global transition temperature $T_c \approx 1.8$ K. Overall, we can conclude that the BKT analysis further supports the conclusion that our thinnest sample is in the 2D limit. However, the manifestation of BKT effects is partly blurred out by the inhomogeneity, which unavoidably comes along with the increase of effective disorder in the thinnest samples. 

\chapter{Fitting of experimental data with Coffey-Clem equation}\label{ap:coffey-clem-fitting-procedure}
\vspace{-0.5cm}
We analysed the experimental $\lambda^{-2}$ vs $T$ data ( Fig.~\ref{fig:lambda-vs-T-fit-Coffey-Clem} ) primarily on Mathematica. We have used the full expression in Eq.~\ref{eq:Coffey_Clem_eqn} to fit the data. As I discussed earlier, we had to choose the temperature dependence of the vortex parameters $\eta$, $\alpha_L$, and $U$ suitable for our system. As we did not know the precise temperature dependence of the parameters, we had to follow a trial-and-error procedure. To check whether the chosen temperature variations are correct, we varied the parameters with the help of the \enquote{Manipulate} function in Mathematica and found the estimated fit to the experimental data. Finally, we used a detailed scheme to find the exact best-fit parameters using the $\chi^2$ method ( Section~\ref{sec:best-fit-error-bar} ).
\begin{figure*}[hbt]
	\centering
	\includegraphics[width=16cm]{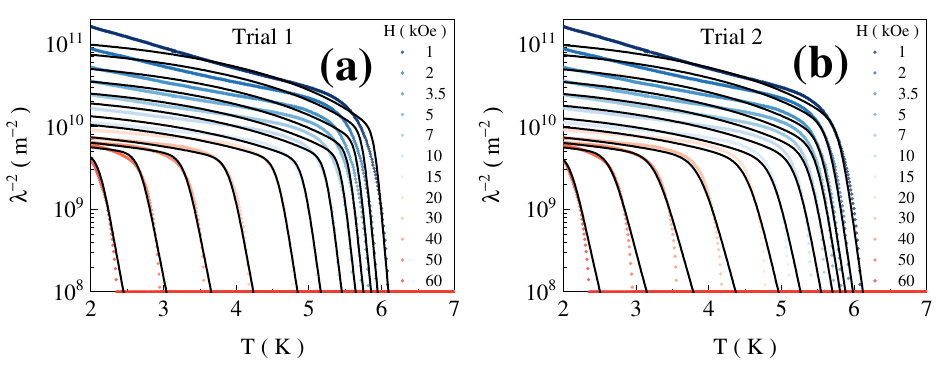}
	\caption[$\lambda^{-2}$ vs $T$ with trial temperature variations of $\alpha_L$ and $U$.]{
		\label{fig:lambda-T-fit-diff-model-appendix} 
		$\lambda^{-2}$ vs $T$ with the trial temperature variations of $\alpha_L$ and $U$: \textbf{(a)} $\alpha_L\sim (1-t^2)^2$ and $U \sim (1-t^2)(1-t^4)^{1/2}$; \textbf{(b)} $\alpha_L\sim (1-t^2)^2$ and $U \sim (1-t^2)(1+t^2)$, where $t=T/T_c$.
	}
\end{figure*}
\section{Fitting with trial temperature variations of \texorpdfstring{$U$}{} and \texorpdfstring{$\alpha_L$}{}}\label{sec:fitting-diff-model}
In this section, I shall discuss some of the trial temperature variations we have considered in order to find a suitable function.
For $\eta$ variation, we have assumed the simple variation mimicking the temperature variation of $H_{c2}$. This might not be true for the low-temperature part of $\delta^{-2}$ vs $T$ data as discussed in section~\ref{sec:delta_inconsistency}.
\par
For $U$ and $\alpha_L$, at first, we have taken the temperature variation from Eqs.~\ref{eq:alphaL-T-variation} and \ref{eq:U-Yeshurun-a0-T-dep} and tried to fit ( Trial 1 in Fig.~\ref{fig:lambda-T-fit-diff-model-appendix}(a) ). But, as seen in Fig.~\ref{fig:lambda-T-fit-diff-model-appendix}, the fit is not that good, especially at low fields. To improve on the fitting, we tried other functions. We found a better fitting ( Trial 2 in Fig.~\ref{fig:lambda-T-fit-diff-model-appendix}(b) ) for the variation of $U$ in Eq.~\ref{eq:U0-T-dep-collective-pinning} while keeping the same variation of $\alpha_L$ as in Eq.~\ref{eq:alphaL-T-variation}. It is evident in Fig.~\ref{fig:lambda-T-fit-diff-model-appendix} that the fit around the transition is better for trial 2. However, in low fields, the increase in $\lambda^{-2}$ could not be captured in either of the above trial variations. Comparisons for the fits with the trial variations are tabulated in the following table.

\begin{table}[hbt]
\centering
\caption{\centering Comparison of Coffey-Clem fits using different trial temperature variations of $U$ and $\alpha_L$.}\label{table:U-alphaL-diff-model}
    \begin{tabular}{|m{0.05\linewidth}|m{0.35\linewidth}|m{0.075\linewidth}|m{0.425\linewidth}|}
    \hline
        Trial & T dependence of $U$ and $\alpha_L$ (Eq. no.) & \centering Fig. No. & \multicolumn{1}{c|}{Remarks}\\
        \hline
        $1.$ & \begin{itemize}[leftmargin=20pt,align=center]
            \item $\displaystyle U\sim(1-t^2)(1-t^4)^{1/2}$ (\ref{eq:U-Yeshurun-a0-T-dep})
            \item $\displaystyle \alpha_L\sim (1-t^2)^2 (\ref{eq:alphaL-T-variation})$
        \end{itemize} & \ref{fig:lambda-T-fit-diff-model-appendix}(a) & \begin{itemize}[leftmargin=20pt,align=center]
            \item Fits well for high-field ($>10$ $kOe$) data.
            \item For low-field data, provides a good fit except at the transition and at low $T$.
        \end{itemize} \\
        \hline
                $2.$ & \begin{itemize}[leftmargin=20pt,align=center]
                    \item $\displaystyle U\sim (1-t^2)(1+t^2)$ (\ref{eq:U0-T-dep-collective-pinning})
                    \item $\displaystyle \alpha_L \sim (1-t^2)^2$ (\ref{eq:alphaL-T-variation})
                \end{itemize} & \ref{fig:lambda-T-fit-diff-model-appendix}(b) & \begin{itemize}[leftmargin=20pt,align=center]
                    \item Fit for high-field data is similar to trial 1.
                    \item For low-field data, the fit is more accurate at the transition, but it fails to capture the behavior at low $T$ similar to trial 1.
                \end{itemize}   \\
                        \hline
                $3.$ & \begin{itemize}[leftmargin=20pt,align=center]
                    \item $\displaystyle U\sim e^{-T/T_0}$ (\ref{eq:U-exp-T-dep})
                    \item $\displaystyle \alpha_L\sim e^{-T/T_0} (\ref{eq:alphaL-exp-T-dep})$
                \end{itemize} & \ref{fig:lambda-vs-T-fit-Coffey-Clem}(b) & Fits reasonably well over the entire range of $T$ ($2-6$ $K$) and field ($1-60$ $kOe$). \\
        \hline
    \end{tabular}
\end{table}

\begin{figure*}[hbt]
	\centering
	\includegraphics[width=16cm]{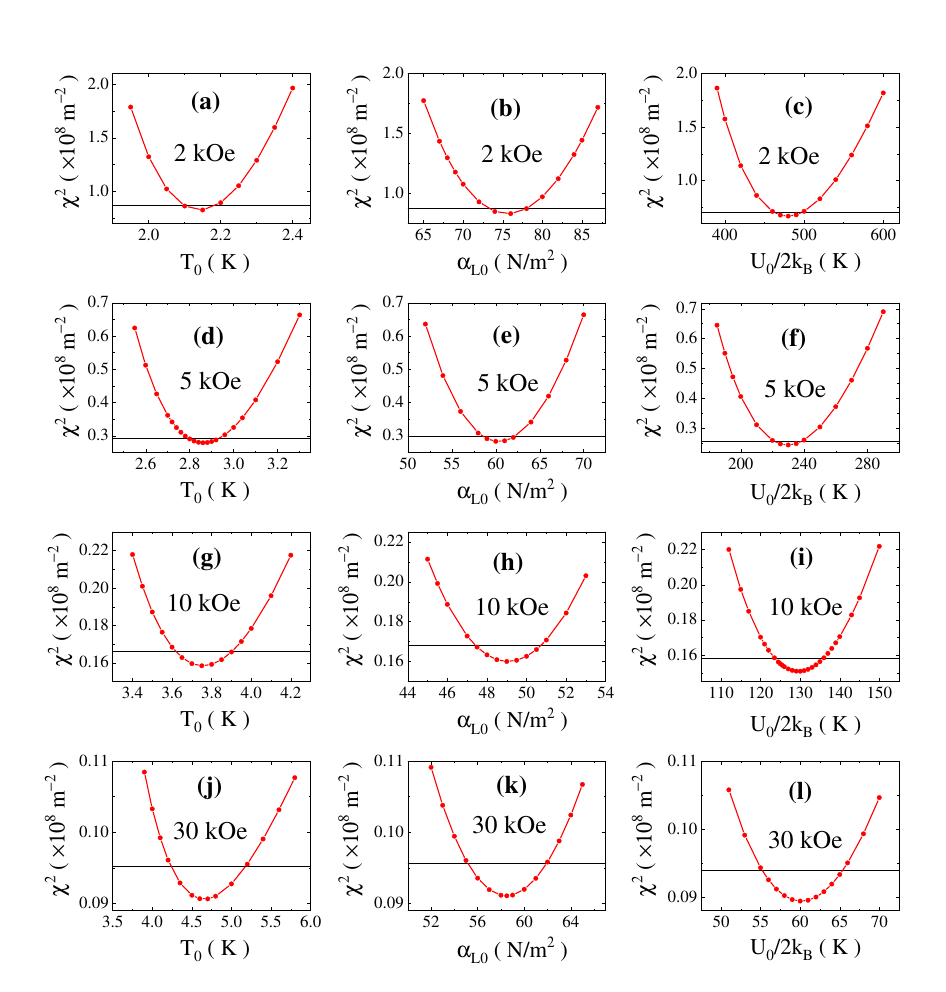}
	\caption[$\chi^2$ as a function of $T_0$, $\alpha_{L0}$ and $U_0/2k_B$ respectively for four representative fields: $2$, $5$, $10$ and $30$ $kOe$.]{
		\label{fig:chi2-vs-T0-alphaL0-U0-error-bar} 
		$\chi^2$ as a function of $T_0$, $\alpha_{L0}$ and $U_0/2k_B$ respectively for four representative fields: \textbf{(a)-(c)} $2$ $kOe$; \textbf{(d)-(f)} $5$ $kOe$; \textbf{(g)-(i)} $10$ $kOe$ and \textbf{(j)-(l)} $30$ $kOe$. The horizontal black line in each of the plots is at 1.05 times the minimum  $\chi^2$ value: the spread of parameter values below the line has been chosen as the error bar of that parameter for the given field.
	}
\end{figure*}
\section{Best fit parameters and error bar}\label{sec:best-fit-error-bar}
In this section, I shall elaborate on the scheme we have followed to find out the best-fit values of the vortex parameters- $U_0$, $\alpha_{L0}$, and $T_0$. The goodness of fit is calculated by the formula:
\begin{equation}
    \chi^2=\frac{1}{N}\frac{\sum(y-y_{fit})^2}{y_{fit}},
\end{equation}
where $y$ is the experimental $\lambda^{-2}$ data in Fig.~\ref{fig:lambda-vs-T-fit-Coffey-Clem} and $y_{fit}$ is calculated using the right-hand side of Eq.~\ref{eq:Coffey_Clem_eqn_simplified}; the normalization factor $1/N$ was to account for the fact that different datasets had a different number of points, $N$. Here, the free parameters are $\alpha_{L0}$, $T_0$ and $U_0/2k_B$. To carry out the error analysis, we used the following protocol: One particular parameter was fixed and the other two were varied freely so as to have the best fit using the \enquote{FindFit} function in Mathematica, and the corresponding $\chi^2$ was calculated. In this procedure, the best fit was determined by the set of values where $\chi^2$ has the minimum value. The results are shown in Fig.~\ref{fig:chi2-vs-T0-alphaL0-U0-error-bar}(a) - (l) for 4 representative fields ( $2, 5, 10,$ and $30$ $kOe$ ).
\begin{figure*}[hbt!]
	\centering
	\includegraphics[width=8cm]{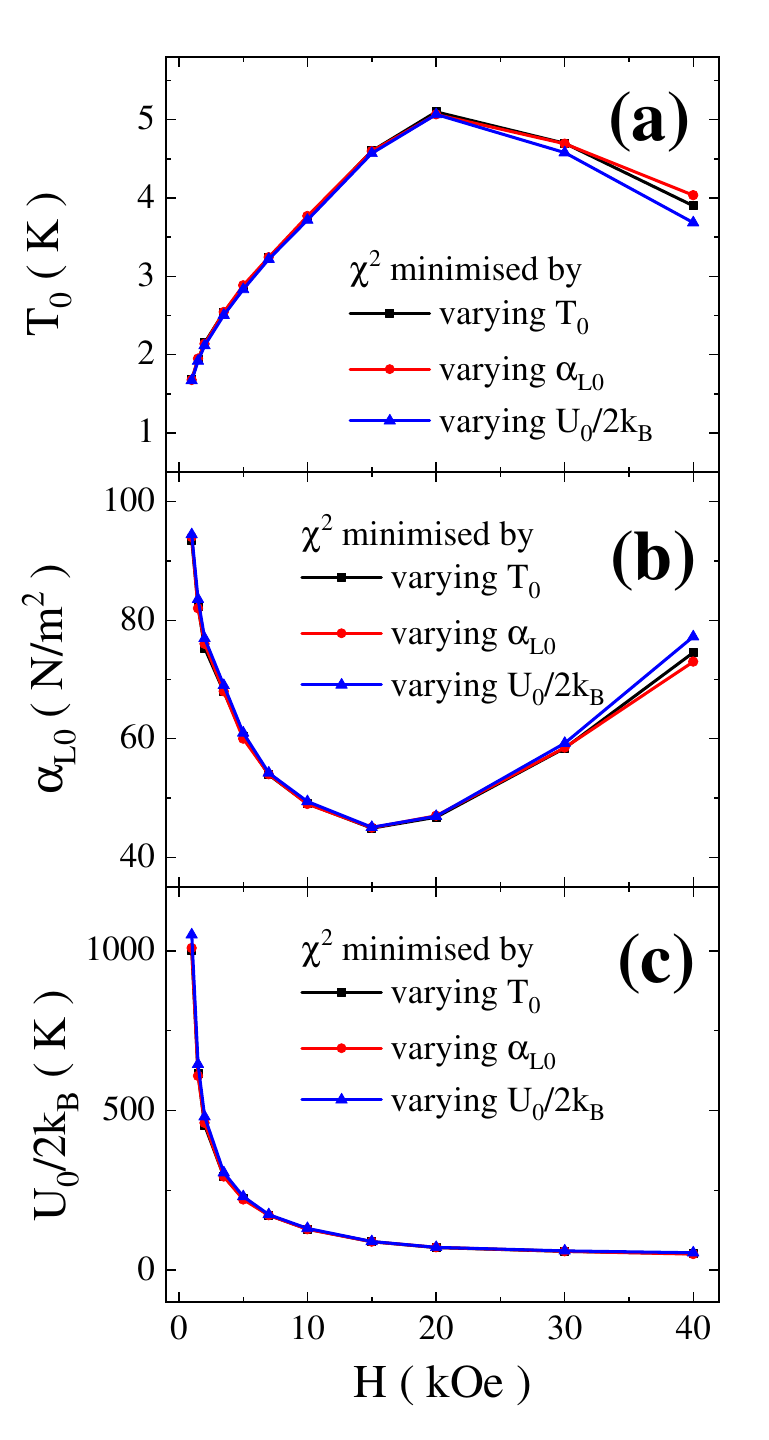}
	\caption[Field dependence of $T_0$, $\alpha_{L0}$ and $U_0/2k_B$ values at the $\chi^2$ minima points.]{
		\label{fig:T0-alphaL0-U0-min-chi2} 
		\textbf{(a)-(c)} Field dependence of $T_0$, $\alpha_{L0}$ and $U_0/2k_B$ values at the $\chi^2$ minima points respectively. In each plot, minimization was done by varying each parameter individually: minima of the  $\chi^2$ vs $T_0$ plots are given by connected black squares, minima of the $\chi^2$ vs $\alpha_{L0}$ plots by connected red circles and minima of the  $\chi^2$ vs $U_0/2k_B$ plots by connected blue triangles.
	}
\end{figure*}
\par
In principle, all three curves should give the same value of $\chi^2$ at the minimum. The slight difference is owing to the fact that the in-built \enquote{FindFit} function minimizes the least-square error instead of $\chi^2$. Nevertheless, the extracted parameter values from the minima of either of these three curves are very close to each other and are shown in Fig.~\ref{fig:T0-alphaL0-U0-min-chi2}. The set of parameters ( $\alpha_{L0}$, $T_0$ and $U_0/2k_B$ ) corresponding to the lowest of the three minima are plotted in Fig.~\ref{fig:T0-alphaL0-U0-min-chi2}. Here we did not include parameters for $50$ and $60$ $kOe$ since the temperature range of data is too small to perform a reliable best fit. Nevertheless, the qualitative variation in these fields is still captured by using extrapolated parameters from lower fields.
\par
The variation of $\chi^2$ for each of the three free parameters, allows us to estimate the error bar. The set of parameter values that do not increase $\chi^2$ beyond $5\%$ of the minimum $\chi^2$ value ( below the horizontal black line in each plot in Fig.~\ref{fig:chi2-vs-T0-alphaL0-U0-error-bar} ), is taken as the error bar for that particular parameter.
        
	\end{appendices}
	
 \cleardoublepage
 \phantomsection
\bibliography{ref.bib}
\end{document}